\documentclass[a4paper,11pt]{article}
\pdfoutput=1

\usepackage{jheppub}

\usepackage{subfig}
\usepackage{xspace}
\usepackage[countmax]{subfloat}
\usepackage{slashed}

\usepackage{comment}

\usepackage{bbm}

\newcommand{\nbar}{{\bar n}}

\newcommand{\ecf}[2]{e_{#1}^{(#2)}} 
\newcommand{\ecfnobeta}[1]{e_{#1}}

\newcommand{\ecflp}[2]{\tilde e_{#1}^{(#2)}}
\newcommand{\ecfop}[2]{\mathbf{E_{#1}}^{(#2)}}

\newcommand{\sja}{n_{sj}}
\newcommand{\sjabar}{\bar{n}_{sj}}
\newcommand{\outj}{B}

\def\log{\text{log}}

\def \thetac {\theta_c}

\def \zs {z_{s}}
\def \zcs {z_{cs}}

\def\be{\begin{equation}}
\def\ee{\end{equation}}

\newcommand{\bnP}{\overline {\mathcal P}}
\newcommand{\w}{\omega}
\newcommand{\cB}{{\mathcal B}}
\newcommand{\bn}{\bar{n}}
\newcommand{\sdt}{\!\cdot\!}
\newcommand{\cP}{{\mathcal P}}
\newcommand{\img}{\mathrm{i}}
\newcommand{\la}{\lambda}
\newcommand{\lp}{\tilde p}

\newcommand{\Dobs}[2]{D_{#1}^{(#2)}} 
\newcommand{\Dobsnobeta}[1]{D_{#1}}

\def\nbar{\bar n}

\newcommand{\eeclp}[2]{\tilde e_{#1}^{(#2)}}

\newcommand{\ecfLa}{e_{2}^{(\alpha)}}
\newcommand{\ecfLb}{e_{2}^{(\beta)}}
\newcommand{\ecfres}{e_{3}^{(\alpha)}}

\newcommand{\sje}{z_{sj}}
\newcommand{\sjtheta}{\theta_{sj}}

\DeclareRobustCommand{\Sec}[1]{Sec.~\ref{#1}}
\DeclareRobustCommand{\Secs}[2]{Secs.~\ref{#1} and \ref{#2}}
\DeclareRobustCommand{\App}[1]{App.~\ref{#1}}
\DeclareRobustCommand{\Tab}[1]{Table~\ref{#1}}

\DeclareRobustCommand{\Fig}[1]{Fig.~\ref{#1}}
\DeclareRobustCommand{\Figs}[2]{Figs.~\ref{#1} and \ref{#2}}
\DeclareRobustCommand{\Eq}[1]{Eq.~(\ref{#1})}
\DeclareRobustCommand{\Eqs}[2]{Eqs.~(\ref{#1}) and (\ref{#2})}
\DeclareRobustCommand{\Ref}[1]{Ref.~\cite{#1}}
\DeclareRobustCommand{\Refs}[1]{Refs.~\cite{#1}}

\newcommand{\Nsub}[2]{\tau_{#1}^{(#2)}}

\newcommand{\pythia}[1]{\textsc{Pythia\xspace #1}}
\newcommand{\madgraph}[1]{\textsc{MadGraph5\xspace #1}}
\newcommand{\fastjet}[1]{\textsc{FastJet\xspace #1}}
\newcommand{\herwig}[1]{\textsc{Herwig\xspace #1}}
\newcommand{\herwigpp}[1]{\textsc{Herwig++\xspace #1}}
\newcommand{\vincia}[1]{\textsc{Vincia\xspace #1}}
\newcommand{\nlojet}{\textsc{NLOJet++}}
\newcommand{\sherpa}{\textsc{Sherpa}}
\newcommand{\ariadne}{\textsc{Ariadne}}

\newcommand{\dire}{\textsc{Dire}}

\usepackage{color}
\definecolor{darkblue}{rgb}{0,0,0.5}
\definecolor{darkred}{rgb}{0.5,0,0}
\definecolor{orange}{rgb}{0.7,0.5,0}

\preprint{MIT--CTP 4681}

\title{Analytic Boosted Boson Discrimination
}

\author{Andrew J. Larkoski,}
\author{Ian Moult,}
\author{and Duff Neill}

\affiliation{Center for Theoretical Physics, Massachusetts Institute of Technology, Cambridge, MA 02139, USA}

\emailAdd{larkoski@mit.edu}
\emailAdd{ianmoult@mit.edu}
\emailAdd{dneill@mit.edu}

\abstract{Observables which discriminate boosted topologies from massive QCD jets are of great importance for the success of the jet substructure program at the Large Hadron Collider. Such observables, while both widely and successfully used, have been studied almost exclusively with Monte Carlo simulations. In this paper we present the first all-orders factorization theorem for a two-prong discriminant based on a jet shape variable, $D_2$, valid for both signal and background jets. Our factorization theorem simultaneously describes the production of both collinear and soft subjets, and we introduce a novel zero-bin procedure to correctly describe the transition region between these limits. By proving an all orders factorization theorem, we enable a systematically improvable description, and allow for precision comparisons between data, Monte Carlo, and first principles QCD calculations for jet substructure observables. Using our factorization theorem, we present numerical results for the discrimination of a boosted $Z$ boson from massive QCD background jets. We compare our results with Monte Carlo predictions which allows for a detailed understanding of the extent to which these generators accurately describe the formation of two-prong QCD jets, and informs their usage in substructure analyses. Our calculation also provides considerable insight into the discrimination power and calculability of jet substructure observables in general.

}

\begin{document} 
\maketitle

\section{Introduction}\label{sec:intro}

The last several years has seen a surge of interest in the field of jet substructure \cite{Abdesselam:2010pt,Altheimer:2012mn,Altheimer:2013yza,Adams:2015hiv}, both as an essential tool for extending new physics searches at the Large Hadron Collider (LHC) into the TeV energy regime, and as an important playground for improving our understanding of high energy QCD, both perturbative and non-perturbative. Of particular phenomenological interest are substructure observables that are sensitive to hard subjets within a jet.  In the highly boosted regime, the hadronic decay products of electroweak-scale particles can become collimated and each appear as a jet in the detector.  Unlike typical massive QCD jets, however, these boosted electroweak jets exhibit a multi-prong substructure that can be identified by the measurement of appropriate observables.  Many such observables have been proposed and studied on LHC simulation or data \cite{CMS:2011xsa,Miller:2011qg,Chatrchyan:2012mec,ATLAS:2012jla,Aad:2012meb,ATLAS:2012kla,ATLAS:2012am,Aad:2013gja,Aad:2013fba,TheATLAScollaboration:2013tia,TheATLAScollaboration:2013sia,TheATLAScollaboration:2013ria,TheATLAScollaboration:2013pia,CMS:2013uea,CMS:2013kfa,CMS:2013wea,CMS-PAS-JME-10-013,CMS-PAS-QCD-10-041,Aad:2014gea,LOCH:2014lla,CMS:2014fya,CMS:2014joa,Aad:2014haa} or used in new physics searches  \cite{CMS:2011bqa,Fleischmann:2013woa,Pilot:2013bla,TheATLAScollaboration:2013qia,Chatrchyan:2012ku,Chatrchyan:2012sn,CMS:2013cda,CMS:2014afa,CMS:2014aka,Khachatryan:2015axa,CMS:1900uua,Khachatryan:2015bma,Aad:2015owa}.

The vast majority of proposed jet substructure observables, however, have been analyzed exclusively within Monte Carlo simulation.  While Monte Carlos play an essential role in the simulation of realistic hadron collision events, they can often obscure the underlying physics that governs the behavior of a particular observable.  Additionally, it is challenging to disentangle perturbative physics from the tuning of non-perturbative physics so as to understand how to systematically improve the accuracy of the Monte Carlo.  Recently, there has been an increasing number of analytical studies of jet substructure observables, including the calculation of the signal distribution for $N$-subjettiness to next-to-next-to-next-to-leading-log order \cite{Feige:2012vc}, a fixed-order prediction for planar flow \cite{Field:2012rw}, calculations of groomed jet masses \cite{Dasgupta:2013ihk,Dasgupta:2013via,Larkoski:2014pca,Dasgupta:2015yua} and the jet profile/ shape \cite{Seymour:1997kj,Li:2011hy,Larkoski:2012eh,Jankowiak:2012na,Chien:2014nsa,Chien:2014zna,Isaacson:2015fra} for both signal and background jets, an analytic understanding of jet charge \cite{Krohn:2012fg,Waalewijn:2012sv}, predictions for fractional jet multiplicity \cite{Bertolini:2015pka}, and calculations of the associated subjet rate \cite{Bhattacherjee:2015psa}.  Especially in the case of the groomed jet observables, analytic predictions informed the construction of more performant and easier to calculate observables.  With the recent start of Run 2 of the LHC, where the phase space for high energy jets only grows, it will be increasingly important to have analytical calculations to guide experimental understanding of jet dynamics.

It is well known that the measurement of observables on a jet can introduce ratios of hierarchical scales appearing in logarithms at every order in the perturbative expansion. Accurate predictions over all of phase space require resummation of these large logarithms to all orders in perturbation theory. While this resummation is well understood for simple observables such as the jet mass \cite{Catani:1991bd,Chien:2010kc,Chien:2012ur,Dasgupta:2012hg,Jouttenus:2013hs}, where it has been performed to high accuracy, a similar level of analytic understanding has not yet been achieved for more complicated jet substructure observables. Jet substructure observables are typically sensitive to a multitude of scales, corresponding to characteristic features of the jet, resulting in a much more subtle procedure for resummation.  

A ubiquitous feature of some of the most powerful observables used for identification of jet substructure is that they are formed from the ratio of infrared and collinear (IRC) safe observables.  Examples of such observables include ratios of $N$-subjettiness variables \cite{Thaler:2010tr,Thaler:2011gf}, ratios of energy correlation functions \cite{Larkoski:2013eya,Larkoski:2014gra,Larkoski:2014zma}, or planar flow \cite{Almeida:2008yp}.  In general, ratios of IRC safe observables are not themselves IRC safe \cite{Soyez:2012hv} and cannot be calculated to any fixed order in perturbative QCD.  Nevertheless, it has been shown that these ratio observables are calculable in resummed perturbation theory and are therefore referred to as Sudakov safe \cite{Larkoski:2013paa,Larkoski:2014wba,Larkoski:2014bia,Larkoski:2015lea}.  Distributions of Sudakov safe observables can be calculated by appropriately marginalizing resummed multi-differential cross sections of IRC safe observables.  An understanding of the factorization properties of multi-differential jet cross sections has been presented in \Refs{Larkoski:2014tva,Procura:2014cba,Larkoski:2015zka} by identifying distinct factorization theorems in parametrically separated phase space regions defined by the measurements performed on the jet.  Combining this understanding of multi-differential factorization with the required effective field theories, all ingredients are now available for analytic resummation and systematically improvable predictions.

As an explicit example, observables that resolve two-prong substructure are sensitive to both the scales characterizing the subjets as well as to the scales characterizing the full jet.  A study of the resummation necessary for describing jets with a two-prong substructure was initiated in \Ref{Bauer:2011uc} which considered the region of phase space with two collinear subjets of comparable energy, and introduced an effective field theory description capturing all relevant scales of the problem.   Recently, an effective field theory description for the region of two-prong jet phase space with a hard core and a soft, wide angle subjet was developed in \Ref{Larkoski:2015zka}, where it was applied to the resummation of non-global logarithms \cite{Dasgupta:2001sh}. Combined, the collinear subjet and soft subjet factorization theorems allow for a complete description of the dominant dynamics of jets with two-prong substructure.

In this paper we will study the factorization and resummation of the jet substructure observable $\Dobsnobeta{2}$ \cite{Larkoski:2014gra}, a ratio-type observable formed from the energy correlation functions.  We will give a detailed effective theory analysis using the language of soft-collinear effective theory (SCET) \cite{Bauer:2000yr,Bauer:2001ct,Bauer:2001yt,Bauer:2002nz} in all regions of phase space required for the description of a one or two-prong jet, and will prove all-orders leading-power factorization theorems in each region. We will then use these factorization theorems to calculate the $\Dobsnobeta{2}$ distribution for jets initiated by boosted hadronic decays of electroweak bosons or from light QCD partons and compare to Monte Carlo simulation.  These calculations will also allow us to make first-principles predictions for the efficiency of the observable $\Dobsnobeta{2}$ to discriminate boosted electroweak signal jets from QCD background jets.

\begin{figure}
\begin{center}
\subfloat[]{\label{fig:intro_plota}
\includegraphics[width= 7.25cm]{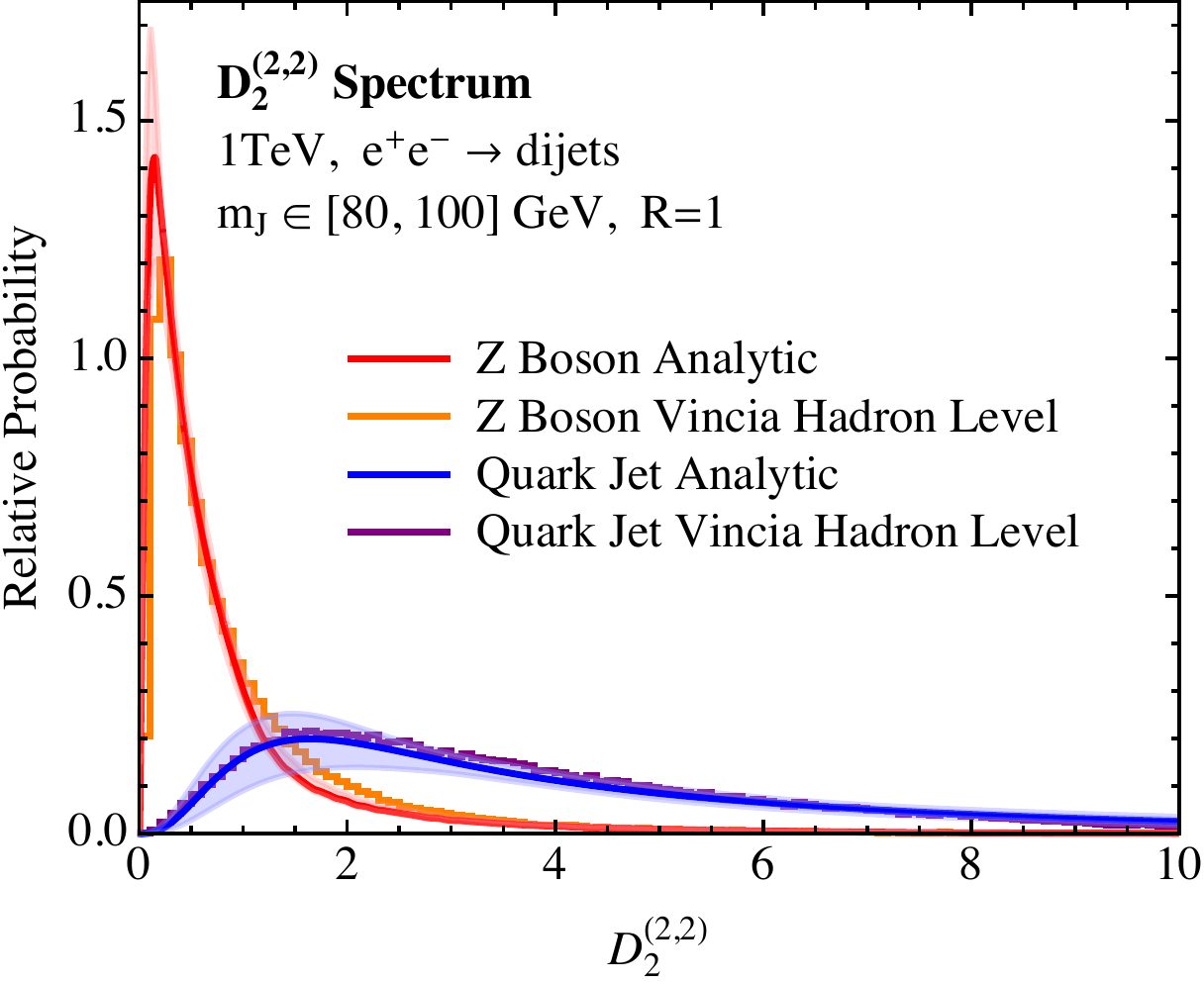}
}
\subfloat[]{\label{fig:intro_plotb}
\includegraphics[width = 7.25cm]{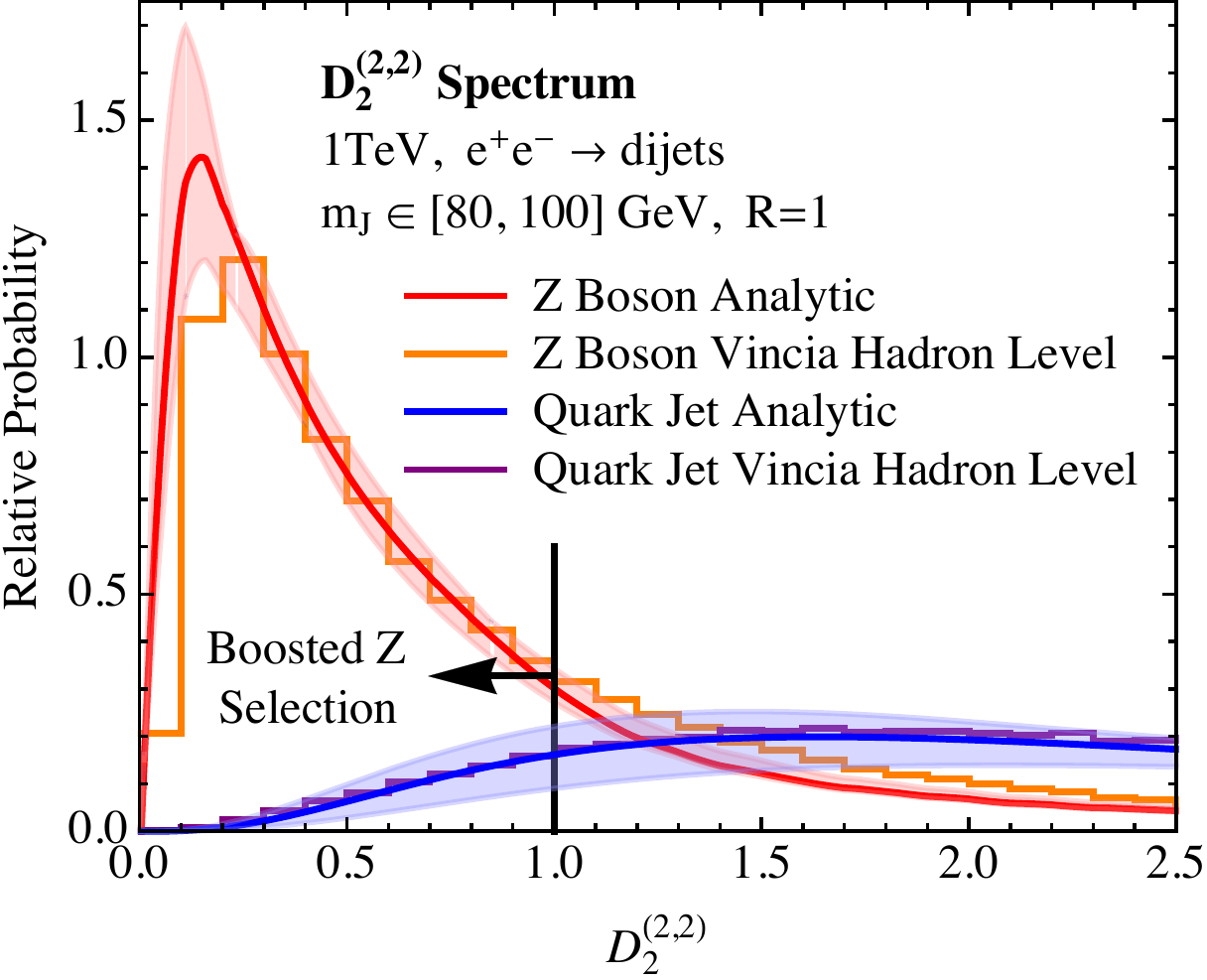}
}
\end{center}
\vspace{-0.2cm}
\caption{Comparison of our analytic calculation with \vincia Monte Carlo predictions for the two prong discriminant, $D_2$. Predictions for both boosted $Z$ bosons and massive QCD jets at a 1 TeV $e^+e^-$ collider are shown. The Monte Carlo is fully hadronized, and non-perturbative effects have been included in the analytic calculation through a shape function. In a) we show the complete distribution, and in b) we zoom in to focus on the region relevant for boosted $Z$ discrimination.
}
\label{fig:intro_plot}
\end{figure}

Our factorized description is valid to all orders in $\alpha_s$, expressing the cross section as a product of field theoretic matrix elements, each of which is calculable order by order in perturbation theory, allowing for a systematically improvable description of the $D_2$ observable. Furthermore, the factorization theorem enables a clean separation of perturbative and non-perturbative physics, allowing for non-perturbative contributions to the observable to be included in the analytic calculation through the use of shape functions \cite{Korchemsky:1999kt,Korchemsky:2000kp}.  In this paper we work to next-to-leading logarithmic (NLL) accuracy to demonstrate all aspects of the required factorization theorems necessary for precision jet substructure predictions. We will see that even at this first non-trivial order, we gain insight into qualitative and quantitative features of the $D_2$ distribution.  While we will give an extensive discussion of our numerical results and comparisons with a variety of Monte Carlo programs in this paper, in \Fig{fig:intro_plot} we compare our analytic predictions for the $D_2$ observable, including non-perturbative effects, for hadronically-decaying boosted $Z$ bosons and QCD jets in $e^+e^-$ collisions with the distributions predicted by the \vincia{} \cite{Giele:2007di,Giele:2011cb,GehrmannDeRidder:2011dm,Ritzmann:2012ca,Hartgring:2013jma,Larkoski:2013yi} Monte Carlo program at hadron level.  Excellent agreement between analytic and Monte Carlo predictions is observed, demonstrating a quantitative understanding of boosted jet observables from first principles.

\subsection{Overview of the Paper}\label{sec:approach}

While there exists a large number of two-prong discriminants in the jet substructure literature, any of which would be interesting to understand analytically, we will use calculability and factorizability as guides for constructing the observable to study in this paper.  This procedure will ultimately lead us to the observable $\Dobsnobeta{2}$ and will demonstrate that $\Dobsnobeta{2}$ has particularly nice factorization and calculability properties.  This approach will proceed in the following steps:
\begin{enumerate}

\item Identify the relevant subjet configurations for the description of a two-prong discriminant.

\item Isolate each of these relevant regions by the measurement of a collection of IRC safe observables.

\item Study the phase space defined by this collection of IRC safe observables, and prove all-orders factorization theorems in each parametrically-defined region of phase space.

\item Identify a two-prong discriminant formed from the collection of IRC safe observables which respects the parametric factorization theorems of the phase space.

\end{enumerate}
A detailed analysis of each of these steps will be the subject of this paper. Here, we provide a brief summary so that the logic of the approach is clear, and so that the reader can skip technical details in the different sections without missing the general idea of the approach.

\begin{figure}
\begin{center}
\subfloat[]{\label{fig:soft_haze}
\includegraphics[width=3.95cm, trim =0 -0.5cm 0 0]{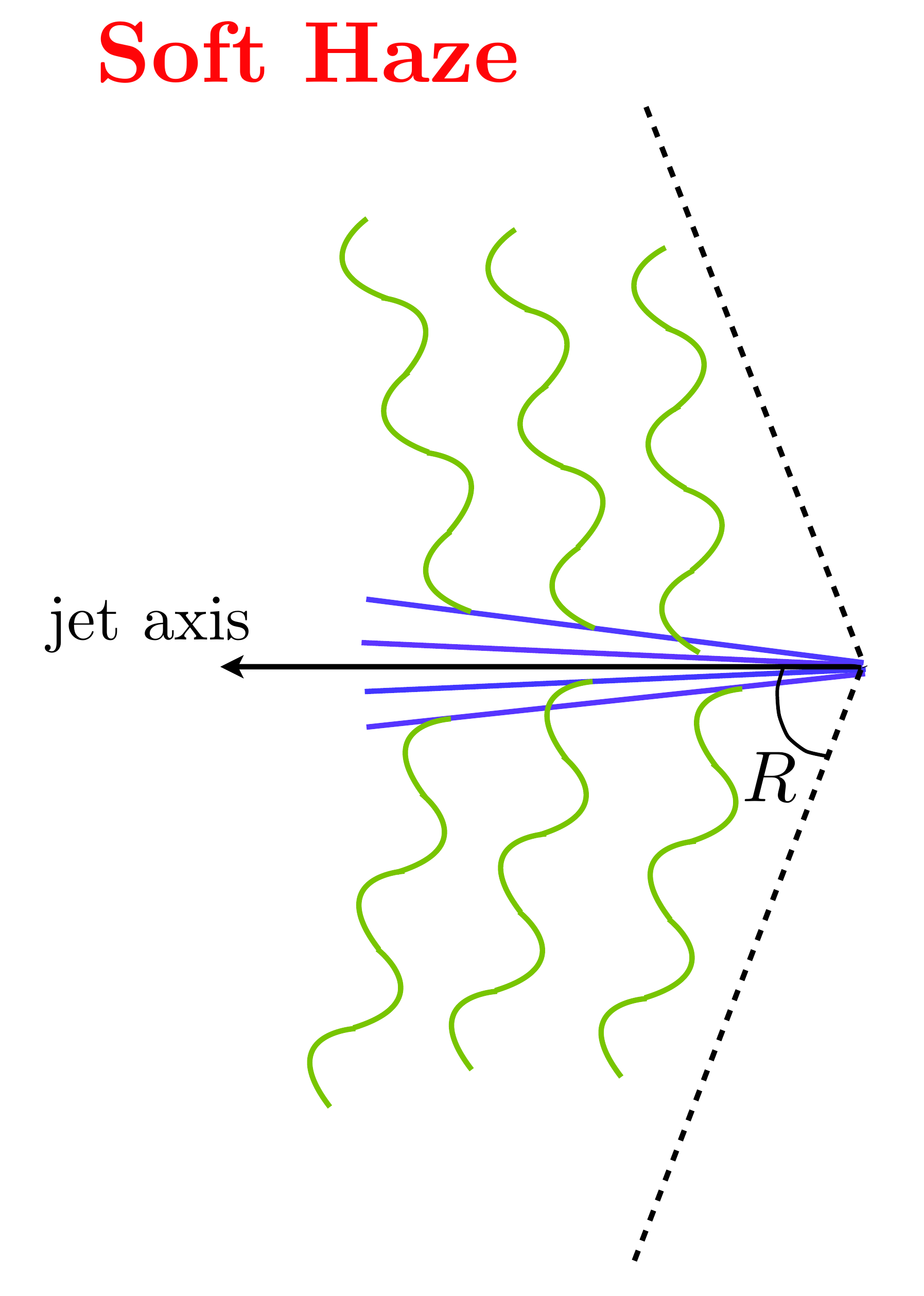}
}\qquad
\subfloat[]{\label{fig:ninja}
\includegraphics[width=4.1cm, trim =0 0 0 0]{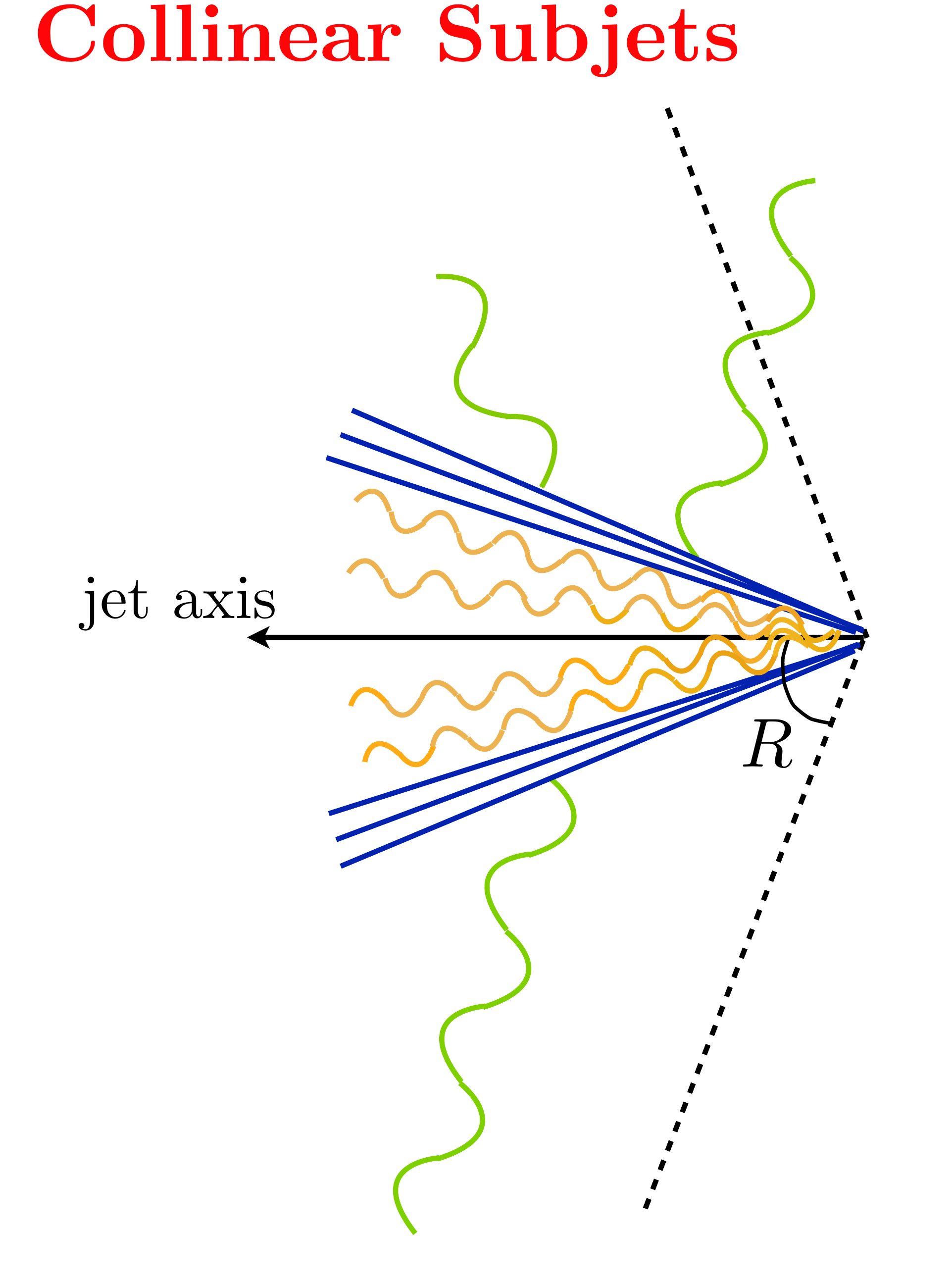}
}\qquad
\subfloat[]{\label{fig:soft_jet}
\includegraphics[width=3.9cm, trim =0 -0.75cm 0 0]{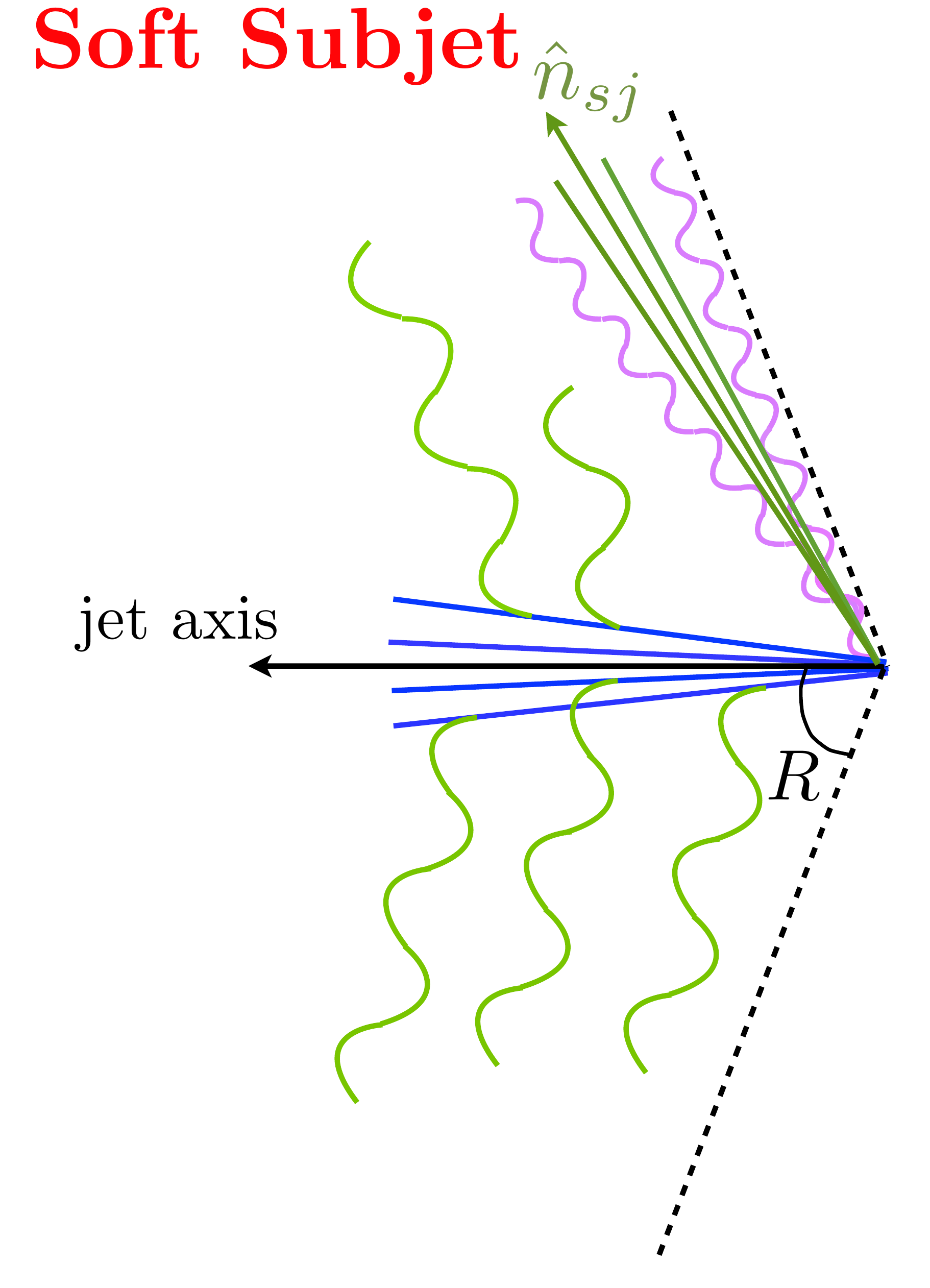}
}
\end{center}
\caption{ Regions of interest for studying the two-prong substructure of a jet. a) Soft haze region in which no subjets are resolved. b) Collinear subjets with comparable energy and a small opening angle. c) Soft subjet carrying a small fraction of the total energy, and at a wide angle from the hard subjet.}
\label{fig:diff_jets}
\end{figure}

The complete description of an observable capable of discriminating one- from two-prong substructure requires the factorized description of the following three relevant subjet configurations, shown schematically in \Fig{fig:diff_jets}:

\begin{itemize}

\item {\bf{Soft Haze}}: \Fig{fig:soft_haze} shows a jet in what we will refer to as the soft haze region of phase space. In the soft haze region there is no resolved subjet, only a single hard core with soft wide angle emissions. This region of phase space typically contains emissions beyond the strongly ordered limit, but is the dominant background region for QCD jets, for which a hard splitting is $\alpha_s$ suppressed.

\item {\bf{Collinear Subjets}}: \Fig{fig:ninja} shows a jet with two hard, collinear subjets. Both subjets carry approximately half of the total energy of the jet, and have a small opening angle. This region of phase space, and its corresponding effective field theory description, has been studied in \Ref{Bauer:2011uc}.

\item {\bf{Soft Subjet}}: \Fig{fig:soft_jet}  shows the soft subjet region of phase space which consists of jets with two subjets with hierarchical energies separated by an angle comparable to the jet radius $R$.  The soft subjet probes the boundary of the jet and we take $R\sim 1$. An effective field theory description for this region of phase space was presented in \Ref{Larkoski:2015zka}.

\end{itemize}

As a basis of IRC safe observables for isolating these three subjet configurations, we use the energy correlation functions \cite{Larkoski:2013eya}, which we define in \Sec{sec:obs_def}. In particular, we will show that the measurement of three energy correlation functions, two 2-point energy correlation functions $\ecf{2}{\alpha}, \ecf{2}{\beta}$, and a 3-point energy correlation function $\ecf{3}{\alpha}$, allows for parametric separation of the different subjet configurations.  While we will focus on the particular case of observables formed from the energy correlation functions, we believe that this approach is more general and could be applied to other IRC safe observable bases. 

With the energy correlation functions as our basis, we study the multi-differential phase space defined by the simultaneous measurement of these observables on a jet in \Sec{sec:phase_space}. Using the power counting technique of \Refs{Larkoski:2014gra,Larkoski:2014zma}, we show that the angular exponents of the energy correlation functions, $\alpha$ and $\beta$, can be chosen such that the different subjet configurations occupy parametrically separated regions of this phase space, and extend to all boundaries of the phase space. This parametric separation allows for each region to be separately described by its own effective field theory. The required effective field theories are described in \Sec{sec:Fact}, and are formulated in the language of SCET.  The formulation in SCET allows us to prove all-orders factorization theorems valid at leading-power in each of the phase space regions, and to resum logarithms to arbitrary accuracy using renormalization group techniques. 

Having understood in detail both the structure of the phase space defined by the IRC safe measurements as well as the factorization theorems defined in each region, we will show in \Sec{sec:def_D2} that this leads unambiguously to the definition of a two-prong discriminant observable which is amenable to factorization. This observable will be a generalized form of $\Dobsnobeta{2}$ \cite{Larkoski:2014gra} which will depend on both angular exponents $\alpha$ and $\beta$.  Calculating the distribution of $\Dobsnobeta{2}$ is accomplished by appropriate marginalization of the multi-differential cross section.  Depending on the phase space cuts that have been made, $\Dobsnobeta{2}$ may or may not be IRC safe itself, and so the marginalization will in general only be defined within resummed perturbation theory.

The outline of this paper is as follows. In \Sec{sec:phase_space} we define the energy correlation functions used in this paper and describe how the different subjet configurations shown schematically in \Fig{fig:diff_jets} can be isolated by demanding parametric relations between the measured values of these observables.  In \Sec{sec:Fact} we discuss the effective field theory descriptions in the different phase space regions, and present the factorization theorems that describe their dynamics. Although some of the relevant effective field theories have been presented elsewhere, we attempt to keep the discussion self-contained by providing a brief review of their most salient features.  All field theoretic definitions of the functions appearing in the factorization theorems, as well as their calculations to one-loop accuracy, are provided in appendices. 
 
 In \Sec{sec:friendly}  we show how the detailed understanding of the multi-differential phase space leads to the definition of $\Dobsnobeta{2}$ as a powerful one- versus two-prong jet discriminant.  In \Sec{sec:sudsafe} we emphasize that without a mass cut, $\Dobsnobeta{2}$ is not IRC safe but is Sudakov safe and whose all-orders distribution exhibits paradoxical dependence on $\alpha_s$.  In \Sec{sec:fixed_order} we study the fixed-order distribution of $\Dobsnobeta{2}$ in the presence of a mass cut to understand its behavior in singular limits. In \Sec{sec:merging} we discuss how the different effective field theories can be consistently merged to give a factorized description of the $\Dobsnobeta{2}$ observable, and introduce a novel zero-bin procedure to implement this merging. 
 
 In \Sec{sec:results} we present numerical results for both signal and background distributions for $\Dobsnobeta{2}$ as measured in $e^+e^-$ collisions and compare our analytic calculation with several Monte Carlo generators. We emphasize many features of the calculation which provide considerable insight into two-prong discrimination, and the ability of current Monte Carlo generators to accurately describe substructure observables. In \Sec{sec:LEP} we discuss numerical results for the $D_2$ observable at $e^+e^-$ collisions at the $Z$ pole at LEP, and demonstrate that being sensitive to correlations between three emissions, the $D_2$ observable can be used as a more differential probe of the perturbative shower for tuning Monte Carlo generators. In \Sec{sec:LHC} we discuss how to extend our calculations to $pp$ collisions at the LHC.  We conclude in \Sec{sec:conc}, and discuss future directions for further improving the analytic understanding of jet substructure.

\section{Characterizing a Two-Prong Jet}
\label{sec:phase_space}

In this section, we develop the framework necessary to construct the all-orders factorization theorems for analytic two-prong discrimination predictions.  We begin in \Sec{sec:obs_def} by defining the energy correlation functions, which we will use to isolate the three subjet configurations discussed in the introduction.  Using the power counting analysis of \Refs{Larkoski:2014gra,Larkoski:2014zma}, we study the phase space defined by measuring the energy correlation functions in \Sec{sec:power_counting}.  Throughout this paper, our proxy for a two-prong jet will be a boosted, hadronically decaying $Z$ boson, but our analysis holds for $W$ or $H$ bosons, as well.

\subsection{Observable Definitions}\label{sec:obs_def}

To distinguish the three different subjet configurations of \Fig{fig:diff_jets} with IRC safe measurements, observables which are sensitive to both one- and two-prong structure are required. Although many possible observable bases exist, in this paper we will use the energy correlation functions \cite{Larkoski:2013eya,Larkoski:2014gra}, as we will find that they provide a convenient basis.

The $n$-point energy correlation function is an IRC safe observable that is sensitive to $n$-prong structure in a jet.  For studying the two-prong structure of a jet, we will need the 2- and 3-point energy correlation functions, which we define for $e^+e^-$ collisions as \cite{Larkoski:2013eya}\footnote{For massive hadrons, there exist several possible definitions of the energy correlation functions depending on the particular mass scheme \cite{Salam:2001bd,Mateu:2012nk}.  The definition in \Eq{eq:ecf_def} is an $E$-scheme definition.  A $p$-scheme definition will be presented in \Sec{sec:LEP} when we discuss the connection to LEP. Since the different definitions are equivalent for massless partons, their perturbative calculations are identical. The different definitions differ only in their non-perturbative corrections.}
\begin{align}\label{eq:ecf_def}
\ecf{2}{\alpha}&= \frac{1}{E_J^2} \sum_{i<j\in J} E_i E_j \left(
\frac{2p_i \cdot p_j}{E_i E_j}
\right)^{\alpha/2} \,, \\
\ecf{3}{\alpha}&= \frac{1}{E_J^3} \sum_{i<j<k\in J} E_i E_j E_k \left(
\frac{2p_i \cdot p_j}{E_i E_j}
\frac{2p_i \cdot p_k}{E_i E_k}
\frac{2p_j \cdot p_k}{E_j E_k}
\right)^{\alpha/2} \,. \nonumber
\end{align}
Here $J$ denotes the jet, $E_i$ and $p_i$ are the energy and four momentum of particle $i$ in the jet and $\alpha$ is an angular exponent that is required to be greater than 0 for IRC safety.  The 4-point and higher energy correlation functions are defined as the natural generalizations of \Eq{eq:ecf_def}, although we will not use them in this paper.

While we will mostly focus on the case of an $e^+e^-$ collider, the energy correlation functions have natural generalizations to hadron colliders, by replacing $E$ by $p_T$ and using hadron collider coordinates, $\eta$ and $\phi$. This definition is given explicitly in \Eq{eq:pp_def_ecf}. At central rapidity, this modification does not change the behavior of the observables, or any of the conclusions presented in the next sections.  Of course, the hadron collider environment has other effects not present in an $e^+e^-$ collider, like initial state radiation or underlying event, that will affect the energy correlation functions.  A brief discussion of the behavior of the energy correlation functions in $pp$ colliders will be given in \Sec{sec:LHC}. Numerical implementations of the energy correlation functions for both $e^+e^-$ and hadron colliders are available in the \texttt{EnergyCorrelator} \fastjet{contrib} \cite{Cacciari:2011ma,fjcontrib}.

\subsection{Power Counting the $\ecf{2}{\alpha}, \ecf{2}{\beta}, \ecf{3}{\alpha}$ Phase Space}\label{sec:power_counting}

With a basis of IRC safe observables identified, we now demonstrate that the measurement of multiple energy correlation functions parametrically separates the three different subjet configurations identified in \Fig{fig:diff_jets}. In particular, the simultaneous measurement of $\ecf{2}{\alpha}$, $\ecf{2}{\beta}$, and $\ecf{3}{\alpha}$ is sufficient for this purpose, and we will study in detail the phase space defined by their measurement.  From this analysis, we will be able to determine for which values of the angular exponents $\alpha$ and $\beta$ the three subjet configurations are parametrically separated within this phase space. The power counting parameters that define ``parametric'' will be set by the observables themselves, as is typical in effective field theory.

We begin by considering how the energy correlation functions can be used to separate one- and two-prong jets. This has been previously discussed in \Ref{Larkoski:2014gra} by measuring $\ecf{2}{\alpha}$ and $\ecf{3}{\alpha}$, but here we consider the phase space defined by $\ecf{2}{\beta}$ and $\ecf{3}{\alpha}$ with $\alpha$ and $\beta$ in general different.  A minimal constraint on the angular exponents, both for calculability and discrimination power, is that the soft haze and collinear subjets configurations are parametrically separated by the measurements.  A power counting analysis of the soft subjet region yields no new constraints beyond those from the soft haze or collinear subjets.

\begin{figure}
\begin{center}
\subfloat[]{\label{fig:unresolved}
\includegraphics[width=6cm]{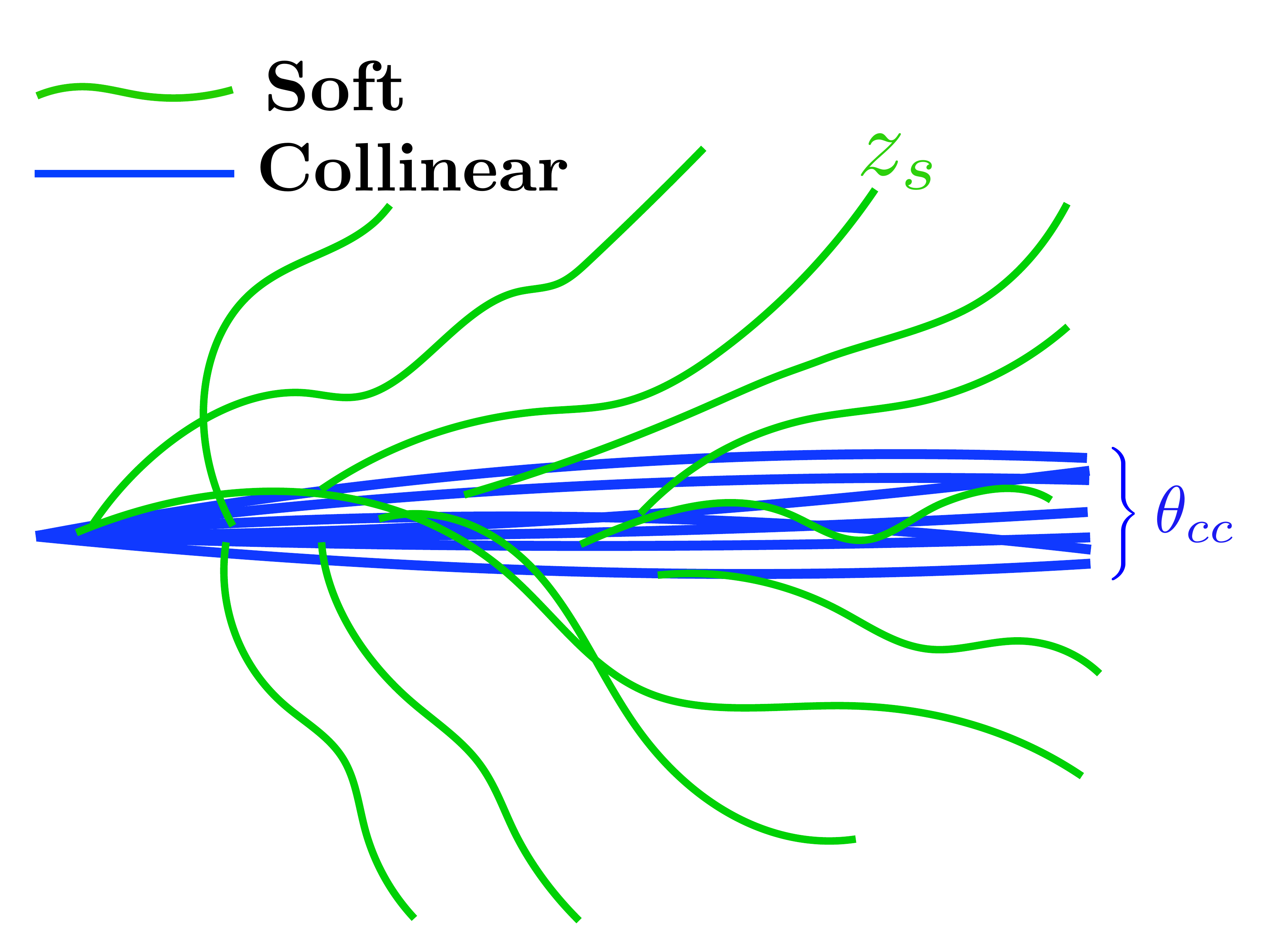}    
}\qquad
\subfloat[]{\label{fig:NINJA}
\includegraphics[width=6cm, trim =0 -0.5cm 0 0]{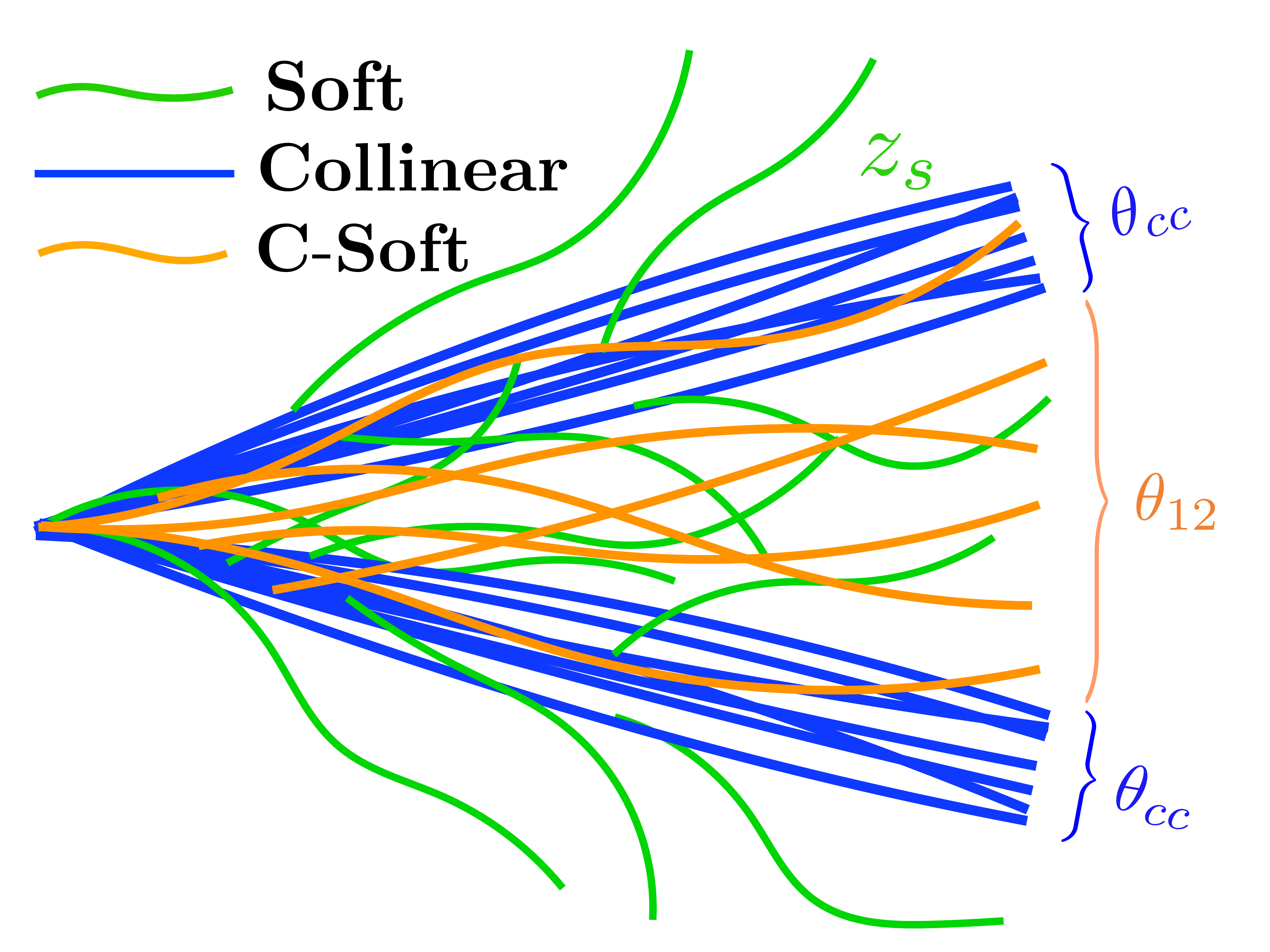}
}
\end{center}
\caption{a) Schematic of a one-prong soft haze jet, dominated by collinear (blue) and soft (green) radiation. The angular size of the collinear radiation is $\theta_{cc}$ and the energy fraction of the soft radiation is $z_s$.  b) Schematic of a jet resolved into two collinear subjets, dominated by collinear (blue), soft (green), and collinear-soft (orange) radiation emitted from the dipole formed by the two subjets.  The subjets are separated by an angle $\theta_{12}$  and the energy fraction of the collinear-soft radiation is $z_{cs}$. 
}
\label{fig:pics_jets}
\end{figure}

The setup for the power counting analysis of the soft haze and collinear subjets configurations is shown in \Fig{fig:pics_jets}, where all relevant modes are indicated. The one-prong jet illustrated in \Fig{fig:unresolved} is described by soft modes with energy fraction $z_s$ emitted at ${\cal O}(1)$ angles, and collinear modes with characteristic angular size $\theta_{cc}$ with ${\cal O}(1)$ energy fraction. The collinear subjets configuration illustrated in \Fig{fig:NINJA} consists of two subjets, each of which carry an ${\cal O}(1)$ fraction of the jet's energy and are separated by an angle $ \theta_{12}\ll1$.  Each of the subjets has collinear emissions at a characteristic angle $ \theta_{cc}\ll  \theta_{12}$, and global soft radiation at large angles with respect to the subjets, with characteristic energy fraction $z_s\ll 1$. In the case of two collinear subjets arising from the decay of a color singlet particle, the long wavelength global soft radiation is not present due to color coherence, but the power counting arguments of this section remain otherwise unchanged. Finally, there is radiation from the dipole formed from the two subjets (called ``collinear-soft'' radiation), with characteristic  angle $ \theta_{12}$ from the subjets, and with energy fraction $z_{cs}$.  The effective theory of this phase space region for the observable $N$-jettiness \cite{Stewart:2010tn} was studied in \Ref{Bauer:2011uc}.\footnote{It is of historical interest to note that the generalization of two-prong event shapes, such as thrust, to event shapes for characterizing three jet structure was considered early on, for example with the introduction of the triplicity event shape \cite{Brandt:1978zm}. However, it was not until more recently, with the growth of the jet substructure field at the LHC, that significant theoretical study was given to such observables.}

We are now able to determine the parametric form of the dominant contributions to the observables $\ecf{2}{\beta}$ and $\ecf{3}{\alpha}$.  In the soft haze region, the dominant contributions to the energy correlation functions are\footnote{It is important to understand that this relationship is valid to an arbitrary number of emissions. When performing the power counting, a summation over all the particles with soft and collinear  scalings in the jet must be considered. However, to  determine  the  scalings  of  the observable,  it  is  sufficient  to consider the scaling of the different types of individual terms in the sum. For example, the three terms contributing to the expression for $\ecf{3}{\alpha}$ arise from correlations between subsets of three collinear particles, one collinear particle and two soft particles, and two collinear particles and a soft particle, respectively. Contributions from other combinations of particles are power suppressed. Because of this simplification, in this paper we will never write explicit summations when discussing the scaling of observables.}
\begin{align}
\ecf{2}{\beta}&\sim \zs+\thetac^\beta\,, \nonumber \\
\ecf{3}{\alpha}&\sim \thetac^{3\alpha}+\zs^2 +\thetac^\alpha \zs\,.
\end{align}
From these parametrics, it is straightforward to show that one-prong jets live in a region of the $\ecf{2}{\beta}, \ecf{3}{\alpha}$ phase space bounded from above and below, whose precise scaling depends on the relative size of the angular exponents $\alpha$ and $\beta$.  The scaling of upper and lower boundaries of the one-prong region of phase space for all $\alpha$ and $\beta$ are listed in \Tab{tab:pc}.  For $\alpha = \beta$, as studied in \Ref{Larkoski:2014gra}, one-prong jets live in the region defined by $\left( \ecf{2}{\beta}\right)^{3} \lesssim \ecf{3}{\beta} \lesssim \left( \ecf{2}{\beta}\right)^{2}$.

\begin{table}[t]
\begin{center}
\begin{tabular}{c|c|c|c|c}
&$\alpha\leq \beta/2$ & $\beta/2 <\alpha < 2\beta/3$ & $ 2\beta/3 \leq \alpha < \beta$ & $\alpha \geq \beta$ \\ 
\hline
upper& $\ecf{3}{\alpha}\sim \left( \ecf{2}{\beta}\right)^{3\alpha/\beta}$ & $\ecf{3}{\alpha} \sim \left (\ecf{2}{\beta} \right )^{\alpha/\beta +1}$&$\ecf{3}{\alpha} \sim \left (\ecf{2}{\beta} \right )^{\alpha/\beta +1}$&$\ecf{3}{\alpha} \sim \left (\ecf{2}{\beta} \right )^{2}$\\
lower  & $\ecf{3}{\alpha}\sim \left( \ecf{2}{\beta}\right)^2$ &$\ecf{3}{\alpha}\sim \left( \ecf{2}{\beta}\right)^{2}$&$\ecf{3}{\alpha}\sim \left( \ecf{2}{\beta}\right)^{3\alpha/\beta}$&$\ecf{3}{\alpha}\sim \left( \ecf{2}{\beta}\right)^{3\alpha/\beta}$
\end{tabular}
\end{center}
\caption{Parametric scaling of the upper and lower boundaries of the one-prong region of the $\ecf{2}{\beta}, \ecf{3}{\alpha}$ phase space as a function of the angular exponents $\alpha$ and $\beta$.
}
\label{tab:pc}
\end{table}

For the collinear subjets configuration, the dominant contributions to the observables $\ecf{2}{\beta}$ and $\ecf{3}{\alpha}$ are
\begin{align}
\ecf{2}{\beta}&\sim \theta_{12}^\beta\,, \nonumber \\
\ecf{3}{\alpha}&\sim \theta_{cc}^{\alpha} \theta_{12}^{2\alpha}+\theta_{12}^\alpha z_s +\theta_{12}^{3\alpha}z_{cs}+z_s^2\,.
\end{align}
The 2-point energy correlation function $\ecf{2}{\beta}$ is set by the angle of the hard splitting, $ \theta_{12}$, and the scaling of all other modes (soft, collinear, or collinear-soft) are set by the $\ecf{3}{\alpha}$ measurement. The requirement
\begin{align}\label{eq:tpb}
z_{cs} \sim \frac{\ecf{3}{\alpha}}{  \left(  \ecf{2}{\beta}  \right)^{3\alpha/\beta} } \ll 1\,,
\end{align}
then implies that the two-prong jets occupy the region of phase space defined by  $\ecf{3}{\alpha} \ll  \left(  \ecf{2}{\beta}  \right)^{3\alpha/\beta}$.  

For optimal discrimination, the one- and two-prong regions of this phase space should not overlap.  Since they are physically distinct, a proper division of the phase space will allow distinct factorizations, simplifying calculations.  Comparing the boundaries of the one-prong region listed in \Tab{tab:pc} with the  upper boundary of the two-prong region from \Eq{eq:tpb}, we find that the one- and two-prong jets do not overlap with the following restriction on the angular exponents $\alpha$ and $\beta$:
\be\label{eq:restriction_1}
 3\alpha \geq 2\beta\,.
\ee
Note that when $\alpha=\beta$ this is satisfied, consistent with the analysis of  \Ref{Larkoski:2014gra}.

\begin{figure}
\begin{center}
\subfloat[]{\label{fig:D2_2D_a}
\includegraphics[width= 6.5cm]{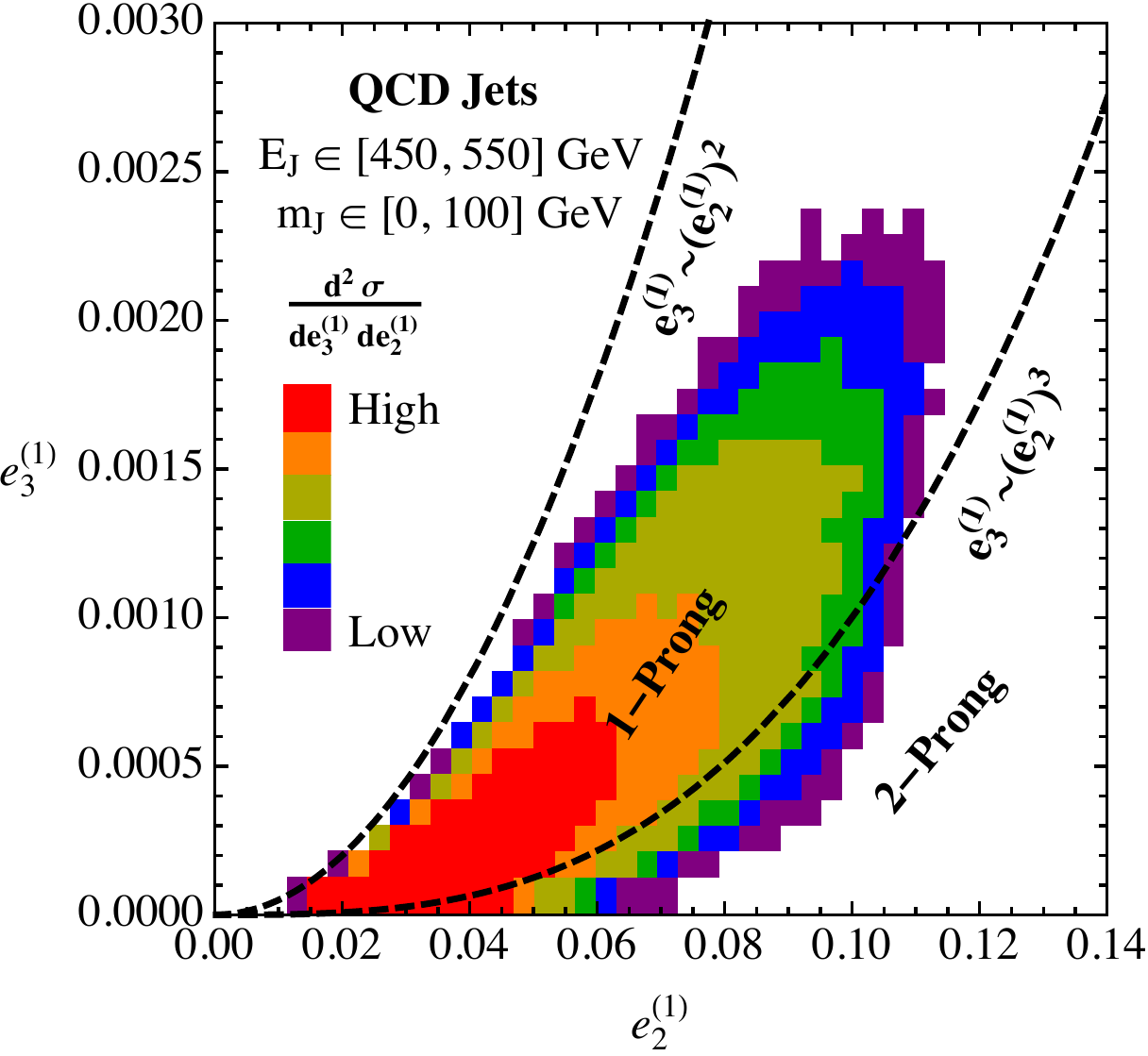}
}
\qquad
\subfloat[]{\label{fig:D2_2D_b}
\includegraphics[width = 6.5cm]{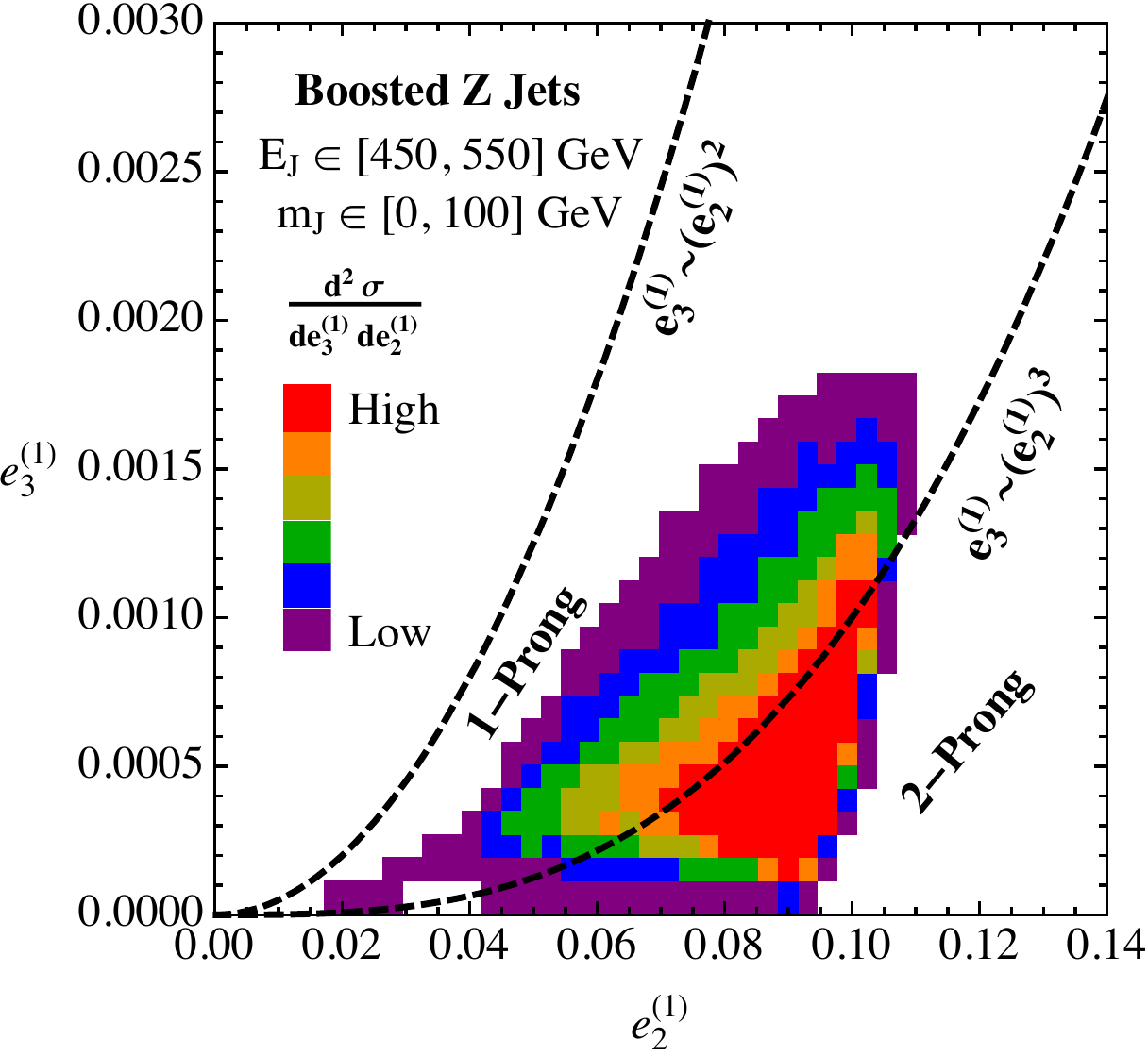}
}
\end{center}
\vspace{-0.2cm}
\caption{Monte Carlo distributions in the $\ecf{2}{1},\ecf{3}{1}$ plane, for QCD quark jets (left) and boosted $Z\to q\bar q$ jets (right). The parametric scalings predicted by the power counting analysis are shown as dashed lines, and the one- and two-prong regions of phase space are labelled, and extend between the parametric boundaries. Note the upper boundary is constrained to have a maximal value of $\frac{1}{2}(\ecf{2}{\alpha})^2=\ecf{3}{\alpha}$.
}
\label{fig:D2_2D}
\end{figure}

Because these power counting arguments rely exclusively on the parametric behavior of QCD in the soft and collinear limits, they must be reproduced by any Monte Carlo simulation, regardless of its shower and hadronization models.  To illustrate the robust boundary between the one- and two-prong regions of phase space predicted in \Eq{eq:restriction_1}, in \Fig{fig:D2_2D}, we plot the distribution in the $\ecf{2}{\beta},\ecf{3}{\alpha}$ plane of jets initiated by light QCD partons and those from boosted hadronic decays of $Z$ bosons as generated in $e^+e^-$ collisions in \pythia{} \cite{Sjostrand:2006za,Sjostrand:2007gs}.  Details of the Monte Carlo generation are presented in \Sec{sec:results}.  QCD jets are dominantly one-pronged, while jets from $Z$ decays are dominantly two-pronged.  We have chosen to use angular exponents $\alpha=\beta=1$ for this plot, as the small value of the angular exponent allows the structure of the phase space to be seen in a non-logarithmic binning.  The predicted behavior persists for all values of $\alpha$ and $\beta$ consistent with \Eq{eq:restriction_1}, while the choice made here is simply for illustrative aesthetics.  On these plots, we have added dashed lines corresponding to the predicted one- and two-prong phase space boundaries to guide the eye.  The one-prong QCD jets and the two-prong boosted $Z$ jets indeed dominantly live in their respective phase space regions as predicted by power counting.

\begin{figure}
\begin{center}
\subfloat[]{\label{fig:tab_pc}
\begin{tabular}{|c|c|}
\hline
Subjet Configuration & Defining Relation\\
\hline
\hline
Soft Haze&   $\left (\ecf{2}{\beta} \right )^{3\alpha/\beta} \lesssim \ecf{3}{\alpha} \lesssim \left (\ecf{2}{\beta} \right )^{2}$   \\
\hline
Collinear Subjets&   $\ecf{2}{\alpha} \sim  \left (\ecf{2}{\beta}   \right) ^{\alpha/\beta}$ and $\ecf{3}{\alpha}\ll \left( \ecf{2}{\beta}\right)^{3\alpha/\beta}$ \\
\hline
Soft Subjet&  $\ecf{2}{\alpha} \sim  \ecf{2}{\beta}  $ and $\ecf{3}{\alpha}\ll \left( \ecf{2}{\beta}\right)^{3\alpha/\beta}$   \\
\hline
\end{tabular}
}\\
\subfloat[]{\label{fig:2pt3ptps}
\includegraphics[width= 5.5cm]{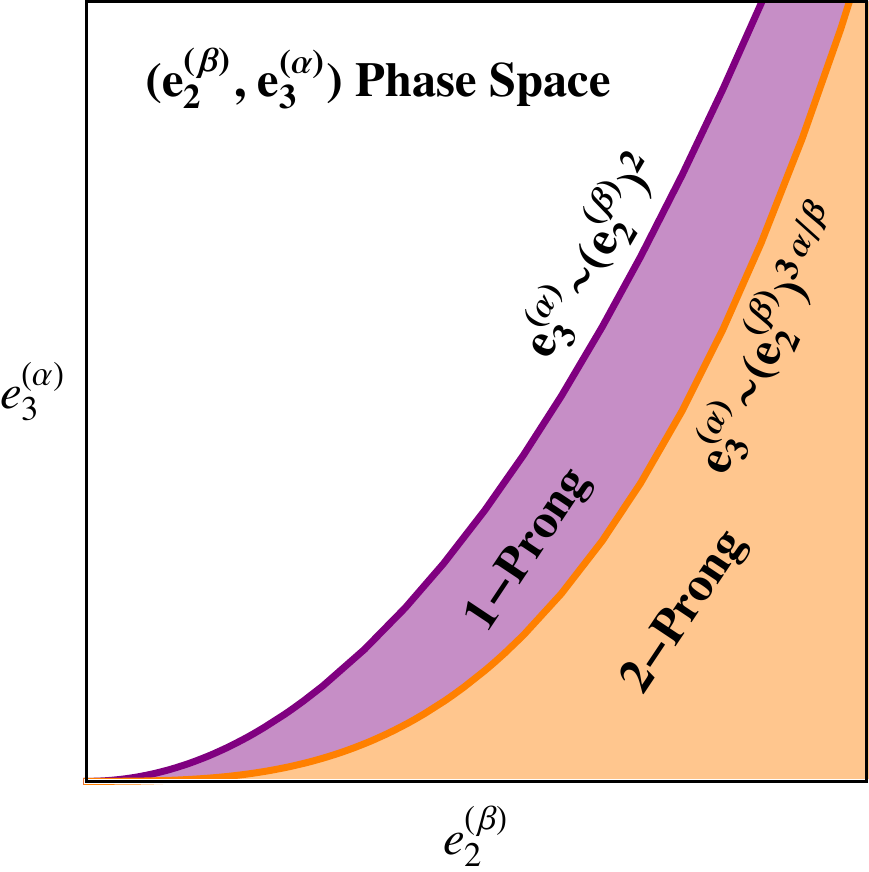}
}
$\qquad$
\subfloat[]{\label{fig:2ptps}
\includegraphics[width= 5.5cm]{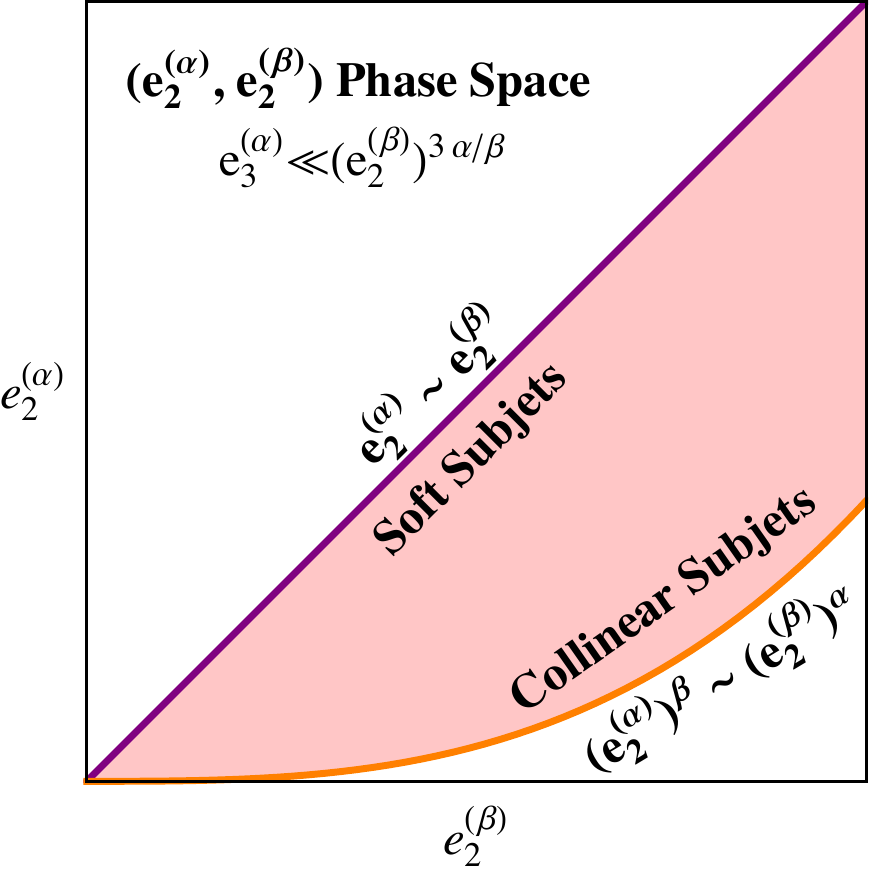}
}
\end{center}
\vspace{-0.2cm}
\caption{ a) Table summarizing the defining relations for the different subjet configurations in terms of the energy correlation functions $\ecf{2}{\alpha}, \ecf{2}{\beta},\ecf{3}{\alpha}$.  b) The one- and two-prong jets regions in the  $\ecf{2}{\beta}, \ecf{3}{\alpha}$ phase space. Jets with a two-prong structure lie in the lower (orange) region of phase space, while jets with a one-prong structure lie in the upper (purple) region of phase space. c) The projection onto the $\ecf{2}{\alpha}, \ecf{2}{\beta}$ phase space in which the soft subjet and collinear subjets are separated. 
}
\label{fig:phasespace}
\end{figure}

The measurement of $\ecf{2}{\beta}$ and $\ecf{3}{\alpha}$ alone is sufficient to separate one- and two prong jets.  However, the two-prong jets can exhibit either collinear subjets or a soft, wide angle subjet. To separate the collinear and soft subjet two-prong jets, we make an additional IRC safe measurement on the full jet.  Following \Ref{Larkoski:2015zka}, in addition to $\ecf{2}{\beta}$ and $\ecf{3}{\alpha}$, we measure $\ecf{2}{\alpha}$, with $\alpha \neq \beta$. In particular, the soft subjet and collinear subjet regions of phase space are defined by the simple conditions
\begin{align}
&\text{Collinear Subjet:} \qquad \ecf{2}{\alpha} \sim  \left (\ecf{2}{\beta}   \right) ^{\alpha/\beta}\,, \\
&\text{Soft Subjet:} \qquad \hspace{0.9cm} \ecf{2}{\alpha} \sim  \ecf{2}{\beta}\,.
\end{align}
For $\alpha \neq \beta$ and $\ecf{2}{\beta}\ll 1$, these two regions are parametrically separated.  Equivalently, in the two-prong region of phase space the measurement of both $\ecf{2}{\alpha}$ and $\ecf{2}{\beta}$ can be used to give IRC safe definitions to the subjet energy fraction and splitting angle, allowing the soft subjet and collinear subjets to be distinguished.  In \Fig{fig:phasespace} we summarize and illustrate the measurements that we make on the jet and the parametric relations between the measured values of the energy correlation functions that define the three phase space regions.  The phase space plots of \Figs{fig:2pt3ptps}{fig:2ptps} were also presented in \Ref{Larkoski:2015zka}.

\subsubsection{Jet Mass Cuts}\label{sec:mass_cuts}

In addition to discriminating QCD jets from boosted $Z$ bosons by their number of resolved prongs, we must also impose a mass cut on the jet to ensure that the jet is compatible with a $Z$ decay.  To include a mass cut in our analysis, for general angular exponents $\alpha$ and $\beta$, we would need to measure four observables on the jet: $\ecf{2}{\alpha}$, $\ecf{2}{\beta}$, $\ecf{3}{\alpha}$ and the jet mass.  This would significantly complicate calculations and introduce new parametric phase space regions that would need to be understood.  To avoid this difficulty, we note that, for our definition of $\ecf{2}{\alpha}$ from \Eq{eq:ecf_def}, if all final state particles are massless, then
\begin{equation}\label{eq:ecf_mass_relation}
\ecf{2}{2} = \frac{m_J^2}{E_J^2} \,,
\end{equation}
where $m_J$ is the mass of the jet.  Therefore, choosing $\beta = 2$ we can trivially impose a mass cut within the framework developed here.  Throughout the rest of this paper, we will set $\beta = 2$ for this reason.  Importantly, from Monte Carlo studies it has been shown that $\beta \sim 2$ provides optimal discrimination power \cite{Larkoski:2013eya,Larkoski:2014gra}, so this restriction does not limit the phenomenological relevance of our results.

Substituting the value $\beta=2$ into the power counting condition of \Eq{eq:restriction_1}, we find that the one- and two-prong regions of phase space are separated if 
\be
\alpha \geq \frac{4}{3}\,.
\ee
To achieve a parametric separation of the one- and two-prong regions of phase space, we will demand that the scalings defining the different regions be separated by at least a single power of $\ecf{2}{\beta}$.  For example, choosing $\alpha=\beta=2$, the scalings of the one-prong and two-prong regions are $\ecf{3}{\alpha}\sim \left( \ecf{2}{\beta}\right)^{3}$ and $\ecf{3}{\alpha}\sim \left( \ecf{2}{\beta}\right)^{2}$, which are parametrically different.  We therefore restrict ourselves to the range of angular exponents
\begin{align}\label{eq:final_constraint_exponents}
\beta=2, \qquad \alpha \gtrsim 2\,.
\end{align}
We expect that for $\alpha<2$ our effective field theory description will begin to break down, while as $\alpha$ is increased above $2$ it should improve.

\section{Factorization and Effective Field Theory Analysis}\label{sec:Fact}

In each region of phase space identified in \Sec{sec:phase_space}, hierarchies of scales associated with the particular kinematic configuration of the jet appear. These include the soft subjet energy fraction $z_{sj}$ in the soft subjet region of phase space, or the splitting angle $\theta_{12}$ of the collinear subjets. Logarithms of these scales appear at each order in perturbation theory, and need to be resummed to all orders to achieve reliable predictions. To perform this resummation, we will prove factorization theorems in each region of phase space by developing an effective field theory description which captures all the scales relevant to that particular region of phase space. These effective field theories are formulated in the language of SCET \cite{Bauer:2000yr,Bauer:2001ct,Bauer:2001yt,Bauer:2002nz}, but include additional modes which are required to describe the dynamics of the scales associated with the jet's particular substructure. Resummation is then achieved by renormalization group evolution within the effective theory.

In this section we discuss each of the effective theories required for a description of the $\ecf{2}{\alpha}, \ecf{2}{\beta}, \ecf{3}{\alpha}$ phase space. For each region of the phase space, we present an analysis of the modes required in the effective field theory description and present the factorization theorem. We also provide a brief discussion of the physics described by each of the functions appearing in the factorization theorem. Field theoretic operator definitions of the functions, as well as their calculation to one-loop accuracy, are presented in appendices.

\subsection{QCD Background}\label{sec:QCD_back}

Three distinct factorization theorems are required to describe the full phase space for massive QCD jets, corresponding to the soft haze, collinear subjets, and soft subjet configurations. Detailed expositions of the factorization theorems for the collinear subjets and soft subjet configurations have been presented in \Refs{Bauer:2011uc,Larkoski:2015zka}, but here we review the important features of the factorization theorems to keep the discussion self-contained.

Throughout this section, all jets are defined using the $e^+e^-$ anti-$k_T$ clustering metric \cite{Cacciari:2008gp,Cacciari:2011ma} with the Winner-Take-All (WTA) recombination scheme \cite{Larkoski:2014uqa,Larkoski:2014bia}. To focus on the aspects of the factorization relevant to the jet substructure, we will present the factorization theorems for the specific case of $e^+e^- \to q \bar q$. The factorization theorem for gluon initiated jets is identical to the quark case, and can be performed using the ingredients in the appendices. The extension to the production of additional jets or $pp$ colliders will be discussed in \Sec{sec:LHC}. 

\subsubsection{Collinear Subjets}\label{sec:ninja}

An effective field theory describing the collinear subjets configuration was first presented in \Ref{Bauer:2011uc} and is referred to as SCET$_+$. We refer the interested reader to \Ref{Bauer:2011uc} for a more detailed discussion, as well as a formal construction of the effective theory. To our knowledge, our calculation is the first, other than that of \Ref{Bauer:2011uc}, to use this effective theory.

\subsubsection*{Mode Structure}\label{sec:ninja_modes}

\begin{figure}
\begin{center}
\subfloat[]{\label{fig:collinear_subjets_a}
\includegraphics[width=5cm]{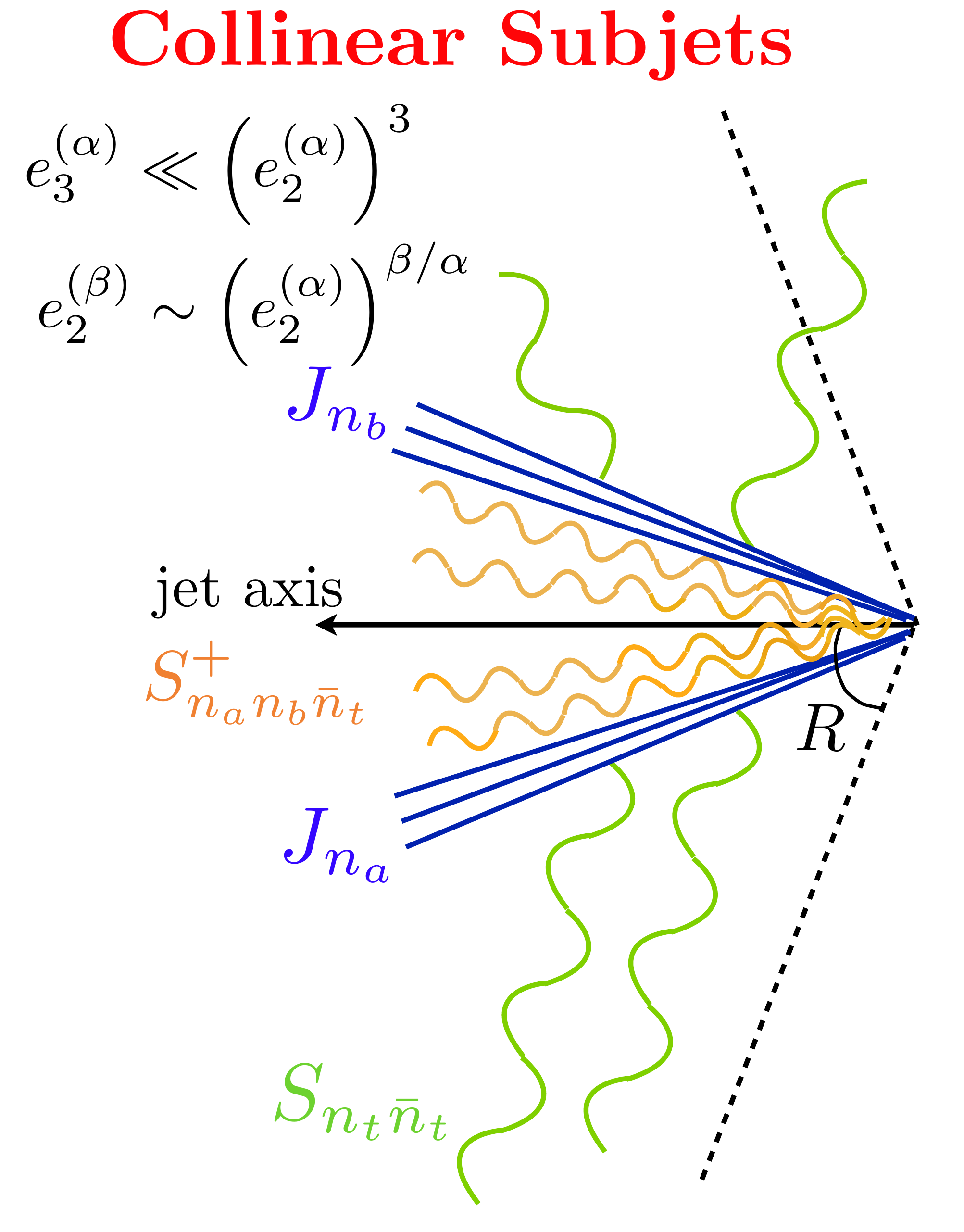}
}
\subfloat[]{\label{fig:collinear_subjets_b}
\includegraphics[width = 10cm]{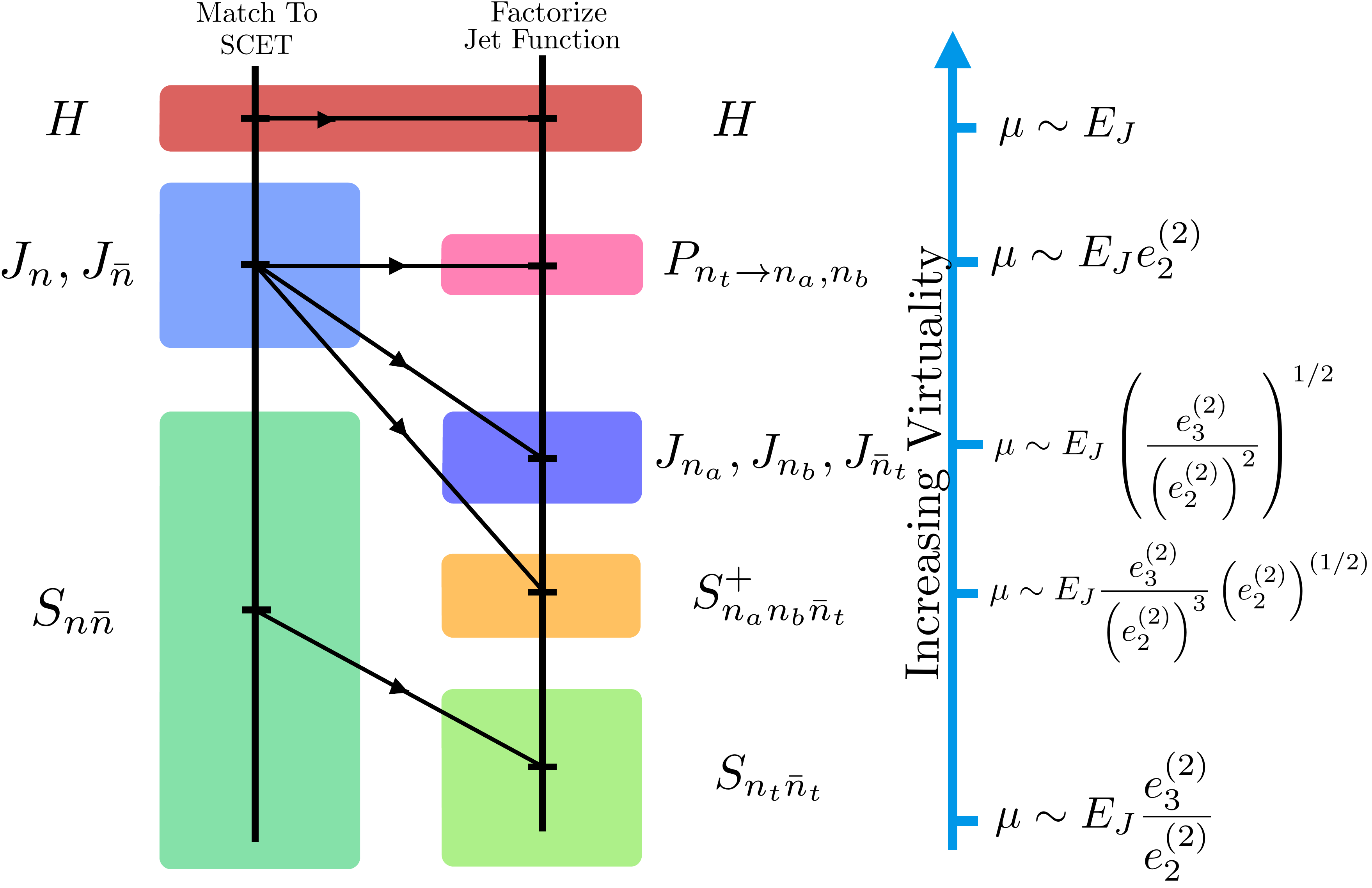}
}
\end{center}
\caption{A schematic depiction of the collinear subjets configuration with dominant QCD radiation and the functions describing its dynamics in the effective field theory is shown in a).   The matching procedure and relevant scales are shown in b), where we have restricted to the case $\alpha=\beta =2$ for simplicity.
}
\label{fig:collinear_subjets}
\end{figure}

The modes of SCET$_+$ are global soft modes, two collinear sectors describing the radiation in each of the collinear subjets, and collinear-soft modes from the dipole of the subjet splitting. These are shown schematically in \Fig{fig:collinear_subjets}. The additional collinear-soft modes, as compared with traditional SCET, are necessary to resum logarithms associated with the subjets' splitting angle. This angle, which is taken to be small, is not resolved by the long wavelength global soft modes.

The parametric scalings of the observables in the collinear subjets region were given in \Sec{sec:power_counting} and are:
\begin{align}\label{eq:collinear_subjets_scaling}
\ecf{2}{\alpha}&\sim \theta_{12}^\alpha \,,  \\
\ecf{2}{\beta}&\sim  \theta_{12}^\beta \,, \\
\ecf{3}{\alpha}&\sim
\theta_{12}^\alpha(\thetac^\alpha \theta_{12}^{\alpha}
+ \zs +\theta_{12}^{2\alpha}\zcs) \,.
\end{align}
Although the measurement of two 2-point energy correlation functions is required to be able to distinguish the soft and collinear subjets, they are redundant in the collinear subjets region from a power counting perspective, due to the relation $\ecf{2}{\beta}\sim \left ( \ecf{2}{\alpha} \right)^{\beta/\alpha}$.  We will therefore always write the scaling of the modes in terms of $\ecf{2}{\alpha}$ and $\ecf{3}{\alpha}$ to simplify expressions.

From \Eq{eq:collinear_subjets_scaling}, we see that $\ecf{2}{\alpha}$ sets the hard splitting scale, while the scalings of all the modes are set by the measurement of $\ecf{3}{\alpha}$. In particular, the scaling of the momenta of the collinear and soft modes are given by
\begin{align}\label{eq:cs_collinear_and_soft}
p_c &\sim E_J\left (\left (\frac{\ecf{3}{\alpha}}{ \left ( \ecf{2}{\alpha} \right )^{2}  }   \right )^{2/\alpha},1,   \left( \frac{\ecf{3}{\alpha}}{ \left ( \ecf{2}{\alpha} \right )^{2}  } \right)^{1/\alpha}   \right )_{n_a\bar n_a, n_b \bar n_b}\,, \\
\label{eq:soft_scale_coll}
p_s &\sim E_J \frac{\ecf{3}{\alpha}}{  \ecf{2}{\alpha}  } \left ( 1,1,1  \right )_{n\bar n} \,,
\end{align}
while the scaling of the collinear-soft mode is given by
\begin{align}\label{eq:cs_cs}
p_{cs} &\sim E_J \frac{\ecf{3}{\alpha}}{ \left ( \ecf{2}{\alpha} \right )^{3}  }    \left (  \left ( \ecf{2}{\alpha} \right)^{2/\alpha},1,    \left ( \ecf{2}{\alpha} \right)^{1/\alpha} \right )_{n\bar n}\,.
\end{align}
Here $E_J$ is the energy of the jet, and the subscripts denote the light-like directions with respect to which the momenta is decomposed.  In the expressions above, the momenta are written in the $(+,-,\perp)$ component basis with respect to the appropriate light-like directions. The subjet directions are labelled by $n_a$ and $n_b$, while the fat jet (containing the two subjets) and the recoiling jet are labelled by $n$ and $\bar n$. The relevant modes and a schematic depiction of the hierarchy of their virtualities is shown in \Fig{fig:collinear_subjets}.  

To have a valid soft and collinear expansion, the scalings of the modes in \Eqs{eq:cs_collinear_and_soft}{eq:cs_cs} imply that
\be
\ecf{2}{\alpha}\sim \left ( \ecf{2}{\beta} \right )^{\alpha/\beta}   \ll 1 \qquad \text{and} \qquad \frac{\ecf{3}{\alpha}}{ \left ( \ecf{2}{\alpha} \right )^{3}  } \sim \frac{\ecf{3}{\alpha}}{ \left ( \ecf{2}{\beta} \right )^{3\alpha/\beta}  } \ll 1 \,.
\ee
This agrees with the boundaries of the phase space found in \Sec{sec:power_counting}.

\subsubsection*{Factorization Theorem}\label{sec:fact_ninja}

In the collinear subjets region of phase space, the values of the 2-point energy correlation functions $\ecf{2}{\alpha}$ and $\ecf{2}{\beta}$ are set by the hard splitting. To leading power, these observables can be used to provide IRC safe definitions of the subjet energy fractions and the angle between the subjets. We therefore write the factorization theorem in terms of $\ecf{2}{\alpha}$, $\ecf{3}{\alpha}$, and the energy fraction of one of the subjets, which we denote by $z$. We further assume that an IRC safe observable, $B$, is measured in the out-of-jet region. Dependence on $B$ enters only into the out-of-jet jet function, and the out-of-jet contribution to the soft function.

The factorization theorem formulated in SCET$_+$ for the collinear subjets region of phase space is given by
\begin{align}\label{eq:NINJA_fact}
\frac{d^3\sigma}{dz\,d\ecf{2}{\alpha}d\ecf{3}{\alpha}}&=\sum_{f,f_a,f_b}H_{n\bar{n}}^{f}J_{\bar{n}}(B) P_{n_t\rightarrow n_a,n_b}^{f\rightarrow f_a f_b}\Big(z;\ecf{2}{\alpha}\Big)\int  de_{3}^{c}de_{3}^{\bar{c}}de_{3}^{s}de_{3}^{cs} \\
& \hspace{-0.5cm} \times\delta\Big(\ecf{3}{\alpha}-e_{3}^c-e_{3}^{\bar{c}}-e_{3}^{s}-e_{3}^{cs}\Big) 
 J_{n_a}^{f_a}\Big(z;e_{3}^{c}\Big)J_{n_b}^{f_b}\Big(1-z;e_{3}^{\bar{c}}\Big)S_{n \bar{n}}\Big(e_3^{s}, B;R\Big)S_{ n_a n_b \bar{n}}^+\Big(e_{3}^{cs}\Big) \,, \nonumber
\end{align}
where we have suppressed the convolution over the out-of-jet measurement, $B$, for simplicity.
Here the $n_a$, $n_b$ denote the collinear directions of the subjets, and we assume that $z\sim 1-z\sim \frac{1}{2}$.  The sum runs over all possible quark flavors that could be produced in an $e^+e^-$ collision. A brief description of the functions entering the factorization theorem of \Eq{eq:NINJA_fact} is as follows:

\begin{itemize}

\item $H_{n\bar{n}}^{f}$ is the hard function describing the underlying short distance process. In this case we consider $e^+e^- \to q \bar q$.

\item $P_{n\rightarrow n_a,n_b}^{f\rightarrow f_a f_b}\Big(z;\ecf{2}{\alpha}\Big)$ is the hard function arising from the matching for the hard splitting into subjets. In this case the partonic channel $f\to f_a f_b$ is restricted to $q\to q g$.

\item  $J_{n_a}^{f_a}\Big(z;e_{3}^{c}\Big)$, $J_{n_b}^{f_b}\Big(1-z;e_{3}^{\bar{c}}\Big)$ are jet functions describing the collinear dynamics of the subjets along the directions $n_a$, $n_b$.

\item  $S_{n \bar{n}}(e_3^{s},B;R)$ is the global soft function. The global soft modes do not resolve the subjet splitting, and are sensitive only to two eikonal lines in the $n$ and $\bar n$ directions. The soft function depends explicitly on the jet radius, $R$.

\item $S_{ n_a n_b \bar{n}}^+\Big(e_{3}^{cs} \Big)$ is the collinear-soft function. The collinear-soft modes resolve the subjet splitting, and hence the function depends on three eikonal lines, namely $n_a, n_b, \bar n$. Although these modes are soft, they are also boosted, and therefore do not resolve the jet boundary, so that the collinear soft function is independent of the jet radius, $R$.

\end{itemize}
This factorization theorem is shown schematically in \Fig{fig:collinear_subjets}, which highlights the radiation described by each of the functions in \Eq{eq:NINJA_fact}, as well as their virtuality scales. The two stage matching procedure onto the SCET$_+$ effective theory, which proceeds through a refactorization of the jet function, is also shown. The fact that the refactorization occurs in the jet function is important in that it implies that it is independent of the global color structure of the event, making it trivial to extend the factorization theorem to events with additional jets. This matching procedure is discussed in detail in \Ref{Bauer:2011uc}.

Operator definitions, and one-loop calculations for the operators appearing in the factorization theorem of \Eq{eq:NINJA_fact}  are given in \App{sec:ninja_app}.


\subsubsection{Soft Subjet}\label{sec:soft_jet}

\begin{figure}
\begin{center}
\subfloat[]{\label{fig:soft_subjets_a}
\includegraphics[width= 5.4cm]{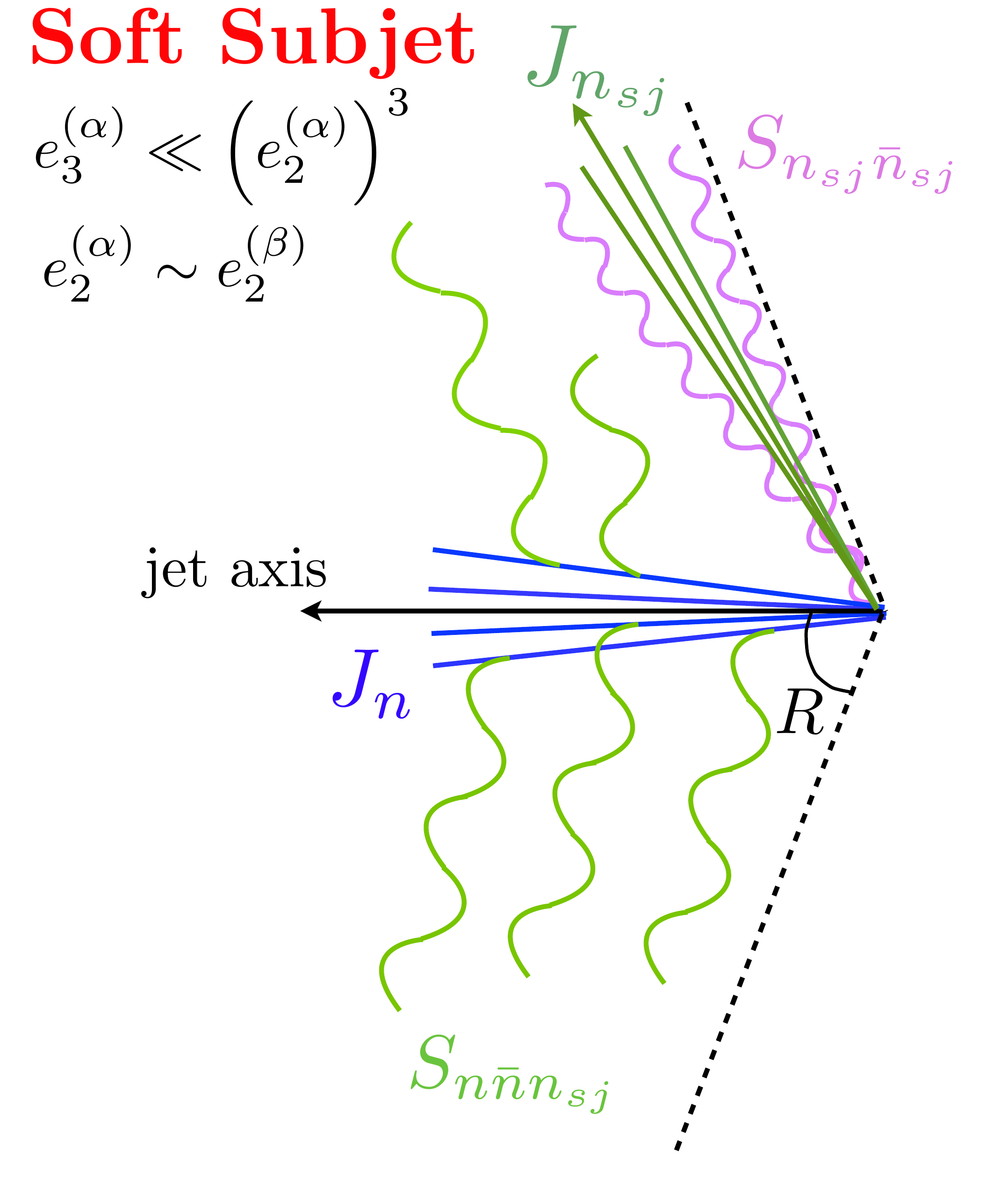}
}
\subfloat[]{\label{fig:soft_subjets_b}
\includegraphics[width =9.5cm]{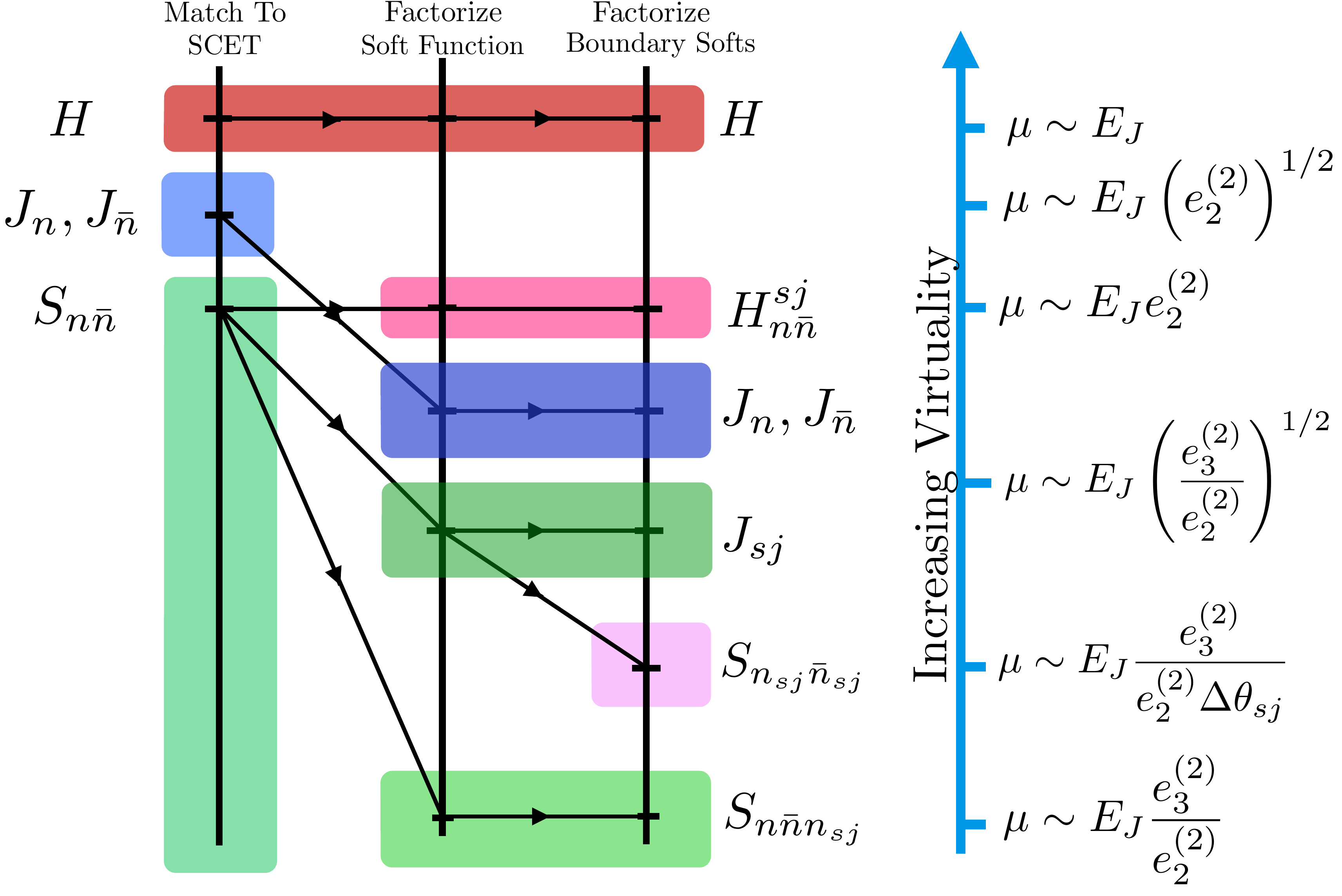}
}
\end{center}
\caption{ A schematic depiction of the soft subjet configuration with dominant QCD radiation and the functions describing its dynamics in the effective field theory is shown in a).   The matching procedure and relevant scales are shown in b), where we have restricted to the case $\alpha=\beta =2$ for simplicity.
}
\label{fig:soft_subjets}
\end{figure}

A factorization theorem describing the soft subjet region of phase space was recently presented in \Ref{Larkoski:2015zka}. In this section we review the basic features of this factorization theorem, but we refer the reader to \Ref{Larkoski:2015zka} for a more detailed discussion.

Unlike for the case of collinear subjets, in the soft subjet configuration, the wide angle soft subjet probes the boundary of the jet. This introduces sensitivity to the details of the jet algorithm used to define the jet, as well as to the measurement made in the region outside the jet. The factorization theorem of \Ref{Larkoski:2015zka} is valid under the assumption that an additive IRC safe observable, $B$, is measured in the out-of-jet region, and that the soft scale associated with this observable, $\Lambda$, satisfies $\Lambda/E_J \ll \ecf{2}{\alpha}$. We will therefore assume that this condition is satisfied throughout this section. However, we will see that the numerical results are fairly insensitive to the details of the choice of scale $\Lambda$. \Ref{Larkoski:2015zka} also used a broadening axis \cite{Larkoski:2014uqa} cone algorithm to define jets, whereas here we use the anti-$k_T$ algorithm, as relevant for phenomenological applications. We will argue that the structure of the factorization theorem is in fact identical in the two cases, to leading power.

\subsubsection*{Mode Structure}\label{sec:soft_jet_modes}

 In the soft subjet region of phase space there are two subjets with an energy hierarchy. We denote the energy of the soft subjet by $z_{sj}$ and the angle from the $n$ axis by $\theta_{sj}$. We also use the notation $\Delta \theta_{sj}=R-\theta_{sj}$ to denote the angle from the soft subjet axis to the jet boundary. The modes of the soft subjet are collinear-soft modes, being both soft and collimated, and we will therefore denote the characteristic angle between them as $\theta_{cs}$. Straightforward power counting can be applied to determine the scaling of the modes for both the energetic jet and the soft subjet. Their contributions to the observable are given by
\begin{align}\label{eq:soft_subjet_scaling}
\ecf{2}{\alpha} &\sim z_{sj}\,, \\ \ecf{2}{\beta} &\sim z_{sj}\,, \\ \ecf{3}{\alpha}&\sim z_{sj} (\theta_c^\alpha +z_{sj} \theta_{cs}^\alpha +z_s)\,.
\end{align}
In the soft subjet region of phase space, we have the relation $\ecf{2}{\alpha} \sim \ecf{2}{\beta}$, and therefore these two observables are redundant from a power counting perspective. We will therefore write the power counting of the modes in terms of $\ecf{2}{\alpha} $ and $\ecf{3}{\alpha} $.

From the contributions to the observables above, we find that the momentum of the collinear and global soft radiation scales like
\begin{align}
p_c &\sim E_J \left ( \left (  \frac{\ecfres}{  \ecfLa }  \right)^{2/\alpha},1,\left (  \frac{\ecfres}{  \ecfLa }  \right)^{1/\alpha}    \right )_{n\bar{n}}\,, \\
p_s &\sim  E_J \,\frac{\ecfres}{  \ecfLa } \left ( 1 ,1,1  \right )_{n\bar{n}}    \,,\nonumber 
\end{align}
where $E_J$ is the energy of the jet and $n$ and $\bar n$ are the light-like directions of the jet of interest and the other jet in the event, respectively.  The soft subjet mode's momentum scales like 
\begin{align}
p_{sj} \sim E_J\, \ecfLa \left ( \left (  \frac{\ecfres}{  \left (\ecfLa\right)^2 }  \right)^{2/\alpha},1,\left (  \frac{\ecfres}{  \left (\ecfLa\right )^2 }  \right)^{1/\alpha}    \right )_{\sja\sjabar}   \,,
\end{align}
in the light-cone coordinates defined by the direction of the soft subjet, $n_{sj}$. These are the complete set of modes defined by the scales set by the measurements of $\ecfLa ,\ecfLb$, and $\ecfres$ alone.

Unlike in the collinear subjet region of phase space there are no collinear-soft modes required in the effective field theory description, since the soft subjet is at a wide angle from the jet axis. However, in this region there is an additional mode, termed a \emph {boundary soft mode} in \Ref{Larkoski:2015zka}, whose appearance is forced by the jet boundary and the energy veto in the region of phase space outside the jet. These modes do not contribute to the $e_2$ observables, but are effectively a collinear-soft mode whose angle with respect to the soft subjet axis is set by the angle to the boundary.  The boundary soft mode's momentum components scale like
\begin{align}\label{eq:boundary_soft_scaling}
p_{bs} &\sim E_J\frac{  \ecfres   }{ \ecfLa   \left(\Delta \sjtheta\right)^\alpha }  \left ( \left(\Delta  \sjtheta\right)^2,1,\Delta  \sjtheta \right )_{\sja\sjabar}   \,, 
\end{align}
written in the light-cone coordinates defined by the soft subjet axis.
The boundary soft modes are required to have a single scale in the soft subjet function. For consistency of the factorization, we must enforce that the soft subjet modes cannot resolve the jet boundary and that the boundary soft modes are localized near the jet boundary.  That is, the angular size of the soft subjet modes, $\theta_{cs}$, must be parametrically smaller than that of the boundary soft modes, namely $\Delta \theta_{sj}$. We therefore find the condition
\begin{equation}\label{eq:applicability_softjet}
\left(\Delta\sjtheta\right)^\alpha \gg  \left(\theta_{cs}\right)^\alpha\sim \frac{\ecfres}{\left (\ecfLa \right )^2}\,, \qquad \text{and}\qquad \Delta\sjtheta \ll 1\,.
\end{equation}
Therefore, the factorization theorem applies in a region of the phase space where the soft subjet is becoming pinched against the boundary of the  jet, but lies far enough away that the collinear modes of the soft subjet do not touch the boundary. A schematic depiction of this region of phase space, along with a summary of all the relevant modes which appear in the factorization theorem is shown in \Fig{fig:soft_subjets}.

In the soft subjet region of phase space, the choice of jet algorithm plays a crucial role, since the soft subjet probes the boundary of the jet. In \Ref{Larkoski:2015zka} the factorization theorem in the soft subjet region of phase space was presented using a broadening axis cone algorithm with radius $R$. We now show that up to power corrections, the factorization theorem in the soft subjet region of phase space is identical with either the anti-$k_T$ or broadening axis cone algorithm. In particular, with the anti-$k_T$ algorithm, the jet boundary is not deformed by the soft subjet, and can be treated as a fixed cone of radius $R$.  This is not true for other jet algorithms, such as such as $k_T$ \cite{Catani:1993hr,Ellis:1993tq} or Cambridge-Aachen \cite{Dokshitzer:1997in,Wobisch:1998wt,Wobisch:2000dk}, where the boundary is deformed by the clustering of soft emissions, a point which has been emphasized elsewhere (see, e.g., \Refs{Appleby:2002ke,Banfi:2005gj,Banfi:2010pa,Kelley:2012kj}).

The validity of the factorization theorem requires the following two conditions, which will put constraints on the power counting in the soft subjet region of phase space. First, the soft subjet must be clustered with the jet axis, rather than with the out-of-jet radiation. This is guaranteed as long as the soft subjet axis satisfies $\theta_{sj} < R$. Second, the radiation clustered with the soft subjet from the out-of-jet region should not distort the boundary of the jet. More precisely, the distortion of the boundary must not modify the value of $\ecf{3}{\alpha}$ at leading power (note that the power counting guarantees that it does not modify $\ecf{2}{\alpha}$). The contribution to $\ecf{3}{\alpha}$ from a soft out-of-jet emission is given by
\be\label{eq:out_jet_estimate}
\ecf{2}{\alpha}   \frac{\Lambda}{E_J}  \ll \ecf{3}{\alpha} \implies  \frac{\Lambda}{E_J}  \ll \frac{  \ecf{3}{\alpha}  }{  \ecf{2}{\alpha}   }\sim\Big(\ecf{2}{\alpha}  \Big)^2\,.
\ee
Since the out-of-jet scale is in principle a free parameter, we can formally enforce this condition in our calculations. Corrections due to a deformation of the jet boundary would enter as power corrections in this region of phase space. The jet boundary therefore acts as a hard boundary of radius $R$, and the factorization theorem is identical to that presented in \Ref{Larkoski:2015zka}.

\subsubsection*{Factorization Theorem}\label{sec:fact_soft_jet}

With an understanding of the precise restrictions on the power counting required for the validity of the soft subjet factorization theorem, we now discuss its structure. Since we have argued that the relevant factorization theorem is identical to that presented in \Ref{Larkoski:2015zka}, we will only state the result.
The factorization theorem in the soft subjet region with the out-of-jet scale satisfying $\Lambda \ll \ecf{2}{\alpha} E_J$, and with jets defined by the anti-$k_T$ jet algorithm, is given by
\begin{align}\label{fact_inclusive_form_1}
&\frac{d\sigma(\outj;R)}{d\ecfLa d\ecfLb d\ecfres }= \\
&\hspace{0.6cm} \int dB_S dB_{J_{\bar n}} \int de_3^{J_n} de_3^{J_{{sj}}} de_3^S  de_3^{S_{sj}}        \delta (B-B_{J_{\bar n}}- B_S)  \delta(\ecfres-e_3^{J_n}-e_3^{J_{{sj}}}-e_3^S-e_3^{S_{sj}}) \nonumber \\
& \hspace{0.6cm}\times H_{n\bar n}(Q^2) H^{sj}_{n\bar{n}}\Big(\ecfLa,\ecfLb\Big)    J_{n}\left(e_3^{J_n}\right)J_{\bar{n}}(B_{J_{\bar n}}) 
 S_{n\bar{n}\sja }\Big(e_3^S;B_{S};R\Big) J_{\sja}\left(e_3^{J_{sj}} \right) S_{\sja\sjabar}\Big(e_3^{S_{sj}};R\Big)\,. \nonumber
\end{align}
In this expression we have explicitly indicated the dependence on the jet boundaries with the jet radius $R$. A brief description of the functions appearing in \Eq{fact_inclusive_form_1} is as follows:
\begin{itemize}
\item $H_{n\bar n}(Q^2)$ is the hard function describing the underlying short distance process. In this case we consider $e^+e^- \to q \bar q$.
\item $H_{n\bar{n}}^{sj}\Big(\ecfLa,\ecfLb\Big) $ is the hard function describing the production of the soft subjet coherently from the initial $q\bar{q}$ dipole, and describes dynamics at the scale set by $\ecfLa,\ecfLb$.
\item $J_{n}\Big(\ecfres\Big)$ is a jet function at the scale $\ecfres$ describing the hard collinear modes of the identified jet along the $n$ direction.
\item  $J_{\bar{n}}(\outj)$ is a jet function describing the collinear modes of the out-of-jet region of the event.
\item $S_{n\bar{n}\sja }\Big(\ecfres;\outj;R\Big)$ is the global soft function involving three Wilson line directions, $n, \bar n, \sja$. The global soft function depends explicitly on both the out-of-jet measurement and the jet radius.
\item $J_{\sja}\Big(\ecfres\Big)$ is a jet function describing the dynamics of the soft subjet modes, which carry the bulk of the energy in the soft subjet.
\item $S_{\sja\sjabar}(\ecfres;R)$ is a soft function describing the dynamics of the boundary soft modes. It depends only on two Wilson line directions $\sja, \sjabar$.
\end{itemize}
These functions, and a schematic depiction of the radiation which they define, are indicated in \Fig{fig:soft_subjets}, along with a schematic depiction of the multistage matching procedure from QCD onto the effective theory, as described in detail in \Ref{Larkoski:2015zka}. Although we will not discuss any details of the matching procedure, it is important to note that it occurs through a refactorization of the soft function, and hence the soft subjet factorization theorem is sensitive to the global color structure of the event, since the soft subjet is emitted coherently from all eikonal lines. This should be contrasted with the case of the collinear subjets factorization theorem, where the matching occurs through a refactorization of the jet function.

In the soft subjet region of phase space, we can relate the variables $\ecfLa\,, \ecfLb$ to the physically more transparent $\sje, \sjtheta$ variables with a simple Jacobian factor, giving the factorization theorem
\begin{align}\label{fact_inclusive_form_2}
&\frac{d\sigma(\outj;R)}{d\sje\, d\sjtheta\, d\ecfres}=\\
&\hspace{0.6cm} \int dB_S dB_{J_{\bar n}} \int de_3^{J_n} de_3^{J_{{sj}}} de_3^S  de_3^{S_{sj}}        \delta (B-B_{J_{\bar n}}- B_S)  \delta(\ecfres-e_3^{J_n}-e_3^{J_{{sj}}}-e_3^S-e_3^{S_{sj}}) \nonumber \\
& \hspace{0.6cm}\times H_{n\bar n}(Q^2) H^{sj}_{n\bar{n}}\left(z_{sj}, \theta_{sj}\right)    J_{n}\left(e_3^{J_n}\right)J_{\bar{n}}(B_{J_{\bar n}}) 
 S_{n\bar{n}\sja }\Big(e_3^S;B_{S};R\Big) J_{\sja}\left(e_3^{J_{sj}} \right) S_{\sja\sjabar}(e_3^{S_{sj}};R)\,. \nonumber
\end{align} 

Operator definitions, and one-loop calculations for the operators appearing in the factorization theorem of \Eqs{fact_inclusive_form_1}{fact_inclusive_form_2}  are given in \App{sec:softjet_app}.

\subsubsection{Soft Haze}\label{sec:soft_haze}

\begin{figure}
\begin{center}
\subfloat[]{\label{fig:soft_hazes_a}
\includegraphics[width= 5cm]{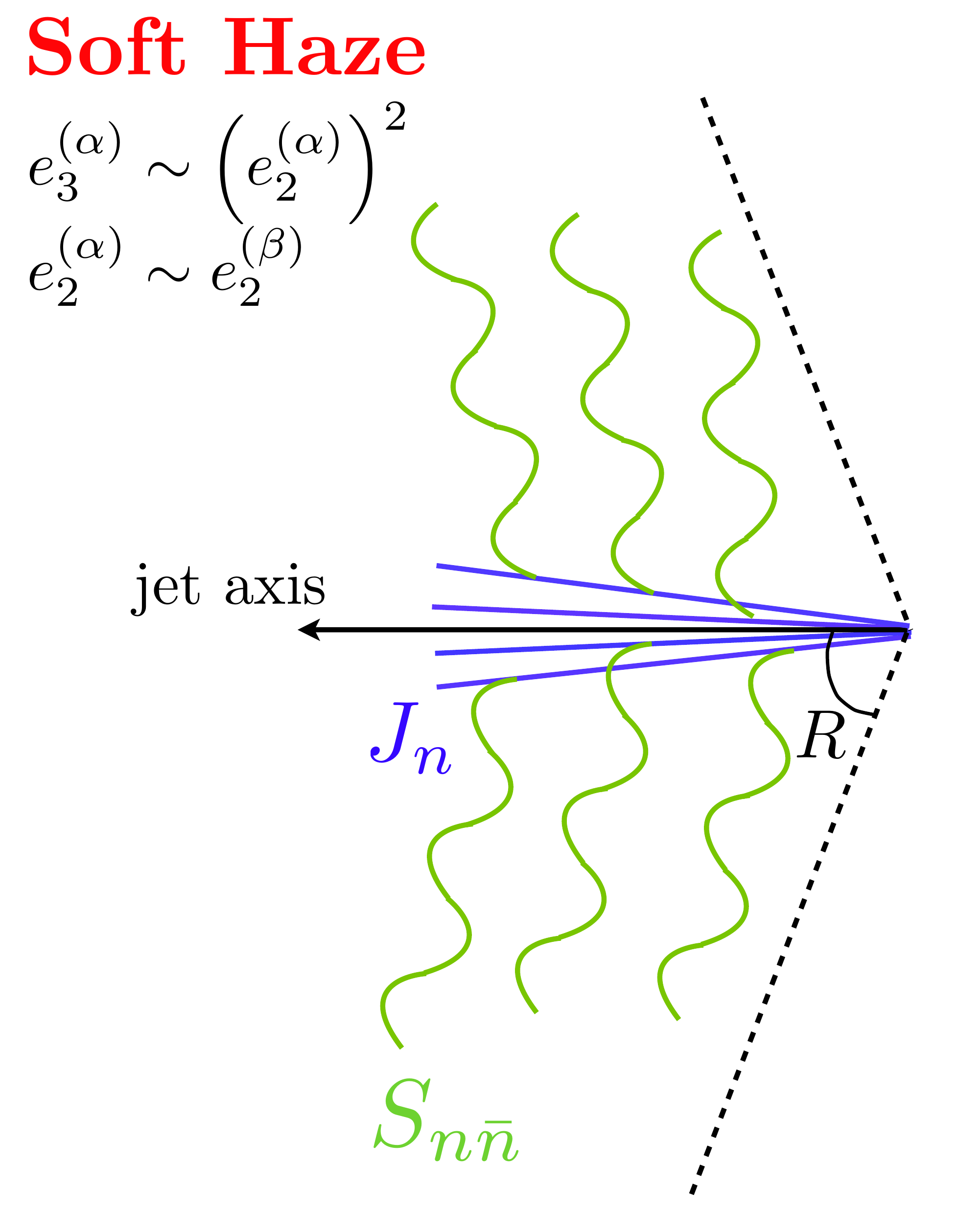}
}
\subfloat[]{\label{fig:soft_hazes_b}
\includegraphics[width = 10cm]{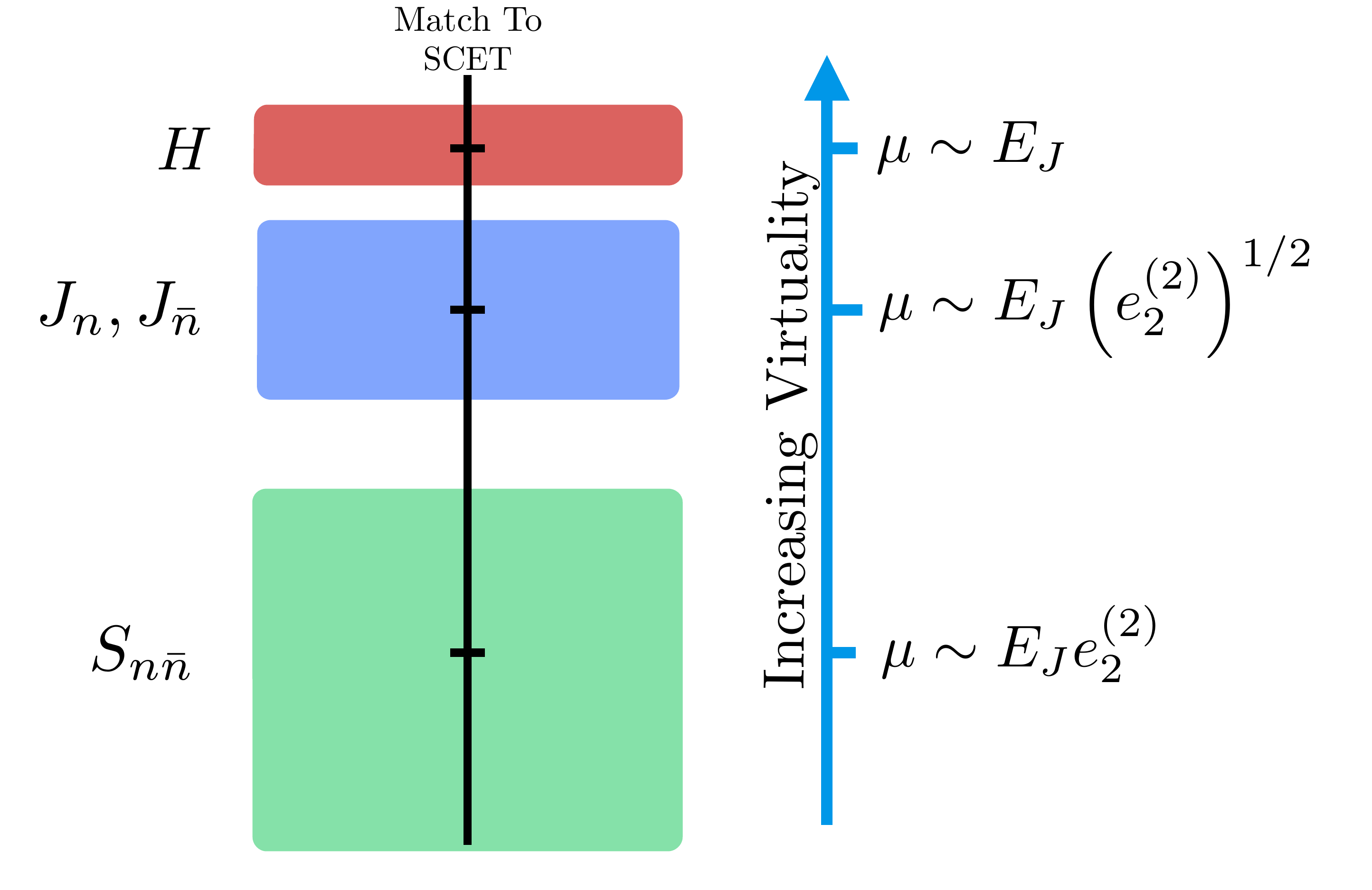}
}
\end{center}
\caption{A schematic depiction of the soft haze configuration where no subjets are resolved, with dominant QCD radiation and the functions describing its dynamics in the effective field theory is shown in a).   The relevant scales in the effective field theory are shown in b), where we have restricted to the case $\alpha=\beta =2$ for simplicity.
}
\label{fig:soft_hazes}
\end{figure}

The soft haze region defines the upper boundary of the $\ecf{2}{\beta},\ecf{3}{\alpha}$ phase space. In this region of phase space  jets consist of a single hard core, with no resolved subjets. A factorization theorem describing this region of phase space has not been presented elsewhere, but can be straightforwardly formulated in standard SCET involving only $n$ and $\bar n$ collinear sectors.

As discussed in \Sec{sec:power_counting}, the power counting in the soft haze region depends sensitively on the relative values of $\alpha$ and $\beta$, and therefore so does the structure of the factorization theorem. Since, from \Eq{eq:final_constraint_exponents}, we restrict ourself to $\alpha \geq \beta$, we will for simplicity only discuss the factorization theorems valid in this case. Factorization theorems for other values of $\alpha$ and $\beta$ can be determined by performing a similar analysis.

\subsubsection*{Mode Structure}\label{sec:soft_modes}

In the soft haze region the observables have the power counting
\begin{align}
\ecf{2}{\alpha}&\sim z_s+ \theta^\alpha_c\,, \\ 
\ecf{2}{\beta}&\sim z_s+ \theta^\beta_c\,, \\ 
\ecf{3}{\alpha}&\sim \zs^2 +\thetac^\alpha \zs+\thetac^{3\alpha}\,,
\end{align}
where we have not yet dropped power suppressed terms. We are interested in the factorization theorem on the upper boundary, with the scaling $\ecf{3}{\alpha} \sim \left (\ecf{2}{\beta} \right )^{2}$.\footnote{There is another parametric choice for the relative scaling of the 2-point energy correlation functions \cite{Larkoski:2014tva}, though it does not extend to the upper boundary of the phase space.  If $(\ecf{2}{\alpha})^\beta\sim (\ecf{2}{\beta})^\alpha$, then the power counting is
\begin{align*}
\ecf{2}{\alpha}&\sim z_s+ \theta^\alpha_c\,, \\ 
\ecf{2}{\beta}&\sim \theta^\beta_c\,, \\ 
\ecf{3}{\alpha}&\sim \zs^2 +\thetac^\alpha \zs\,,
\end{align*}
with both 2-point correlation functions dominated by collinear physics.  For $\alpha>\beta$, this region has the scaling $\ecf{3}{\alpha} \sim \left (\ecf{2}{\beta} \right )^{2\alpha/\beta}$ which does not extend to the upper boundary.} We now assume $\alpha > \beta$. In this case, dropping power suppressed terms, the appropriate power counting is
\begin{align}
\ecf{2}{\alpha}&\sim z_s\,, \\ 
\ecf{2}{\beta}&\sim z_s+ \theta^\beta_c\,, \\ 
\ecf{3}{\alpha}&\sim \zs^2\,.
\end{align}
It is also interesting to consider the case $\alpha=\beta$ because in the soft haze region it is not necessary to measure two different 2-point energy correlation functions, unlike in the two-prong region of phase space.
In the case that $\alpha=\beta$, we have instead,
\begin{align}\label{eq:power_count_softhaze_alpha}
\ecf{2}{\alpha}&\sim z_s+ \theta^\alpha_c\,, \\ 
\ecf{3}{\alpha}&\sim \zs^2 +\thetac^\alpha \zs\,,
\end{align}
where the second term in the expression for $\ecf{3}{\alpha}$ is no longer power suppressed. This will modify the factorization theorem between the two cases.

In both cases, the scaling of the modes is then given by
\begin{align}
p_c &\sim E_J \left (\left (\ecf{2}{\beta}\right )^{2/\beta}, 1, \left (\ecf{2}{\beta}\right )^{1/\beta}\right )_{n \bar n}, \\ p_s &\sim \ecf{2}{\beta}E_J \left (1,1,1\right)_{n \bar n},
\end{align}
with $\beta=\alpha$ in the second case.
Here $E_J$ is the energy of the jet and the subscripts denote the light-like directions with respect to which the momenta is decomposed.
This scaling should be recognized as the usual power counting of the collinear and soft modes for the angularities with angular exponent $\beta$ \cite{Ellis:2010rwa,Larkoski:2014tva}.

\subsubsection*{Factorization Theorem}\label{sec:fact_soft_haze}

The factorization theorem in the soft haze region of phase space can now be straightforwardly read off from the power counting expressions of the previous sections.  We state it both for the case $\alpha=\beta$ and $\alpha > \beta$. For $\alpha>\beta$, we have
\begin{align} \label{eq:fact_soft_haze}
\hspace{-0.5cm}\frac{d\sigma}{d\ecf{2}{\alpha}d\ecf{2}{\beta}d\ecf{3}{\alpha}}&=H_{n\bar{n}}(Q^2)    J_{\bar{n}}(B)\int de_2^{c}de_2^{s}    \delta\left(\ecf{2}{\beta}-  e_2^c  -e_2^{s}\right)  J_{n}\left(e_{2}^{c}\right)S_{n \bar{n}}\left(e_2^{s},\ecf{2}{\alpha},\ecf{3}{\alpha},R,\outj\right)\,,
\end{align}
where we have suppressed the convolution over the out-of-jet measurement $B$, to focus on the structure of the in-jet measurements.
For $\alpha=\beta$, the factorization theorem takes an interesting form\footnote{When calculating the tail of the $D_2$ distribution, one might be tempted to marginalize over $\ecf{2}{\beta}$ in \Eq{eq:fact_soft_haze}. This na\"ive marginalization does not yield the correct result. Rather, if one started the derivation of the factorization theorem with only the measurements of $\ecf{2}{\alpha}$ and  $\ecf{3}{\alpha}$ imposed, so that all possible $\ecf{2}{\beta}$ configurations are integrated over, then \Eq{eq:fact_soft_haze2} would be obtained. Thus \Eq{eq:fact_soft_haze2} is the correct marginalization over $\ecf{2}{\beta}$ in \Eq{eq:fact_soft_haze}.}
\begin{align} \label{eq:fact_soft_haze2}
\frac{d\sigma}{d\ecf{2}{\alpha}d\ecf{3}{\alpha}}&=H_{n\bar{n}}(Q^2)    J_{\bar{n}} (B)\int de_2^{c}de_2^{s} de_3^{s}   \delta\left(\ecf{2}{\alpha}-e_2^{c}-e_2^{s}\right) \delta \left ( \ecf{3}{\alpha}- e_2^c \,e_2^{'s}-e_3^{s}  \right)    \\  
&\hspace{8cm} \times J_{n}\left(e_{2}^{c}\right)S_{n \bar{n}}\left(e_2^{s},e_2^{'s},e_3^{s},R,\outj\right)\,, \nonumber
\end{align}
where again the convolution over $B$ has been suppressed.
A brief description of the functions appearing in the factorization theorems is as follows:
\begin{itemize}

\item $H_{n\bar n}\left(Q^2\right)$  is the hard function describing the underlying short distance process. In this case we consider $e^+e^- \to q \bar q$.

\item $J_{\bar{n}} (B)$ is the jet function describing the collinear modes for the recoiling jet.

\item $J_{n}\left(e_{2}^{c} \right)$ is the jet function describing the collinear modes for the jet in the $n$ direction.

\item $S_{n \bar{n}}\left(e_2^{s},e_{2}^{'s},e_3^{s},R,\outj\right)$ and $S_{n \bar{n}}\left(e_2^{s},\ecf{2}{\alpha},\ecf{3}{\alpha},R,\outj\right)$ are soft functions describing the global soft radiation from the $n \bar n$ dipole.  In the case of $\alpha=\beta$, an additional energy correlation, $e_{2}^{'s}$, is measured on the soft radiation, with an angular factor of $2\alpha$, which multiplies against the collinear contribution to $\ecf{2}{\alpha}$ when contributing to $\ecf{3}{\alpha}$. These also carry the jet algorithm constraints denoted by $R$, and any out-of-jet measurements $\outj$.

\end{itemize}
These functions, and a schematic depiction of the radiation which they define are indicated in \Fig{fig:soft_hazes}.  In \App{app:softhaze}, we give operator definitions of these functions and the leading-power expression for the $\ecf{3}{\alpha}$ measurement operator in the soft function.

There are several interesting features about the factorization theorems of \Eqs{eq:fact_soft_haze}{eq:fact_soft_haze2}. First, the soft functions are multi-differential, in that they require the simultaneous measurement of multiple quantities. Such multi-differential jet and soft functions have been discussed in detail in \Ref{Larkoski:2014tva,Procura:2014cba}. One other interesting feature of the factorization theorem of \Eq{eq:fact_soft_haze2}, for the case of equal angular exponents, is the appearance of the product structure in the $\delta$-function defining the value of $\ecf{3}{\alpha}$. This product structure follows from the power counting of \Eq{eq:power_count_softhaze_alpha} which describes the properties of the 3-point energy correlation function in the soft and collinear limits. It is important to note that this product form does not violate soft-collinear factorization, since only the knowledge of the total $\ecf{2}{\alpha}$ of the soft or collinear sector is required.

The soft contribution to the 3-point energy correlation is first non-vanishing with two real emissions. Therefore at one-loop, the factorization theorem of \Eq{eq:fact_soft_haze} reduces exactly to the factorization theorem for the multi-differential angularities studied in \Refs{Larkoski:2014tva,Procura:2014cba}, whereas the factorization theorem of \Eq{eq:fact_soft_haze2} reduces to the factorization theorem for a single angularity. In this paper, we will not perform the two-loop calculation necessary to obtain a non-trivial contribution to the three point energy correlation function. Instead, we will obtain an approximation to the cross section in this region by taking a limit of our factorization theorems in the two-prong region of phase space.  This is possible, because as we will show in \Sec{sec:fixed_order} by studying the fixed order distributions for the observable $D_2$, there is no fixed order singularity in the soft haze region of phase space in the presence of a mass cut. This implies that the resummation is not needed to regulate a fixed order singularity. This will be discussed in \Sec{sec:soft_haze_match}. The field theoretic definitions of the functions appearing in the factorization theorem of  \Eq{eq:fact_soft_haze} as well as power expansions of the measurement operators are collected in \App{app:softhaze}. However, because of the fact that we do not explicitly use the results of the soft haze factorization theorem in our calculation, we simply refer the reader to \Refs{Larkoski:2014tva,Procura:2014cba} for the calculations of the one-loop functions relevant to the factorization theorems of \Eqs{eq:fact_soft_haze}{eq:fact_soft_haze2}, and leave for future work the full two-loop calculation.

\subsubsection{Refactorization of the Global Soft Function}\label{sec:refac_soft}

In each of the factorization theorems required for the description of QCD background jets, namely the collinear subjets, soft subjet, and soft haze factorization theorems, there is a global soft function, which is sensitive to both the in-jet measurement of the energy correlation functions,  as well as the out-of-jet measurement $B$. To ensure that all large logarithms are resummed by the renormalization group evolution, we must perform a refactorization of the soft function \cite{Fleming:2007xt,Ellis:2010rwa,Kelley:2011aa,Chien:2012ur,Jouttenus:2013hs}. This ensures that the only logarithms which appear in a given soft function that are sensitive to both in-jet and out-of-jet scales are true non-global logarithms (NGLs) \cite{Dasgupta:2001sh}, which first appear at two-loop order in the calculation of a particular soft function.\footnote{It is important to emphasize that throughout this section we refer to the NGLs which appear in the soft function of a given factorization theorem, and the order in $\alpha_s$ at which they will appear in this particular soft function. Because we combine distinct factorization theorems, some of which include hard splitting functions, or eikonal emission functions, this order is in general distinct from the order at which they will appear in the total cross section, which can be different for each factorization theorem. This combination of the factorization theorems is completely independent from the refactorization of the soft function in a particular factorization theorem.}  Here we focus on the refactorization of the soft subjet and collinear subjets factorization theorems of \Secs{sec:ninja}{sec:soft_jet}, which will be used in our numerical calculation. For both of these factorization theorems, we can write the soft function to all orders in $\alpha_s$ as
\begin{equation}
S\left(   \ecf{3}{\alpha}, B;R,\mu\right)=S^{(\text{out})}  \left( B;R,\mu  \vphantom{ \ecf{3}{\alpha}}  \right)S^{(\text{in})}\left(   \ecf{3}{\alpha};R,\mu\right)S_{\text{NGL}}\left(   \ecf{3}{\alpha}, B;R\right)\,,
\end{equation}
where we have explicitly indicated the renormalization scale $\mu$ dependence \cite{Hornig:2011tg}.  The non-global part of the soft function $S_{\text{NGL}}\left(   \ecf{3}{\alpha}, B;R\right)$ is first non-trivial at two-loop order, beyond the accuracy to which we explicitly calculated the soft functions in this paper. Furthermore, the anomalous dimension of the soft function factorizes to all orders in perturbation theory as
\begin{align}
\gamma_{S}  \left(   \ecf{3}{\alpha}, B;R;\mu \right)=\gamma^{(\text{out})}_{S}  \left( B;R;\mu  \vphantom{ \ecf{3}{\alpha}} \right)   +\gamma^{(\text{in})}_{S}  \left(   \ecf{3}{\alpha};R;\mu \right)  \,,
\end{align}
and therefore the renormalization group kernels factorize as well. Briefly, this occurs because renormalization group consistency relates the soft anomalous dimension to the sum of all the other anomalous dimensions, each of which can be associated with the in-jet or out-of-jet contributions.\footnote{As discussed in \Ref{Jouttenus:2013hs} there is some ambiguity in how the hard function, for example, is associated with the in-jet or out-of-jet anomalous dimensions, but this does not affect the above argument.}

 While similar refactorizations of the global soft function have been discussed previously, and used in numerical calculations (see especially \Ref{Jouttenus:2013hs} for a detailed discussion), we will discuss it here for completeness.  The refactorization of the global soft function plays a role in our numerical results and is particularly important in appropriately separating scales in the global soft function of the soft subjet factorization theorem of \Sec{sec:soft_jet}.  In \Ref{Larkoski:2015zka} the structure of the one-loop calculation of the soft subjet factorization theorem was discussed in detail, with a particular focus on the dependence on the angle $\Delta \theta_{sj}$ between the soft subjet and the boundary. There it was found that the while the out-of-jet soft function contained dependence on the angle between the soft subjet and the boundary, $\Delta \theta_{sj}$, this dependence vanishes in the in-jet contribution to the soft function due to a zero bin subtraction. Renormalization group consistency is achieved since the $\Delta \theta_{sj}$ dependence associated with the in-jet region is carried by  the boundary soft function. Therefore, the refactorization of the global soft function for the soft subjet factorization theorem allows the soft function to be separated into a piece with $\Delta \theta_{sj}$ dependence, and a piece with no $\Delta \theta_{sj}$ dependence, and is crucial for resumming all large logarithms associated with this scale.  The one-loop anomalous dimensions, split into out-of-jet and in-jet contributions, as well as canonical scales for both the in-jet and out-of-jet soft functions are given in \App{sec:ninja_app}, \App{sec:softjet_app}, and \App{sec:soft_subjet_cbin}. Further details of this refactorization, and in particular a discussion on the dependence on $\Delta \theta_{sj}$ is also given.

For completeness, we also give the final refactorized expressions for the factorization theorems for the collinear subjets and soft subjet factorization theorems that will be used when presenting numerical results. For the collinear subjets factorization theorem, we have 
 \begin{align}\label{eq:NINJA_fact_refac}
\frac{d^3\sigma}{dz\,d\ecf{2}{\alpha}d\ecf{3}{\alpha}}&=\sum_{f,f_a,f_b}H_{n\bar{n}}^{f}(Q^2)     P_{n_t\rightarrow n_a,n_b}^{f\rightarrow f_a f_b}\Big(z;\ecf{2}{\alpha}\Big)    \int dB_S dB_{J_{\bar n}}        \int  de_{3}^{c}de_{3}^{\bar{c}}de_{3}^{s}de_{3}^{cs} \\
& \hspace{-0.5cm} \times     \delta (B-B_{J_{\bar n}}- B_S)    \delta\Big(\ecf{3}{\alpha}-e_{3}^c-e_{3}^{\bar{c}}-e_{3}^{s}-e_{3}^{cs}\Big)     \nonumber\\
& \hspace{-0.5cm} \times J_{\bar{n}}(B_{J_{\bar n}} )  J_{n_a}^{f_a}\Big(z;e_{3}^{c}\Big)J_{n_b}^{f_b}\Big(1-z;e_{3}^{\bar{c}}\Big)    S^{(\text{out})}_{n \bar{n}}\Big(B_S;R\Big)   S^{(\text{in})}_{n \bar{n}}\Big(e_3^{s};R\Big)    S_{ n_a n_b \bar{n}}^+\Big(e_{3}^{cs}\Big) \,, \nonumber
\end{align}
 while for the soft subjet factorization theorem, we have
 \begin{align}\label{fact_inclusive_form_2_refac}
\frac{d\sigma(\outj;R)}{d\sje\, d\sjtheta\, d\ecfres}&= H_{n\bar n}(Q^2) H^{sj}_{n\bar{n}}\left(z_{sj}, \theta_{sj}\right)   \int dB_S dB_{J_{\bar n}} \int de_3^{J_n} de_3^{J_{{sj}}} de_3^S  de_3^{S_{sj}}     \\
&\hspace{-0.5cm}  \times      \delta (B-B_{J_{\bar n}}- B_S)  \delta(\ecfres-e_3^{J_n}-e_3^{J_{{sj}}}-e_3^S-e_3^{S_{sj}}) \nonumber \\
& \hspace{-0.5cm}\times     J_{n}\left(e_3^{J_n}\right)J_{\bar{n}}(B_{J_{\bar n}}) 
 S^{(\text{out})}_{n\bar{n}\sja }\Big(B_{S};R\Big)     S^{(\text{in})}_{n\bar{n}\sja }\Big(e_3^S;R\Big)     J_{\sja}\left(e_3^{J_{sj}} \right) S_{\sja\sjabar}(e_3^{S_{sj}};R)\,. \nonumber
\end{align} 
In this form, each function in \Eqs{eq:NINJA_fact_refac}{fact_inclusive_form_2_refac} contains logarithms of a single scale, which can be resummed through renormalization group evolution.

\subsection{Boosted Boson Signal}\label{sec:signal_fact}

\begin{figure}
\begin{center}
\subfloat[]{\label{fig:boosted_boson_a}
\includegraphics[width=6.2cm]{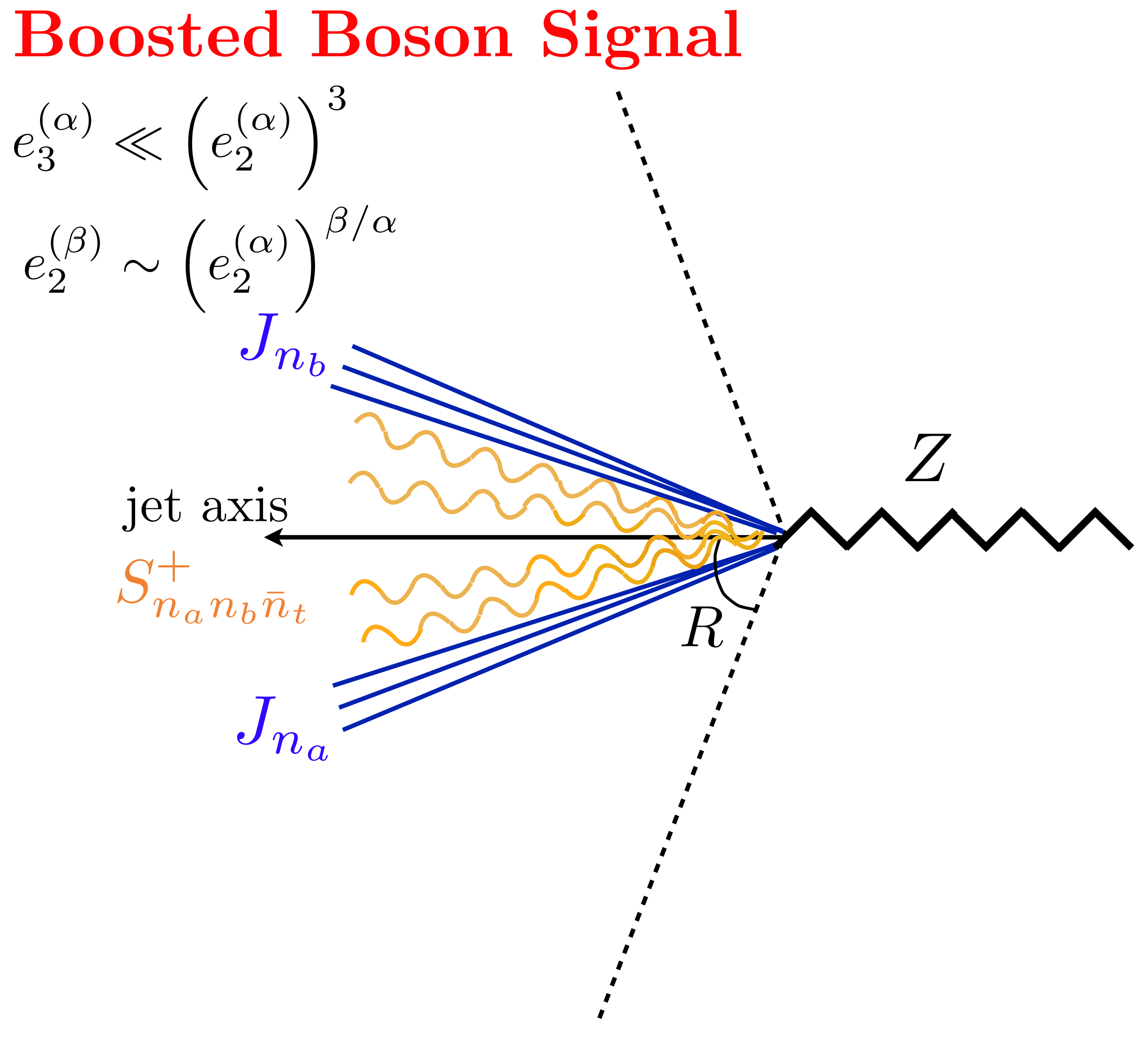}
}
\subfloat[]{\label{fig:boosted_boson_b}
\includegraphics[width = 8.5cm]{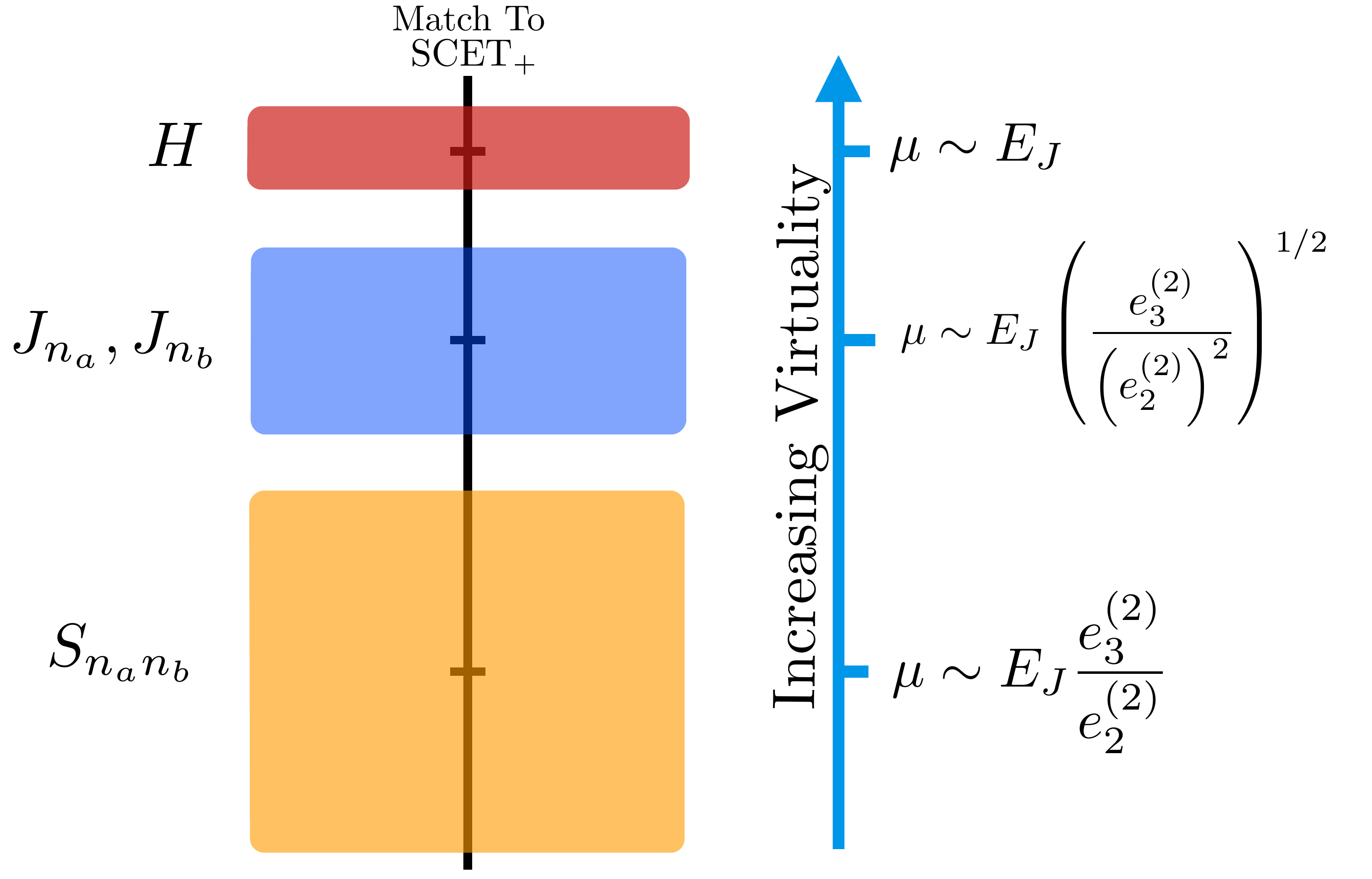}
}
\end{center}
\caption{A schematic depiction of the boosted $Z$ boson configuration with dominant QCD radiation and the functions describing its dynamics in the effective field theory is shown in a).  The relevant scales, ordered in virtuality, are summarized in b), where we have restricted to the case $\alpha=\beta =2$ for simplicity.
}
\label{fig:boosted_boson}
\end{figure}

In this section we discuss the effective field theory and factorization theorem relevant for the hadronically-decaying boosted boson signal. For concreteness, we will consider the case of a boosted $Z$ boson decaying to a massless $q\bar q$ pair; however, the extension to other color-neutral boosted particles is trivial. We will work in the narrow width approximation, setting the width of the $Z$ boson $\Gamma_Z=0$. Corrections to this approximation are trivial to implement, as they do not modify the structure of the factorization, and are expected to have a minimal effect.

A factorization theorem for the $N$-subjettiness observable $\Nsub{1,2}{\beta}$ \cite{Stewart:2010tn,Thaler:2010tr,Thaler:2011gf} measured on boosted $Z$ jets was presented in \Ref{Feige:2012vc}. This factorization theorem was obtained by boosting an appropriately chosen $e^+e^-$ event shape. A factorization theorem can also be formulated using the SCET$_+$ effective theory,\footnote{Here we have slightly extended the usage of the SCET$_+$ nomenclature beyond that which it was originally used in \Ref{Bauer:2011uc}. In particular, in the case of the signal distribution, there are no global soft modes, and the matching to the effective theory proceeds in quite a different way than for the case of a two prong QCD jet as originally considered in \Ref{Bauer:2011uc}. Nevertheless, because the effective theory contains a collinear-soft mode, we will refer to it as SCET$_+$. } where the collinear-soft mode, which was described in \Sec{sec:ninja}, corresponds to the boosted soft mode of the $e^+e^-$ event shape. We will take this second approach, as it is in line with the general spirit of this paper, of developing effective field theory descriptions of jet substructure configurations. However, the approach of relating to boosted $e^+e^-$ event shape variables is useful for relating results to higher order calculations known in the literature. Despite the fact that the factorization for the energy correlation functions in the signal region follows straightforwardly from that of \Ref{Feige:2012vc}, or from the SCET$_+$ factorization theorem of \Sec{sec:ninja},  we will discuss it here for completeness.

We assume the process $e^+e^-\to ZZ \to q \bar q l\bar l$, where $l$ is a lepton to avoid having to describe additional jets, although the extension to two hadronically-decaying $Z$ bosons is trivial. The factorization theorem is then similar to that presented in \Sec{sec:ninja}, however, there are no global soft modes since the $Z$ is a color singlet. The scaling of the collinear and collinear-soft modes are identical to those given in \Sec{sec:ninja}, so we do not repeat them here.  The factorization theorem is given by
\begin{align}\label{eq:signal_fact}
\frac{d\sigma}{dz\,d\ecf{2}{\alpha}d\ecf{3}{\alpha}}&=H(Q^2) P_{n\rightarrow n_a,n_b}^{Z\rightarrow q \bar q}\Big(z;\ecf{2}{\alpha}\Big)\int  de_{3}^{c}de_{3}^{\bar{c}}de_{3}^{s}de_{3}^{cs} \\
& \hspace{-0.5cm} \times \delta\Big(\ecf{3}{\alpha}-e_{3}^c-e_{3}^{\bar{c}}-e_{3}^{cs}\Big) 
 J_{n_a}^{q}\Big(z;e_{3}^{c}\Big)J_{n_b}^{\bar q}\Big(1-z;e_{3}^{\bar{c}}\Big)S_{ n_a n_b }^+\Big(e_{3}^{cs}\Big) \,. \nonumber
\end{align}
As with the factorization theorem in \Sec{sec:ninja}, we have chosen to write the factorization theorem in terms of $\ecf{2}{\alpha}$, $\ecf{3}{\alpha}$, and the energy fraction of one of the subjets, $z$. A brief description of the functions appearing in \Eq{eq:signal_fact} is as follows:
 \begin{itemize}
 \item $H(Q^2)$ is the hard function describing the production of the on-shell $Z$ bosons in an $e^+e^-$ collision. It also includes the leptonic decay of the $Z$ boson. Following \Ref{Feige:2012vc} we assume that the $Z$ boson is unpolarized and so its decay matrix element is flat in the cosine of the boost angle. Non-flat distributions corresponding to some particular decay or production mechanism are straighforward to include.
 \item $P_{n\rightarrow n_a,n_b}^{Z\rightarrow q\bar q}\Big(z;\ecf{2}{\alpha}\Big)$ describes the decay of the on-shell $Z$ boson into a $q\bar q$ pair with momenta along the $n_a$ and $n_b$ axes.  
 \item  $J_{n_a}^{q}\Big(z;e_{3}^{c}\Big)$, $J_{n_b}^{\bar q}\Big(1-z;e_{3}^{\bar{c}}\Big)$ are the jet functions describing the collinear radiation associated with the two collinear subjets.
 \item  $S_{ n_a n_b }^+\Big(e_{3}^{cs}\Big)$ is the collinear-soft function describing the radiation from the $q\bar q $ dipole formed by the two collinear subjets.
 \end{itemize}
The basic structure of the factorization theorem, and the radiation described by the different functions, as well as their scalings, are shown schematically in \Fig{fig:boosted_boson}. Operator definitions, and one-loop calculations for the operators appearing in the factorization theorem of \Eq{eq:signal_fact}  are given in \App{sec:signal_app}. Because the collinear soft modes are boosted, the collinear soft function does not require a refactorization, as was necessary for the global soft functions, in \Sec{sec:refac_soft}.

It is important to emphasize the distinction between our treatment of a boosted $Z$ jet, where we presented a single factorization theorem, and a massive QCD jet, where three distinct factorization theorems were required. While it is obvious that the soft haze region does not exist for a boosted $Z$ jet, the soft subjet region does. However, unlike the case of a massive QCD jet, where the soft subjet region is enhanced by a factor of $1/z_{sj}$ from the eikonal emission factor, no such enhancement exists for the $Z$ decay. Indeed, it was shown in \Ref{Feige:2012vc} that the effect of the jet boundary, which would arise from the soft subjet configuration, is power suppressed by $1/Q$. While it would be potentially interesting to analytically study the jet radius dependence for the signal distribution using the soft subjet factorization theorem, this is beyond the scope of this paper. We will therefore neglect jet radius effects and write the factorization theorem in \Eq{eq:signal_fact} with no $R$ dependence.  

The factorization theorem of \Eq{eq:signal_fact} provides an accurate description of the boosted boson signal in the two-prong region of phase space, where $\ecf{3}{\alpha} \ll \left ( \ecf{2}{\alpha} \right )^{3}$. However, to be able to compare the signal and background distributions, a valid description of the region $\ecf{3}{\alpha} \gtrsim \left ( \ecf{2}{\alpha} \right )^{3}$ is also required. Unlike for the case of a massive QCD jet, where this region is described by the soft haze factorization theorem, for a boosted $Z$ boson, an accurate description of this region requires matching to the fixed order $Z\to q\bar q g$ matrix element. Since the boost of the $Z$ boson is fixed, this corresponds to a hard gluon emission from the $q\bar q$ dipole. In the numerical results shown throughout the paper, we have performed this matching to fixed order, directly within the SCET$_+$ effective theory. The fixed order cross section for $\Dobs{2}{\alpha,\beta}$ onto which the result of the factorization theorem was matched, was calculated numerically by boosting the leading order $e^+e^-\to q\bar q g$ matrix element and performing a Monte Carlo integration. This allows for the consideration of general angular exponents $\alpha$ and $\beta$ in which case the required integrals are difficult, if not impossible, to evaluate analytically.

\section{A Factorization Friendly Two-Prong Discriminant}\label{sec:friendly}

The approach to two-prong discrimination taken in this paper is to use calculability and factorizability constraints to guide the construction of an observable. Having understood in detail the structure of the $\ecf{2}{\alpha}, \ecf{2}{\beta}, \ecf{3}{\alpha}$ phase space, along with the effective field theories describing each parametric region, we now show how a powerful two-prong discriminant, $D_2$, emerges from this analysis naturally.  After defining the $D_2$ observable, we discuss some of its interesting properties, and show that the factorization theorems of \Sec{sec:Fact} can be combined to give a factorized description of the observable over the entire phase space.

\subsection{Defining $D_2$}\label{sec:def_D2}

The goal of boosted boson discrimination is to define observables which distinguish between one- and two-prong jets. As a simplification, we will take the view that both collinear and soft subjets should be treated as two-pronged by the discriminant, while soft haze jets should be treated as one-pronged.  Treating both the collinear and soft subjets as two-pronged immediately implies that a marginalization over the soft subjet and collinear subjet factorization theorems will need to be performed to obtain a prediction for the two-prong discriminant. This will be discussed in \Sec{sec:merging}. 
A more sophisticated observable could take advantage of the different fraction of signal and QCD jets in the soft subjet and collinear subjets regions of phase space, and we will give a simple example of such an observable in \Sec{sec:2insight}. 

\begin{figure*}[t]
\begin{center}
\subfloat[]{\label{fig:D2_contours}
\includegraphics[width=6.5cm]{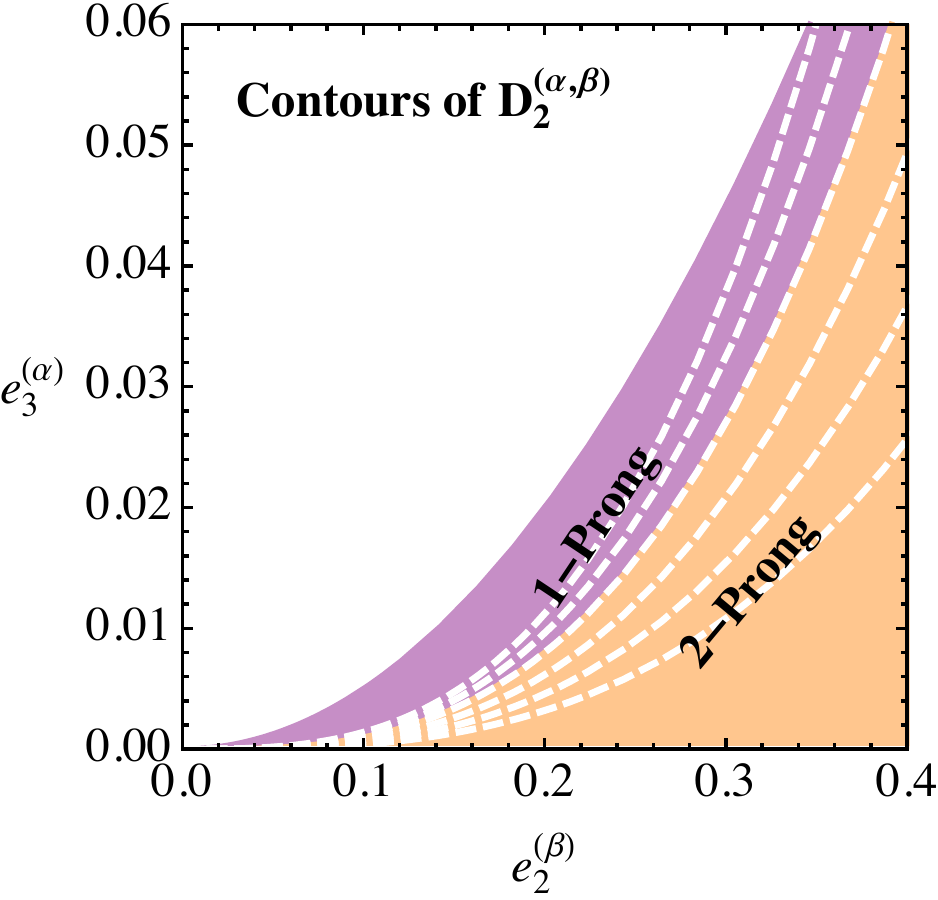}
}\ 
\subfloat[]{\label{fig:D2_plot}
\includegraphics[width=7.5cm,trim =0 0cm 0 0]{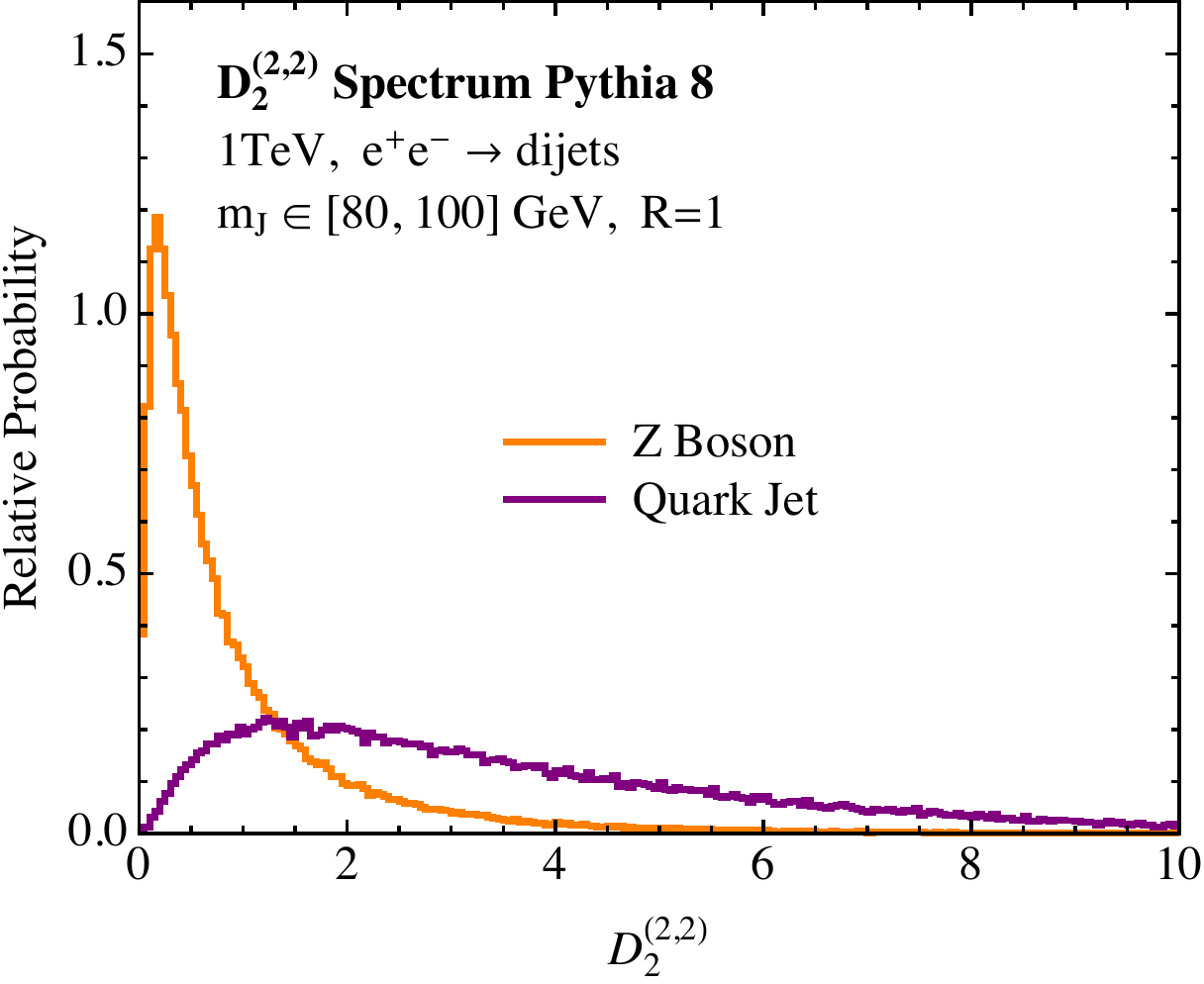}
}
\end{center}
\caption{ a) Contours of the observable $D_2$ in the $\ecf{2}{\beta},\ecf{3}{\alpha}$ plane. b) Sample $D_2$ spectra for boosted $Z$ bosons and QCD jets, generated in Monte Carlo. Angular exponents $\alpha=\beta =2$ have been used. 
}
\label{fig:C2vD2}
\end{figure*}

We will consider discriminants, which we denote $\Dobs{2}{\alpha, \beta}$, which parametrize a family of contours in the $\ecf{2}{\beta},\ecf{3}{\alpha}$ plane, as shown schematically in \Fig{fig:C2vD2}. Such observables can be calculated by marginalizing the double differential cross section \cite{Larkoski:2013paa}
\be \label{eq:marginalization}
\frac{d\sigma}{d\Dobs{2}{\alpha, \beta}}=\int d\ecf{2}{\beta} d\ecf{3}{\alpha} \delta \left(\Dobs{2}{\alpha, \beta}-\Dobs{2}{\alpha, \beta}(\ecf{2}{\beta},\ecf{3}{\alpha}) \right)   \frac{d^2 \sigma}{d\ecf{2}{\beta} d\ecf{3}{\alpha}}\,.
\ee
For the observable $\Dobs{2}{\alpha, \beta}$ to be calculable using the factorization theorems of \Sec{sec:Fact}, the curves over which the marginalization is performed in \Eq{eq:marginalization} must lie entirely in a region of phase space in which there is a description in terms of a single effective field theory (up to the marginalization over the collinear and soft subjets). Stated another way, the contours of $\Dobs{2}{\alpha, \beta}$ must lie either entirely in the one-prong region of phase space, or entirely in the two-prong region of phase space. This condition is also natural from the perspective that $\Dobs{2}{\alpha, \beta}$  provide good discrimination power, a point which has been emphasized in \Refs{Larkoski:2014gra,Larkoski:2014zma}. If the contours do not respect the parametric scalings of the phase space, the marginalization cannot be performed within a single effective field theory. A more sophisticated interpolation between the different effective field theories, along the lines of \Refs{Larkoski:2014tva,Procura:2014cba} is then required.

In \Sec{sec:phase_space}, a power counting analysis was used to show that for $3\alpha/\beta > 2$, the one- and two-prong regions of phase space are parametrically separated, with the contour separating them scaling as $\ecf{3}{\alpha} \sim \left(\ecf{2}{\beta}\right)^{3\alpha/\beta}$. This implies that, parametrically, the optimal two-prong discriminant formed from $\ecf{2}{\beta}$ and $\ecf{3}{\alpha}$ is
\begin{equation}\label{eq:D2_gen}
\Dobs{2}{\alpha, \beta}= \frac{\ecf{3}{\alpha}}{(\ecf{2}{\beta})^{3\alpha/\beta}}\,.
\end{equation}
This extends the definition of \Ref{Larkoski:2014gra}, which considered the observable $\Dobs{2}{\alpha, \alpha}$, with equal angular exponents.
To simplify our notation, we will often not explicitly write the angular exponents $\alpha$ and $\beta$, referring to the observable simply as $D_2$.

The $D_2$ observable takes small values for a two-prong jet and large values for a one-prong jet.  Its contours in the $\ecf{2}{\beta},\ecf{3}{\alpha}$ phase space are shown schematically in \Fig{fig:C2vD2}, along with illustrative Monte Carlo generated spectra for both boosted $Z$ jets and massive QCD jets in $e^+e^-$ collisions. A more detailed discussion of the discrimination power of $D_2$, as well as the details of the Monte Carlo generation, will be given in \Sec{sec:results}.

\subsection{Sudakov Safety of $D_2$}\label{sec:sudsafe}

One interesting feature of the $D_2$ observable is that it is not IRC safe without an explicit cut on $\ecf{2}{\beta}$. For every value of $D_2$, the contour over which the double differential cross section is marginalized passes through the origin of the phase space, where the soft and collinear singularities are located. This feature is shown in \Fig{fig:D2_contours}. At every fixed order in perturbation theory, this gives rise to an ill-defined (divergent) cross section. However, a resummed calculation of the double differential cross section regularizes the singular region of phase space, and leads to a finite distribution for the $D_2$ observable. This property is referred to as Sudakov safety \cite{Larkoski:2013paa,Larkoski:2015lea}.  Because Sudakov safe observables are not calculable in fixed order perturbation theory, they do not generically have an $\alpha_s$ expansion, and we will show that the $D_2$ spectrum exhibits a particularly interesting dependence on $\alpha_s$.

The regularization of the fixed order singularity in the double differential cross section is achieved by the all orders resummation of logarithmically enhanced terms in the perturbative expansion. In the effective field theory description, this resummation is achieved by renormalization group evolution, and its properties are therefore determined by the form of the SCET anomalous dimensions. To illustrate how the $\alpha_s$ dependence arises from the structure of the renormalization group evolution in SCET, we consider the soft subjet factorization theorem of \Sec{sec:soft_jet} in the leading logarithmic (LL) approximation. The cusp pieces of the anomalous dimensions for the different functions appearing in the factorization are given in Laplace space by (see \App{sec:softjet_app})  
\begin{align}
\mu\frac{d}{d\mu}\log\, H^{sj}_{n\bar{n}}(\sje,\sja,\mu) &=-\frac{\alpha_s C_A}{\pi}\log \left[\frac{\mu^2}{Q^2} z_{sj}^{-2}\right]\,,\\
\mu\frac{d}{d\mu}\log\, J_{\sja }\Big(\eeclp{3}{\alpha}\Big) &=-2\frac{\alpha_s C_A}{\pi(1-\alpha)}\log \left[\eeclp{3}{\alpha}\frac{\mu^\alpha}{Q^\alpha}  z_{sj}^{2-\alpha}     \right] \,,\\
\mu\frac{d}{d\mu}\log\,  S_{\sja \,\sjabar }\Big(\eeclp{3}{\alpha};R\Big)&=\frac{\alpha_s C_A}{\pi(1-\alpha)}\log \left[\eeclp{3}{\alpha}\frac{\mu}{Q} z_{sj}\right]\,,\\
\mu\frac{d}{d\mu}\log\,  S_{\sja \,n\,\bar{n}}\Big(\eeclp{3}{\alpha},B;R\Big)&=\frac{\alpha_s C_A}{\pi(1-\alpha)}\log \left[\eeclp{3}{\alpha}\frac{\mu}{Q} z_{sj}\right] \,,
\end{align}
where we have used $\eeclp{3}{\alpha}$ to denote the Laplace conjugate to $\ecf{3}{\alpha}$, and we have kept only IR scales in the logs. Furthermore, we have kept only the terms proportional to $C_A$ so as to resum only the physics associated with the soft subjet.
The hard matching coefficient for the soft subjet production is given by the tree level eikonal emission factor
\begin{align}
H^{sj(\text{tree})}_{n\bar{n}}(\sje,\sja)&=\frac{\alpha_s C_F}{\pi\sje}\frac{n\cdot \bar{n}}{n\cdot\sja\,\sja\cdot\bar{n}}\,.
\end{align}
Solving the renormalization group equations, and running all functions to the hard scale $Q$, we then find that in the soft subjet region of phase space the multi-differential cross section can be written to LL accuracy as
\begin{align}\label{eq:LL_sudakov}
&\frac{d\sigma}{d\ecf{3}{\alpha}    dz_{sj} d\theta_{sj}  }= - \frac{\alpha_s^2 C_F C_A}{\alpha\pi^2}\frac{4}{n\cdot\sja\,\sja\cdot\bar{n}}\frac{\log \left[\ecf{3}{\alpha}z_{sj}^{-2}\right]}{\sje\ecf{3}{\alpha}} e^{-\frac{\alpha_s}{\pi}\frac{C_A}{\alpha}\log^2\left[
\ecf{3}{\alpha}z_{sj}^{-2}
\right]}\,,
\end{align}
exhibiting a familiar Sudakov form.

A complete calculation of the $D_2$ spectrum requires marginalizing over both the soft subjet and collinear subjet configurations, which we discuss in \Sec{sec:merging}. However, to demonstrate the $\alpha_s$ behavior in the simplest manner, we will consider just the soft subjet effective theory. In particular, we will fix the angle of the soft subjet, but allow it to be arbitrarily soft, so as to probe the singular region of phase space. The result is then representative of the contribution from the soft subjet region of phase space. An exactly analogous behavior occurs for the contribution from the collinear subjets region of phase space.

Fixing $\theta_{sj}$ to satisfy $n \cdot n_{sj}=1/2$ (and therefore $\bar n\cdot n_{sj}=3/2$), and restricting to $\alpha=\beta$ for simplicity, the 2-point energy correlation function in the soft subjet region of phase space is simply
\begin{equation}
\ecf{2}{\alpha} = z_{sj} \,.
\end{equation}
The corresponding $D_2$ distribution is then obtained by marginalizing the multi-differential cross section of \Eq{eq:LL_sudakov}
\begin{align}
\frac{d\sigma}{dD_2}&=\int dz_{sj}\, d\theta_{sj}\, d\ecf{3}{\alpha}\, \delta \left ( D_2 -\frac{ \ecf{3}{\alpha}  }{(\ecf{2}{\alpha})^3   }   \right )    \frac{d\sigma}{d\ecf{3}{\alpha}    dz_{sj} d\theta_{sj}  }\\
&
\hspace{1cm}
\to\int dz_{sj} \,d\ecf{3}{\alpha}\, \delta \left ( D_2 -\frac{ \ecf{3}{\alpha}  }{z_{sj}^3   }   \right )    \frac{d\sigma}{d\ecf{3}{\alpha}    dz_{sj} d\theta_{sj}  }  \nonumber \,,
\end{align}
where, in the second line, we have fixed $\theta_{sj}$ and so we do not integrate over it.  Inserting the multi-differential cross section and fixing $\theta_{sj}$, we then have
\begin{align}\label{eq:sudsaferes}
\frac{d\sigma^{sj}}{dD_2}&=-\frac{16}{3}\frac{\alpha_s^2 C_F C_A}{\alpha\pi^2}\int_0^{1} dz_{sj}   \frac{\log \left[D_2z_{sj}\right]}{D_2\sje } e^{-\frac{\alpha_s}{\pi}\frac{C_A}{\alpha}\log^2\left[
D_2 z_{sj}
\right]}\\
&=\frac{8}{3}\frac{\alpha_s C_F}{\pi}\frac{e^{-\frac{\alpha_s}{\pi}\frac{C_A}{\alpha} \log^2 D_2}}{D_2} \,, \nonumber
\end{align}
where the $sj$ superscript denotes that this is representative of a contribution from the soft subjet region of phase space.  Importantly, because the soft subjet is defined by requirements on IRC safe measurements, the cross section in \Eq{eq:sudsaferes} is a well-defined and in principle measurable quantity.

The $\alpha_s$ dependence in this distribution of $D_2$ is very surprising.  Because $D_2$ is defined with respect to the 3-point energy correlation function, one would na\"ively expect that $D_2$ only makes sense for a jet with at least three partons.  Indeed, if we make an explicit cut on $z_{sj}$, for example, then $D_2$ is IRC safe, and first non-zero for a jet with three partons at ${\cal O}(\alpha_s^2)$.  However, because $D_2$ without a cut on $z_{sj}$ is not IRC safe, this intuition fails, and in a fascinating way.  By resumming the large logarithms of $z_{sj}$ to all orders and then marginalizing, the $D_2$ distribution calculated in \Eq{eq:sudsaferes} actually starts at ${\cal O}(\alpha_s)$!  Including emissions to all orders has effectively generated a non-trivial distribution for $D_2$ at one order {\it lower} in $\alpha_s$ than when it is first, na\"ively, non-zero.  Other examples of Sudakov safe observables in the literature have expansions in $\sqrt{\alpha_s}$ \cite{Larkoski:2013paa,Larkoski:2015lea} or are even independent of $\alpha_s$ \cite{Larkoski:2014wba,Larkoski:2014bia,Larkoski:2015lea}.  To our knowledge, $D_2$ is the first example of a Sudakov safe observable for which all-orders resummation reduces the order in $\alpha_s$ when the observable's distribution is first non-zero.\footnote{For observables that do not have universal behavior in the ultraviolet \cite{Larkoski:2015lea}.}  We re-emphasize that though the distribution of $D_2$ in \Eq{eq:sudsaferes} is a Taylor series in $\alpha_s$, it is impossible in purely fixed-order perturbation theory to systematically calculate it.

\subsection{Fixed-Order $D_2$ Distributions with a Mass Cut}\label{sec:fixed_order}

\begin{figure}
\begin{center}
\subfloat[]{\label{fig:fixed_order_a}
\includegraphics[width=6.8cm]{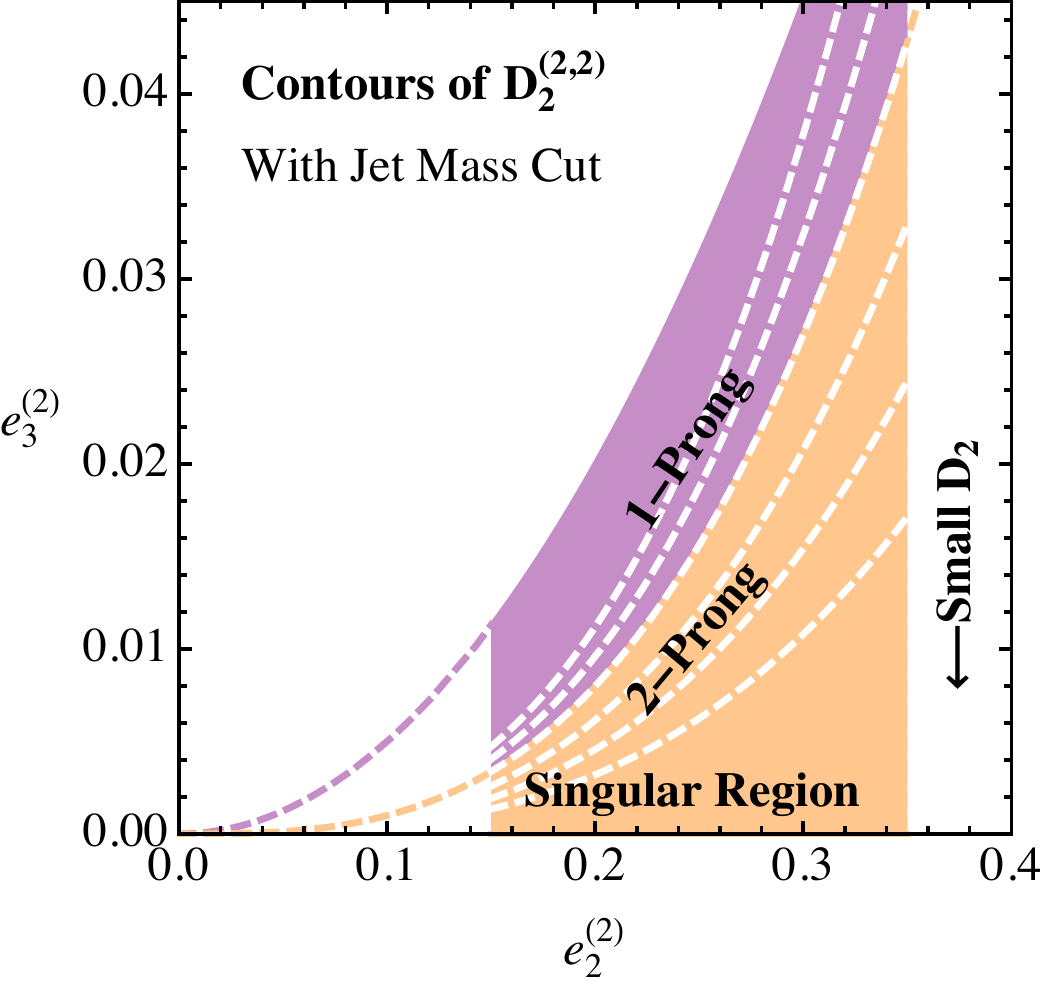}
}\ 
\subfloat[]{\label{fig:fixed_order_b}
\includegraphics[width=7.15cm, trim =0 0cm 0 0]{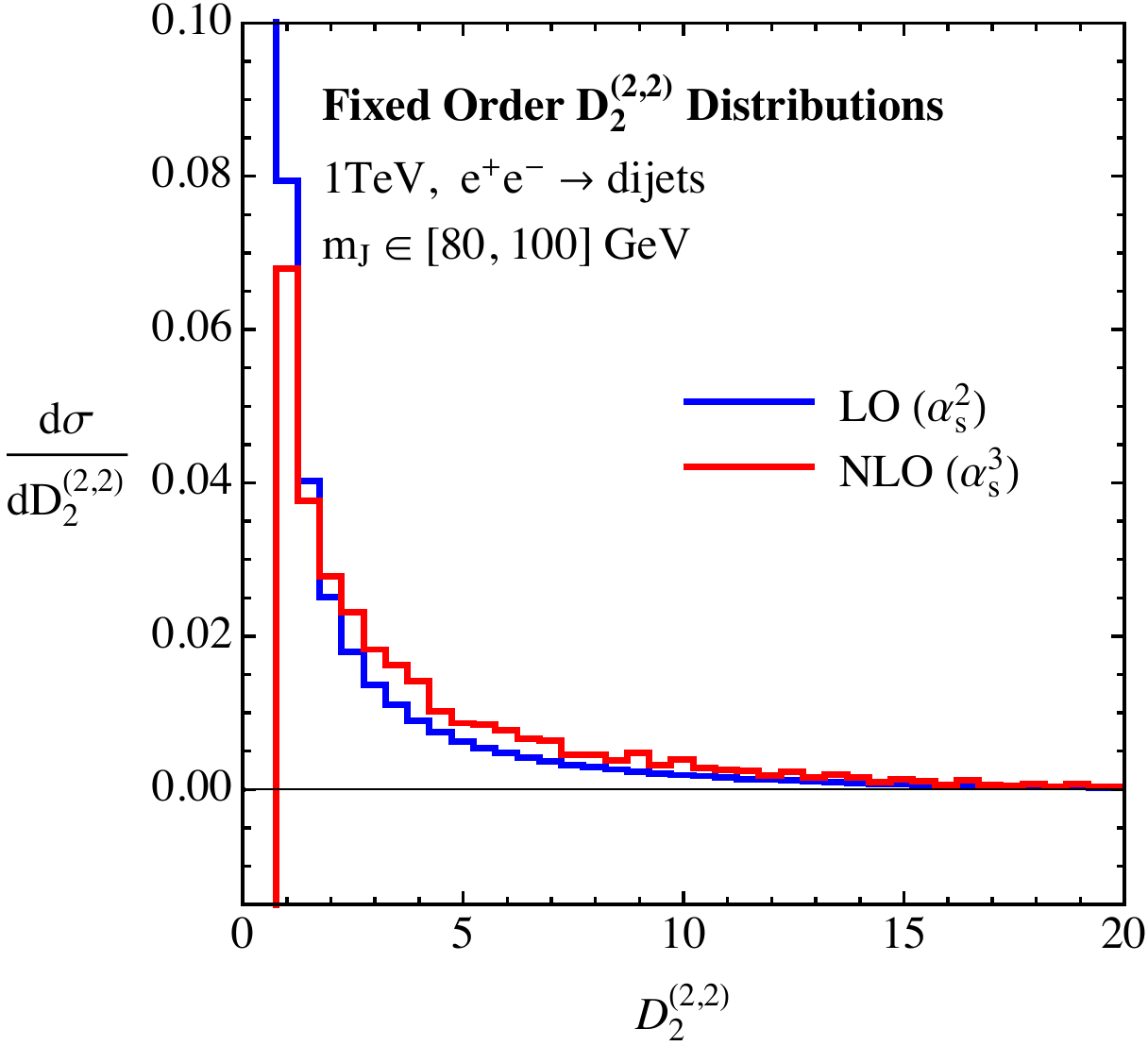}
}
\end{center}
\caption{ a) A schematic depiction of the $\ecf{2}{2}, \ecf{3}{2}$ phase space in the presence of a mass cut, along with contours of the $D_2$ observable.  b) Leading order (through $\alpha_s^2$) and next-to-leading order (through $\alpha_s^3$) distributions for the $D_2$ observable in the presence of a mass cut as measured on hemisphere jets in $e^+e^-$ collisions. 
}
\label{fig:fixed_order}
\end{figure}

Although $D_2$ is not IRC safe without a cut on $\ecf{2}{\beta}$, leading to its interesting Sudakov safe behavior, in experimental analyses a jet mass cut will be always be applied. We will therefore be most interested in this case. In \Fig{fig:fixed_order_a} we show a schematic depiction of the $\ecf{2}{\alpha}, \ecf{3}{\beta}$ phase space in the presence of a mass cut for $\alpha=\beta =2$, along with contours of the $D_2$ observable. As is indicated in the figure, the mass cut removes the origin of the phase space, making $D_2$ IRC safe and calculable in fixed-order perturbation theory. It is therefore interesting to study the singularity structure of the fixed-order perturbative expansion of $D_2$ in the presence of a mass cut.

In \Fig{fig:fixed_order_b} we show both the leading order $(\alpha_s^2)$ (LO) and the next-to-leading order $(\alpha_s^3)$ (NLO)  fixed-order distributions of the $D_2^{(2,2)}$ observable as measured on the most energetic hemisphere jet in $e^+e^- \to $ dijets events at $1$ TeV center of mass energy, and with a jet mass cut of $m_J\in [80,100]$ GeV, in anticipation of our application to boosted $Z$ boson discrimination. However, the detailed range of the mass cut window is irrelevant to the arguments of this section. \nlojet~\cite{Nagy:1997yn,Nagy:1998bb,Nagy:2001fj,Nagy:2001xb,Nagy:2003tz} was used to generate the distributions. The fixed-order $D_2$ distribution diverges at small values, and its sign in this region flips order-by-order, characteristic of the Sudakov region. This behavior makes clear the necessity of resummation in the small $D_2$ region. However, importantly, there is no divergence or other structure at large values of $D_2$. Instead, the distribution exhibits a tail extending to large values both at LO and NLO, and this behavior is expected to persist to higher orders.  This long tail arises from the fact that the upper boundary of the phase space is parametrically far, of distance $\sim1/\ecf{2}{\alpha}$, from the two-prong region of phase space. A schematic depiction of the singularity structure in the $\ecf{2}{2}, \ecf{3}{2}$ phase space is shown in \Fig{fig:fixed_order_a}.  The observation that a fixed-order singularity exists only at small values of $D_2$ is important for the resummation of the observable in the presence of a mass cut. In particular, while resummation in the soft subjet and collinear subjet factorization theorems are necessary to regulate a fixed-order singularity, the soft haze factorization theorem presented in \Sec{sec:soft_haze} is not.

The fixed-order behavior of the $D_2$ observable is in some ways much more similar to that of a traditional jet or event shape than might na\"ively be expected. However, there are some important differences. In particular, a mass cut of $80 < m_J <100 $ GeV has been applied, which is comparable to the location of the Sudakov peak in the mass for a jet of energy $500$ GeV. Therefore, unlike in the case of a traditional jet shape, where there is a transition from a region where resummation is important to a far tail region where a fixed order calculation provides an accurate description, in this case, for all values of $D_2$, there is an overall Sudakov suppression due to the mass cut, in addition to the divergence at small values of $D_2$. This is however, a small effect in the fixed order distribution compared to the divergence at smaller values, and most importantly, does not require regularization, as it is regulated by the mass cut.

\subsection{Merging Factorization Theorems}\label{sec:merging}

A complete description of the $D_2$ observable for background jets requires combining the three factorization theorems presented in \Sec{sec:Fact}. This involves both the merging of the soft subjet and collinear subjets factorization theorems, which must be performed before the marginalization over the $D_2$ contours, as well as the matching between the small $D_2$ description of the resolved two-prong region and large $D_2$ description of the unresolved region.  We will discuss how the matching is accomplished for these two cases in turn.

\subsubsection{Merging Soft and Collinear Subjets}\label{sec:soft_NINJA_match}

The region of phase space in which two subjets are resolved by the measurement is described by two distinct factorization theorems.  These two regions of phase space are separated by the measurement of the two 2-point energy correlation functions, $\ecf{2}{\alpha},\ecf{2}{\beta}$. However, in the calculation of $D_2$, both regions are treated as two-pronged, and the additional 2-point energy correlation function must be marginalized over. Since each effective theory can only be used within its regime of validity, a merged description, valid in both the soft subjets and collinear subjets region of phase space, is required. To accomplish this, we introduce a novel procedure for merging the two factorization theorems.

At a fixed $\ecf{3}{\alpha}$, the soft subjet and collinear subjets fill out the $\ecf{2}{\alpha}, \ecf{2}{\beta}$ phase space, which was shown in \Fig{fig:2ptps}. This phase space has also been studied in the context of two angularities measured on a single jet in \Refs{Larkoski:2014tva,Procura:2014cba}. In this case factorization theorems involving only collinear and soft modes exist on the boundaries of phase space, and an additional collinear-soft mode is required in the bulk of phase space. New logarithms exist in the bulk of the phase space, so called $k_T$ logarithms \cite{Larkoski:2014tva}, which can either be captured by the additional collinear-soft mode proposed in \Ref{Procura:2014cba}, or by the interpolation procedure of \Ref{Larkoski:2014tva}. In this case, the factorization theorems involving only the collinear and soft modes do not extend beyond the boundaries of the phase space, and they cannot be directly matched onto one another, as this would neglect the resummation of the $k_T$ logarithms, which are not present in either factorization theorem. We will now argue that the case of interest in this paper, namely of two resolved subjets, is different. In particular, the soft subjet and collinear subjets factorization theorems extend from the boundaries of phase space, and already contain all the modes required for a description in the bulk of the phase space. In particular no additional modes exist in the bulk region of the phase space. This implies in particular that a description of the entire phase space region can be obtained by a proper merging of the collinear subjets and soft subjet factorization theorems, which is the approach that we will take.

\begin{figure}
\begin{center}
\subfloat[]{\label{fig:soft_collinear_match_a}
\includegraphics[width= 6.25cm]{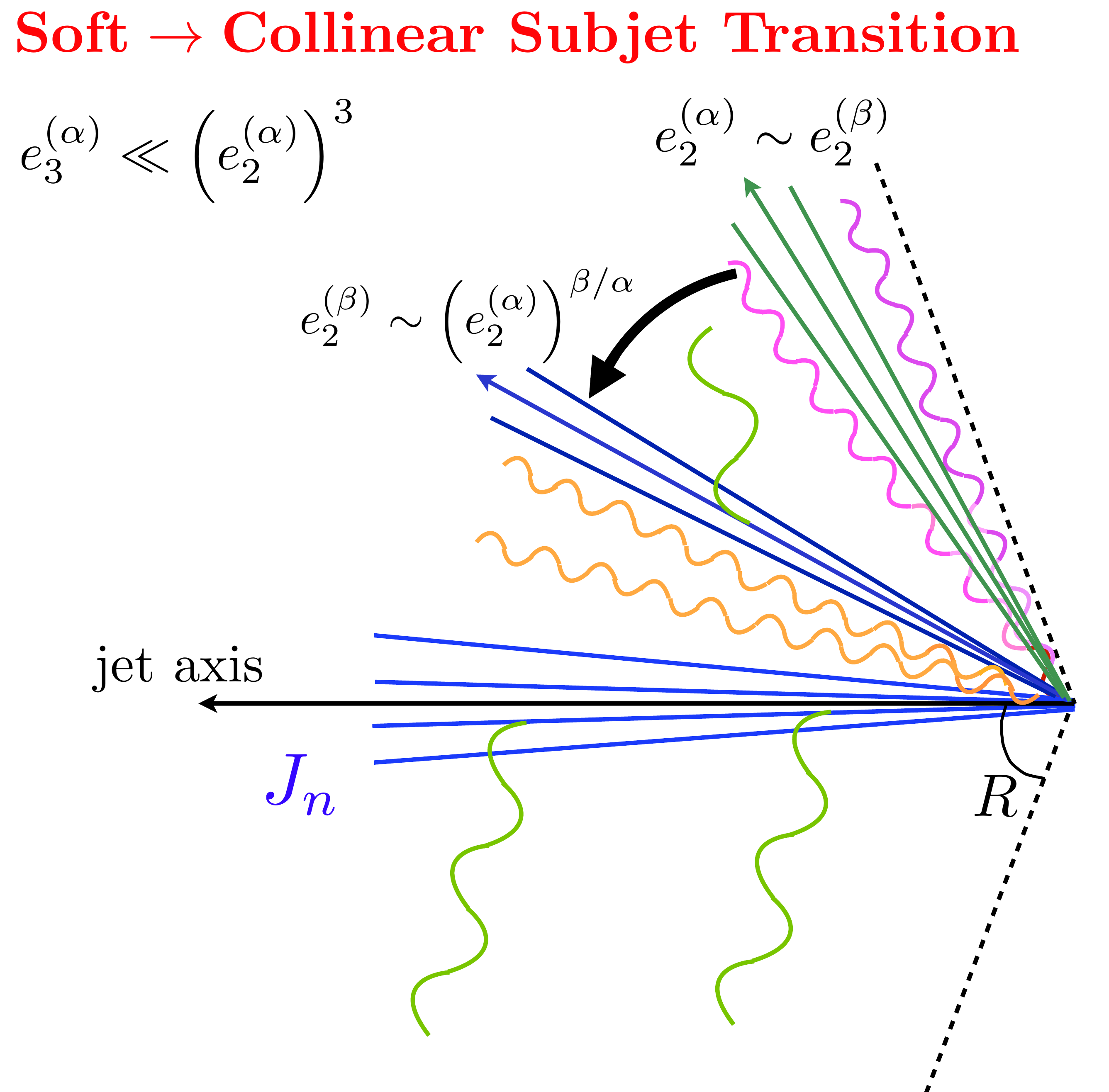}
}
\ 
\subfloat[]{\label{fig:soft_collinear_match_b}
\includegraphics[width = 7.3cm]{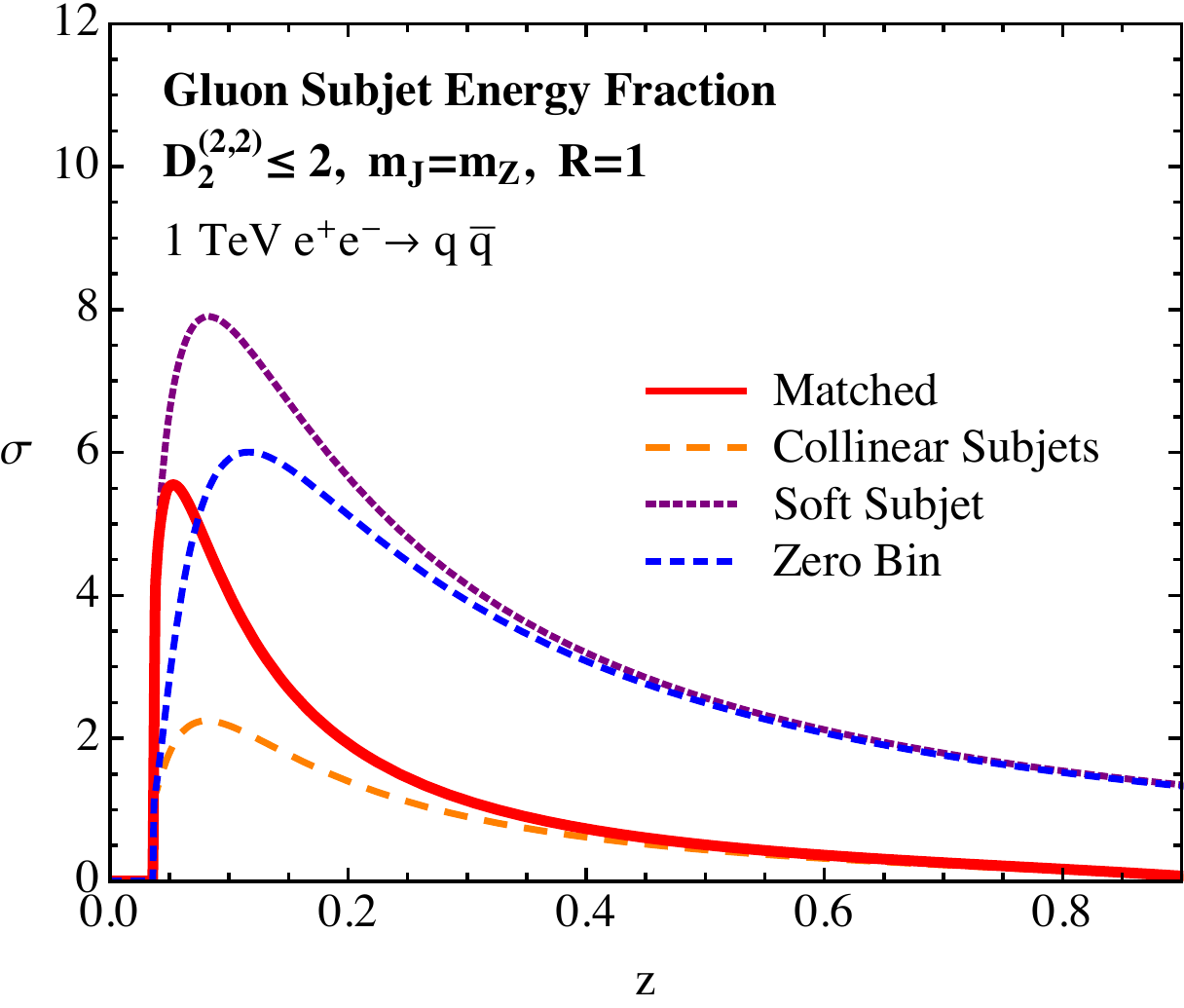}
}
\end{center}
\vspace{-0.2cm}
\caption{ a) A schematic depiction of the transition between the soft subjet and collinear subjets regions of phase space.   b) Distribution of the energy fraction of the gluon subjet as predicted by the collinear subjets effective theory, the soft subjet effective theory, and the merged description. The collinear zero bin of the soft subjet is also shown.
}
\label{fig:soft_collinear_match}
\end{figure}

To see that no additional modes are present in the bulk of the phase space, it is sufficient to look for modes which transition between the modes present in the effective theory descriptions in the soft subjet and collinear subjets regions of phase space, and which contribute at leading power. When transitioning from the collinear subjets region of phase space to the soft subjet region of  phase space, as is shown schematically in \Fig{fig:soft_collinear_match_a}, the collinear modes of one of the jets become the soft subjet and boundary soft modes of the soft subjet factorization theorem. On the other hand, the collinear-soft modes transition to the global soft modes. However, one could possibly be concerned that there exist additional modes which appear as collinear-soft modes on the boundary of phase space where the collinear subjets exist, but which transition to soft subjet modes instead of global soft modes. However, one can immediately see that such modes cannot exist, since the energy fraction of the soft subjet modes is set by the $\ecfnobeta{2}$ measurement, while the energy fraction of the collinear-soft modes is set by the $\ecfnobeta{3}$ measurement. Since $e_3$ is fixed, and the transition is occurring only in the $\ecf{2}{\alpha}, \ecf{2}{\beta}$ phase space, such modes cannot exist. This implies that all contributing modes already exist in either the soft subjet, or collinear subjets factorization theorems. This is a crucial difference from the case of the double differential angularities, which in some sense simplifies the analysis. Since no additional modes exist in the bulk of the phase space, the factorization theorems can be extended from the boundaries, and can be matched onto each other. This will allow for the resummation of all large logarithms. We will now discuss in more detail our implementation of this matching, after which we will see that our argument, presented here based on power counting, for the absence of additional modes, is explicitly realized through our merging procedure.

This suggests then the procedure we will use for interpolating between the collinear subjets and soft subjet factorization theorem, as sketched in \Ref{Larkoski:2015zka}, where the soft subjet factorization theorem was originally introduced. It proceeds by implementing a zero bin subtraction \cite{Manohar:2006nz} in factorization theorem space (the meaning of this will become clear shortly) to remove double counting in the overlapping region between the effective theories. This is a non-trivial and novel example of the zero bin procedure, and demonstrates the general utility of its approach. 

Recall that in a standard SCET factorization, the cross section is written as a convolution of a jet function, which describes the collinear physics, and a soft function, which describes the soft physics. To achieve this mode separation without introducing a double counting, the soft limit of the jet function must be subtracted, which is referred to in the literature as a zero bin subtraction. Here we extend this approach to the case of two distinct factorization theorems  which describe different regions of a multi-differential phase space, the soft subjet and collinear subjets effective field theories, but which overlap in the bulk of the two-prong phase space. It is important that here we only focus on the two-prong region of phase space; the matching to the one-prong region of phase space will be discussed in \Sec{sec:soft_haze_match}.  To perform the matching in the two-prong region of phase space, inspired by the zero-bin procedure,  we will write the cross section as a sum of the contributions from the soft subjet factorization theorem and the collinear subjets factorization theorem, with a zero bin contribution to remove the overlap between the effective theories. Explicitly, we write
\begin{align}\label{eq:matched}
\sigma=\left( \sigma_{sj}-\sigma_{sj}|_{cs} \right)+\sigma_{cs}\,,
\end{align}
where we have suppressed that at this stage the cross section is still differential in the kinematics of the subjets, so that our notation is not overly cumbersome.  The cross section in the soft subjet or collinear subjets regions of phase space are denoted by $sj$ and $cs$ subscripts, respectively.
Here the zero bin contribution, which removes the double counting, is given by $\sigma_{sj}|_{cs}$. Explicitly, $\sigma_{sj}|_{cs}$ is obtained by taking the limit of the soft subjet factorization theorem in the power counting of the collinear subjets factorization theorem. The anomalous dimensions and one-loop matrix elements for the collinear zero bin of the soft subjet factorization theorem are given in \App{sec:soft_subjet_cbin}. Each of the three contributions to the cross section given in \Eq{eq:matched} are associated with their own factorization theorem. However, the contributions to the cross section with the clearest physical interpretation are $\sigma_{cs}$ and the combined term $\left( \sigma_{sj}-\sigma_{sj}|_{cs} \right)$, which we will refer to the as the zero bin subtracted soft subjet contribution. It is the contribution which can be interpreted over the entire phase space as the contribution from a soft subjet, and all logarithms contained in this expression are of soft scales.

We specifically subtract the collinear-bin of the soft subjet factorization, and not the soft-bin of the collinear factorization. This is due to the need to cancel the contributions from the boundary soft modes of the soft subjet factorization in the collinear region. Since no analogous mode to the boundary softs is found in the collinear resummation, any soft expansion would miss this contribution, resulting in a logarithm being resummed in an inappropriate \emph{collinear} region of phase space. This is in contrast to what happens when comparing the two subtractions \emph{in the soft region}. So long as one uses the relative transverse momentum of the subjets as the splitting scale of the collinear factorization, the collinear-bin of the soft subjet does match the soft-bin of the collinear factorization in the soft region. This is the result of the merging of various soft scales. In the soft jet collinear-bin, the expanded boundary softs and global soft scales naturally merge, and in the soft-bin of the collinear jets, the global softs and collinear-softs also naturally merge in the soft region. This can be explicitly verified with the canonical scales given in \App{sec:canonical_merging_scales}. Thus the collinear-bin of the soft subjet is the appropriate subtraction throughout phase space, to remove double counting at all points.

\begin{figure}
\begin{center}
\subfloat[]{\label{fig:soft_collinear_match_a_pt6}
\includegraphics[width= 7.0cm]{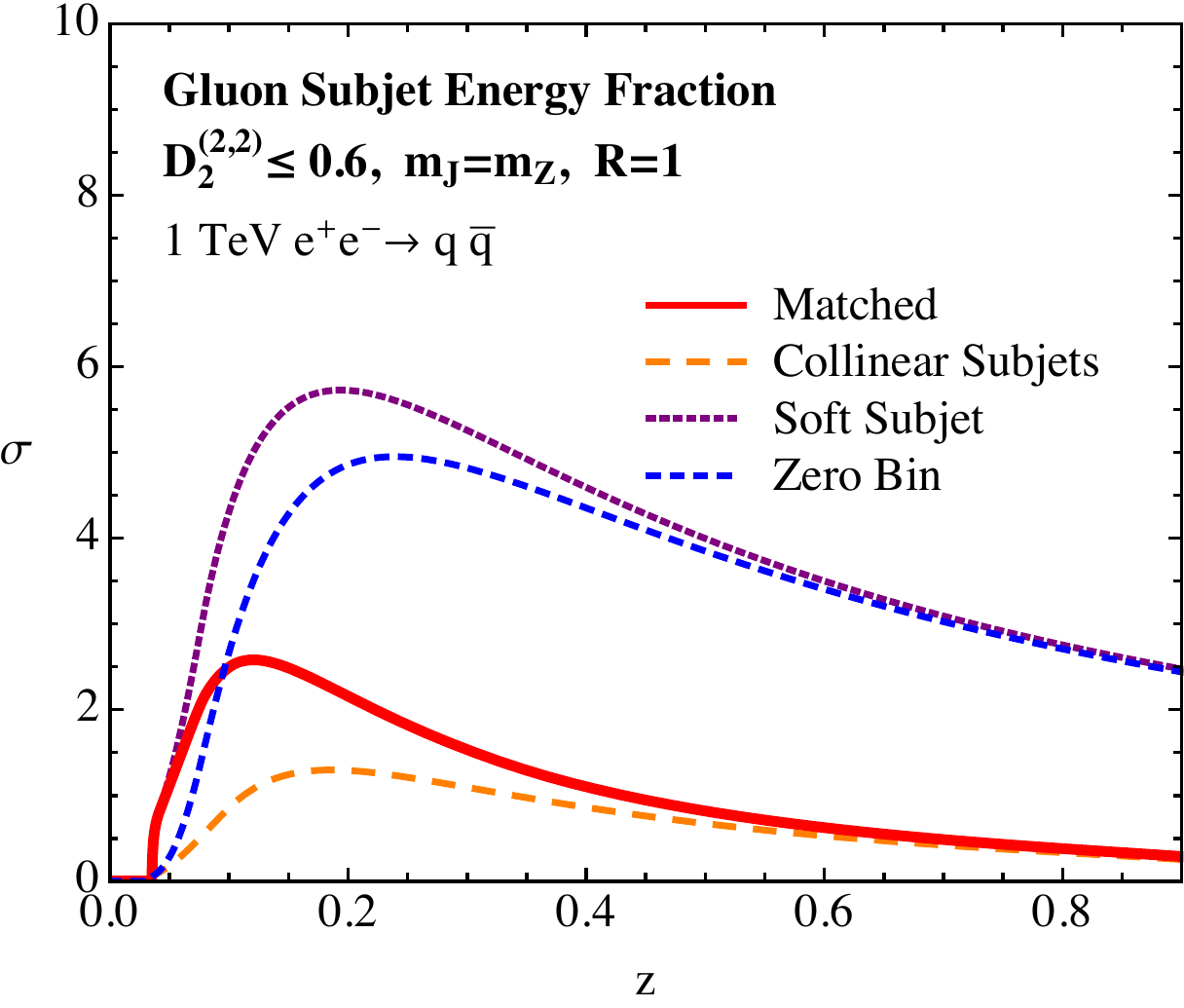}
}
\ 
\subfloat[]{\label{fig:soft_collinear_match_b_pt6}
\includegraphics[width = 6.9cm]{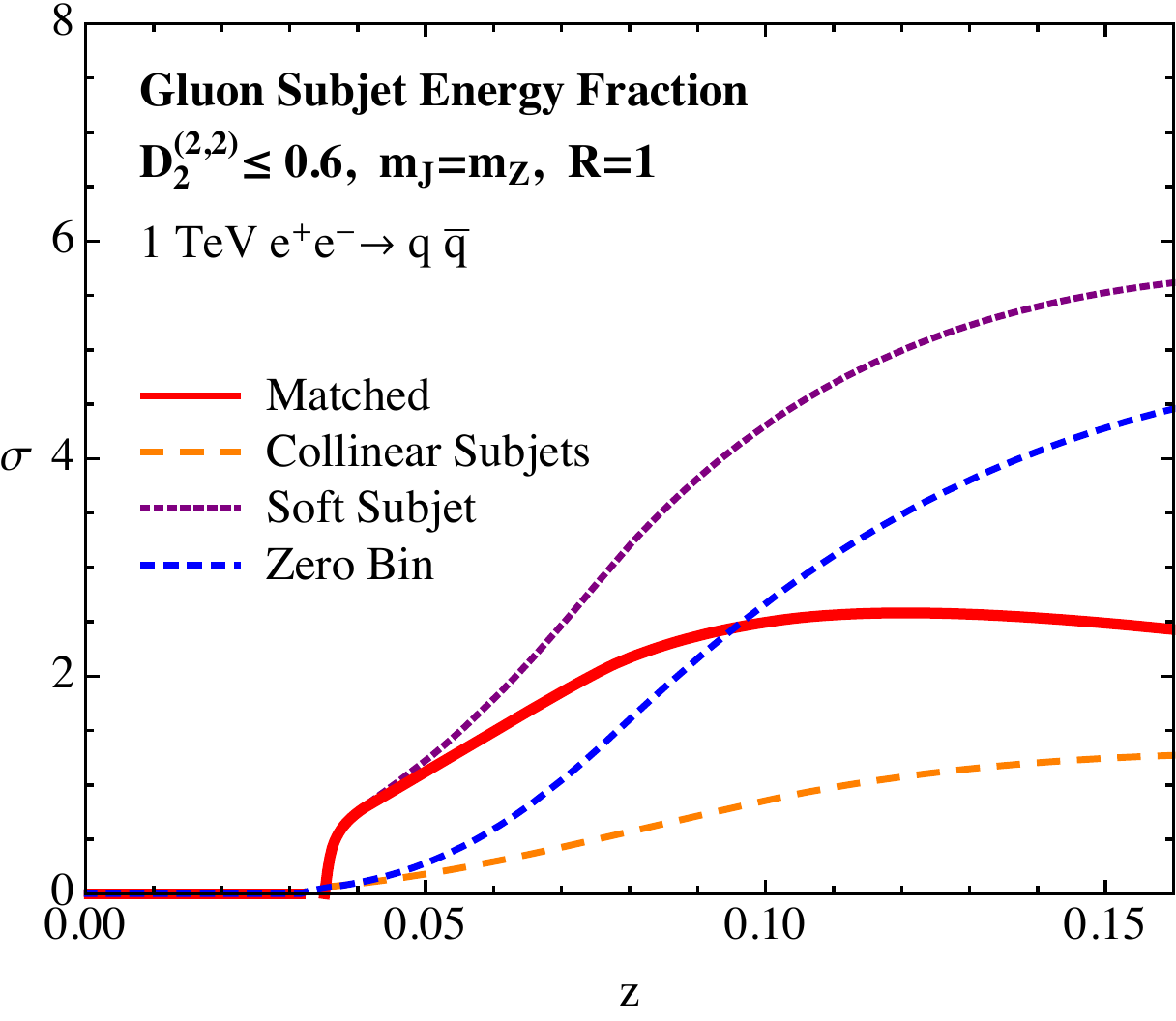}
}
\end{center}
\vspace{-0.2cm}
\caption{ a) Distribution of the energy fraction of the gluon subjet as predicted by the collinear subjets effective theory, the soft subjet effective theory, the collinear zero bin, and the matched description. A zoomed version at small $z$ is shown in b).
}
\label{fig:soft_collinear_match_pt6}
\end{figure}

Having defined our merging procedure, implemented through the zero bin, we can now revisit our argument for the absence of additional modes, previously given by power counting,  which can be verified from an explicit calculation. Taking the collinear-bin of the soft subjet factorization, and the soft-bin of the collinear subjet factorization, one finds identical fixed order expressions, as well as a one-to-one mapping of the anomolous dimensions between these two re-expanded factorizations. With the merging of the soft scales in the ``bins''  of the primary factorizations as one enters the soft region then implies they are numerically equivalent. No new logarithms appear in the bulk of phase space, unlike the case of two angularities \cite{Larkoski:2014tva}. This emphasizes that the collinear-soft region is a genuine overlap between the factorizations, with no new structures not already found in the factorizations.

To see visually the effect that this matching has, it is interesting to look at the distribution of the energy fraction of the one of the subjets. In \Fig{fig:soft_collinear_match_b}, we plot the distribution of the gluon subjet's energy fraction as computed in the collinear subjets and soft subjet factorization theorems, as well as the energy spectrum for the matched cross section of \Eq{eq:matched} and zero bin contribution. The energy spectrum is cumulative $D_2 \leq 2$, which is the majority of the two-prong region, and for simplicity we have fixed the jet mass $m_J=m_Z$. The matched contribution smoothly interpolates between the spectrum for the collinear subjets at large values of $z$, where the collinear subjets factorization theorem is valid, and captures all logarithms of the splitting angle, and that for the soft subjet factorization theorem at small values of $z$, accurately resumming large logarithms of $z$. It is also important to note that for large $z$, the zero bin contribution matches exactly onto the soft subjet contribution, removing its contribution in this region.  One can also see that the collinear-bin of the soft subjets cancels the collinear contribution to the soft region, up to power corrections, as argued above. We find that the collinear subjets provides a good description over a large range of values, with the soft subjet factorization theorem only required at small values of $z$. 

In \Fig{fig:soft_collinear_match_a_pt6}, we show the energy spectra at cumulative $D_2 \leq 0.6$, along with a zoomed version at small values of $z$, in \Fig{fig:soft_collinear_match_b_pt6}. This figures makes clear that our matched prediction, computed using our zero-bin approach, reproduces correctly the behavior of the collinear subjets at large values of $z$, and the soft subjet factorization theorem at small values of $z$. In particular, in \Fig{fig:soft_collinear_match_b_pt6}, we see that below $z\sim 0.05$, the soft subjet and matched predictions are indistinguishable.

Although we will not study this case explicitly in this paper, we have also performed the matching for gluon jets, where the dominant contribution comes from $g \to gg$ splitting. This case is somewhat interesting due to the fact that the Bose symmetry of the final gluons guarantees that the $z$ distribution is symmetric about $z=0.5$, leading to peaks in the $z$ distribution due to soft singularities at both $z=0$ and $z=1$. Nevertheless, the same matching procedure works identically in this case, and this procedure could therefore also be straightforwardly applied for studying substructure in gluon jets, as would be required for a complete calculation at the LHC.

We have shown here the matched subjet energy spectra for the particular choice of jet radius $R=1$ at a center of mass energy of $1$ TeV for quark jets, as this is the particular case that we will focus on throughout the rest of the paper. However, we have investigated the properties of the matching away from these parameters. It is important to note that our procedure for merging factorization theorem must be carefully treated at small $R$. This manifests itself as a breakdown in the zero bin procedure. In particular, for a fixed value of $\ecf{2}{\alpha}$, if $R$ is small, then the power counting $\ecf{2}{\alpha}\sim z_{sj}$ is invalidated. In other words, for small $R$ there does not exist a region of phase space which contributes to $\ecf{2}{\alpha}$ for which $z_{sj}$ is sufficiently small that the soft subjet expansion is valid. 

We can bound the specific $R$ that eliminates the soft subjet region by considering the minimum energy fraction accessible to a subjet at a fixed $\ecf{2}{\alpha}$:
\begin{align}\label{eq:min_z}
z_{\min}\approx \frac{\ecf{2}{\alpha}}{\left(2\, \text{sin} \frac{R}{2}\right)^{\alpha}}\,.
\end{align}
As a necessary condition for a soft subjet, one must fulfill the condition:
\begin{align}\label{eq:min_z_sj}
z_{\min}\sim \ecf{2}{\alpha} \rightarrow 1\sim \left(2\, \text{sin} \frac{R}{2}\right)^{\alpha}\,,
\end{align}
and so $R\sim 1$ for the soft subjet to contribute.  To eliminate the soft subjet then requires $R \ll 1$ and to still have valid collinear subjet regions requires that $R$ and $\ecf{2}{\alpha}$ are related as:
\begin{align}\label{eq:no_sj_R}
1\gg R^\alpha\gg \ecf{2}{\alpha}.
\end{align}
Finally, one should distinguish a fixed mass jet from a fixed $\ecf{2}{\alpha}$. In the case $\alpha=2$, since $\ecf{2}{2}= \frac{m_J^2}{E_J^2}$, by varying $E_J$ or $R$, we can open or close the soft subjet region. 

This appears in the zero bin by the fact that the zero bin subtraction is greater in all regions than the soft subjet, leading to a negative total cross section. We find numerically that this occurs for $R< 0.5$ for the case of $m_J=90$ GeV, and $Q=1$ TeV. This value depends fairly sensitively on $m_J$ and $Q$, or equivalently $\ecf{2}{\alpha}$. In this case, only the collinear subjets factorization theorem should be used, and it is valid throughout the entire available phase space. In this paper we focus primarily on the case of fat jets, defined with $R=1$, and therefore it is necessary to perform the matching between the soft subjet region and the collinear subjets region for jets of energy $500$ GeV. However, in \Sec{sec:R_dependence}, we perform a brief survey of different $R$ values, comparing our analytic predictions with distributions from Monte Carlo generators. A more phenomenological study of the importance of the matching for different physics processes of interest for an $e^+e^-$ collider, the LHC, or even a possible $100$ TeV collider, where even higher boosts can be achieved, would be interesting, but is well beyond the scope of our initial investigation and can be straightforwardly treated using our techniques.

While we have used a zero bin procedure to perform the matching between the collinear subjets and soft subjet factorization theorems, it is also possible to develop a dedicated effective field theory valid when the soft subjet becomes collinear. This effective field theory is related to our zero bin contribution, and has been developed in \Refs{Pietrulewicz:2016nwo}. While we believe that this approach is nice in principle, for the observable $D_2$, we find that such an effective field theory has a vanishing region of validity, as can be seen from the zero bin contribution in \Fig{fig:soft_collinear_match_b}, and \Figs{fig:soft_collinear_match_a_pt6}{fig:soft_collinear_match_b_pt6}. We therefore believe that our use of the zero bin, as generalized to distinct factorization theorems, represents a natural approach to the merging of the distinct factorization theorems. However, we acknowledge that this is an observable dependent statement, and there may be cases where there is a sufficiently large region of overlap between the soft subjet and collinear subjets effective theories, and in this case it might prove useful to have a separate effective field theory description which is valid in the case that the soft subjet becomes collinear.

\subsubsection{Matching Resolved to Unresolved Subjets}\label{sec:soft_haze_match}

An important feature of the $D_2$ observable is that its contours respect the parametric scaling of the phase space, as emphasized in \Fig{fig:C2vD2}. This implies that the marginalization over the contours defining the observable can be performed at small $D_2$ entirely within the merged effective theory of \Sec{sec:soft_NINJA_match}, and at large $D_2$ within the soft haze effective field theory. Hence the matching between these two different descriptions can be performed at the level of the $D_2$ distribution instead of at the level of the double differential cross section, which is a great simplification, and primary feature of the $D_2$ observable.

The soft haze factorization theorem presented in \Sec{sec:soft_haze} first contributes to the shape of the $D_2$ distribution at two emissions, the first order at which $\ecf{3}{\alpha}$ can be non-zero (technically at next-to-next-to-leading logarithmic prime order, NNLL$'$, in the logarithmic counting). Since our focus is on an initial investigation of the factorization properties of two-prong discriminants, the necessary two-loop calculation is beyond the scope of this paper.  Na\"ively, this implies that since the merged effective field theory describing the two-prong region of phase space is only valid for $D_2 \lesssim 1$, our predictions should not be extended beyond $D_2 \lesssim 1$. However, we will argue that because of the structure of fixed order singularities for the $D_2$ observable, extending our two-prong factorization theorems to large $D_2$ will provide an accurate description of the $D_2$ distribution for a wide range of $E_J$ and $R$.

As shown in \Sec{sec:fixed_order}, there does not exist a fixed order singularity at large $D_2$. In particular, this implies that if extended into this region, the factorization theorems valid at small $D_2$ will not diverge. Furthermore, one in fact expects that they provide a reasonable description of the shape. They contain both an overall Sudakov factor for the $\ecf{2}{\beta}$ scale of the jet, and also provide a description of the internal structure of the jet in terms of splitting functions (in the case of the collinear subjets factorization). While the splitting function does not exactly reproduce the matrix elements in the soft haze factorization theorem, it provides a good description of them. We believe that this is a consistent approach which suffices for this initial investigation. 

Perhaps the most important fixed order correction not captured in the subjet factorition for $D_2$ is simply the endpoint of the distribution, which arises from the kinematic boundaries of the phase space. Since we will normalize our distributions to $1$, in order to compare to the Monte Carlo generators, the height of the peak is correlated with the endpoint. Matching to the soft haze region would give the resummed distribution the correct endpoint in the tail, and thus can shift the peak up in general. This endpoint is sensitive to the specific $R$ and $E_J$ of the jet, as well as to the values of the angular exponents $\alpha$ and $\beta$. Recall that since the Monte Carlo generators respect momentum conservation, they always terminate their distributions before the physical endpoint of the spectrum. We will also see how this disagreement in the tail region changes as a function of $R$ and $E_J$ in \Secs{sec:R_dependence}{sec:jet_energy} respectively.  However, for the case of dijets produced at a center of mass energy of $1$ TeV, with a jet mass cut of $80< m_J <100$ GeV, as is relevant for boosted boson discrimination, and on which we primarily focus throughout this paper, we will see that this discrepancy in the tail region is minimal, and we will find good agreement between our analytic calculations and the Monte Carlo predictions. It would of course be interesting to perform the complete two-loop calculation in the soft haze region of phase space; however, we believe that this would have a minor effect for a substantial range of parameter space. Nevertheless, the proper inclusion of this region of phase space would also be interesting from a resummation perspective, as it would require matching between two distinct factorization theorems involving a different number of resolved jets, instead of the more familiar case of matching a resummed distribution to a fixed order calculation. We leave further investigations of this to future work.

\section{Numerical Results and Comparison with Monte Carlo}\label{sec:results}

We now present numerical results for signal and background distributions for the $D_2$ observable in $e^+e^-$ collisions. We give a detailed comparison with Monte Carlo, at parton level in Secs.~\ref{sec:MC} through \ref{sec:jet_energy} and including hadronization in \Sec{sec:Hadronization}. We then study the discrimination power of $D_2$ analytically in \Sec{sec:ROC}, and comment on the optimal choice of angular exponents. In \Sec{sec:2insight} possible observables which go beyond $D_2$, and separately resolve the soft subjet, and collinear subjets region of phase space, and how these could be used for possible improvements to boosted boson discrimination.

Throughout this section we use \fastjet{3.1.2} \cite{Cacciari:2011ma} and the \texttt{EnergyCorrelator} \fastjet{contrib} \cite{Cacciari:2011ma,fjcontrib} for jet clustering and analysis. All jets are clustered using the $e^+e^-$ anti-$k_T$ metric \cite{Cacciari:2008gp,Cacciari:2011ma} using the WTA recombination scheme \cite{Larkoski:2014uqa,Larkoski:2014bia}, with an energy metric.\footnote{We thank Jesse Thaler for use of a preliminary version of his code for WTA in $e^+e^-$ collisions. This code is now available in the \fastjet{contrib}.}

\subsection{Comparison with Parton-Level Monte Carlo}\label{sec:MC}

Previous studies of boosted boson discrimination with ratios of IRC safe jet observables have relied entirely on Monte Carlo simulations. While the implementation of both the perturbative shower and hadronization are well-tuned to describe simple event-wide observables, jet substructure observables probe significantly more detailed correlations.  For the particular case of observables sensitive to two-prong structure, their discrimination power is sensitive to the description of massive QCD jets in the phase space region where the jets are dominated by a resolved splitting.  One might na\"ively expect that this region of phase space is sensitive to the implementation of the parton shower model, and we will see that this is indeed the case.

 While a comparison to recent LHC data on jet substructure observables (for example: \cite{ATLAS:2012am,Aad:2013gja,TheATLAScollaboration:2013tia,TheATLAScollaboration:2013qia,CMS:2014fya,CMS:2014joa}) is possible, the lack of analytic calculations means that it is difficult to disentangle perturbative from non-perturbative effects. In this section we compare the results of our analytic calculation for $D_2$ with a number of Monte Carlo generators at parton level, focusing in particular on the small $D_2$ region.\footnote{One should always be wary of comparisons of Monte Carlo generators at parton level which employ different hadronization models.  Our comparisons at parton level presented in this section are to set the stage for fully hadronized comparisons in the following section. However, we take the view that a parton shower should achieve, to the greatest extent possible, a clean separation between perturbative and non-perturbative physics, and therefore should provide an accurate description of observables both at parton and hadron level.} This allows for a detailed probe of the simulation of two-prong jets in QCD by the perturbative shower (for a discussion of some other variables, see \Ref{Fischer:2014bja,Fischer:2015pqa}).  A large number of implementations of the perturbative shower exist, and are implemented in popular Monte Carlo generators (for reviews, see e.g. \cite{Buckley:2011ms,Skands:2012ts,Seymour:2013ega,Gieseke:2013eva,Hoche:2014rga}). Some examples include \pythia{} \cite{Sjostrand:2006za,Sjostrand:2007gs}, a $p_T$-ordered dipole shower; \vincia{} \cite{Giele:2007di,Giele:2011cb,GehrmannDeRidder:2011dm,Ritzmann:2012ca,Hartgring:2013jma,Larkoski:2013yi}, \sherpa{} \cite{Gleisberg:2003xi,Gleisberg:2008ta}, \ariadne{} \cite{Lonnblad:1992tz}, and \dire{} \cite{Hoche:2015sya}, dipole-antenna showers; and \herwigpp{} \cite{Marchesini:1991ch,Corcella:2000bw,Corcella:2002jc,Bahr:2008pv}, an angular-ordered dipole shower.\footnote{\herwigpp{} also has the option for a dipole-antenna shower implementation \cite{Platzer:2011bc} though we will not use it here.}

As representative of these different Monte Carlo shower implementations, we will use the following Monte Carlo generators throughout this section:
\begin{itemize}
\item \pythia{8.205}
\item \vincia{1.2.01} with a $p_T$-ordered shower
\item \vincia{1.2.01} with a virtuality-ordered shower
\item \herwigpp{2.7.1}
\end{itemize}
All Monte Carlos were showered with default settings except for the caveats listed below and requiring two-loop running of $\alpha_s$ in the CMW scheme \cite{Catani:1990rr,Dokshitzer:1995ev} with $\alpha_s(m_Z)=0.118$.  The different shower evolution variables within the \vincia{} Monte Carlo enables a study of their effects.  For background distributions, we generate $e^+e^-\to $ dijets at 1 TeV center of mass energy and study the highest energy $R=1.0$ anti-$k_T$ jet in the event.  For signal distributions in \pythia{} and \vincia{}, we generate $e^+e^-\to ZZ$ events with both $Z$s decaying hadronically.  For \herwig{}, the fixed-order signal distributions are generated in \madgraph{2.1.2} \cite{Alwall:2014hca} and showered in \herwig{}.  All jets are required to have a mass in the window $m_J\in[80,100]$ GeV.  In all plots shown in this section, hadronization has been turned off in all Monte Carlos.  Fixed-order matching was also turned off in \vincia{}.

\begin{figure}
\begin{center}
\subfloat[]{\label{fig:D2_ee_bkg}
\includegraphics[width= 7.15cm]{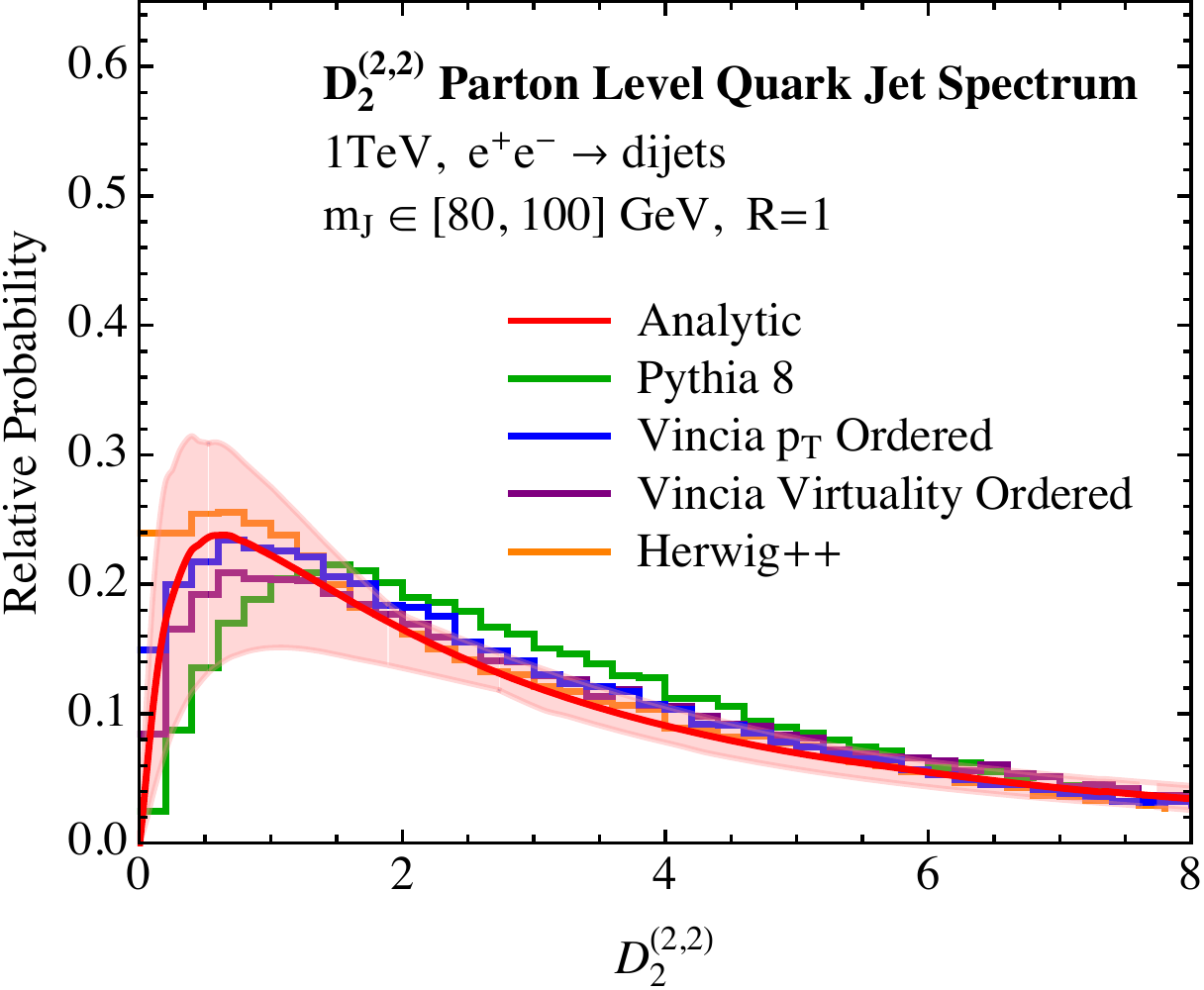}
}
\ 
\subfloat[]{\label{fig:D2_ee_sig}
\includegraphics[width = 7.15cm]{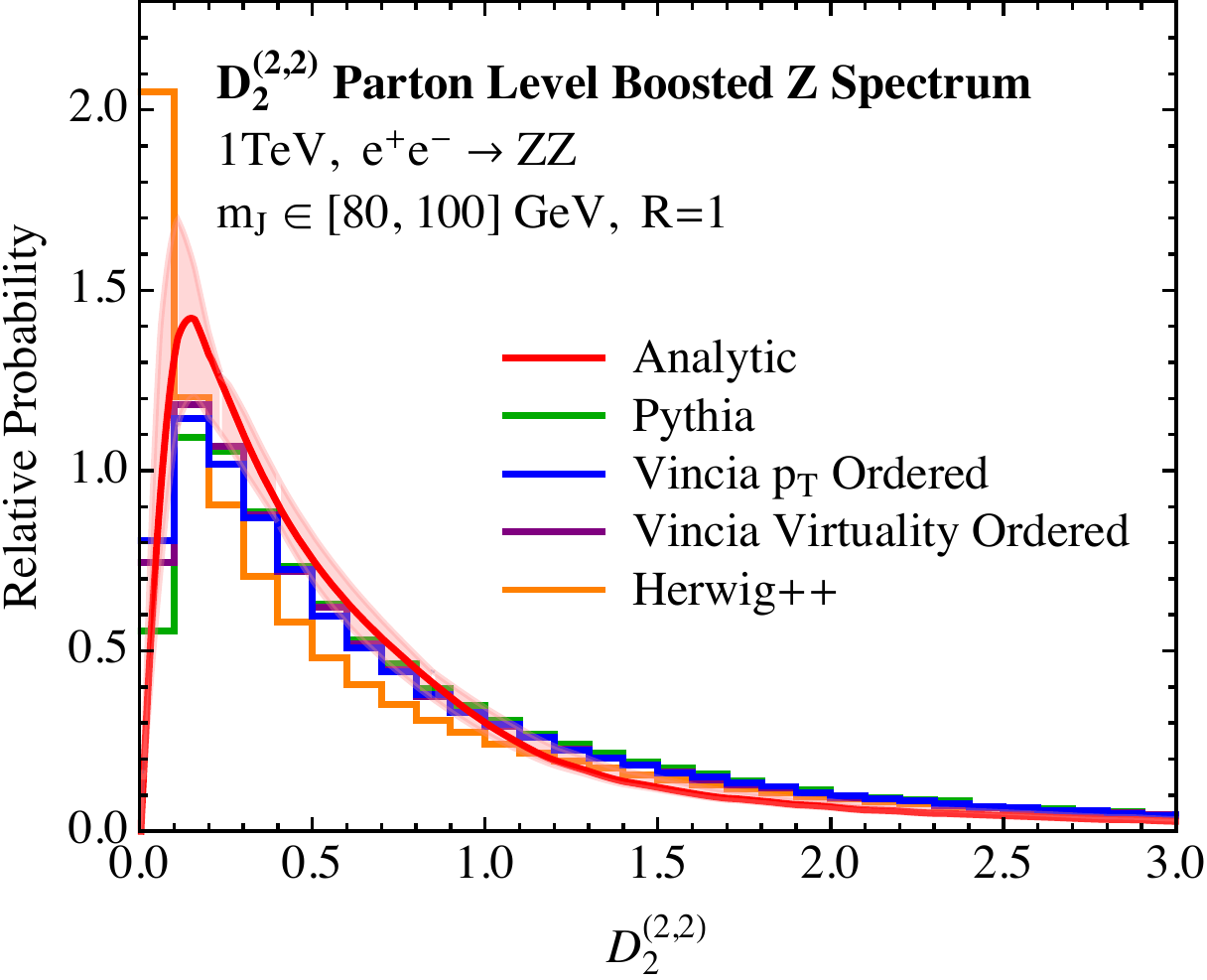}
}
\end{center}
\vspace{-0.2cm}
\caption{A comparison of our analytic prediction for $\Dobs{2}{2,2}$ compared with the parton-level predictions of the \pythia{}, \vincia{} and \herwig{} Monte Carlos. a) The $D_2$ distributions as measured on QCD background jets.  b) The $D_2$ distributions as measured on boosted $Z$ boson signal jets. The solid line is the central value of our analytic calculation and the shaded bands are representative of perturbative scale variations.  The pinch in the scale variations is a consequence of unit normalizing the distributions.
}
\label{fig:MC_compare}
\end{figure}

\Fig{fig:MC_compare} compares our analytic prediction for the $\Dobs{2}{2,2}$ spectrum to the parton-level Monte Carlo simulations in both background (\Fig{fig:D2_ee_bkg}) and signal (\Fig{fig:D2_ee_sig}) samples. The details of the scale variations used to make the uncertainty bands will be explained in \Sec{sec:scales}, but the pinch in the uncertainties should not be taken as physical.  The pinch comes from unit normalizing the distributions, and is common in analyses in which scale variations are applied to normalized distributions (see, e.g., \Refs{Ellis:2010rwa,Dasgupta:2012hg,Jouttenus:2013hs}).  All Monte Carlos have similar distributions as measured on signal jets, though \herwig{} is more peaked at small values than the other generators.  Our analytic prediction, shown with perturbative scale variation, agrees well with the Monte Carlo generators. On background jets, however, the distributions are distinct, especially at small values of $D_2$.  Small $D_2$ is the region where the jet has a two-prong structure, but unlike for signal jets, for background jets that structure is not generated by a hard matrix element. In the case of collinear subjets, it is generated by a hard splitting function, while for a soft subjet, it is generated by an eikonal emission.  In the Monte Carlos, small $D_2$ is the region that is most sensitive to the cutoff effects and other infrared choices.  As we will show in following sections, by adjusting unphysical infrared scales, differences between the Monte Carlos at parton level can be reduced and essentially eliminated.


For reference, in \App{app:twoemissionMC} we show a collection of $\ecf{2}{2}$ distributions at both parton and hadron level for each of the different Monte Carlo generators. Since $\ecf{2}{2}$, which is related to the jet mass by \Eq{eq:ecf_mass_relation}, is set by a single emission, the agreement between the different generators, particularly at parton level, is significantly better than for the $D_2$ observable. This further emphasizes the fact that the $D_2$ observable offers a more differential probe of the perturbative shower, going beyond the one emission observables on which Monte Carlo generators have primarily been tuned.

In the following sections we will study the partonic $D_2$ distributions in more detail.  We will restrict ourselves to comparing and contrasting $p_T$-ordered \vincia{} and \pythia{} for a few reasons.  First, as exhibited in \Fig{fig:D2_ee_bkg}, these Monte Carlos represent the largest spread in their predicted $D_2$ spectra.  \herwig{}, while it performs very similarly to \vincia{}, has a different hadronization model than \pythia{} and \vincia{}.  So, directly comparing \pythia{} and \vincia{} minimizes any implicit hadronization effects 
when comparing the Monte Carlos at parton level. There are still differences due to the cutoff of the perturbative shower, which will be discussed in \Sec{sec:scales}.

\subsection{Monte Carlos and Perturbative Scale Variation}\label{sec:scales}

The fact that, in particular, the $p_T$-ordered \vincia{} distribution for $D_2$ as measured on background agreed with our calculation while the \pythia{} distribution disagreed in the small $D_2$ region can be understood and quantified further.  The bulk of the disagreement between our analytic calculation and \pythia{}, illustrated in \Fig{fig:D2_ee_bkg}, occurs near the peak of the $D_2$ distribution.  It is well-known that for many observables perturbative uncertainties tend to be significant in the peak region of the distribution.  Therefore, it is possible that the difference between the $p_T$-ordered \vincia{} and \pythia{} $D_2$ distributions can fully be explained by large perturbative uncertainties.  In this section, we will show that by adjusting the cutoff of the parton level shower, the differences between \vincia{} and \pythia{} can be significantly reduced.

To estimate perturbative uncertainties in our resummed analytic calculation, the standard procedure is to vary the scales that appear in the calculation by factors of 2.  This is at the very least a proxy for the sensitivity of the cross section on these scales.  Because our factorization theorems contain many functions, as well as merging of distinct factorization theorems, in principle there are numerous scales that could be varied, a complete analysis of which is beyond the scope of this paper.  A complete list of the variations considered as well as the resummation procedure can be found in \App{sec:canonical_merging_scales}, while here we only summarize. In all factorizations theorems, we vary the subjet splitting scales, the in-jet soft radiation scales, the out-of-jet soft radiation scales, as well as where the freeze-out for the Landau pole occurs in the running of $\alpha_s$. We do not separately vary the scale in the soft subjet factorization theorem and the collinear zero bin to ensure that the zero bin subtraction is implemented correctly. The scale variation band for the total cross section is then taken as the combined band for all possible combinations of these scale variations. The soft subjet cross section displays a particular sensitivity to the out-of-jet scale setting, since the running between the boundary soft modes and the out-of-jet modes forces the soft subjet energy spectrum to vanish at the jet boundary,\footnote{As explained in \Ref{Larkoski:2015zka}, this is connected with the buffer region of \Ref{Dasgupta:2002bw}.} though the fixed order cross section probes the soft divergence in this region. Thus we also consider several different schemes for handling the out-of-jet scale setting. We believe that our scale variation bands are representative, and this is supported by the agreement with the Monte Carlo.

\begin{figure}
\begin{center}
\includegraphics[width= 7.2cm]{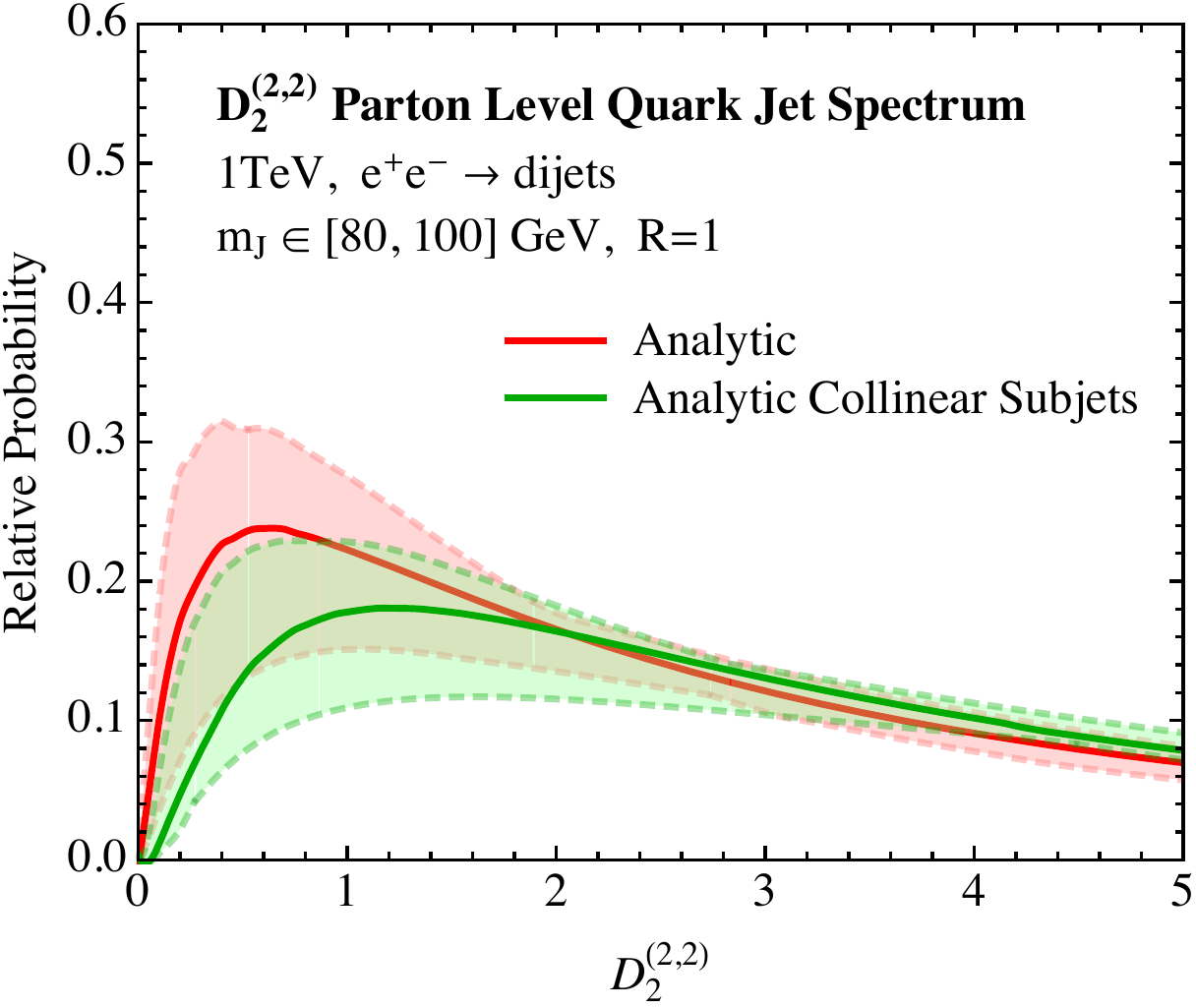}
\end{center}
\caption{
Analytic prediction for the $D_2$ distribution for background QCD jets including the envelope of the perturbative scale variation, as compared with an analytic calculation including just the collinear subjets region of phase space. The effect of the soft subjet region of phase space is clearly visible at small values of $D_2$.
}
\label{fig:D2_scale}
\end{figure}

Having understood the perturbative uncertainty bands, we now discuss in more detail the discrepancy between the different Monte Carlo generators arising at small values of $D_2$, as exemplified by the difference between the $p_T$-ordered \vincia{} and \pythia{} distributions. To understand the origin of this discrepancy, we begin by understanding the effect of the soft subjet region of phase space in our analytic calculation. This is possible due to our complete separation of the phase space using the energy correlation functions. In particular, because we have formulations of distinct factorization theorems in the soft subjet and collinear subjets regions of phase space, we can make an analytic prediction for the contribution arising just from the collinear subjets region of phase space. In \Fig{fig:D2_scale} we show a comparison of the $D_2$ distribution for background QCD jets as computed using our complete factorization theorem, incorporating both the soft subjet and collinear subjets region of phase space, as compared with the calculation incorporating only the collinear subjets region of phase space. Comparing the two curves, we are able to understand the effect of the soft subjet region of phase space. In particular, we see that the soft subjet has a considerable effect on the distribution at small values of $D_2$, giving rise to a more peaked distribution, with the peak at smaller values of $D_2$, as compared to the result computed using only the collinear subjets region of phase space. Although the perturbative error bands are large, the systematic effect of the soft subjet region of phase space is clear. 

One further feature of the $D_2$ distributions, which is made clear by \Fig{fig:D2_scale}, is that the full $D_2$ distribution is not the result of a single Sudakov peak, and therefore our intuition about the behavior of different orders in the perturbative expansion, and the behavior of scale variations from traditional event shapes fails. In particular, while it is generically the case for traditional event shape distributions that lower order resummed results overshoot in the peak region, it is not at all clear that this behavior should be true for $D_2$, and indeed it is not observed. Instead, the contribution from the collinear subjets alone is expected to undershoot the peak of the $D_2$ distribution, since it does not incorporate the soft subjet region of phase space. The final contribution is then obtained as a superposition of two distinct Sudakov peaks, and can therefore behave quite differently from traditional event shapes.

\begin{figure}
\begin{center}
\subfloat[]{\label{fig:cutoff_Rpt5}
\includegraphics[width= 7.2cm]{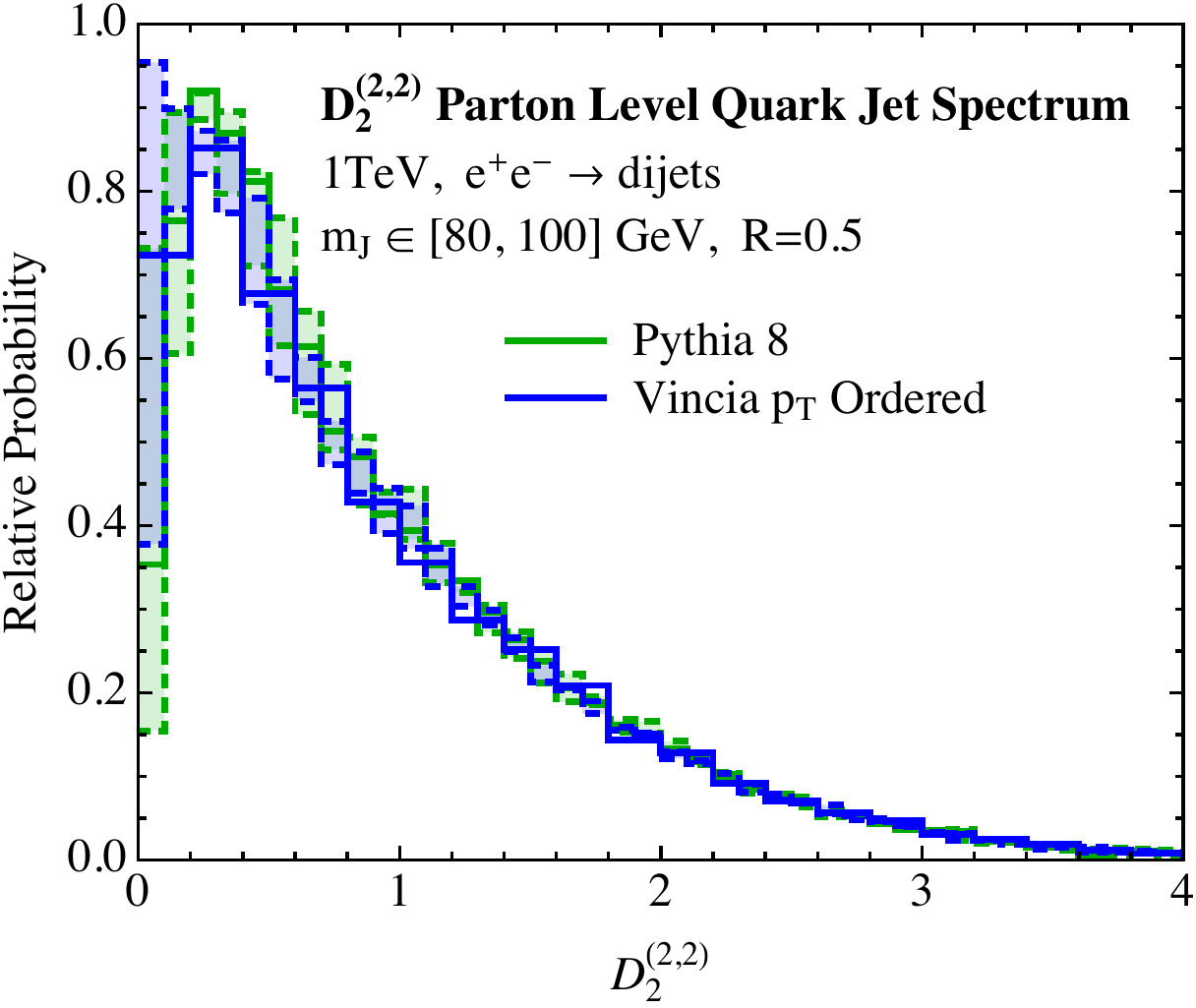}
}
\ 
\subfloat[]{\label{fig:cutoff_Rpt7}
\includegraphics[width = 7.2cm]{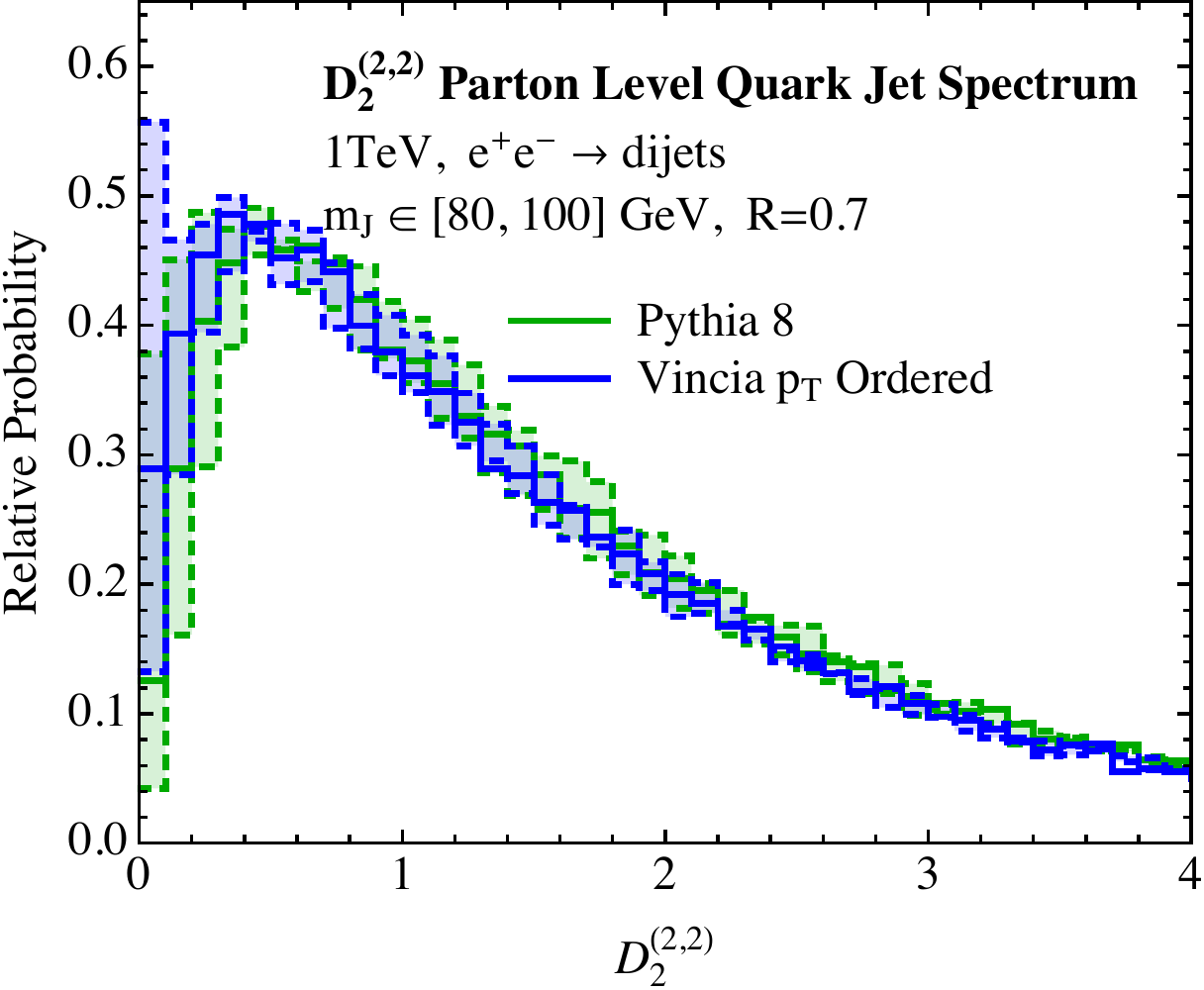}
}\\
\subfloat[]{\label{fig:cutoff_R1}
\includegraphics[width= 7.2cm]{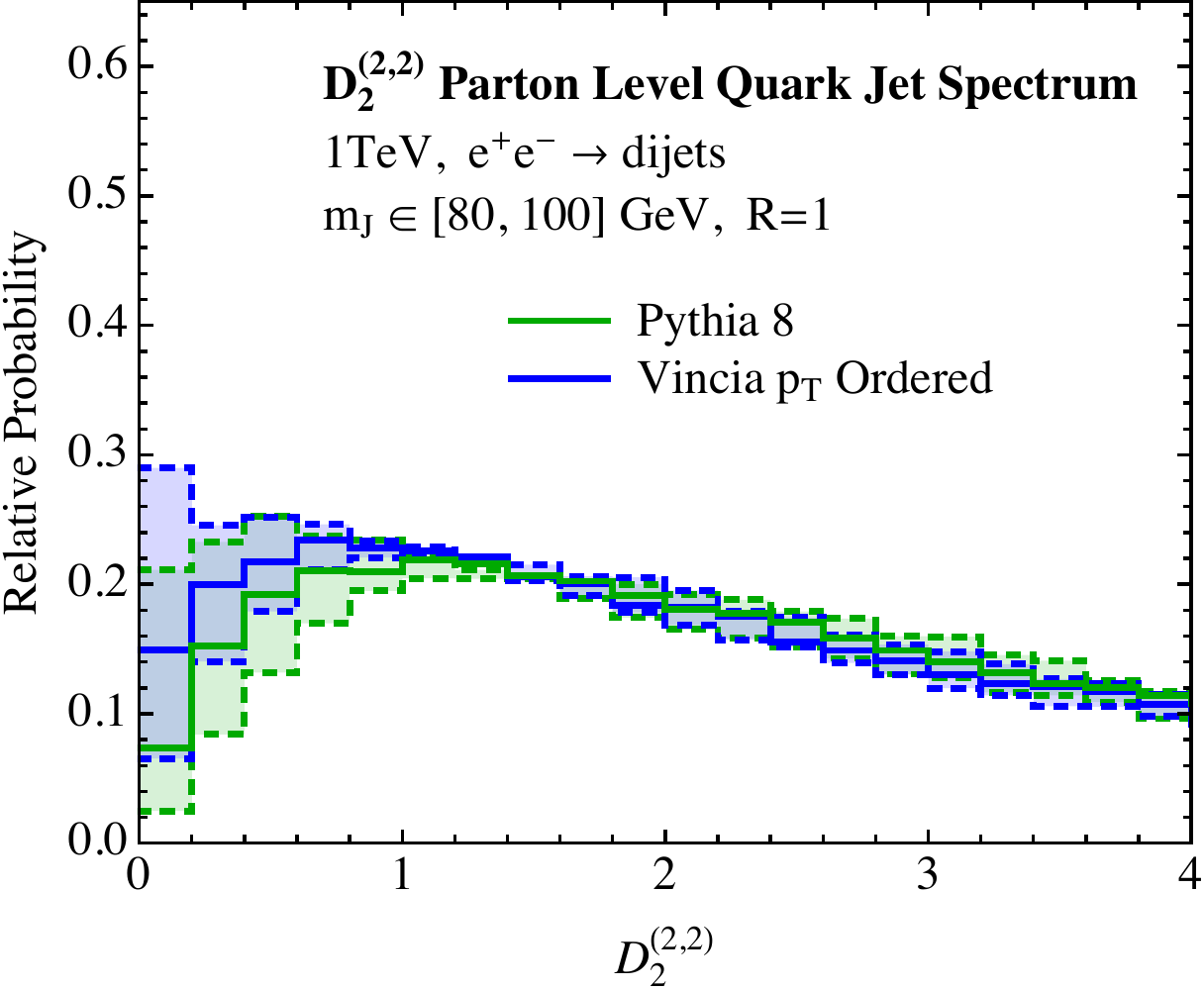}
}
\ 
\subfloat[]{\label{fig:cutoff_R1pt2}
\includegraphics[width = 7.2cm]{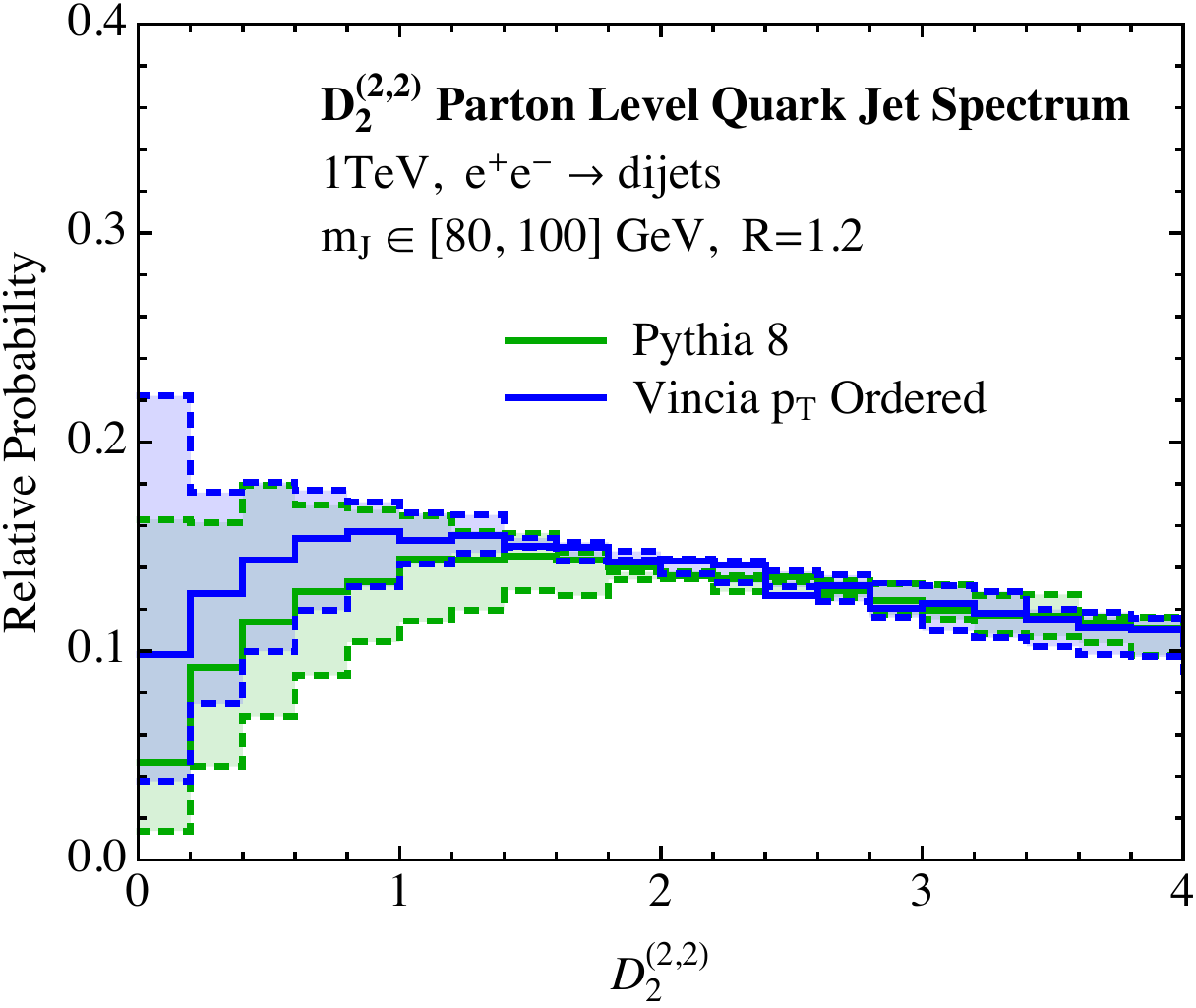}
}
\end{center}
\caption{
Comparison of the $D_2$ distribution for background QCD jets in \pythia{} and $p_T$-ordered \vincia{} for different jet radii. In each plot, the central value is obtained using a shower cutoff of $0.8$ GeV, and the uncertainty bands are generated by varying this cutoff between $0.4$ GeV, and $1.2$ GeV. 
}
\label{fig:cutoff_test}
\end{figure}

Monte Carlo descriptions of the perturbative shower should provide a similar description of collinear physics, but can differ in their description of soft wide angle radiation. Some of these differences were discussed in \Sec{sec:MC}.  As discussed earlier, because \vincia{} is a dipole-antenna shower, it should accurately describe both the hard collinear and soft wide-angle regions of phase space.  
Because small values of $D_2$ are sensitive to both collinear and soft physics, the fact that the \pythia{} distribution at small $D_2$ is distinct suggests that its description of soft wide-angle physics is the reason.\footnote{Part of the reason for why \pythia{} seems to not correctly describe the soft, wide-angle region of phase space may be due to the fact that while it uses kinematics of dipoles in its shower, it still uses the Altarelli-Parisi splitting functions as an approximation of the squared matrix element.  The dipole and its emission is then boosted to the appropriate frame, which may over-populate the soft wide-angle region of phase space as compared to the eikonal matrix element. We thank Torbj\"orn Sj\"ostrand and Peter Skands for detailed discussions of this point.}  The difference observed in our analytic calculation arising from the soft subjet region of phase space is similar to that observed between the $p_T$-ordered \vincia{} and \pythia{} Monte Carlo distributions. It is therefore interesting to investigate whether the difference in Monte Carlo distributions can arise exclusively from different descriptions of wide angle soft radiation.  

We will show that for the $D_2$ observable and jet samples we studied, most of the difference can be accounted for by differences in the treatment of unphysical infrared scales at parton level in the Monte Carlos.
Since we perform this comparison at parton level, there is some ambiguity in effects due to the perturbative cutoff of the shower, and those arising from different descriptions of wide angle soft radiation.  In particular, the Monte Carlos will in general have different low-scale $p_T$ cutoffs at which the perturbative parton shower is terminated.  Varying this scale can potentially greatly increase or decrease the number of soft emissions because the value of $\alpha_s$ in this region is large.  In particular, for the versions of \pythia{} and \vincia{} that were use to generate events in \Fig{fig:MC_compare}, the cutoff in \pythia{} is $0.4$ GeV, while the cutoff in \vincia{} is $0.8$ GeV. Indeed, these are the default values for these showers.  Therefore, we expect that the \pythia{} parton shower produces more soft emissions than \vincia{}, which would increase the value of $D_2$, and potentially also contribute to the observed difference.

To attempt to disentangle the effects of the shower cutoff from differences in the modeling of soft radiation, in \Fig{fig:cutoff_test}, we consider Monte Carlo predictions of the $D_2$ distribution as measured on QCD jets, with different jet radii, namely $R=0.5,0.7,1.0,1.2$. By using different jet radii, we can control the importance of the soft subjet region of phase space. With small jet radii, the soft subjet region of phase space does not exist, while it becomes increasingly important as the jet radius is increased.  Analytic predictions for the $D_2$ distribution for different values of the jet radius, $R$, will be given in \Sec{sec:R_dependence}. Here we compare the $p_T$-ordered \vincia{} and \pythia{} Monte Carlo. To generate the central values of the curves, we have used a cutoff of $0.8$ GeV, and the uncertainty bands are generated by varying this cutoff from $0.4$ GeV to $1.2$ GeV, to understand its effect. From \Fig{fig:cutoff_test} we see that while there is a relatively large uncertainty band from varying the perturbative cutoff of the shower, they do overlap for all jet radii studied.  This suggests that the dominant difference between the $D_2$ distributions from \vincia{} and \pythia{} is due to emissions at a scale near the parton shower cutoff.


This analysis also shows some of the difficulties in disentangling perturbative from non-peturbative effects, and the importance of having analytic calculations and precise theoretical control of different phase space regions to do so. However, by measuring sufficiently many observables on a jet, we are able to isolate distinct phase space regions and study in detail the extent to which Monte Carlo parton showers reproduce the physics in the different regions.  $D_2$, or similar jet substructure observables, could therefore be powerful tools for tuning Monte Carlos, both to formally-accurate perturbative calculations, as well as data. In the remaining sections of the paper, we will use the default shower cutoffs in the Monte Carlo generators, as was done in \Fig{fig:MC_compare}, and will not show uncertainty bands on our Monte Carlo distributions from varying this parameter.

\subsection{Analytic Jet Radius Dependence}\label{sec:R_dependence}

As demonstrated in the previous section, the region of small $D_2$ is a sensitive probe of the dominant soft or collinear structure in the jet.  It is therefore interesting to study the jet radius dependence of $D_2$ analytically, because the relative size of soft subjet and collinear subjets contributions to $D_2$ will depend on the jet radius.  At large jet radius, as shown earlier, the soft subjet region is an important contribution at small $D_2$, but as the jet radius decreases, the collinear subjets should dominate.  In this section, we will study the jet radius dependence of $D_2$ and compare our analytic calculation to Monte Carlo. This will also demonstrate that our analytic calculation accurately describes the $R$ dependence of the $D_2$ distribution.  As in the previous section, we will restrict this study to $p_T$-ordered \vincia{} and \pythia{} showers, and will take the jet radius to be $R=0.5,0.7,1.0,1.2$, which are representative of a wide range of values of experimental interest. Larger values of $R$ can be straightforwardly studied with our approach, but are of less phenomenological interest.  It is expected that for smaller values of $R$ logarithms of $R$ may become numerically important \cite{Seymour:1997kj,Tackmann:2012bt,Dasgupta:2014yra}, so we do not consider them here.

\begin{figure}
\begin{center}
\subfloat[]{\label{fig:Rdep_a}
\includegraphics[width= 7.0cm]{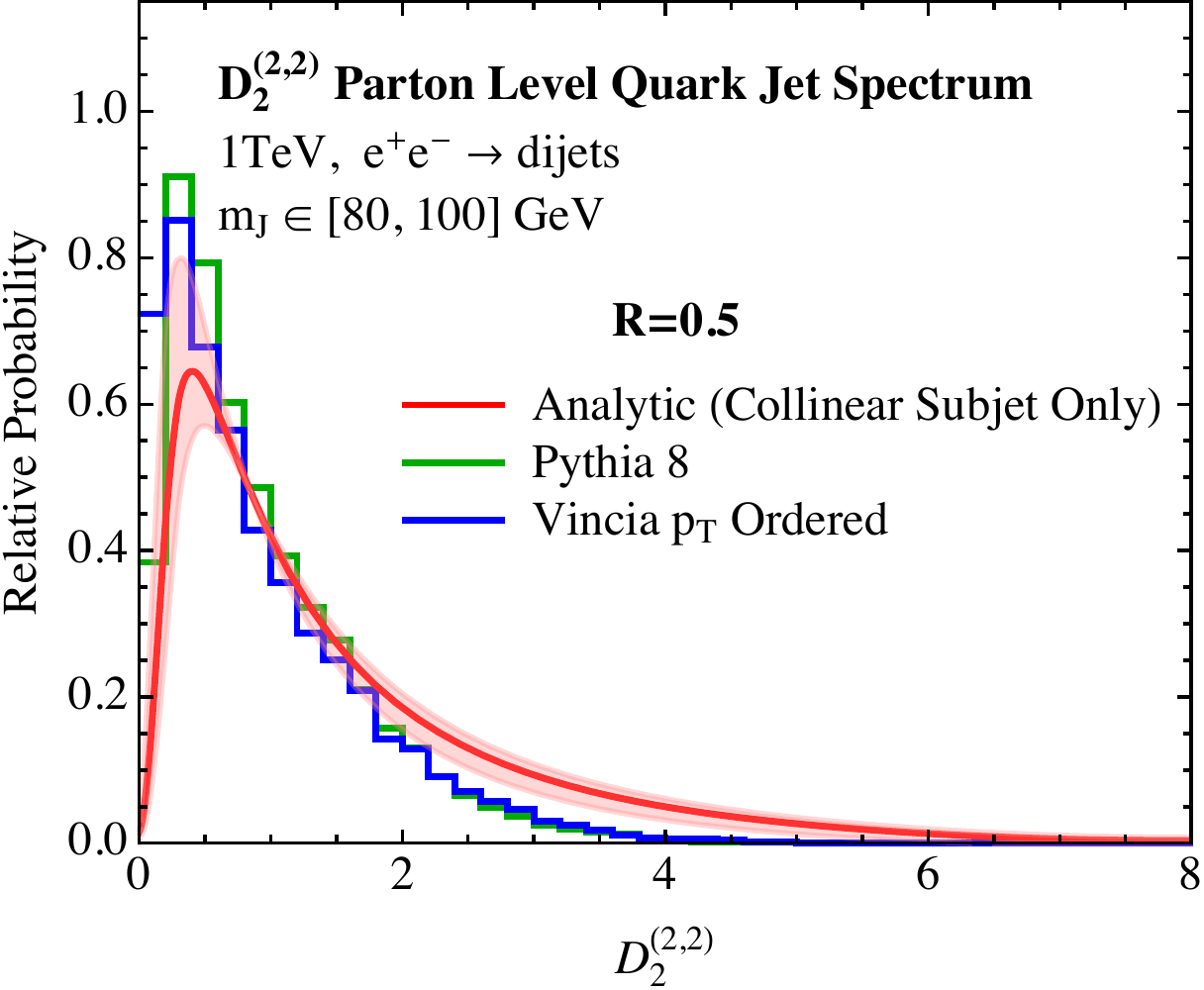}
}
\subfloat[]{\label{fig:Rdep_b}
\includegraphics[width = 7.0cm]{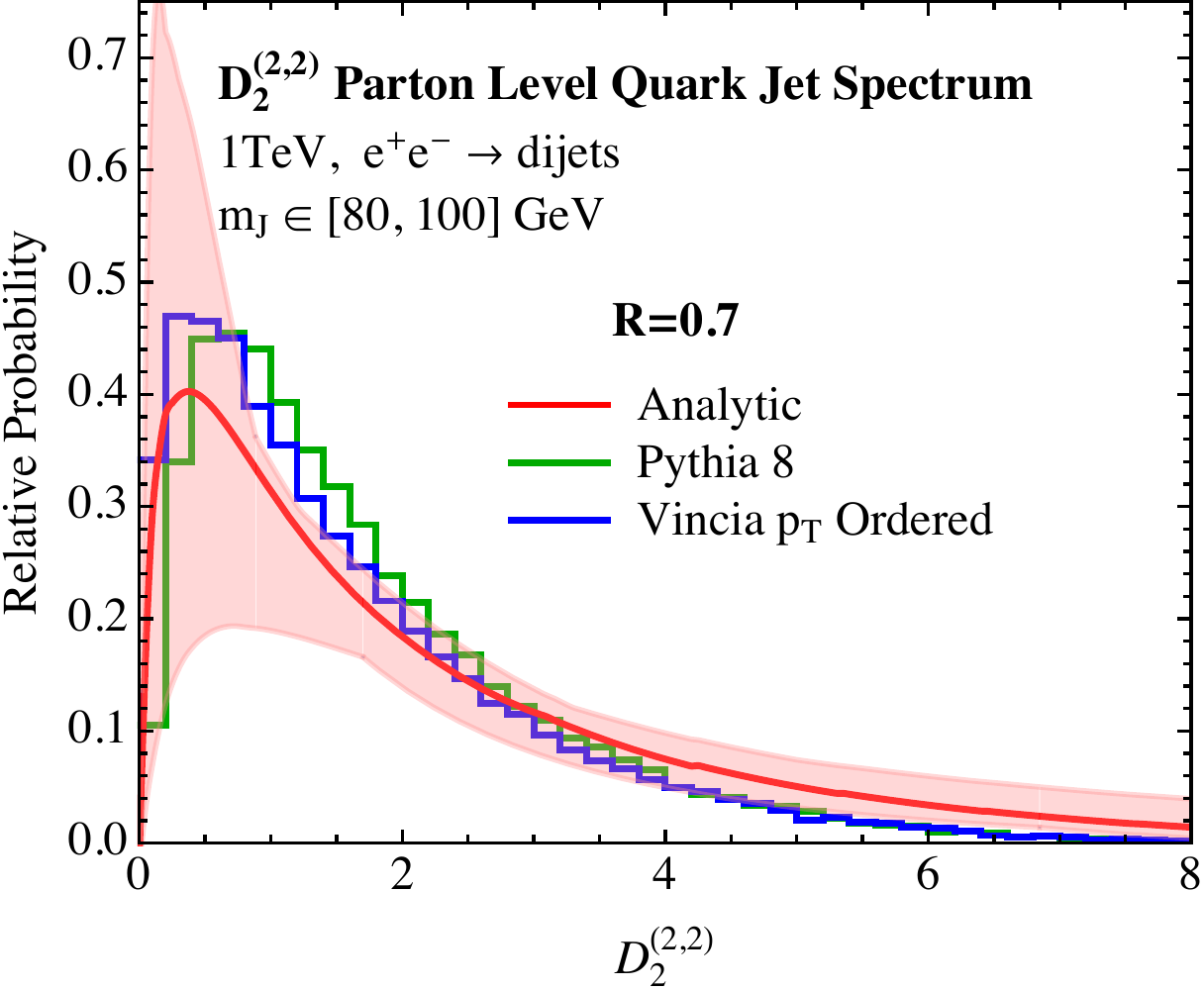}
}\\
\subfloat[]{\label{fig:Rdep_c}
\includegraphics[width= 7.0cm]{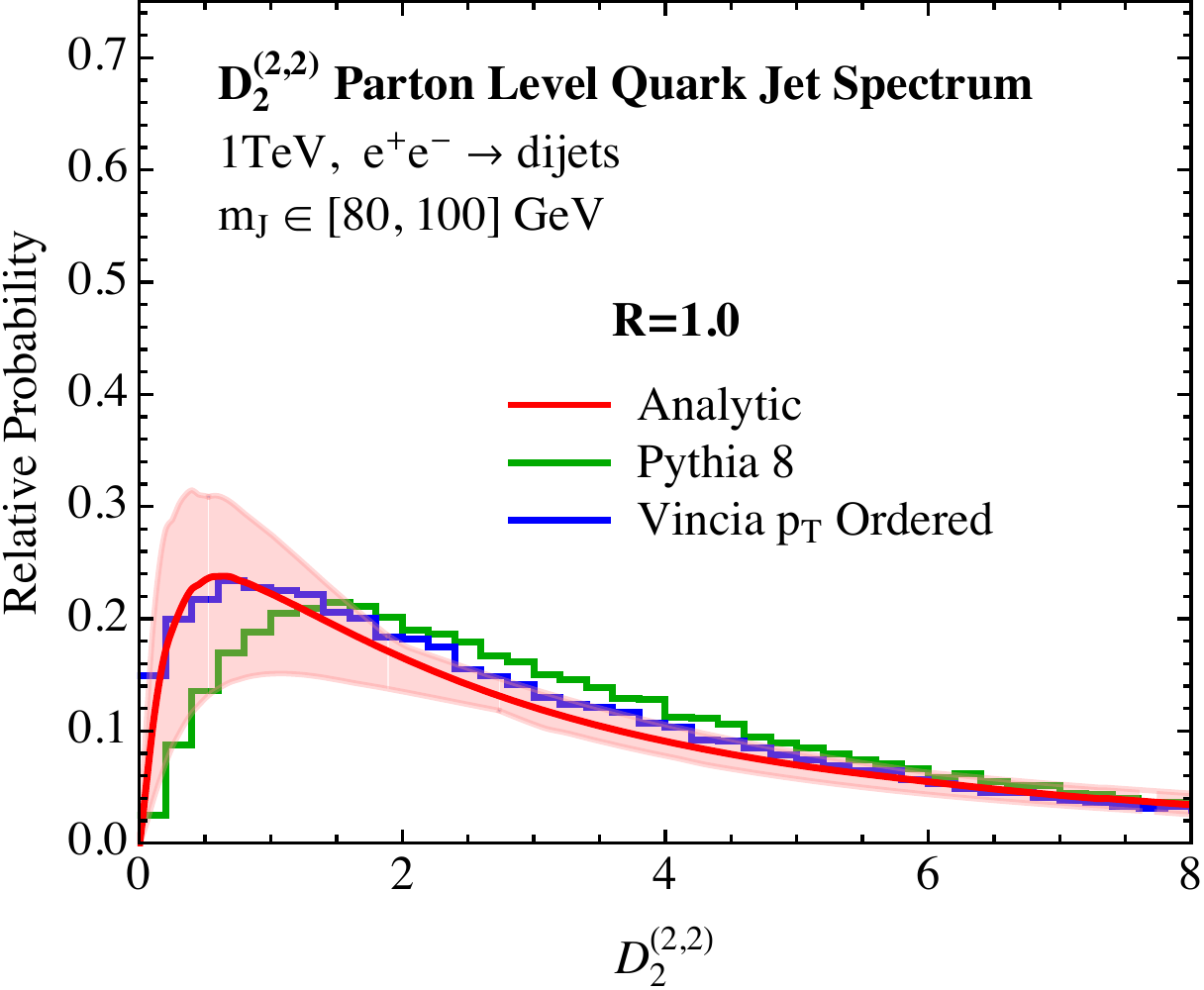}
}
\subfloat[]{\label{fig:Rdep_d}
\includegraphics[width = 7.0cm]{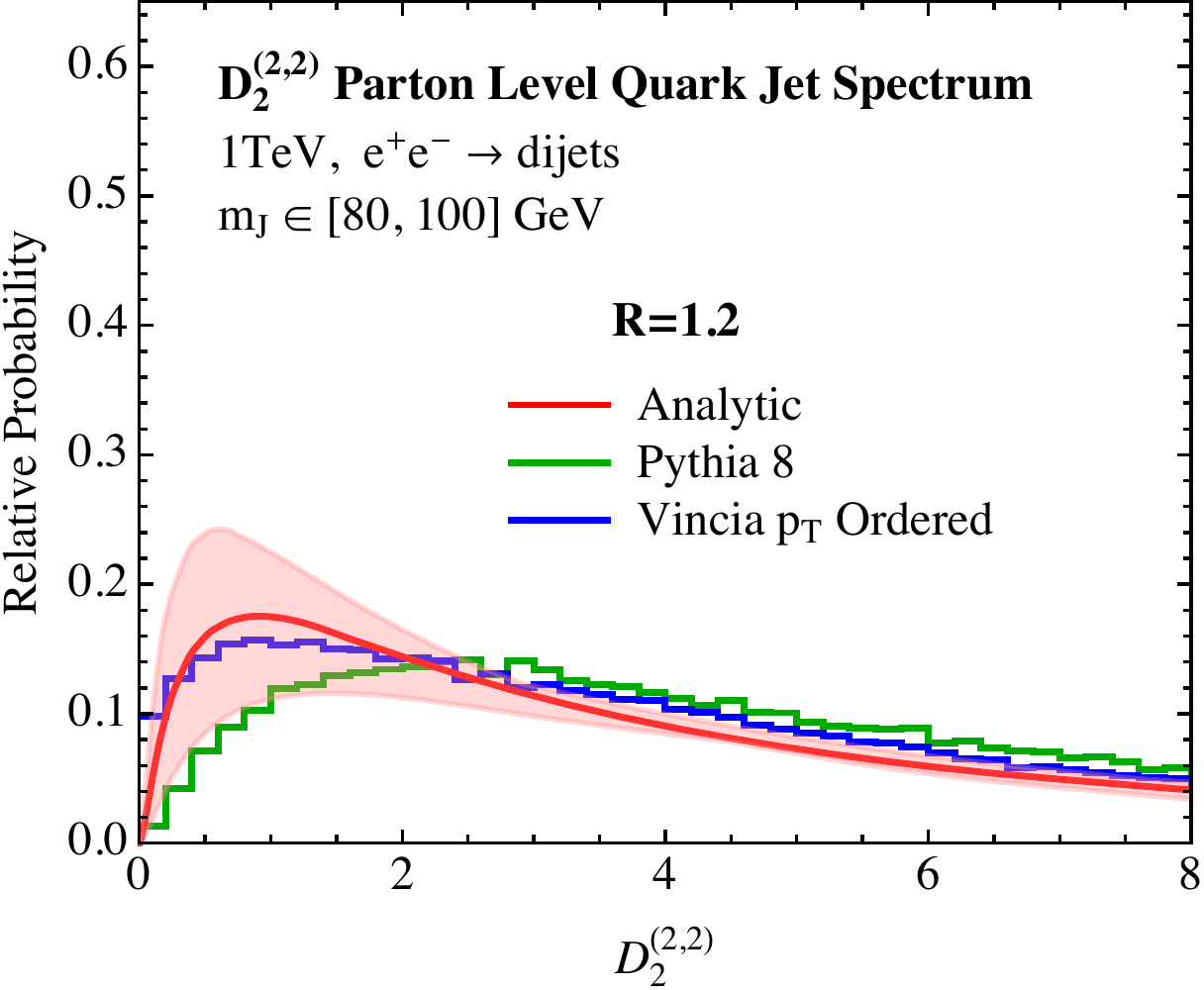}
}
\end{center}
\caption{ Comparison of QCD background $D_2$ distributions from $p_T$-ordered \vincia{} and \pythia{} to our analytic prediction as a function of the jet radius, $R$. The values $R=0.5,0.7,1.0,1.2$ are shown in Figures a)-d), respectively. In the analytic prediction for $R=0.5$, only the collinear subjets factorization theorem is used, while for all other values of the jet radius the analytic calculation includes contributions from both the collinear subjets and soft subjet factorization theorems. 
The pinch in the scale variations is a consequence of unit normalizing the distributions.
}
\label{fig:R_dependence}
\end{figure}

Comparisons of parton level Monte Carlo results from both $p_T$-ordered \vincia{} and \pythia{} to our analytic calculation are shown in \Fig{fig:R_dependence}. Since we scan over a range of jet radii, perturbative uncertainties for each $R$ value are not as extensively explored as earlier with $R=1$, and are only meant as a rough estimate of the perturbative uncertainty.  Our focus here is simply to show that the scaling behavior with $R$ between our analytic calculation and the Monte Carlos agree. 
There is excellent agreement between the Monte Carlo results and our analytic calculations over the entire range of $R$ values.
For $R\gtrsim 1$, there is some 
disagreement in the position of the peak of the distribution between the generators, though, as shown earlier, this can be accounted for by adjusting the shower cutoffs. 
In the peak region, hadronization will play an important role, smearing out differences between Monte Carlos.
The effect of hadronization, and its implementation in our analytic calculation, will be discussed in \Sec{sec:Hadronization}.

For jet radii of $R=0.7,1.0,1.2$ our analytic calculation consists of both collinear subjets and soft subjet contributions.  For $R=0.5$, however, we only include the contribution from collinear subjets, which is guided by our matching procedure between the collinear subjets and soft subjet factorization theorems, as discussed in \Sec{sec:soft_NINJA_match}.  For a fixed jet mass, as the value of $R$ is decreased, the region of validity of the soft subjet factorization theorem vanishes rapidly.  For jet masses in the range $80 < m_J < 100$ GeV, and $Q=1$ TeV, we find that between $R=0.7$ and $R=0.5$  the region of validity of the soft subjet rapidly shrinks to zero, and there should not be a transition between the collinear subjets factorization theorem and the soft subjet factorization theorem. Because of this, for the value of $R=0.7$, our perturbative error bands are more extensive, and are taken as the envelope of curves both that include the matched soft jet, and curves that do not. While this is certainly over conservative in the error estimate, we have included this to emphasize this point. This feature is also seen explicitly in the plots of \Fig{fig:R_dependence}, where the region of disagreement between the different Monte Carlo generators is squeezed towards zero. A similar effect occurs as the energy (or $p_T$) of the jet is increased with a fixed jet mass, which will be discussed in \Sec{sec:jet_energy}.

Throughout the remainder of this paper, we will study the case $R=1$ exclusively, because both collinear subjets and soft subjet regions of phase space must be included and that radius is relevant to a large number of jet substructure studies using fat jets.

\subsection{Analytic Jet Energy Dependence}\label{sec:jet_energy}

In addition to studying the dependence on the jet radius as a probe of the importance of the soft subjet and of the Monte Carlo description of the shower, it is also interesting to study the dependence of the $D_2$ distribution on the energy of the jet, with a fixed mass cut. For highly energetic jets, one expects that the soft subjet will play a negligible role, as the region of validity of the soft subjet factorization theorem shrinks as the energy of the jet is increased, as long as the mass of the jet is kept fixed. On the other hand, since we keep the jet radius used in the clustering fixed, the angular separation of the collinear particles decreases with energy, but the phase space for wide angle global soft radiation increases considerably.  This radiation is present both in the collinear subjets and soft haze factorization theorems. It is also of course present in the soft subjet factorization theorem, although we have argued that we expect this to give a small contribution. Studying the jet energy dependence therefore probes the behavior of the generators in a fashion complementary to the $R$ dependence.

In this section, we study the perturbative $D_2$ distribution for center of mass energies ranging from $500$ GeV to $2$ TeV, for a fixed jet radius of $R=1$, and with a fixed mass cut of $80<m_J<100$ GeV. This region of energies covers the majority of the phenomenologically interesting phase space available at the LHC. We will also perform a more detailed study at LEP energies in \Sec{sec:LEP}. For our resummation, we require (amongst other things), that $\ecf{2}{\alpha} \ll 1$. For the case of $\alpha=2$ for which we will be most interested, this corresponds to the assumption $\ecf{2}{2} = m_J^2/E_J^2 \ll 1$. For a mass cut around the $Z$ pole mass, this expansion is valid throughout the range of energies we consider.   The case when $\ecf{2}{2}\lesssim 1$, but not parametrically so, is outside the scope of this paper.

\begin{figure}
\begin{center}
\subfloat[]{\label{fig:Edep_a}
\includegraphics[width= 7.0cm]{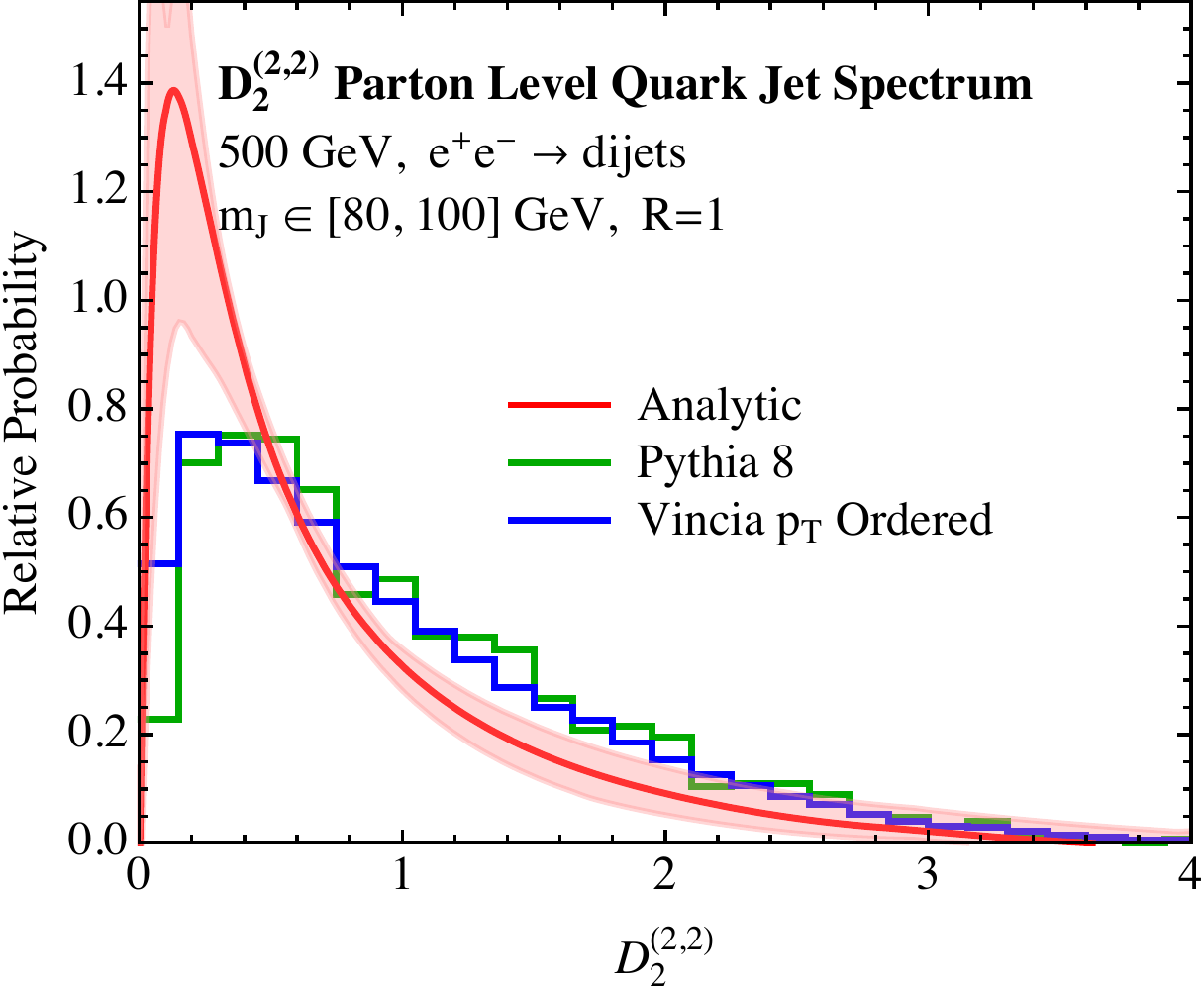}
}
\subfloat[]{\label{fig:Edep_b}
\includegraphics[width = 7.0cm]{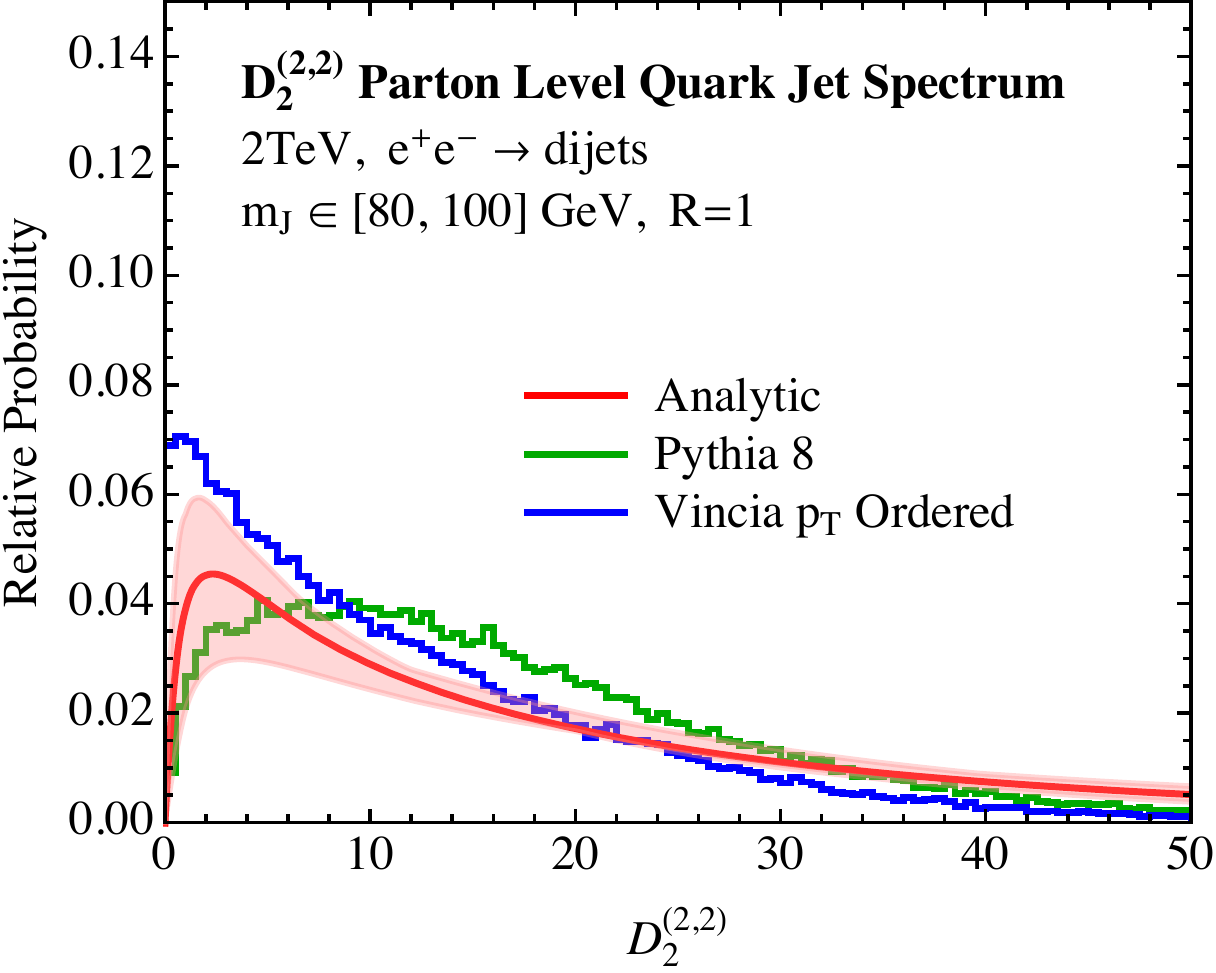}
}
\end{center}
\caption{ Comparison of QCD background $D_2$ distributions from $p_T$-ordered \vincia{} and \pythia{} to our analytic prediction as a function of the jet energy, $E_J$. The values $E_J=500$ GeV, and $2$ TeV are shown in Figures a) and b), respectively. A jet radius of $R=1$ is used for all values of the jet energy.
The pinch in the scale variations is a consequence of unit normalizing the distributions.
}
\label{fig:E_dependence}
\end{figure}

In \Fig{fig:E_dependence} we show distributions for the $D_2$ observable as obtained from Monte Carlo simulation, and compared with our analytic calculation. As in \Sec{sec:R_dependence}, we restrict to $p_T$-ordered \vincia{} and \pythia{} at parton level. The perturbative scale variations for each energy value are less extensively explored and are only meant to provide a rough estimate of the perturbative uncertainty.  The evolution of the difference between the \vincia{} and \pythia{} generators is again quite fascinating, with the discrepancy between the generators increasing significantly with energy, to the point that at $2$ TeV the qualitative shape of the distributions doesn't agree. In particular, the behavior at small $D_2$ is completely different between the two generators, with \vincia{} having a large peak, which is not present in \pythia{}.

As discussed in \Sec{sec:scales}, this discrepancy between the generators is dominantly due to differences in the treatment of the parton shower cutoff.  As the energy is increased with a fixed jet mass and jet radius, emissions that contribute to $D_2$ are forced to have smaller and smaller energy.
As evidence that this is indeed the cause of the discrepancy, we have checked that the conclusions of \Sec{sec:R_dependence} remains true at higher energy, as long as the jet radius is taken to scale as $R \sim 2m_J/p_T$, so that it constrains the wide angle soft radiation. For example, for $R=0.2$ at $2$ TeV, we find excellent agreement between the $D_2$ distributions as generated by \pythia{} and \vincia{}.\footnote{We include this plot in \App{app:twoemissionMC}, \Fig{fig:D2_app_R021}, for reference.} 
Because the fact that the disagreement is so large between the generators, and is arising from the modeling of soft radiation, this may be an excellent observable to study soft radiation and color coherence in parton showers.

As a reference, in \App{app:twoemissionMC} we show distributions of the $\ecf{2}{2}$ observable, measured at both $500$ GeV, and $2$ TeV for both the \vincia{} and \pythia{} Monte Carlos, and at both parton and hadron level. Unlike for the $D_2$ observable, since $\ecf{2}{2}$ is set by a single emission, excellent agreement is observed for the $\ecf{2}{2}$ observables between \pythia{} and \vincia{} both at parton level, emphasizing that $D_2$ offers a more differential probe of the perturbative shower than single emission observables.

Our analytic predictions at $2$ TeV, as shown in \Fig{fig:E_dependence}, are intermediate between the \pythia{} and \vincia{} results. They exhibit a peaked structure at small values of $D_2$, but not to the extent seen in the \vincia{} distribution. We believe that this is largely due to the normalization of the distributions, and the fact that we do not match to fixed order in the tail of the distribution. Since this tail becomes longer at higher energies, a larger disagreement in the peak region is also seen. 
On the other hand, at $500$ GeV, our analytic prediction has a large peak. This is evidence that because the $D_2$ spectrum is much more sharply peaked at $500$ GeV, higher order resummation may be more important in the peak region. However, the relatively good agreement between analytics and Monte Carlo shows that our factorization theorem is able to accurately capture the energy dependence over a large range of energies.

It is important to note that hadronization will remove some of the discrepancies in the $D_2$ distributions between the \vincia{} and \pythia{} generators, especially at high energies, where it will smear out the peak at low values of $D_2$. While this improves qualitatively the behavior of the distributions, discrepancies in the shape still remain. This will be discussed in detail in \Sec{sec:Hadronization}, along with its incorporation into our analytic calculation.  For comparison to precision analytic calculations and interpreting data, it is vital that Monte Carlo generators provide accurate descriptions of both the perturbative and non-perturbative aspects of QCD jets, and not compensating for perturbative discrepancies by the tuning of non-perturbative parameters. This is especially important for disentangling non-perturbative effects from perturbative effects, the latter of which should in principle be under much better control, and for extracting reliable information about non-perturbative QCD from jet physics.

Throughout the rest of the paper we will focus on jets with radius $R=1$ at a center of mass energy of $1$ TeV.

\subsection{Impact of Hadronization}\label{sec:Hadronization}

Hadronization plays an important role in a complete description of any jet observable, and a description of non-perturbative effects, preferably from a field-theoretic approach, is required to compare with experimental data. An advantage of the factorization approach taken in this paper is that it allows for a clean separation of perturbative and non-perturbative physics. Non-perturbative effects enter the factorization theorems presented in \Sec{sec:Fact} through the soft function, which describes the dynamics of soft radiation, both perturbative and non-perturbative, between the jets. For a large class of additive observables, the treatment of non-perturbative physics in the soft function is well-understood, and can be incorporated using shape functions \cite{Korchemsky:1999kt,Korchemsky:2000kp,Bosch:2004th,Hoang:2007vb,Ligeti:2008ac}. Shape functions have support over a region of size $\Lambda_{\text{QCD}}$, and are convolved with the perturbative soft function. In the tail region of the distribution, where the observable is dominated by perturbative emissions, they reduce to a shift.  For a large class of observables, this shift is determined by a universal \cite{Akhoury:1995sp,Dokshitzer:1995zt} non-perturbative parameter multiplied by a calculable, observable dependent number \cite{Dokshitzer:1995zt,Lee:2006fn,Lee:2007jr}. Similar shape functions have also been used to incorporate the effects of pile-up and the underlying event at hadron colliders \cite{Stewart:2014nna}.

The effect of non-perturbative physics on multi-differential cross sections has not been well-studied. For the double differential cross section of two angularities,  \Ref{Larkoski:2013paa} considered using uncorrelated shape functions for each angularity individually, but it is expected that a complete description would require a shape function incorporating non-perturbative correlations between observables. For the particular case of the $D_2$ observable, we will argue that a single parameter shape function can be used to accurately describe the dominant non-perturbative effects, and in particular, that a study of multi-differential shape functions with non-perturbative correlations, is not required.  Of course, to justify the use of a shape function requires the observable in question to be infrared and collinear safe.  Therefore, we will only consider non-perturbative corrections to $D_2$ in the presence of a mass cut on the jet.

In \Sec{sec:fixed_order} we performed a study of the fixed order singular structure of the $D_2$ observable in the presence of a jet mass cut. Importantly, we showed that $D_2$ only has a singularity at $D_2=0$, with its behavior at all other values regulated by the mass cut. Non-perturbative corrections to the $D_2$ observable will play an important role only when the soft scale becomes non-perturbative, which as just argued, for a perturbative mass cut of the form studied in this paper, only occurs as $D_2 \to 0$. Recall that the $D_2$ observable is defined as 
\begin{equation}
\Dobs{2}{\alpha, \beta}= \frac{\ecf{3}{\alpha}}{(\ecf{2}{\beta})^{3\alpha/\beta}}\,,
\end{equation}
which is not additive. However, in the two-prong region of phase space, namely $D_2\to 0$, the value of $\ecf{2}{\beta}$ is set to leading power by the hard splitting, and so $D_2$ effectively reduces to an additive observable. In this region of phase space the description of non-perturbative effects in terms of a shape function can therefore be rigorously justified from our factorization theorem, and it can be applied directly to the $D_2$ distribution.  For large values of $D_2$, it is not additive, and the use of a shape function cannot be formally justified. However, in this region, a shape function is not required, as any singular behavior is regulated by a mass cut. We therefore will use a shape function that falls off exponentially at large values of $D_2$.  We believe that this is a self-consistent approach until non-perturbative corrections to multi-differential cross sections are better understood.

In the two-prong region of phase space, we have shown that two distinct factorization theorems, namely the soft subjet and collinear subjets, are required, and in \Sec{sec:soft_NINJA_match} we showed how these two descriptions can be merged to provide a complete description of the two-prong region of phase space. Importantly, the two factorization theorems describing the two-prong region of phase space have soft functions with different numbers of Wilson lines. The collinear subjets soft function is a two-eikonal line soft function, while the soft subjet soft function has three eikonal lines. Since the shape function describes the non-perturbative contribution to the soft function, in general we should allow for two distinct shape functions, with independent parameterizations. The zero-bin merging procedure in \Sec{sec:soft_NINJA_match} would then be performed on the non-perturbative cross sections, after convolution with the appropriate shape function. However, at the level of perturbative accuracy which we work, and because we will simply be extracting our shape function parameters by comparing to Monte Carlo, the use of distinct parameterizations of different shape functions for both the soft subjet and collinear subjets soft functions would introduce many redundant parameters. To simplify the situation in this initial investigation, we will choose to use the same parametrization of the shape function, and the same non-perturbative parameters for both soft functions. This allows for the non-perturbative corrections to be described by a single parameter, and as we will see provides an excellent description of the Monte Carlo data. Because we use the same shape function for both the soft subjet, and collinear subjets soft functions, it also implies that the shape function can be applied after the zero bin merging procedure, namely, directly at the level of the $D_2$ distribution.

As a simple parametrization of a shape function for $D_2$, we follow \Ref{Stewart:2014nna} and consider
\begin{align}\label{eq:shape_func}
F(\epsilon)=\frac{4\epsilon}{\Omega_D^2}e^{-2\epsilon/\Omega_D}\,,
\end{align}
where $\epsilon$ is the energy and $\Omega_D\sim \Lambda_{\text{QCD}}$ is a non-perturbative scale. Note that while we will use the same value of $\Omega_D$ for the signal and background distributions, it will have very different effects on the two distributions, which will arise naturally from the power counting in the different factorization theorems, as will be shown in this section. The function of \Eq{eq:shape_func} satisfies the required properties that it is normalized to $1$, has a finite first moment $\Omega_D$, vanishes at $\epsilon=0$, and falls off exponentially at high energies \cite{Hoang:2007vb}. More general bases of shape functions are discussed in \Ref{Ligeti:2008ac}, although we find that the single parameter shape function of \Eq{eq:shape_func} is sufficient to describe the dominant effects of hadronization.  

As discussed above, we will use the shape function of \Eq{eq:shape_func} for both the collinear subjets and soft subjets factorization theorems, with the same value of $\Omega_D$ in both cases. Because we have enforced this simplification to reduce the number of parameters, it is then most interesting to focus on $\Omega_D$ for the collinear subjets factorization theorem, which has two eikonal lines. In this case, we can show that we can relate the $\Omega_D$ parameter to universal non-perturbative parameters appearing in $e^+e^-\to$ dijet factorization theorems, which have been measured in experiment. Therefore, throughout the rest of this section, we will focus on deriving scaling relations for $\Omega_D$, assuming we are working in the collinear subjets factorization theorem. Again, we wish to emphasize that this is merely a simplification we have made to reduce the number of parameters, and a more general treatment could be performed, but we will see that with only the single $\Omega_D$, with properties derived assuming the collinear subjets factorization theorem, excellent agreement with Monte Carlo is observed.

The effect of non-perturbative physics as modeled by the shape function is very different for background or signal distributions.  For background, when $D_2$ is small, the contribution to $\ecf{3}{\alpha}$ from a non-perturbative soft emission is
\begin{equation}
\left.\ecf{3}{\alpha}\right|_\text{np}\sim \frac{\epsilon}{E_J} \ecf{2}{\alpha}\,,
\end{equation}
where $\epsilon$ is the energy of the non-perturbative emission and $E_J$ is the energy of the jet, as shown in \Eq{eq:soft_scale_coll}.  The non-perturbative contribution to $D_2$ is therefore
\begin{equation}
\left.\Dobs{2}{\alpha,\beta}\right|_\text{np} = \frac{\left.\ecf{3}{\alpha}\right|_\text{np}}{(\ecf{2}{\beta})^{3\alpha/\beta}} \sim \frac{\epsilon}{E_J}\frac{ \ecf{2}{\alpha}}{(\ecf{2}{\beta})^{3\alpha/\beta}} \,.
\end{equation}
In terms of the shape function, the non-perturbative distribution of $D_2$ for background jets can then be written as a convolution:\footnote{In this initial investigation we do not include a gap in our shape function, which would implement a minimum hadronic energy deposit, as expected physically \cite{Hoang:2007vb}. Such gapped shape functions, and their associated renormalon \cite{Beneke:1998ui} ambiguity \cite{Gardi:2000yh} have been studied for arbitrary angular exponents \cite{Hornig:2009vb}, and could be straightforwardly incorporated in our analysis. However, we observe excellent agreement with our single parameter shape function, which we therefore find to be sufficient for our purposes.   }
\begin{equation}\label{eq:shape_bkgd}
\frac{d\sigma_{\text{np}}  }{d\Dobs{2}{\alpha, \beta}}=\int_0^\infty d\epsilon \,   F(\epsilon)\,\frac{d\sigma_{\text{p}} \left ( \Dobs{2}{\alpha, \beta}-\frac{\epsilon}{E_J}\frac{\ecf{2}{\alpha}}{(\ecf{2}{\beta})^{3\alpha/\beta}} \right)  }{d\Dobs{2}{\alpha, \beta}}  \,,
\end{equation}
where $\sigma_{\text{np}}$ and $\sigma_{\text{p}}$ denote the non-perturbative and perturbative cross sections, respectively.

We can estimate the scale at which the global softs of the collinear subjets factorization theorem become non-perturbative from the scaling of the modes given in \Eq{eq:cs_collinear_and_soft}. Rewriting this scaling in terms of the center of mass energy of the $e^+e^-$ collision, $Q$, and $D_2$, we find that the global soft scale of the collinear subjets factorization theorem has virtuality
\begin{align}
\mu_S=2^3\, D_2\, m_Z \left(   \frac{m_Z}{Q} \right)^3\,,
\end{align}
where we have assumed a jet mass, $m_J=m_Z$, as relevant for boosted $Z$ discrimination. Taking $\Lambda_{\text{QCD}}=500$ MeV, we find that the global soft scale enters the non-perturbative regime at $D_2 \simeq 1$.   

Restricting to $\beta = 2$, in the collinear subjets region of the background jet phase space, the non-perturbative distribution of $\Dobs{2}{\alpha, 2}$ is then
\begin{equation}\label{eq:shape_function_eq1}
\frac{d\sigma_{\text{np}}  }{d\Dobs{2}{\alpha, 2}}=\int_0^\infty d\epsilon \,   F(\epsilon)\,\frac{d\sigma_{\text{p}} \left ( \Dobs{2}{\alpha, 2}-2^{\alpha -2}\frac{\epsilon}{E_J}\frac{E_J^{2\alpha}}{  m_J^{2\alpha}  } \right)  }{d\Dobs{2}{\alpha, 2}}  \,,
\end{equation}
where we have used 
\begin{equation}
\ecf{2}{2} = \frac{m_J^2}{E_J^2} \,,
\end{equation}
and that, in the collinear subjets region of phase space,
\begin{equation}
\ecf{2}{\alpha} \simeq 2^{\alpha-2}\left(
\ecf{2}{2}
\right)^{\alpha/2}\,.
\end{equation}
Because we consider fixed-energy jets with masses in a narrow window, $\ecf{2}{2}$ is just a number and can be removed by appropriate change of variables.  Making this change, we then have
\begin{equation}\label{eq:np_background}
\frac{d\sigma_{\text{np}}  }{d\Dobs{2}{\alpha, 2}}=\int_0^\infty d\epsilon \,   F(\epsilon)\,\frac{d\sigma_{\text{p}} \left ( \Dobs{2}{\alpha, 2}-\frac{\epsilon}{E_J} \right)  }{d\Dobs{2}{\alpha, 2}}  \,,
\end{equation}
where the non-perturbative parameter in the shape function is effectively modified to
\begin{equation}
\tilde \Omega_D =  2^{\alpha - 2}\frac{    \Omega_D }{ \frac{m_J^{2\alpha}}{E_J^{2\alpha}}  }\,.
\end{equation}

The non-perturbative parameter $\Omega_D$ still has implicit dependence on the angular exponent $\alpha$.  Because the global soft modes have the lowest virtuality and can only resolve the back-to-back soft Wilson lines in the $n$ and $\bar n$ directions, we can use the results of \Refs{Lee:2006fn,Lee:2007jr} to extract the $\alpha$ dependence.  By the boost invariance of the soft function\footnote{This boost invariance holds strictly only for a soft function with no jet algorithm restrictions. However, since we are considering fat jets close to hemispherical, we expect corrections to the boost invariance of the soft function to be small.} along the $n-\bar n$ directions and the form of the observable $\ecf{3}{\alpha}$ as measured on soft particles, it follows that $\Omega_D$ takes the form
\begin{equation}\label{eq:alpha_np}
\Omega_D = \frac{3}{2\alpha-1}\Sigma \,,
\end{equation}
where $\Sigma$ is a universal non-perturbative matrix element of two soft Wilson lines and all dependence on $\alpha$ has been extracted.\footnote{In this section we ignore the effects of hadron masses, and their associated power corrections of $\mathcal{O}(m_H/Q)$, where $m_H$ is the mass of a stable hadron in the jet. While these power corrections can also be incorporated through the shape function, in general, they break the universality of the non-perturbative matrix element, $\Sigma$ \cite{Salam:2001bd,Mateu:2012nk}. In particular, \Eq{eq:alpha_np} is no longer in general true, for a $\Sigma$ that is independent of the angular exponent $\alpha$ \cite{Salam:2001bd,Mateu:2012nk}. This depends on the precise definition of the energy correlation functions for massive particles. However, the value of $\Sigma$ can still be extracted from dijet event shapes in the same universality class as a particular angularity \cite{Mateu:2012nk}. Furthermore, $\Omega_D$ has a scale dependence from renormalization group evolution, $\Omega_D=\Omega_D(\mu)$, although this dependence is logarithmic, and is therefore small compared to our uncertainties. We will discuss briefly the impact of hadron masses and the renormalization group evolution of $\Omega_D$ in \Sec{sec:LEP}, and in \App{app:shape_RGE}.}  We have normalized the matrix element such that the coefficient is unity for $\alpha = 2$. We will shortly discuss the extent to which the values of $\Omega_D$ we obtain from comparison with the parton shower agree with the known values of this universal non-perturbative matrix element.

For signal jets, the lowest virtuality mode in the jet are the collinear-soft modes. Unlike the global soft modes of the collinear subjets factorization theorem, which did not resolve the substructure of the jets, allowing us to relate the non-perturbative parameter appearing in the shape function to that appearing in dijet event shapes, the collinear soft modes in the signal factorization theorem resolve the jet substructure. However, since the decaying boson is a color singlet, there are still only two eikonal lines present in the factorization theorem. Boost invariance of the soft function will therefore again allow us to relate the non-perturbative parameter for the signal distribution to that appearing in dijet event shapes. This is similar to the argument used in \Ref{Feige:2012vc} to calculate the signal distribution for $2$-subjettiness.

A non-perturbative collinear-soft emission contributes to $\ecf{3}{\alpha}$ as
\begin{equation}
\left.\ecf{3}{\alpha}\right|_\text{np}\sim \frac{\epsilon}{E_J} (\ecf{2}{\alpha})^3\,,
\end{equation}
where now $\epsilon$ is the energy of the non-perturbative collinear-soft emission, as shown in \Eq{eq:cs_cs}.  The non-perturbative contribution to $D_2$ for signal jets is therefore
\begin{align}
\left.\Dobs{2}{\alpha,\beta}\right|_\text{np} &= \frac{\left.\ecf{3}{\alpha}\right|_\text{np}}{(\ecf{2}{\beta})^{3\alpha/\beta}} \sim \frac{\epsilon}{E_J}\frac{ (\ecf{2}{\alpha})^3}{(\ecf{2}{\beta})^{3\alpha/\beta}} \\
&\simeq 2^{3(\alpha-\beta)}\frac{\epsilon}{E_J}\nonumber \,,
\end{align}
where in the second line we have used the parametric relationship between $\ecf{2}{\alpha}$ and $\ecf{2}{\beta}$ in the collinear subjets region.
Convolving with the shape function, the non-perturbative distribution for signal jets is then
\begin{equation}\label{eq:np_signal}
\frac{d\sigma_{\text{np}}  }{d\Dobs{2}{\alpha, \beta}}=\int_0^\infty d\epsilon \,   F(\epsilon)\,\frac{d\sigma_{\text{p}} \left ( \Dobs{2}{\alpha, \beta}-2^{3(\alpha-\beta)}\frac{\epsilon}{E_J} \right)  }{d\Dobs{2}{\alpha, \beta}}  \,.
\end{equation}

It is important to note how the different scales for the soft radiation in the case of the signal and background jets leads to different behavior of the $D_2$ distributions after hadronization. In particular, from \Eqs{eq:np_background}{eq:np_signal} one can determine the shift in the first moment of the $D_2$ distribution caused by hadronization, which we will denote by $\Delta_D$. Restricting to the case $\alpha=\beta=2$ for simplicity, we find that for the background distribution, 
\begin{equation}\label{eq:bkg_shift_np}
\Delta_D=\frac{\Omega_D}{ E_J \left(  \frac{m_J}{E_J} \right)^4}\,,
\end{equation}
whereas for the signal jets, we have
\begin{equation} \label{eq:signal_nonpert}
\Delta_D=\frac{\Omega_D}{ E_J }\,.
\end{equation}
Since $\Omega_D$ should be of the scale $1$ GeV, we see that for signal jets, the shift in the first moment due to hadronization is highly suppressed, and behaves differently than a traditional event shape due to the boost factor, while for background jets, since $m_J \ll E_J$, the effect of hadronization is significant. We will see that both of these features, which are consequences of the power counting of the dominant modes, are well reproduced in the Monte Carlo simulations.

\begin{figure}
\begin{center}
\subfloat[]{\label{fig:D2_ee_hadbkg_a}
\includegraphics[width= 7.25cm]{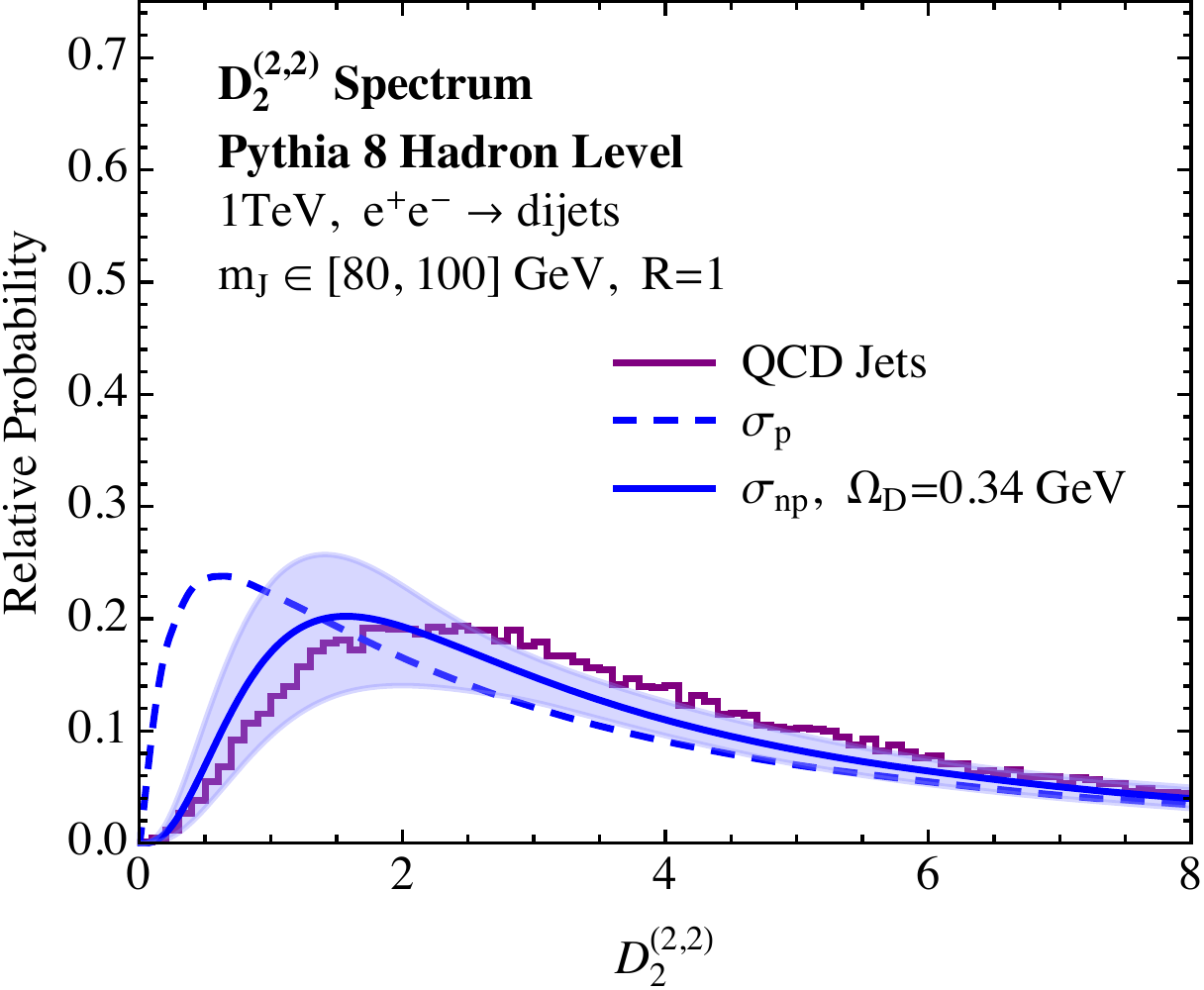}
}
\subfloat[]{\label{fig:D2_ee_hadbkg_b}
\includegraphics[width= 7.25cm]{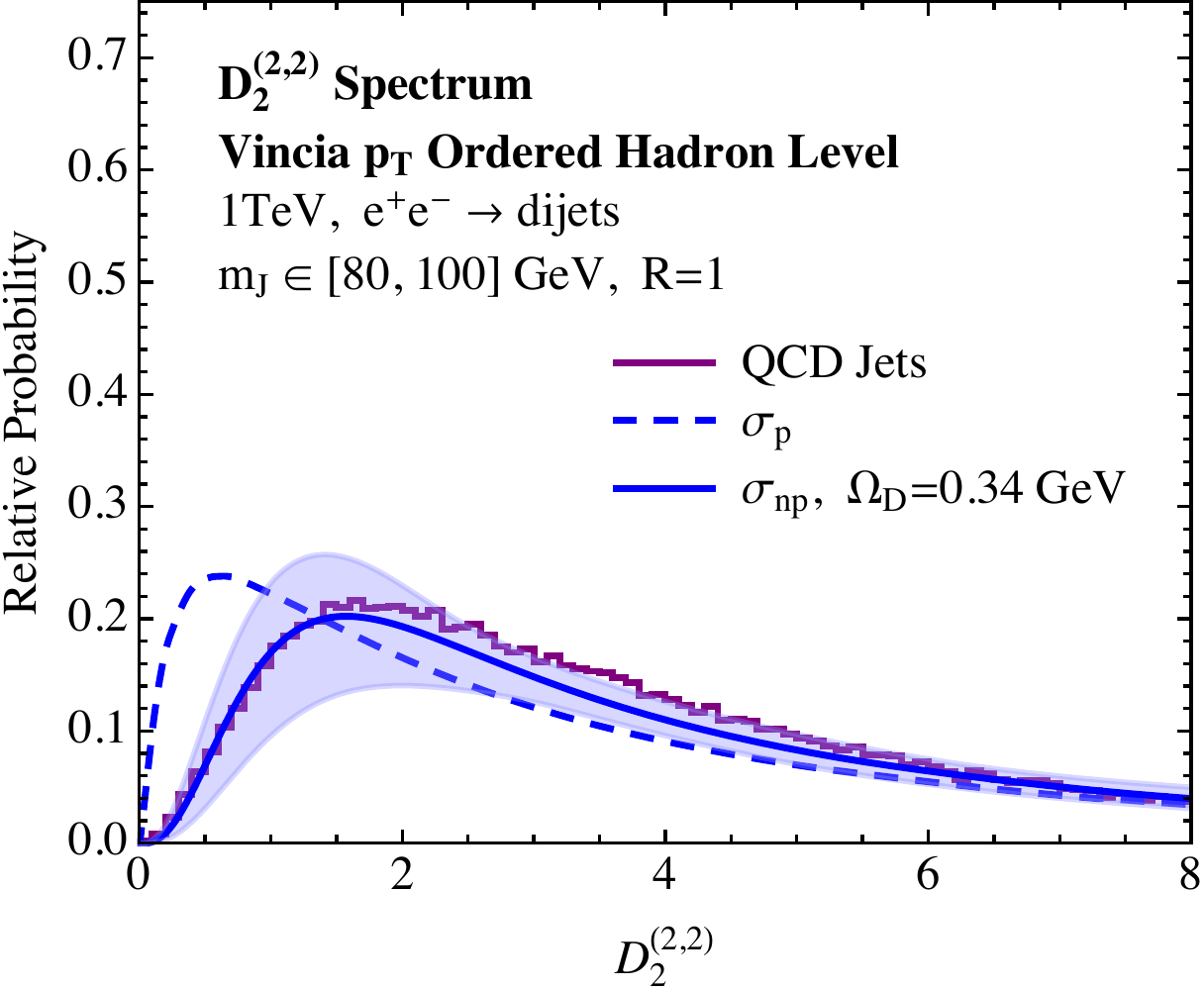}
}\\
\subfloat[]{\label{fig:D2_ee_hadbkg_c}
\includegraphics[width= 7.25cm]{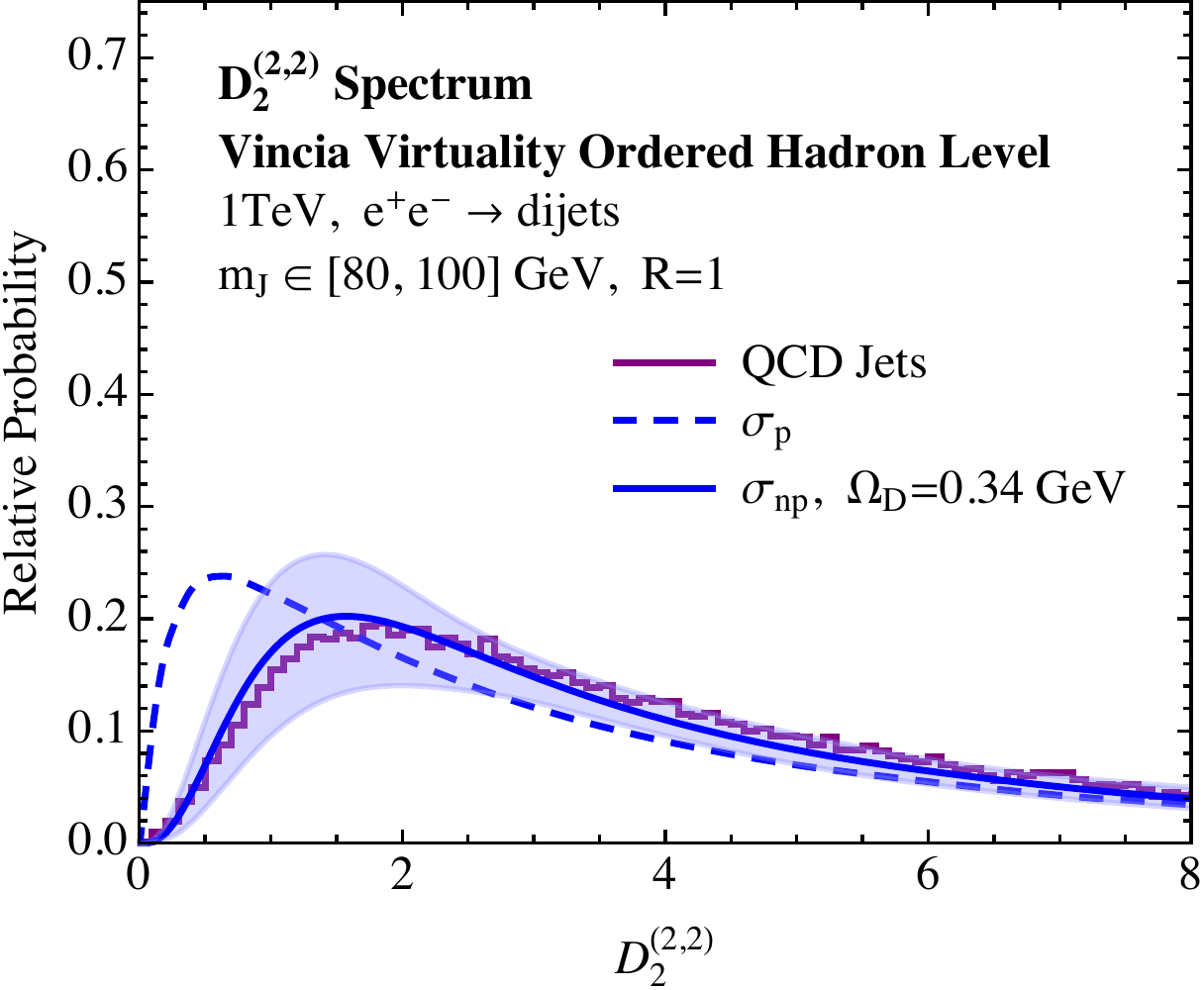}
}
\subfloat[]{\label{fig:D2_ee_hadbkg_d}
\includegraphics[width= 7.25cm]{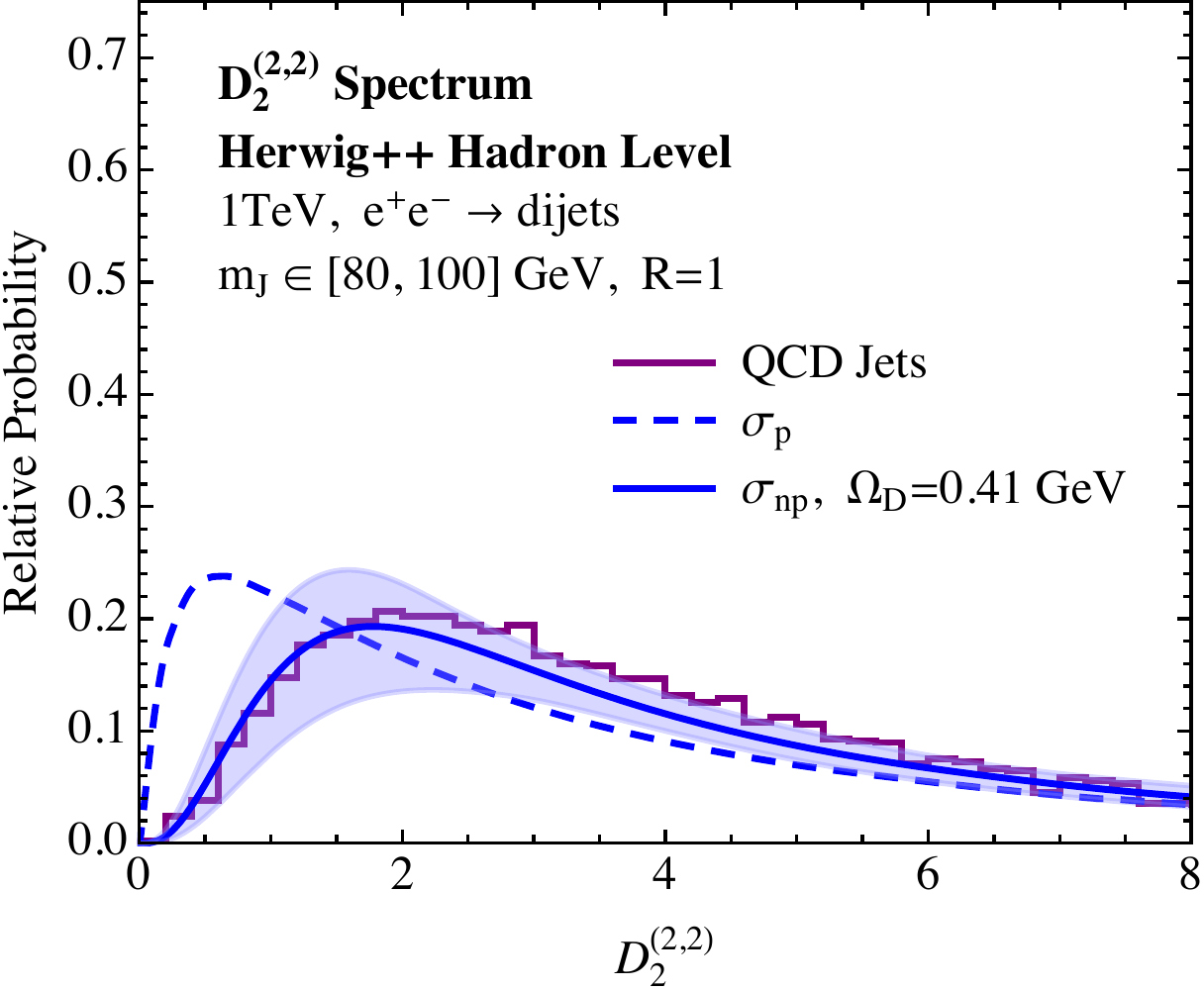}
}
\end{center}
\vspace{-0.2cm}
\caption{A comparison of the $\Dobs{2}{2,2}$ distributions for background QCD jets from our analytic prediction and the various hadron-level Monte Carlos.  $\sigma_p$ denotes the parton level perturbative prediction for the distribution and $\sigma_{np}=\sigma_p\otimes F_D$ is the perturbative prediction convolved with the non-perturbative shape function.  The values of the non-perturbative parameter $\Omega_D$ used are also shown.
}
\label{fig:D2_hadr}
\end{figure}

\begin{figure}
\begin{center}
\subfloat[]{\label{fig:D2_ee_hadsig_a}
\includegraphics[width = 7.25cm]{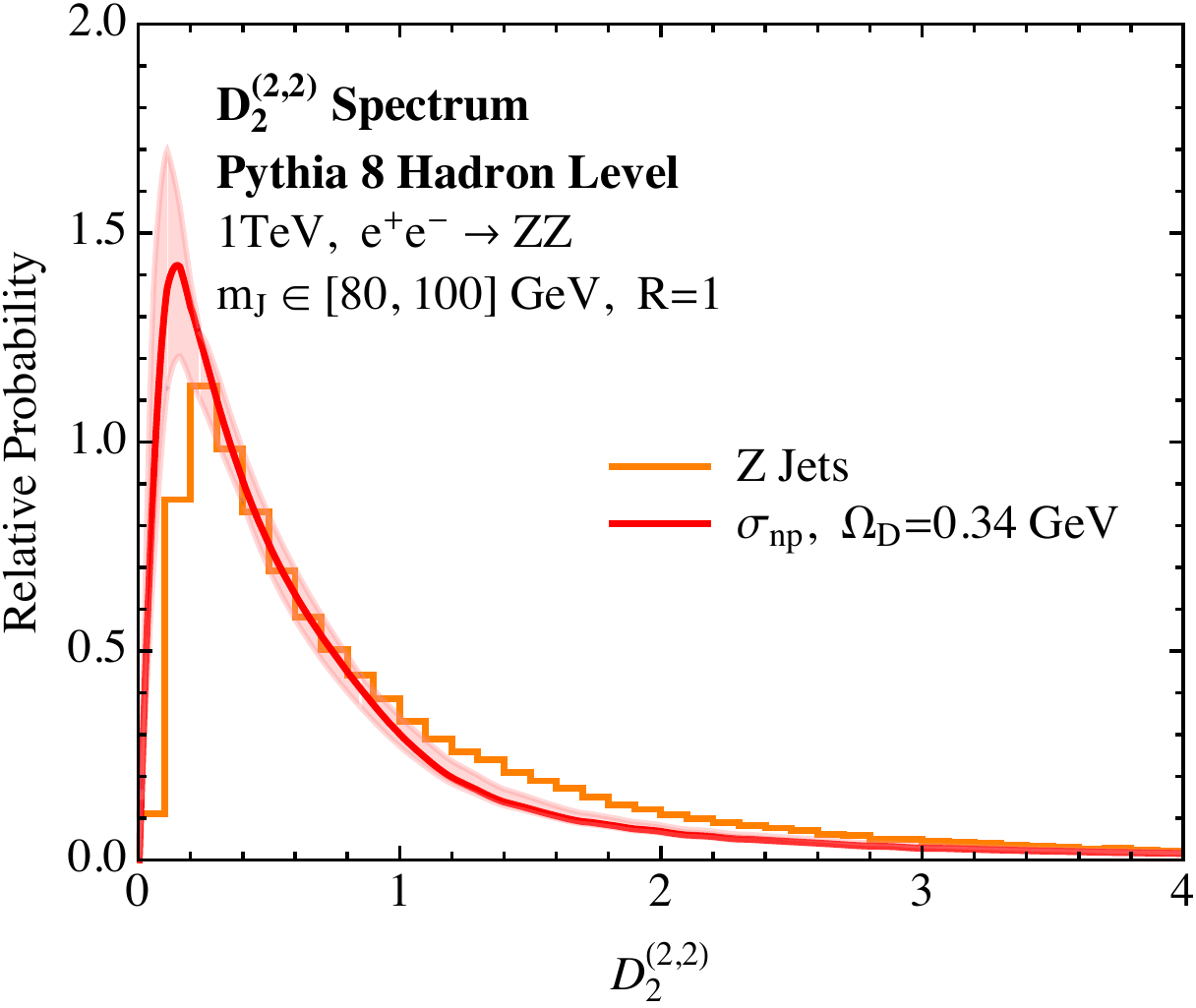}
}
\subfloat[]{\label{fig:D2_ee_hadsig_b}
\includegraphics[width = 7.25cm]{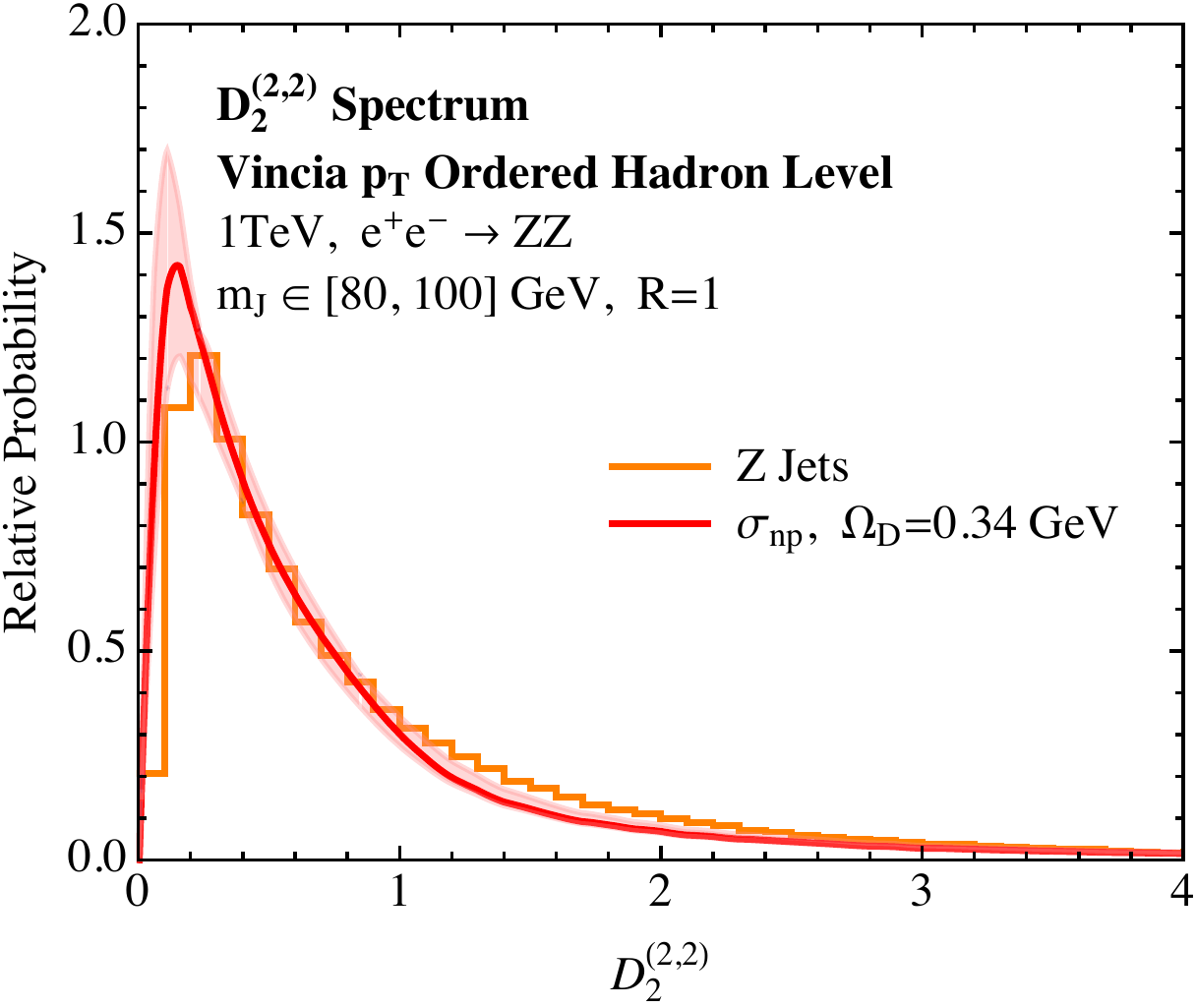}
}\\
\subfloat[]{\label{fig:D2_ee_hadsig_c}
\includegraphics[width = 7.25cm]{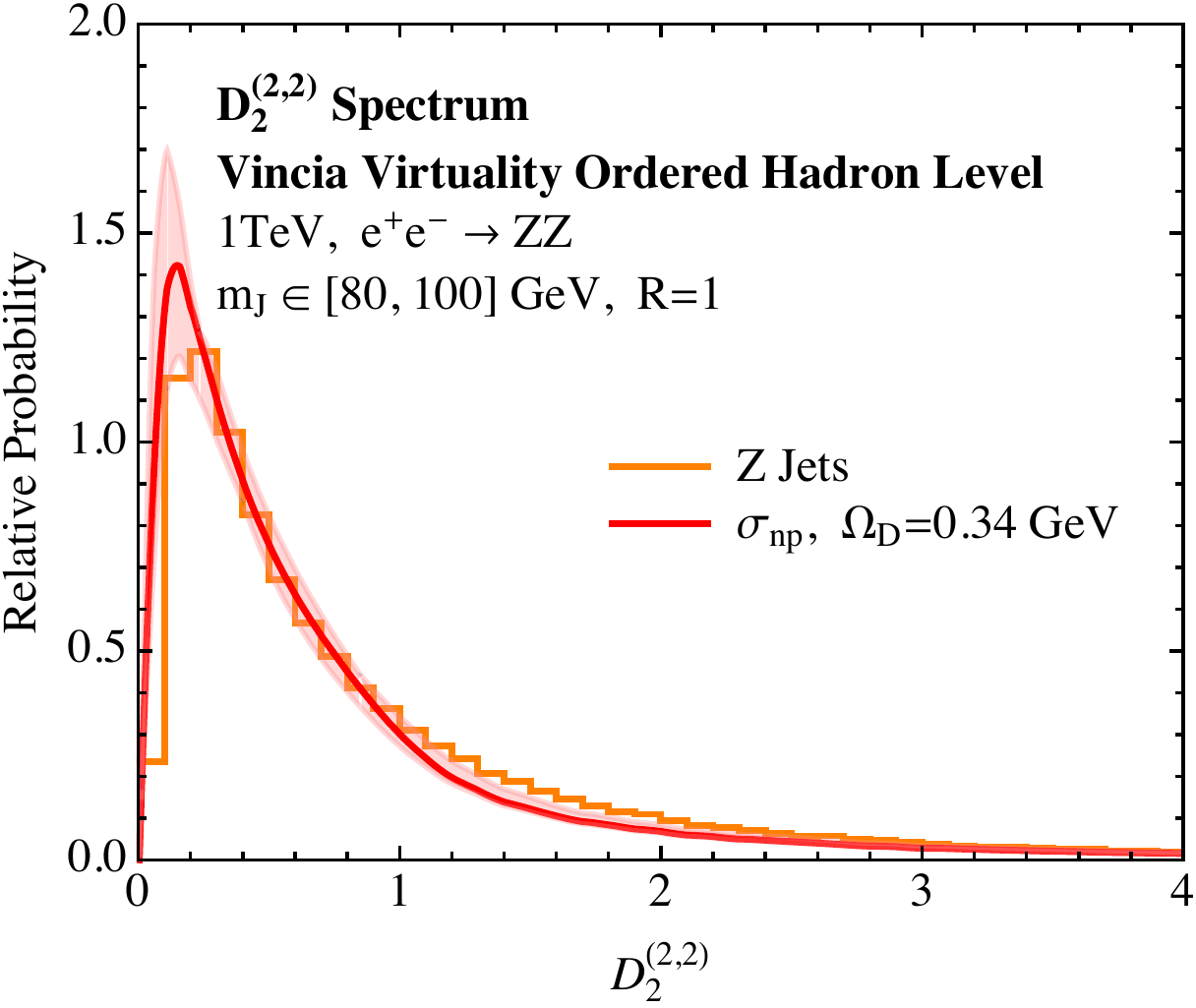}
}
\subfloat[]{\label{fig:D2_ee_hadsig_d}
\includegraphics[width = 7.25cm]{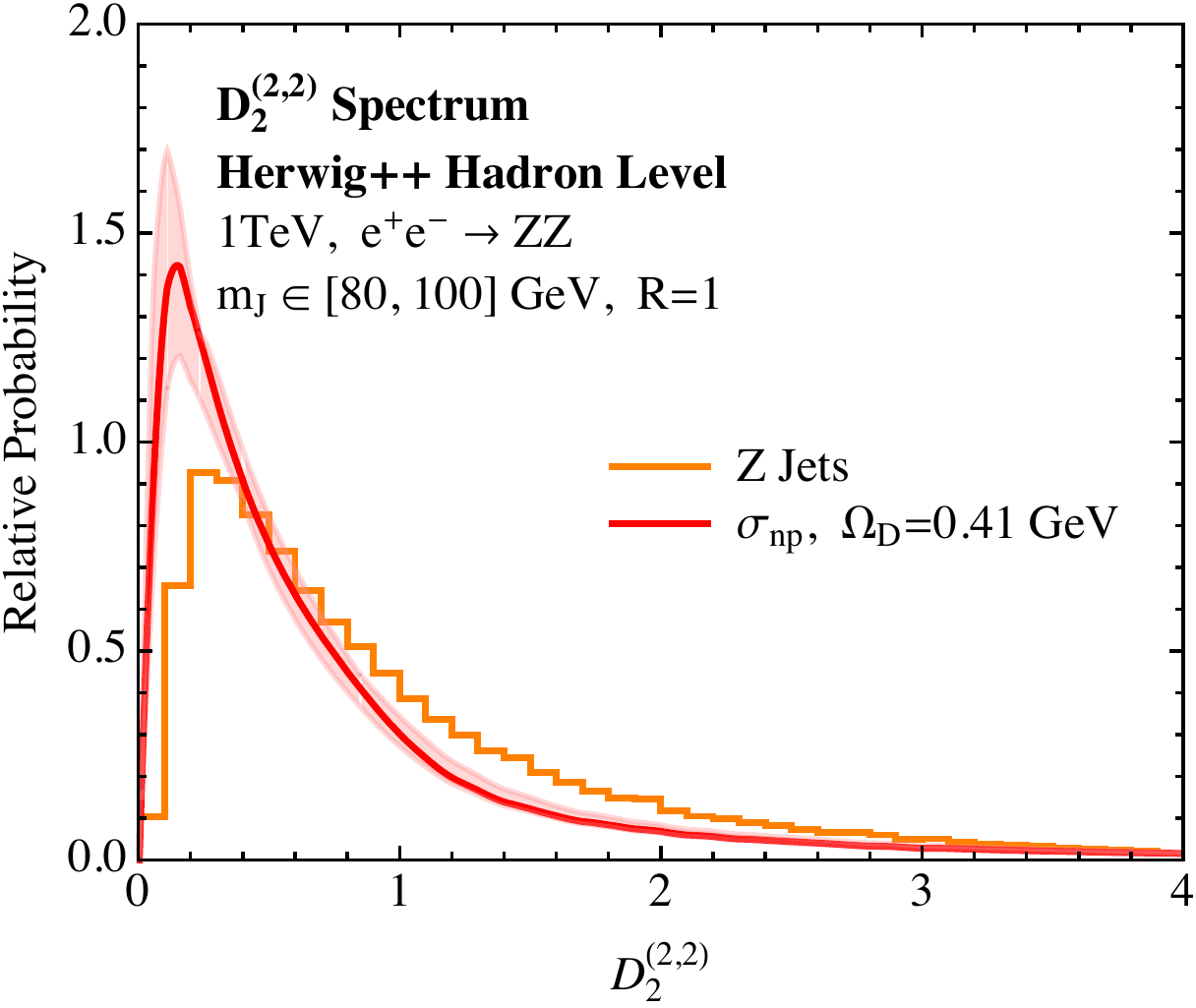}
}
\end{center}
\vspace{-0.2cm}
\caption{
A comparison of the $\Dobs{2}{2,2}$ distributions for signal boosted $Z$ jets from our analytic prediction and the various hadron-level Monte Carlos.  $\sigma_p$ denotes the parton level perturbative prediction for the distribution and $\sigma_{np}=\sigma_p\otimes F_D$ is the perturbative prediction convolved with the non-perturbative shape function, although for the signal this has a negligible effect.  The values of the non-perturbative parameter $\Omega_D$ used are also shown.
}
\label{fig:D2_hadr_signal}
\end{figure}

Comparisons between the hadron-level distributions of $\Dobs{2}{2,2}$ from our analytic calculations and the Monte Carlos are presented in \Fig{fig:D2_hadr} for background and \Fig{fig:D2_hadr_signal} for signal jets.  For background distributions, we compare our perturbative calculation convolved with the shape function, as defined in \Eq{eq:shape_bkgd}. Both \vincia{} and \pythia{} use the same hadronization model, but \herwigpp{} uses a distinct hadronization model, and therefore we allow for a different shape parameter, $\Omega_D$, for the two cases. For the case of \pythia{} and \vincia{}, 
we choose to extract the value of $\Omega_D$ by fitting to the hadronized distribution for $p_T$ ordered \vincia{}. However, we will shortly discuss the level of ambiguity in $\Omega_D$ arising from this extraction.  For jets with an energy of 500 GeV and mass of 90 GeV, we find that the choice $\Omega_D = 0.34\pm 0.03$ GeV provides the best agreement of our perturbative calculation with $p_T$ ordered \vincia{}, while $\Omega_D=0.41 \pm 0.03$ GeV provides the best agreement with \herwigpp{}. The errors assigned here come only from the fitting itself, and are due to the statistical uncertainties of the Monte Carlo distributions due to the finite width of the histogram bins.  These errors do not take into account any other uncertainties; for example, whether one should perform the fit to hadron level \vincia{} or \pythia{}. This level of agreement between the non-perturbative parameters extracted from \pythia{} and \herwigpp{} is comparable to more detailed studies, such as \Ref{Mateu:2012nk}.  A comparison of the distributions of \Fig{fig:D2_hadr} before and after hadronization shows that hadronization has a considerable effect on the background distributions, particularly at small values of $D_2$, as expected from \Eq{eq:bkg_shift_np}. This effect, which in the Monte Carlos is realized through tuned hadronization models,  is well described by the single parameter shape function. Importantly, as discussed above, if different shape parameters were used for the collinear subjets and soft subjets factorization theorems, they would be nearly degenerate in the fit at the level of perturbative accuracy that we work, which is why we have made the simplification of working with a single non-perturbative shape parameter.

We have argued that the non-perturbative parameter $\Omega_D$ in the collinear subjets factorization theorem can be related to a universal non-perturbative matrix element of two soft Wilson lines.  Such non-perturbative matrix elements appear in the factorization theorems of a large class of $e^+e^-$ event observables, and has therefore been measured from data at LEP.\footnote{An extremely large literature exists on such measurements, and their theoretical interpretation, to which we cannot do justice in this brief section. We refer the reader to, for example, \Refs{Korchemsky:1999kt,Korchemsky:2000kp,Achard:2004sv,Gehrmann:2009eh,Abbate:2010xh,Abbate:2012jh,Hoang:2014wka,Hoang:2015hka} and references therein.} While the value of $\Omega_D$ that we have determined for the two parton showers is by no means precise, it is interesting to compare our value with those extracted from precision studies of $e^+e^-$ collider observables which have been performed in the literature. Using the particular case of $\alpha=\beta=2$, and converting to our normalization, a recent extraction of the non-perturbative parameter from an N$^3$LL$'$ analysis of the $C$-parameter event shape using LEP data, and including power corrections and hadron mass effects \cite{Salam:2001bd,Mateu:2012nk}, gives a value of $\Omega_D = 0.28$ GeV \cite{Hoang:2014wka,Hoang:2015hka}. This agrees well with our values extracted through comparison with Monte Carlo.
Going forward, with the goal of increasing both the precision and understanding of jet substructure, the ability to relate the dominant non-perturbative corrections to the $D_2$ observable to known non-perturbative parameters measured in $e^+e^-$ is a valuable feature, and that further study on the non-perturbative corrections to multi-differential cross sections is of great importance.

\begin{figure}
\begin{center}
\subfloat[]{\label{fig:D2_ee_hadbkg_a2}
\includegraphics[width= 7.25cm]{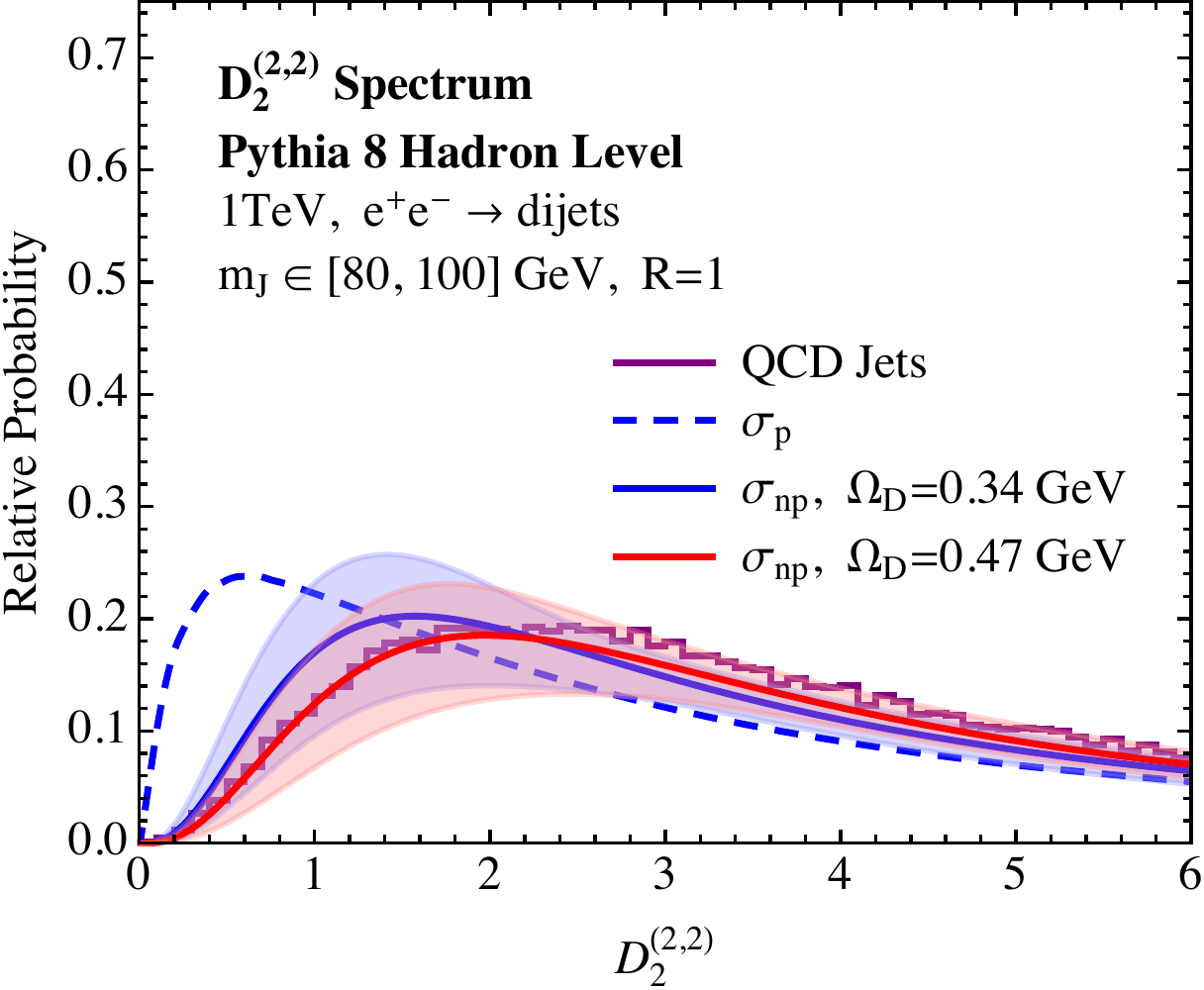}
}
\subfloat[]{\label{fig:D2_ee_hadbkg_b2}
\includegraphics[width= 7.25cm]{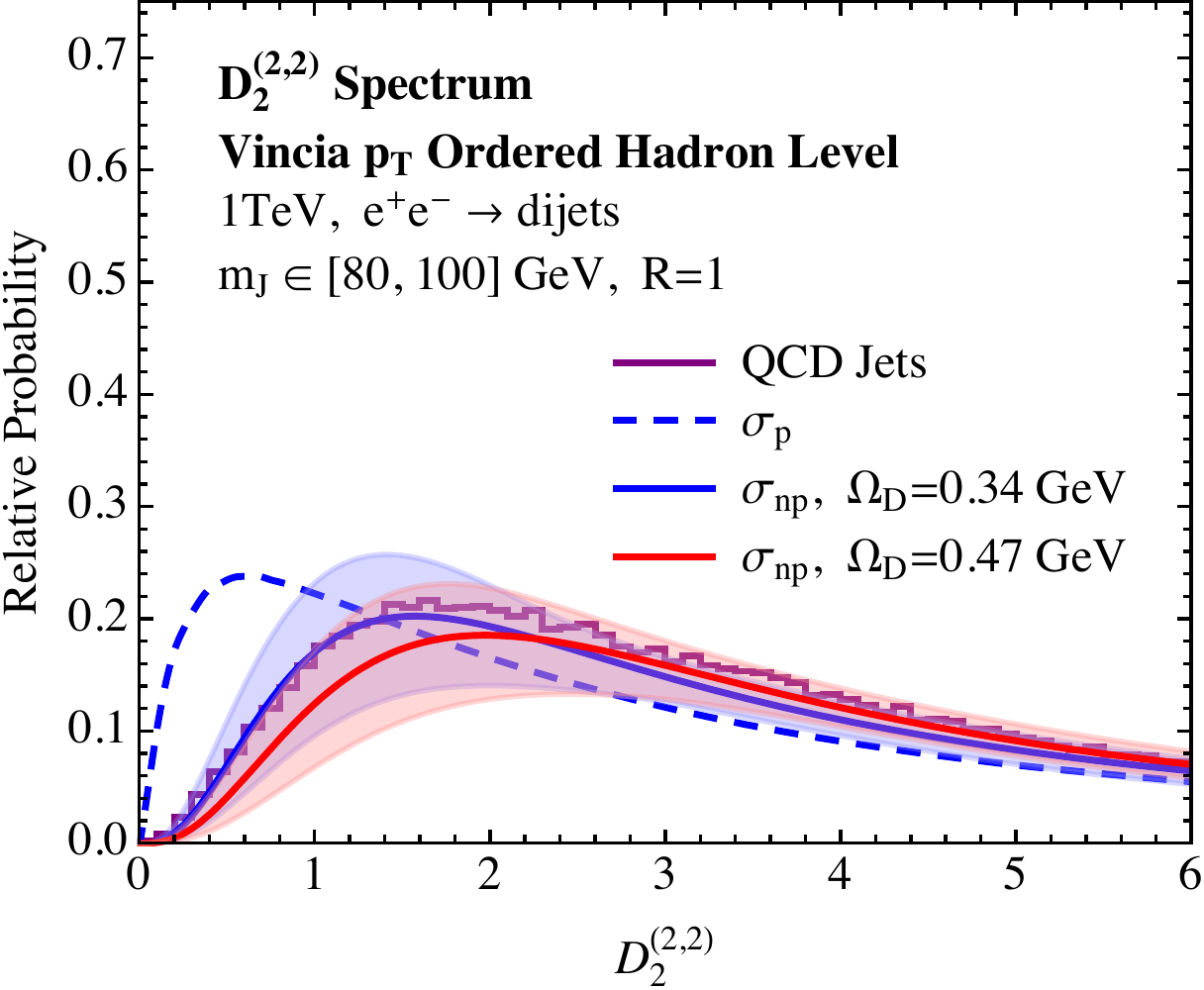}
}
\end{center}
\vspace{-0.2cm}
\caption{A comparison of the $\Dobs{2}{2,2}$ distributions for background QCD jets from our analytic prediction and \pythia{} and $p_T$ ordered \vincia{} Monte Carlos in a) and b).  Analytic predictions for different values of the non-perturbative shape parameter $\Omega_D$ are shown. 
}
\label{fig:D2_hadr_doubleshape}
\end{figure}

Many of the features of the background distributions which were present before hadronization in \Fig{fig:D2_ee_bkg} persist after convolution with the shape function.  However, they are greatly reduced, and they become difficult to disentangle from modifications to the non-perturbative shape parameter at the order we work. In particular, from \Fig{fig:D2_hadr}, we see that for the choices of $\Omega_D$ that we have used, both \vincia{} showers agree well with our analytic calculation.  On the other hand, with the chosen value of $\Omega_D$,
the $D_2$ distribution in \pythia{} is systematically pushed to higher values as compared with our calculation.

To try and asses the extent to which this can be accommodated for by adjusting the value of $\Omega_D$, in \Fig{fig:D2_hadr_doubleshape} we show plots of both \pythia{} and \vincia{} with $p_T$ ordering compared with our analytic results for two different values of the shape parameter. The values $\Omega_D=0.34$ GeV and $\Omega_D=0.47$ GeV were chosen to give best agreement with the \vincia{} and \pythia{} distributions, respectively. This figure makes clear that the disagreement between the $D_2$ distributions as generated by the two Monte Carlo generators can largely be remedied by using different values of the non-perturbative parameter. We note also that the effect of changing the non-perturbative parameter is of course similar to that of changing the perturbative cutoff of the shower, as was discussed in \Sec{sec:scales}, making it difficult to disentangle these two effects. 

This plot also gives a feel for the extent to which $\Omega_D$ can be varied before significant disagreement is seen between the analytic calculation and a given Monte Carlo distribution. Performing the perturbative calculation to higher accuracy would help to resolve some of these ambiguities in the value of the shape parameter, by reducing the perturbative uncertainty on the shape of the distribution, as well as its normalization. Throughout the rest of this paper, when comparing our analytic predictions with \vincia{} or \pythia{}, we will use the value $\Omega_D=0.34$ GeV as obtained from our fit to hadron level $p_T$-ordered \vincia{}.  
However, one should keep in mind the level of sensitivity to this parameter. In particular, for the application of boosted $Z$ discrimination, we will see that the discrimination power of the observable will depend sensitively on the shape of the $D_2$ distribution below the peak, and will therefore exhibit great sensitivity to the value of the shape parameter.

For the signal distributions, shown in \Fig{fig:D2_hadr_signal}, we use the same choice of non-perturbative parameters as for the background distributions. From \Eq{eq:signal_nonpert}, we have seen that for the jets with $E_J=500$ GeV, the non-perturbative shift is expected to be of the order $1/500$, and is therefore completely negligible to the level of accuracy that we work, and the equality of the non-perturbative parameters between the signal and background distributions is not tested. For the signal distributions, we see excellent agreement between the theory prediction and all the Monte Carlo generators. Due to the sharp peak in the distribution, we expect higher order resummation is necessary to provide a more accurate description right in the peak region, where the perturbative uncertainty in our calculation becomes large. Due to the fact that the distributions are normalized, this uncertainty also manifests itself in the tail of the distribution. It is known how to calculate the signal distribution to higher accuracy \cite{Feige:2012vc}, and so we do not consider this issue further here.  The effect of the shape function on our analytic results are consistent with all of the Monte Carlos, whose signal $D_2$ distribution is changed only slightly (i.e., only in the lowest bins) after hadronization.

We conclude this section by emphasizing how the choice of variable can greatly facilitate comparisons with Monte Carlos. An important feature of the $D_2$ observable is that it cleanly separates phase space regions dominated by different physics. In particular, it separates the region of phase space where a subjet is formed from that where no subjet is formed, as well as separating the regions of phase space where hadronization is important from those where it plays a minor role. This enables these effects to be cleanly disentangled, and provides a sensitive probe of their modeling. We therefore believe that the observable $D_2$ could play an important role in the tuning of Monte Carlo generators for jet substructure studies, and could be used to complement some of the observables proposed and studied in \Refs{Fischer:2014bja,Fischer:2015pqa}.\footnote{Note that \Refs{Fischer:2014bja,Fischer:2015pqa} used the observable $C_2$, also formed from the energy correlation functions, which was proposed in \Ref{Larkoski:2013eya}. Unlike $D_2$, $C_2$ does not cleanly separate the two-prong region of phase space from the one-prong region of phase space. A detailed discussion of this point can be found in \Ref{Larkoski:2014gra}. The clean separation of the one- and two-prong regions of phase space is the essential feature of the $D_2$ observable, which allows for its precise theoretical calculation and its sensitivity to the shower implementation. } Furthermore, the observable $D_3$ \cite{Larkoski:2014zma}, which is sensitive to three-prong substructure within a jet also provides a clean separation of two- and three-prong regions, and could be used to provide an even more detailed understanding of jet substructure and the perturbative shower evolution.

\subsection{Analytic Boosted $Z$ Discrimination with $D_2$}\label{sec:ROC}

In this section, we use our analytic calculation, combined with the non-perturbative shape functions of \Sec{sec:Hadronization}, to make complete predictions for the discrimination power of $\Dobs{2}{2,2}$ for hadronically-decaying boosted $Z$ bosons versus QCD quark jets at an $e^+e^-$ collider.  We present comparisons of our calculation to the results of fully hadronized \pythia{}, \vincia{}, and \herwig{} Monte Carlos.  Here, we also present Monte Carlo results from scanning over a range of values for the angular exponent $\alpha$ that is consistent with our factorization theorem. Analytic results for boosted boson discrimination were also presented recently in \Ref{Dasgupta:2015yua} for groomed mass taggers, as well as an analytic study of the optimal parameters.

\begin{figure}
\begin{center}
\subfloat[]{\label{fig:D2_disc_py}
\includegraphics[width= 7.25cm]{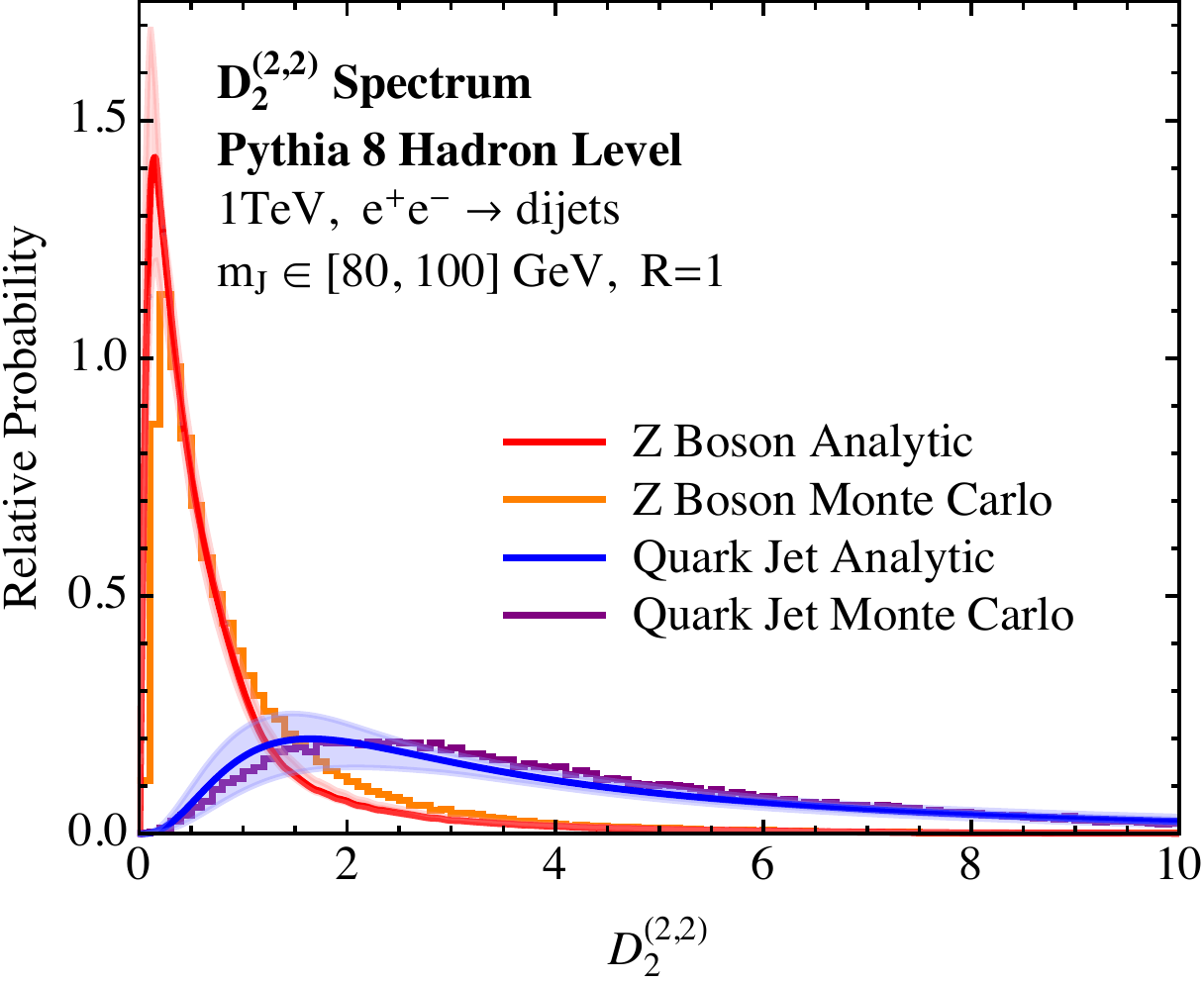}
}
\subfloat[]{\label{fig:D2_disc_vinp}
\includegraphics[width = 7.25cm]{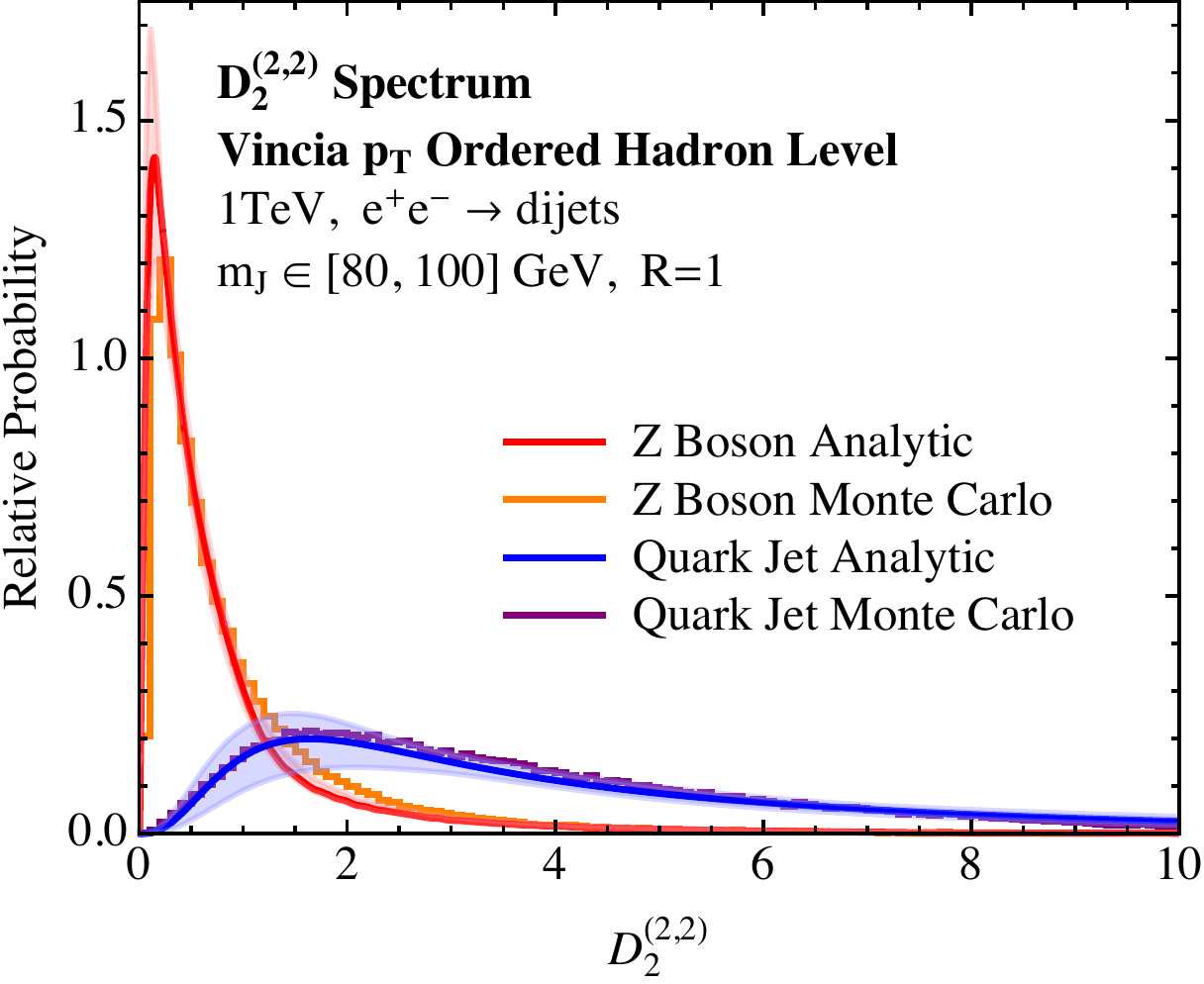}
}\\
\subfloat[]{\label{fig:D2_disc_vinv}
\includegraphics[width= 7.25cm]{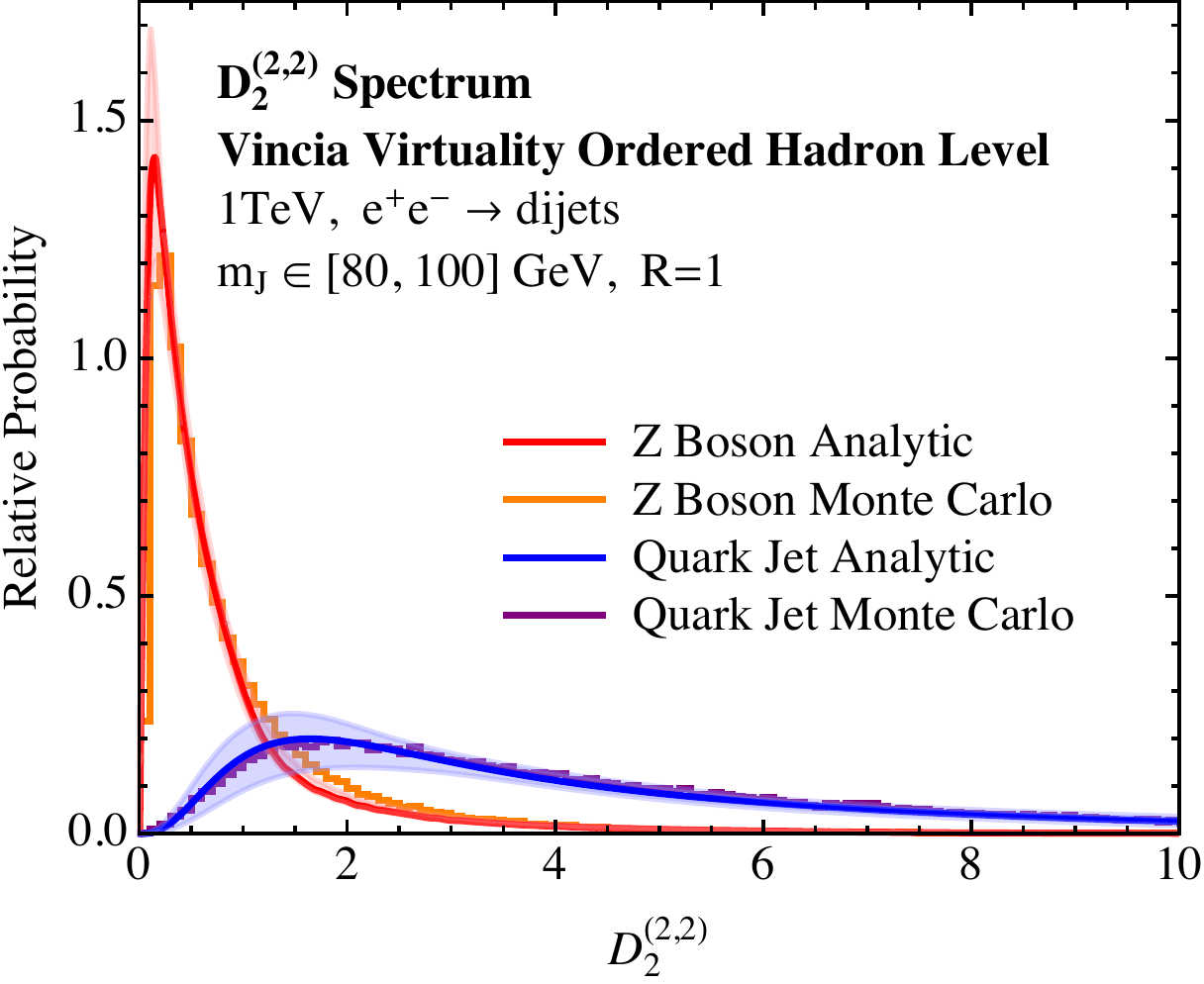}
}
\subfloat[]{\label{fig:D2_disc_her}
\includegraphics[width = 7.25cm]{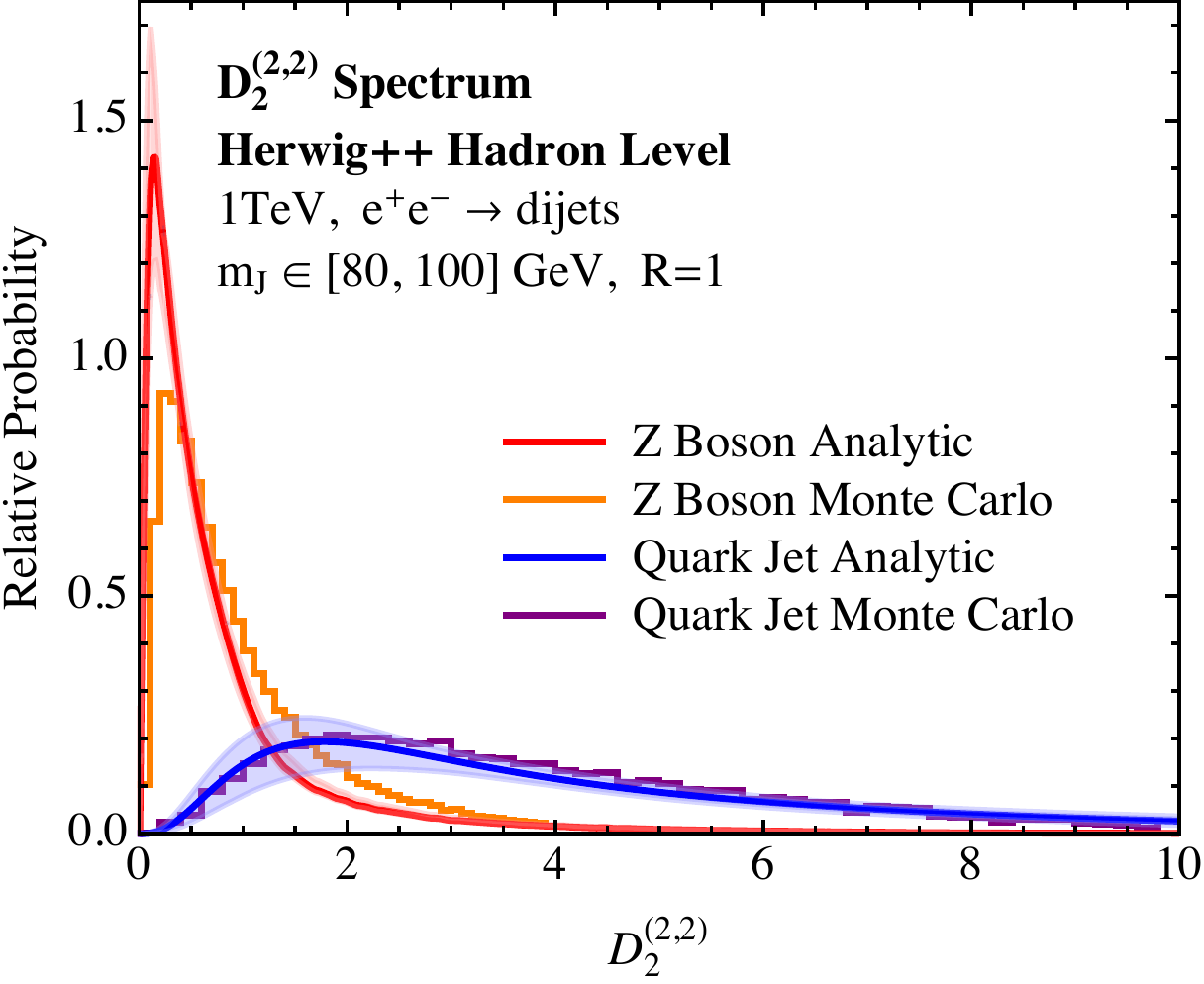}
}
\end{center}
\vspace{-0.2cm}
\caption{
A comparison of signal and background $\Dobs{2}{2,2}$ distributions for the four different Monte Carlo generators and our analytic calculation, including hadronization. Here we show the complete distributions, including the long tail for the background distribution. Although we extend the factorization theorem beyond its naive region of applicability into the tail, excellent agreement with Monte Carlo is found.
}
\label{fig:D2_disc}
\end{figure}

\begin{figure}
\begin{center}
\subfloat[]{\label{fig:D2_disc_py_zoom}
\includegraphics[width= 7.25cm]{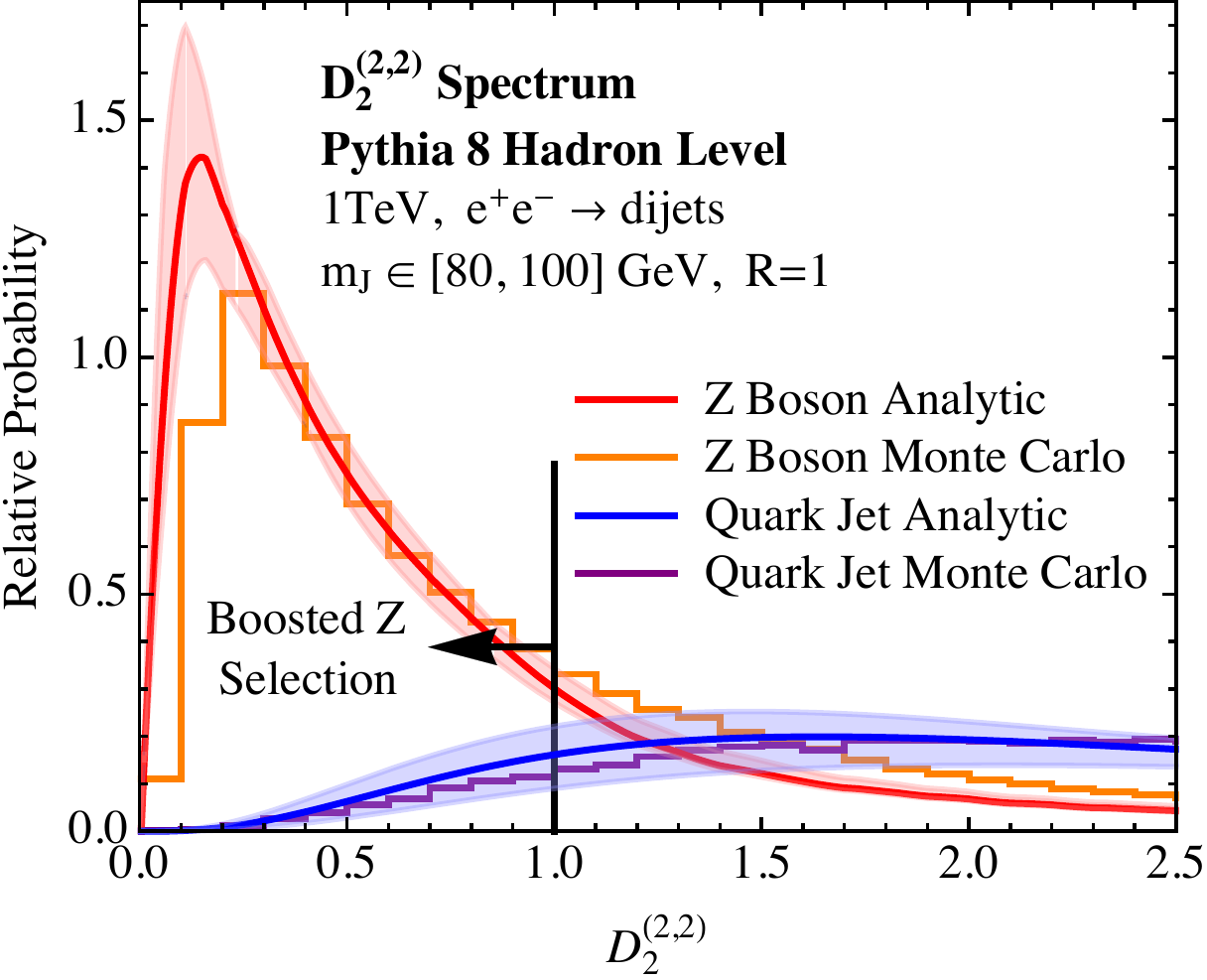}
}
\subfloat[]{\label{fig:D2_disc_vinp_zoom}
\includegraphics[width = 7.25cm]{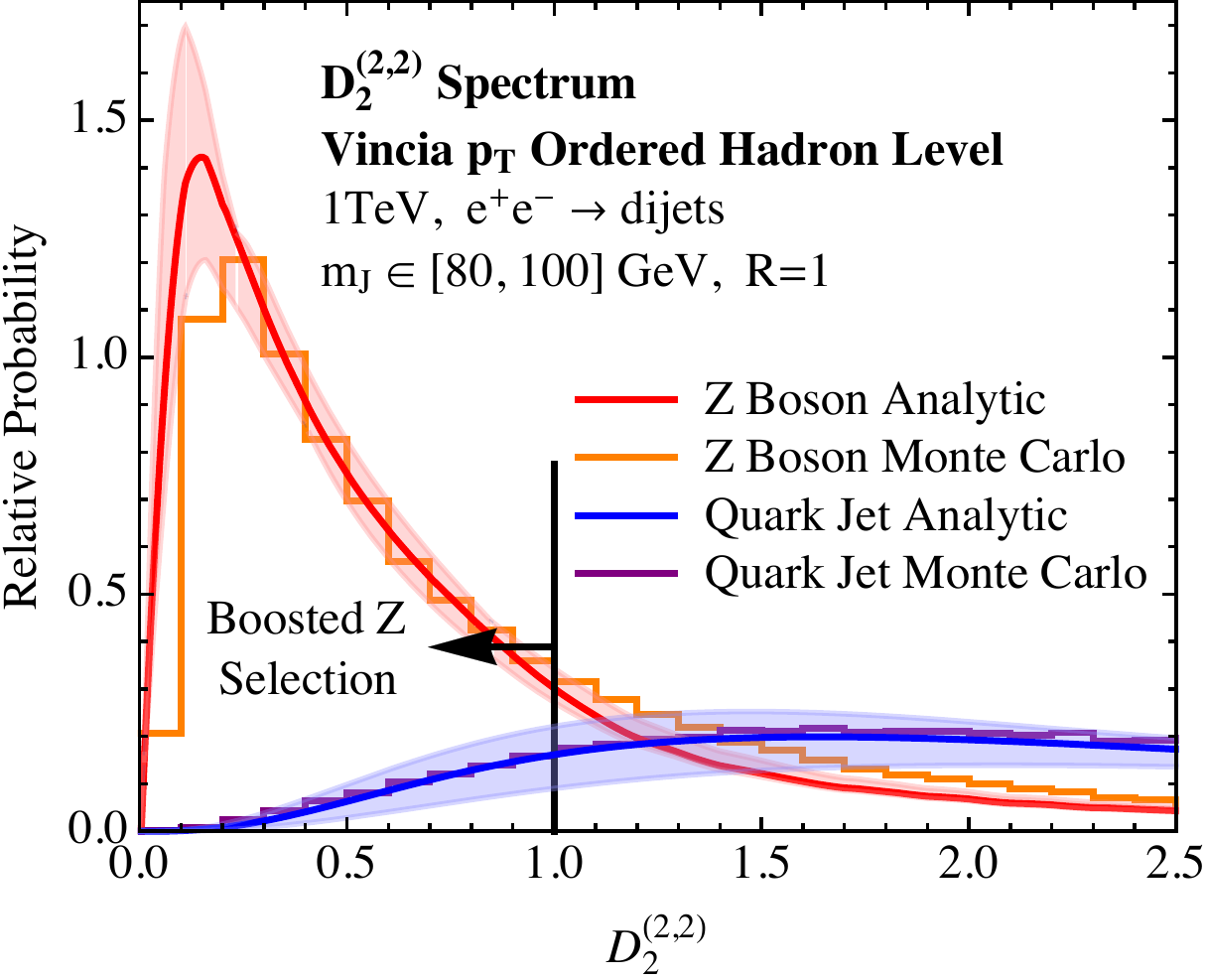}
}\\
\subfloat[]{\label{fig:D2_disc_vinv_zoom}
\includegraphics[width= 7.25cm]{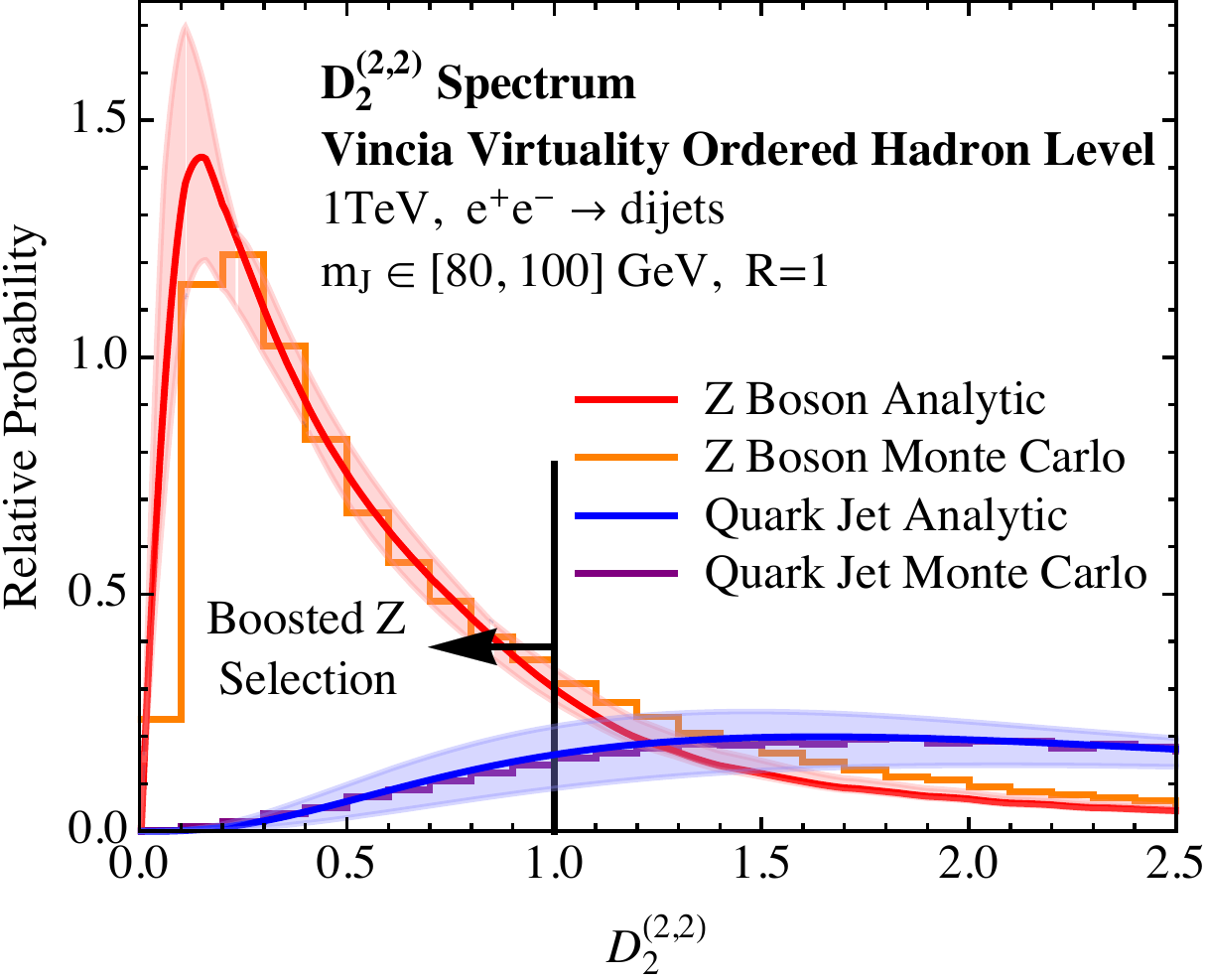}
}
\subfloat[]{\label{fig:D2_disc_her_zoom}
\includegraphics[width = 7.25cm]{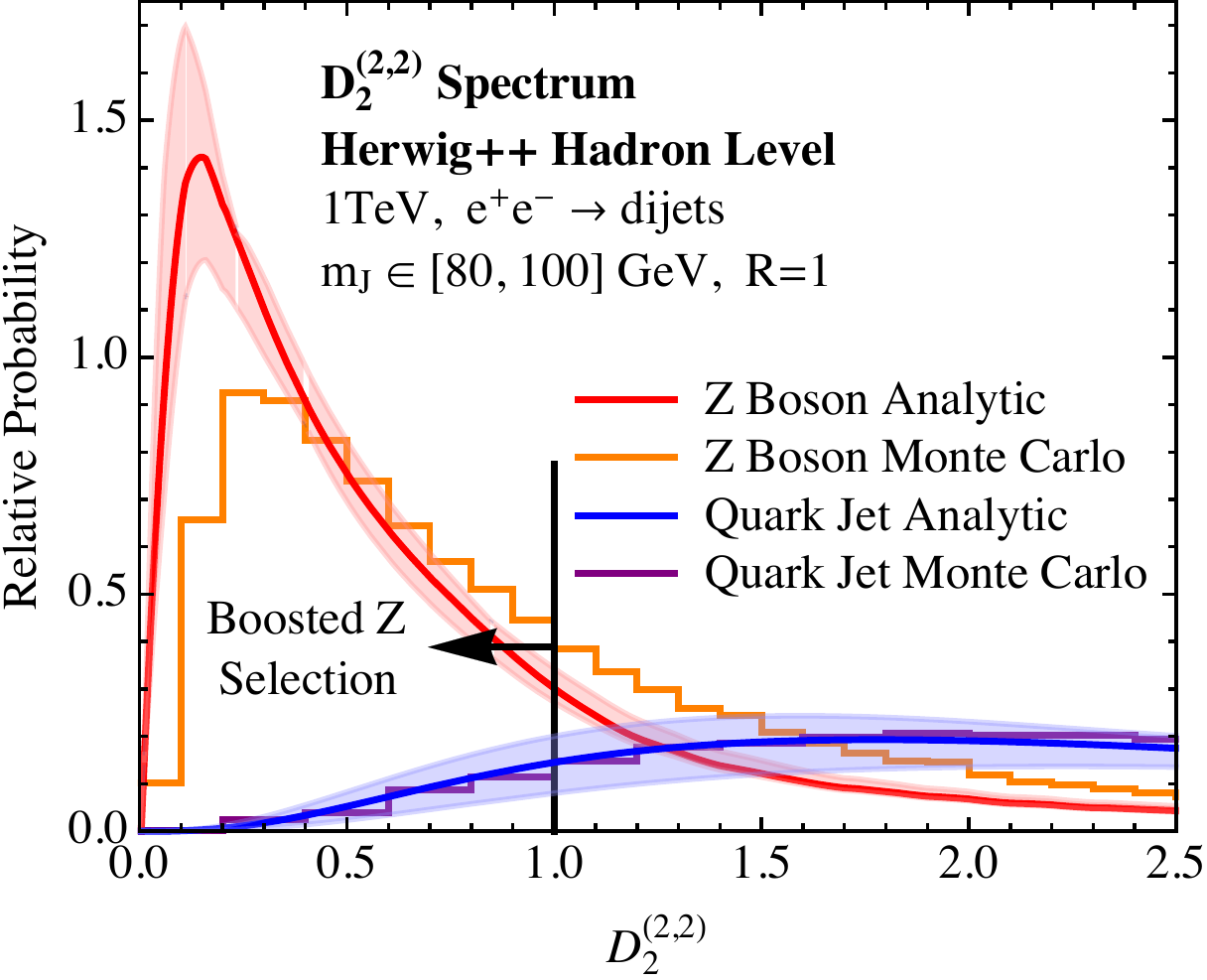}
}
\end{center}
\vspace{-0.2cm}
\caption{
A comparison of signal and background $\Dobs{2}{2,2}$ distributions for the four different Monte Carlo generators and our analytic calculation, including hadronization. Here we show a zoomed in view of the distributions at small $D_2$, along with a representative cut that could be used to select a relatively pure sample of boosted $Z$ bosons. Relevant cuts for boosted $Z$ discrimination are to the left of the perturbative peak for the background distributions.
}
\label{fig:D2_disc_zoom}
\end{figure}

In \Figs{fig:D2_disc}{fig:D2_disc_zoom} we overlay the distributions for $\Dobs{2}{2,2}$ as measured on signal and background for each Monte Carlo sample, and compare with our analytical calculations including the non-perturbative shape function contributions. \Fig{fig:D2_disc} shows the complete $D_2$ distributions, including the long tail of the background distribution, while \Fig{fig:D2_disc_zoom} shows a zoomed in version, focusing on small values of $D_2$, as is most relevant for signal versus background discrimination. A representative cut on the $D_2$ distribution, as could be used to select a relatively pure sample of boosted $Z$ bosons, is also indicated. In general, the agreement between the Monte Carlos, for both signal and background distributions, and our calculation is impressive. This holds true both for the overall shape of the distributions, including the long tail of the background distribution, and for the detailed shape at small values of $D_2$. It is also important to note that the perturbative uncertainties remain under control, even in the small $D_2$ region, as seen in \Fig{fig:D2_disc_zoom}. The uncertainty bands do not incorporate variations in the non-perturbative parameter $\Omega_D$. There are however, some small deviations between the analytic predictions and the Monte Carlo distributions.  The background distribution in \pythia{} is pushed to slightly higher values than our calculation.  This implies that the signal versus background discrimination power as predicted with \pythia{} will be overestimated.  The most conservative prediction for the signal versus background discrimination power is from \herwig{}, whose background distribution is nearly identical to our calculation.  That \pythia{} tends to be optimisitic and \herwig{} tends to be pessimistic with respect to discrimination power has been observed in several other jet substructure analyses \cite{Larkoski:2013eya,Larkoski:2014gra,Larkoski:2014zma,Aad:2014gea}.

An important feature of the $D_2$ distributions, made clear by \Fig{fig:D2_disc_zoom}, is that in the region of interest relevant for boosted $Z$ discrimination, the background distribution is deep in the non-perturbative regime. Therefore, although the perturbative uncertainties are small, the effect of the shape function, and variations of the non-perturbative parameter $\Omega_D$, is large.  Estimates of the uncertainties due to the form of the shape function, or the use of more complicated functional forms, along the lines of \Ref{Ligeti:2008ac} are well beyond the scope of this paper.  An advantage of our factorization approach is that we are able to achieve a clean separation of perturbative and non-perturbative effects, and demonstrate relations between the non-perturbative matrix elements appearing in our factorization theorems and non-perturbative matrix elements which have been measured with other event shapes, by using their field theoretic definitions. This separation is essential for understanding discrimination performance in the non-perturbative region, which we see is required for jet substructure studies related to boosted boson discrimination.  Importantly, though, $D_2$ seems to take advantage of the different hadronization corrections to signal and background jets, and the overlap of the signal and background regions of $D_2$ decreases significantly in going from parton-level to fully hadronized jets.

\begin{figure}
\begin{center}
\subfloat[]{\label{fig:D2_roca}
\includegraphics[width= 7.2cm]{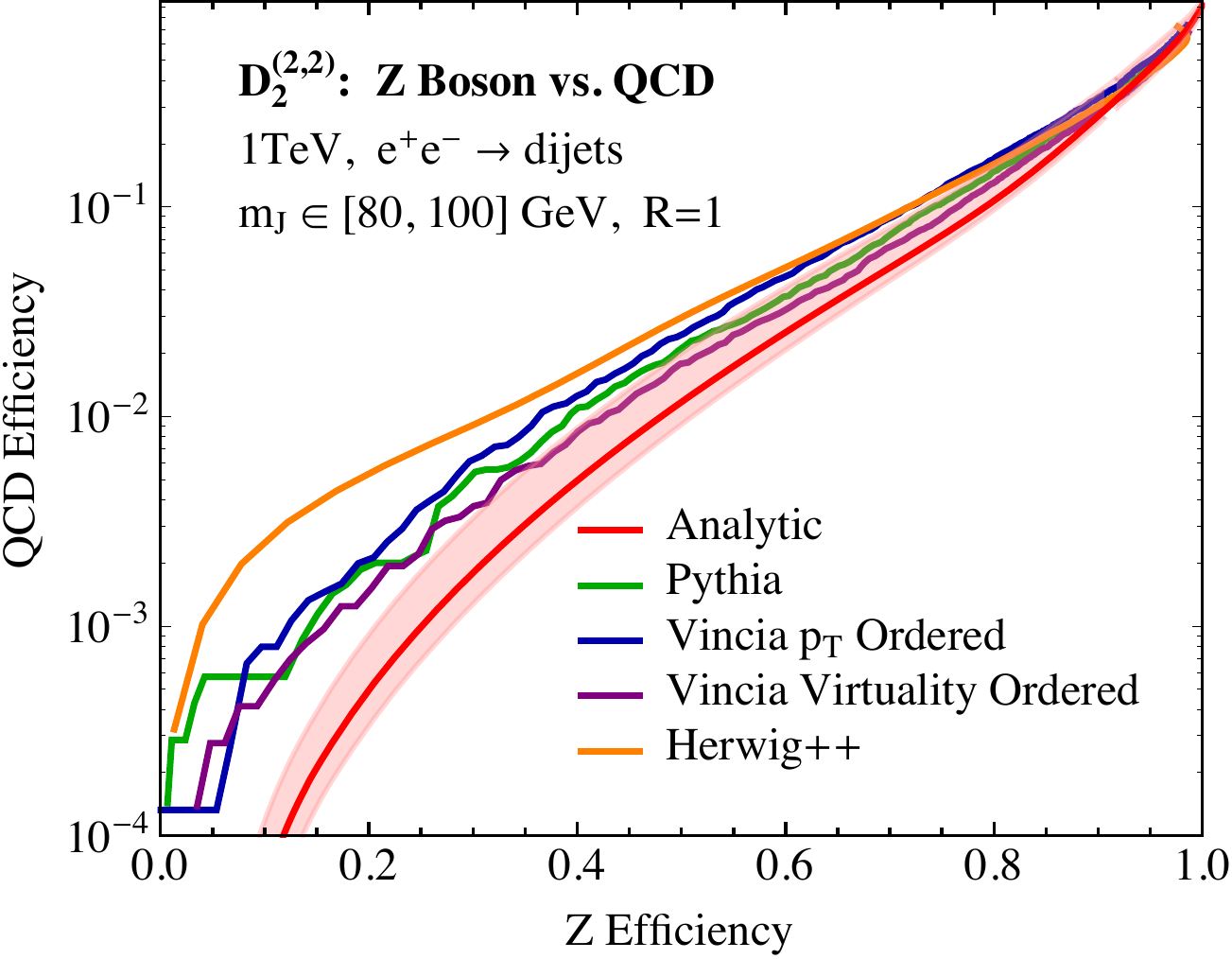}
}
\ 
\subfloat[]{\label{fig:D2_rocb}
\includegraphics[width = 7.0cm]{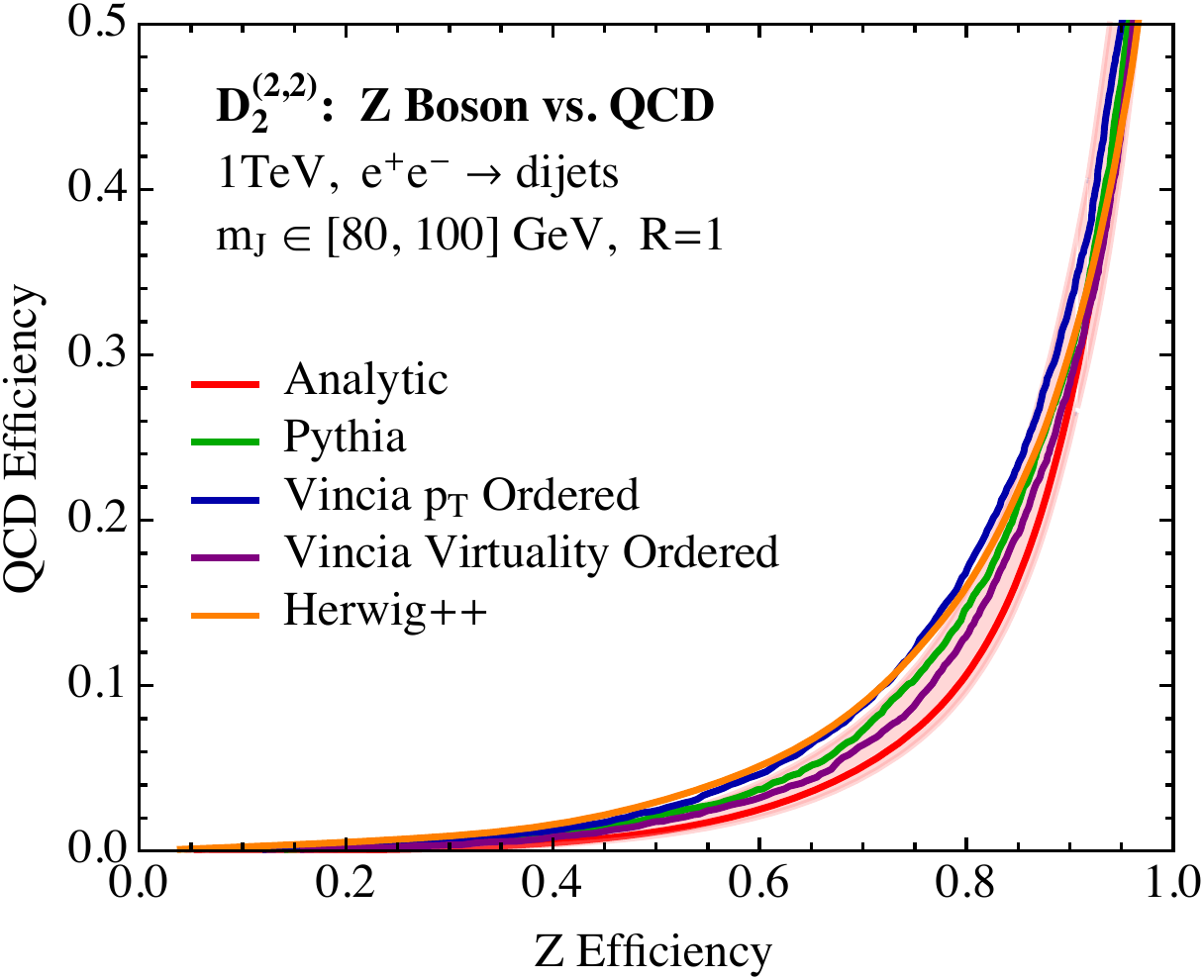}
}\end{center}
\caption{ Signal vs.~background efficiency curves for $\Dobs{2}{2,2}$ for the Monte Carlo samples as compared to our analytic prediction on a a) logarithmic scale plot and b) linear scale plot.  The band of the analytic prediction is representative of the perturbative scale uncertainty.
}
\label{fig:ROC}
\end{figure}

In \Fig{fig:ROC}, we have used these raw distributions to produce signal versus background efficiency curves (ROC curves) by making a sliding cut in $D_2$.  The ROC curve from each Monte Carlo sample as well as our analytic prediction from our calculated signal and background distributions are shown in both logarithmic plot and linear plot in \Figs{fig:D2_roca}{fig:D2_rocb}, respectively. The band around our analytic prediction should be taken as representative of the signal versus background efficiency range from varying the perturbative scales.\footnote{Note that ROC curves only make sense for normalized distributions, and therefore the envelopes from scale variation cannot be used. Instead, ROC curves are generated from normalized signal and background distributions made with a variety of scale choices, with scales varied separately in the signal and background distributions. We then take the envelope of these ROC curves to generate the uncertainty bands for the ROC curves. } For the analytic predictions, we use $\Omega_D=0.34$, as obtained from our fit to the $p_T$ ordered \vincia{} shower.  Consistent with the distributions in \Fig{fig:D2_disc}, the Monte Carlos are in qualitative agreement with our analytic prediction for the ROC curve.  In general, our analytic prediction seems to give an optimistic prediction for the discrimination power, however, this is driven by the fact that our resummed prediction for the signal distribution is more peaked. It would be interesting to perform the NNLL resummation for the signal, which should significantly reduce the uncertainty in the signal calculation, particularly in the peak region, where the perturbative uncertainties in our present calculation are quite large. Because of the fact that the distributions are normalized, an improved behavior in the peak of the distribution could also improve the agreement in the tail of the signal distribution, which is currently systematically low, due to the fact that the peak is systematically high. This could enable a conclusive understanding as to the discrepancy between the different Monte Carlo generators for both signal and background distributions. In particular, our analytic calculations suggest that the \herwigpp{} generator provides pessimistic predictions for the discrimination power of the $D_2$ observable due to the underestimation of the peak height for the signal distribution, and it would be interesting to understand this further. Due to the importance of analytically understanding the discrimination power of jet substructure observables, such a calculation is well motivated. For the case of $\alpha=\beta=2$, the required perturbative components could be obtained following relations to $e^+e^-$ event shapes as were used in \Ref{Feige:2012vc}.

\begin{figure}
\begin{center}
\subfloat[]{\label{fig:D2_roca2}
\includegraphics[width= 7.2cm]{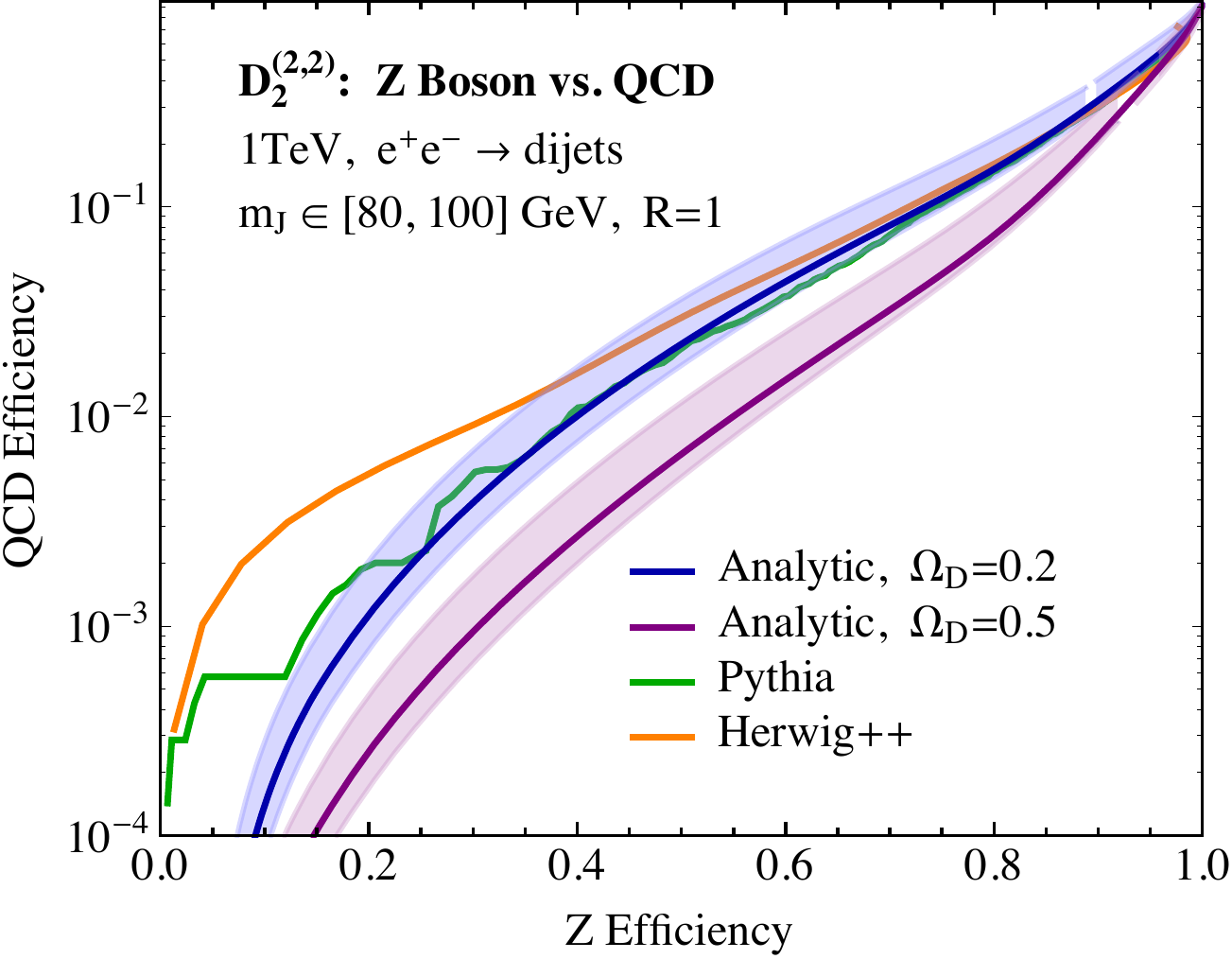}
}
\ 
\subfloat[]{\label{fig:D2_rocb2}
\includegraphics[width = 7.0cm]{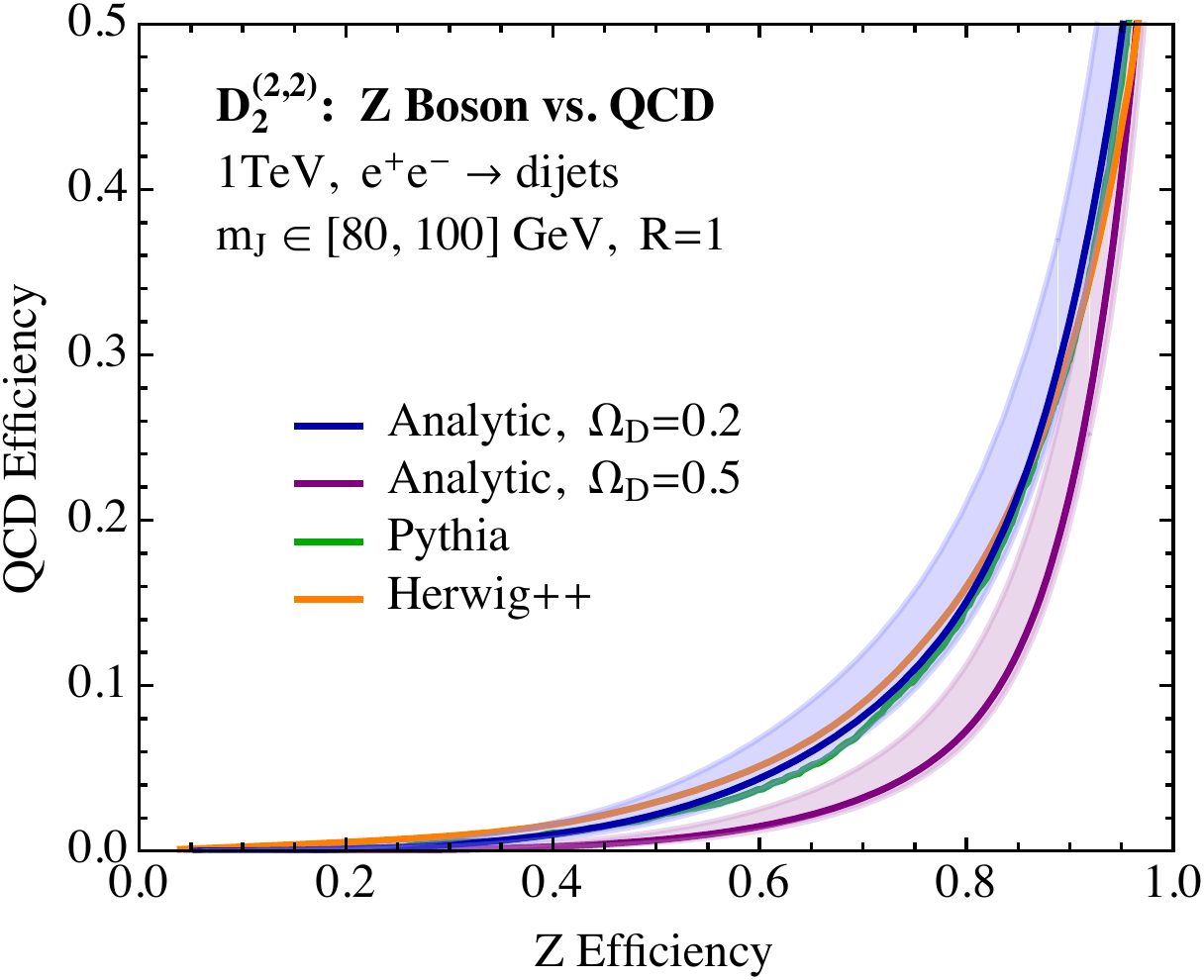}
}\end{center}
\caption{ Signal vs.~background efficiency curves for $\Dobs{2}{2,2}$ for the Monte Carlo samples as compared to our analytic prediction for two different values of the non-perturbative shape parameter, chosen by varying our central value by $\pm0.15$ GeV. Results are shown on a  logarithmic scale in a) and a linear scale in b).  Perturbative scale uncertainties are also shown.
}
\label{fig:ROC_doubleshape}
\end{figure}

One feature made clear by the linear ROC curve in \Fig{fig:D2_rocb} is the increase in perturbative uncertainty with increasing $Z$ efficiency. As emphasized earlier, this is due to the fact that for the region of interest for $Z$ discrimination, one is probing values of $D_2$ which are below the peak of the background distribution, and therefore in the non-perturbative regime. As the $Z$ efficiency is increased, one enters the peak region of the background distribution, where the perturbative uncertainty is largest, causing a corresponding increase in the uncertainty band for the ROC curve. However, we do not include uncertainties due to the non-perturbative parameter $\Omega_D$ or from the shape function, in \Fig{fig:D2_rocb}, which are the dominant sources of uncertainty in this region.

To demonstrate that is indeed the case, in \Fig{fig:ROC_doubleshape} we show ROC curves in both linear and log scales for two different values of the non-perturbative shape parameter. The values of $\Omega_D$ where chosen by varying our central value of $\Omega_D=0.34$ GeV by $\pm 0.15$ GeV (and rounding to nice numbers). We have also shown the distributions from the \herwigpp{} and \pythia{} generators as representative of the ROC curves generated by the Monte Carlo generators. This figure makes clear that in the region of efficiencies of interest for boosted $Z$ tagging, one is extremely sensitive to the $D_2$ distribution in the deeply non-perturbative region, and this uncertainty swamps the perturbative uncertainty. To be able to improve the accuracy in this region will require detailed comparisons with Monte Carlo, data, and analytic calculations, to allow for a clean separation of the non-perturbative parameter from perturbative modifications to the shape of the distribution.

\begin{figure}
\begin{center}
\subfloat[]{\label{fig:D2_50sig}
\includegraphics[width= 7.2cm]{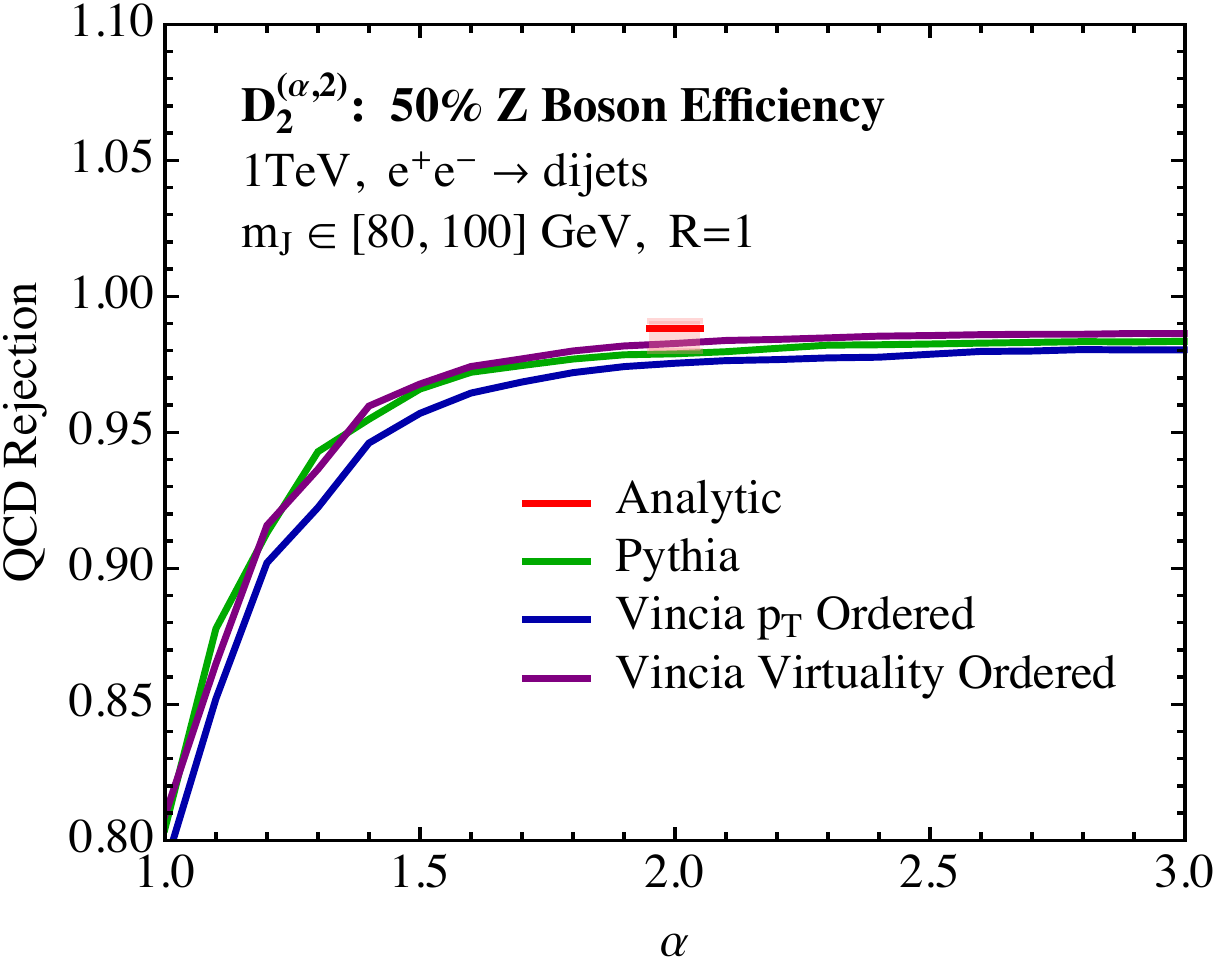}
}
\ 
\subfloat[]{\label{fig:D2_75sig}
\includegraphics[width = 7.0cm]{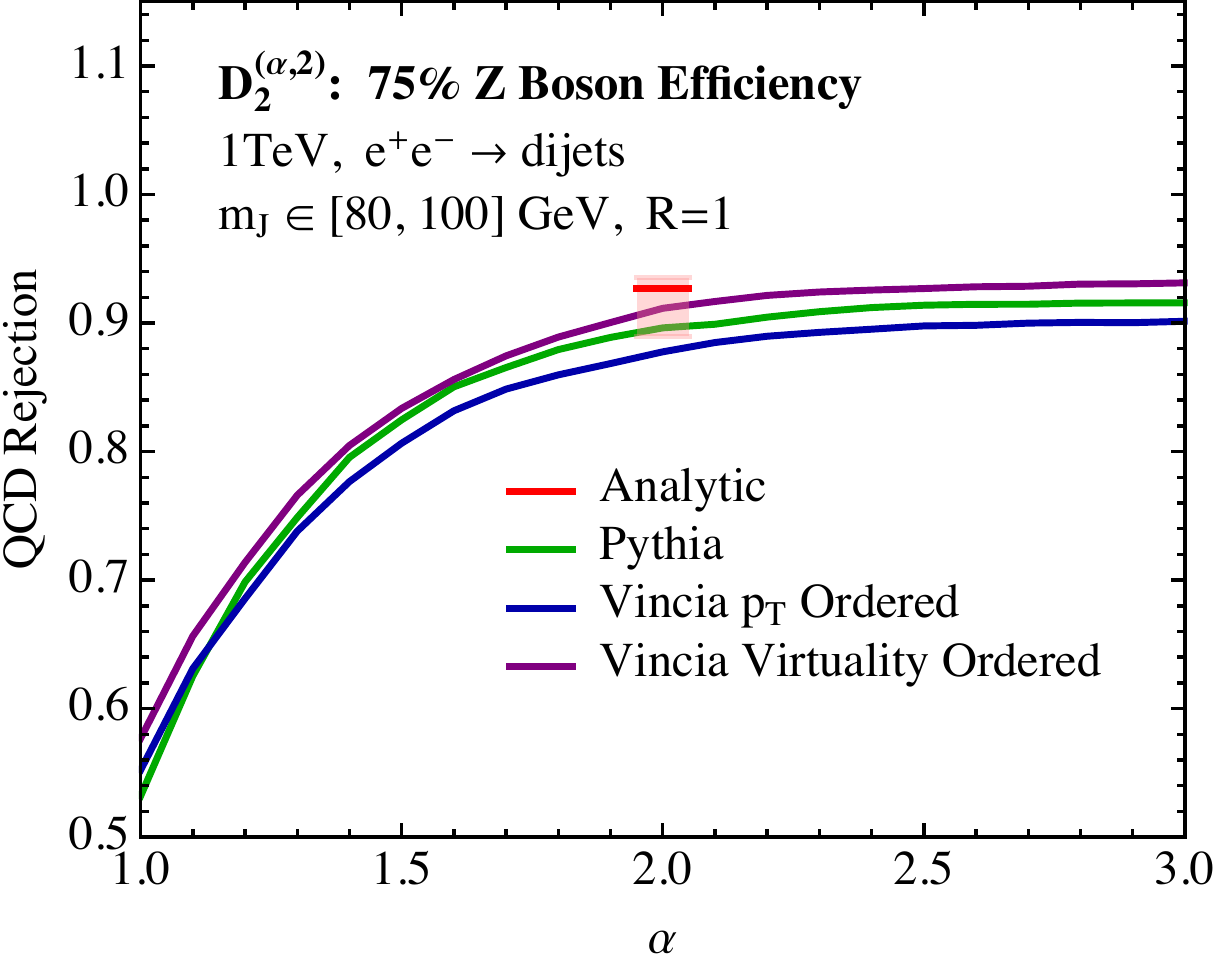}
}\end{center}
\caption{Background rejection rate at fixed a) 50\% and b) 75\% signal efficiency as a function of the angular exponent of the 3-point energy correlation function in $D_2$, and a comparison to our analytic prediction for $\alpha=2$.
}
\label{fig:effs}
\end{figure}

To further understand the discrimination power of the $D_2$ observable, in \Figs{fig:D2_50sig}{fig:D2_75sig} we show the background rejection rate at 50\% and 75\% signal efficiency as a function of $\alpha$, the angular exponent of the 3-point energy correlation function in $D_2$.  Below about $\alpha=4/3$, all rejection rates dramatically decrease as $\alpha$ decreases, while above about $\alpha = 4/3$, the QCD rejection rate in all Monte Carlo samples is impressively flat.  This is consistent with our power counting analysis of the $\ecf{2}{2},\ecf{3}{\alpha}$ phase space plane in \Sec{sec:mass_cuts} and is a powerful verification that the Monte Carlos respect the parametric dynamics of QCD.

Although our factorization theorem is valid in the region $\alpha\gtrsim 2$, for $\beta=2$, in \Figs{fig:D2_50sig}{fig:D2_75sig}  we have only shown the analytic prediction for the value $\alpha=2$, where we find that it agrees well with the Monte Carlo results, as expected from the agreement of the distributions and ROC curves. For $\alpha>2$, while our prediction for the background distribution remains accurate (indeed our power counting becomes more valid in this region), the signal distribution becomes extremely sharply peaked, which is difficult to describe, and sensitive to normalization. Due to the fact that this region is also of less phenomenological interest, both because the large angular exponent makes the observable sensitive to pile up contamination, and because both power counting and Monte Carlo analyses indicate that optimal performance is achieved for $\alpha=2$, we have decided not to focus on this region. It would be potentially interesting to see if higher order resummation would be sufficient to describe the sharply peaked signal distribution in this region, as well as to test the universality of the non-perturbative power corrections.

One further interesting feature of \Figs{fig:D2_50sig}{fig:D2_75sig} is the correspondence between the perturbative scale variations, and the spread in the curves from the different Monte Carlo generators, which agree well at both 50\% and 75\%. For the case of $p_T$-ordered \vincia{} as compared with virtuality ordered \vincia{}, this correspondence is precise, as the difference between the Monte Carlos can be viewed as a scale variation, and identical hadronization models are used.

\subsection{Discrimination in the Two-Prong Regime}\label{sec:2insight}

\begin{figure}
\begin{center}
\subfloat[]{\label{fig:broadening_a}
\includegraphics[width= 7.25cm]{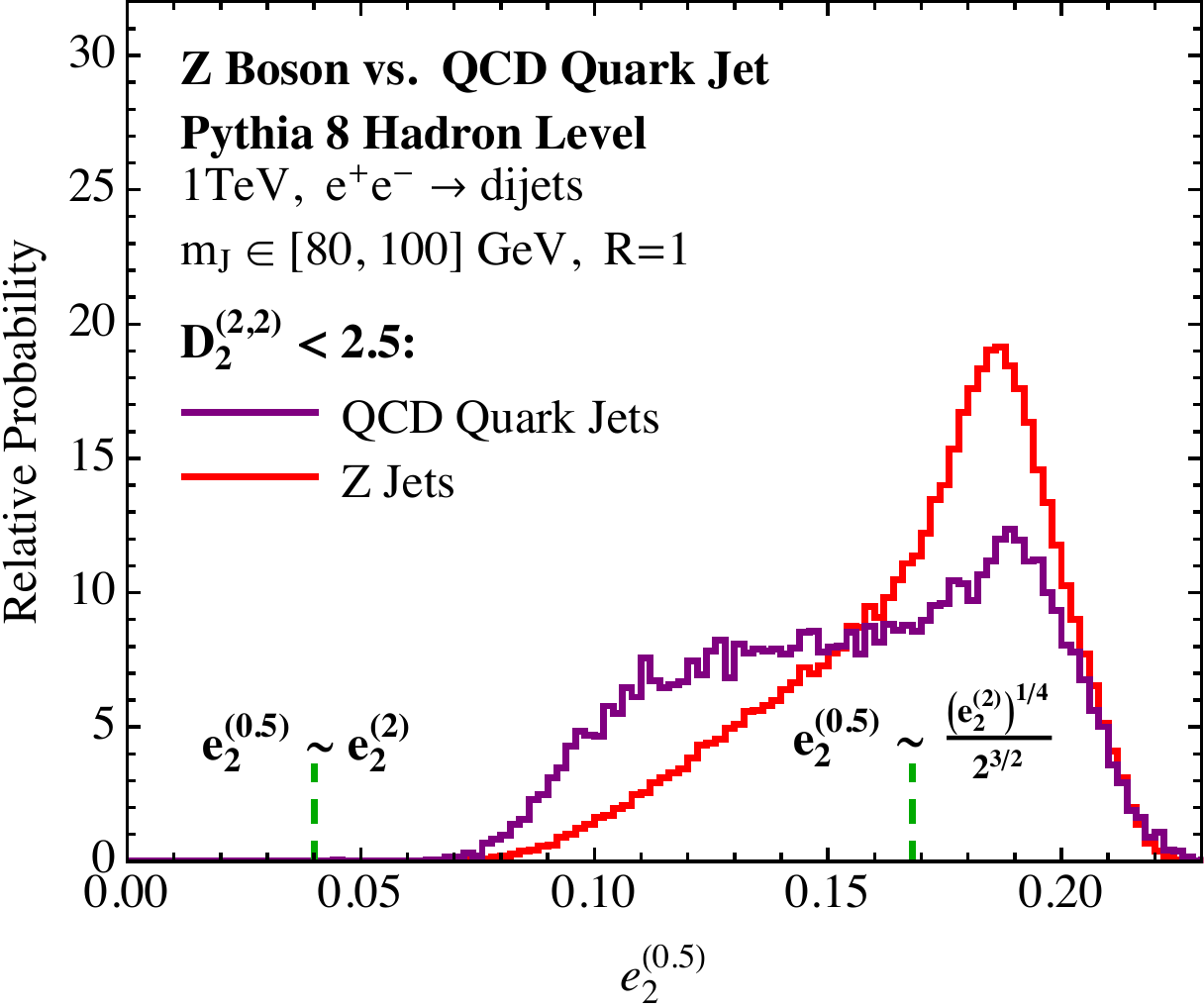}
}
\subfloat[]{\label{fig:broadening_b}
\includegraphics[width= 7.25cm]{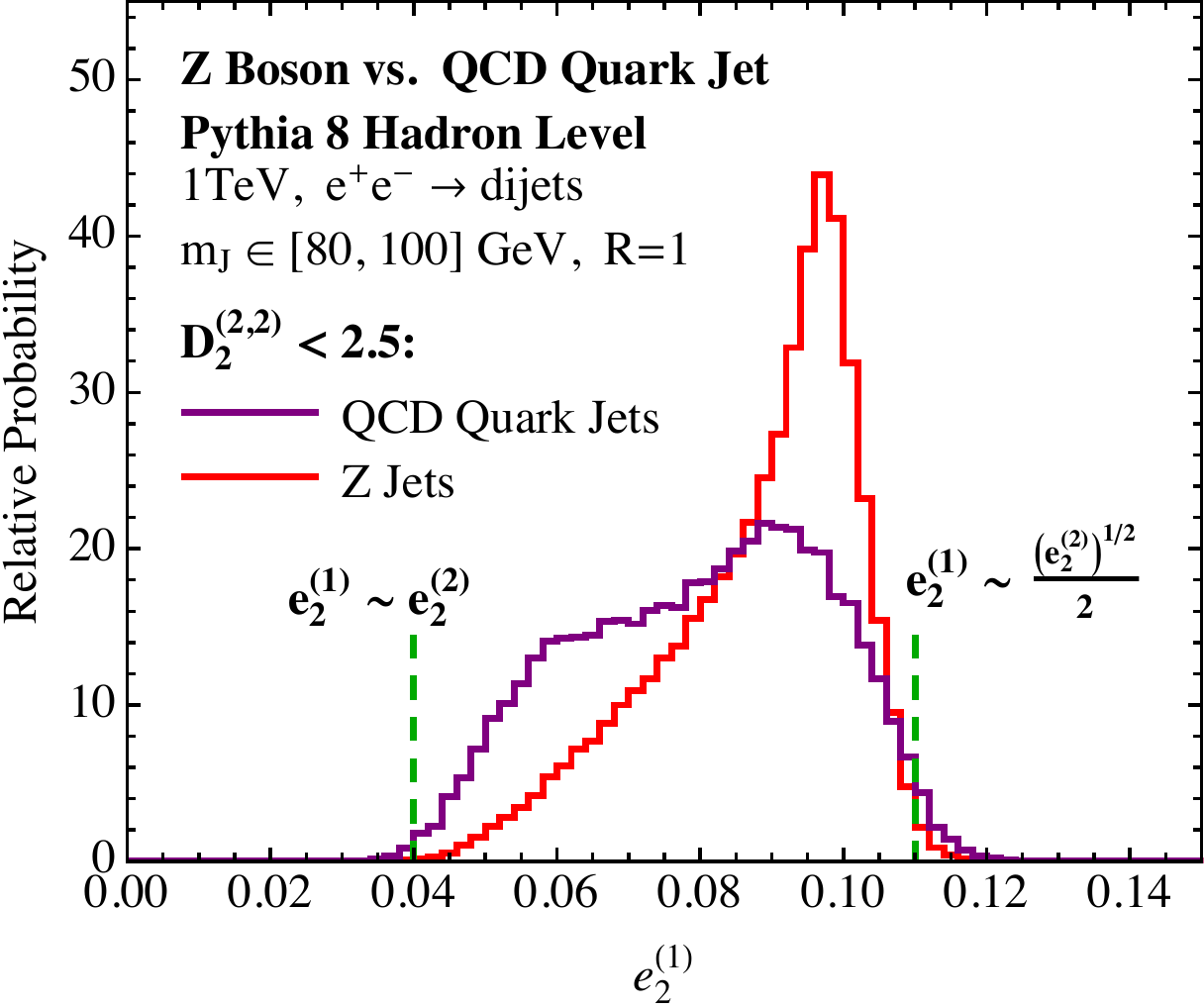}
}
\end{center}
\caption{Distributions for $\ecf{2}{0.5}$ (left) and $\ecf{2}{1.0}$ (right) from the signal and background \pythia{} Monte Carlo samples.  In addition to the mass cut $m_J\in [80,100]$ GeV, these jets are also required to have $\Dobs{2}{2,2}<2.5$ to guarantee that these jets are dominated by two-prong structure.  The parametric boundaries of $\ecf{2}{\beta}$ from \Eq{eq:e2bounds} are shown with the green dashed lines.
}
\label{fig:D2_ee_further}
\end{figure}

Throughout this paper, we have emphasized that the discrimination of boosted hadronically decaying $Z$ bosons (or $W$ or $H$ bosons) from massive QCD jets is effectively a problem of discriminating one- from two-prong jets.  We have demonstrated that the observable $D_2$ is powerful for this goal.  However, in the formulation of our factorization theorem for calculating the distribution of $D_2$, we needed to perform additional 2-point energy correlation function measurements on the jet to separate contributions from soft subjet and collinear subjets contributions to background.  While indeed the signal jets are dominantly two-pronged, we further know that those prongs are dominantly collinear, and do not have parametrically different energies.  Therefore, we are able to further discriminate signal from background jets in the two-prong region of phase space by exploiting additional measurements that can isolate the soft subjet and collinear subjet configurations.  A detailed analysis of this is beyond the scope of this paper, but here, we will demonstrate in Monte Carlo that such a procedure is viable.

To investigate this, we measure the observable $\Dobs{2}{2,2}$ on jets on which a tight mass window cut has been applied.  Other angular exponents for $D_2$ can be used also, but here we only measure $D_2$ to define two-prong jets.  We restrict to the two-prong region of phase space by requiring that $\Dobs{2}{2,2} < 2.5$.  Then, on the jets that pass these cuts, we measure two, 2-point energy correlation functions, $\ecf{2}{2}$ and $\ecf{2}{\beta}$, where $\beta < 2$.  As discussed in \Sec{sec:Fact}, the measurement of the two 2-point energy correlation functions provides an IRC safe definition of the subjets' energy fractions and splitting angle.  Because we make a tight mass cut on the jets, $\ecf{2}{2}$ is essentially fixed, and only $\ecf{2}{\beta}$ is undefined.  We will study the distribution of $\ecf{2}{\beta}$ for both signal and background jets in this region of phase space.

For a fixed value of $\ecf{2}{2}$ and $\beta < 2$, $\ecf{2}{\beta}$ is parametrically bounded as
\begin{equation}\label{eq:e2bounds}
\ecf{2}{2} \lesssim \ecf{2}{\beta} \lesssim 2^{\beta-2}(\ecf{2}{2})^{\beta / 2}\,.
\end{equation}
In the two-prong region, the lower bound is set by the soft subjet while the upper bound is set by collinear subjets.  Therefore, $\ecf{2}{\beta}$ for signal jets will peak near $2^{\beta-2}(\ecf{2}{2})^{\beta / 2}$, while background QCD jets will fill out the full range.  We illustrate this in \Fig{fig:D2_ee_further} on the hadronized \pythia{} sample with the appropriate cuts applied.  We show plots of the distributions of $\ecf{2}{0.5}$ and $\ecf{2}{1}$ on both signal and background jets and have added dotted lines to denote the parametric upper and lower boundaries.  As expected, signal peaks near the upper boundary and background fills out the entire allowed region and so this additional information could be used for discrimination.  For the very small values of $\beta=0.5$, an $\mathcal{O}(1)$ drift is observed with respect to the parametric boundaries, while for $\beta=1$, the parametric boundaries are extremely well respected. 

This demonstrates a simple example of an observable which goes beyond the simple one vs. two prong picture of jet substructure, asking more differential questions about the subjets themselves. In particular, it could be used both to further improve the discrimination power of boosted boson discriminants, and to study in detail the QCD properties of subjets.

\section{Looking Back at LEP}\label{sec:LEP}

In this section, we consider the $D_2$ distribution for QCD jets in $e^+e^-$ collisions at the $Z$ pole at LEP, for which a large amount of data exists.   While the use of $D_2$ for boosted boson discrimination is not possible, nor relevant, at LEP, this will emphasize the sensitivity of $D_2$ as a probe of two-prong structure in jets. We will suggest the importance of using variables sensitive to two emissions off of a primary quark in tuning Monte Carlo generators to LEP data.

Our definition of the energy correlation functions in \Eq{eq:ecf_def} makes implicit assumptions about the treatment of hadron masses, which we have ignored to this point.  The definition given there is an $E$-scheme treatment of hadron masses \cite{Salam:2001bd,Mateu:2012nk}, but we could equally well define $p$-scheme energy correlation functions as:
\begin{align}\label{eq:ecf_pscheme}
\ecf{2}{\beta}&=\frac{1}{E_J^2}\sum_{ i<j\in J} |\vec p_{i}|\,   |\vec p_{j}| \left[2(1-\cos\theta_{ij})\right]^{\beta/2}\,, \\
\ecf{3}{\beta}&=\frac{1}{E_J^3}\sum_{ i<j<k \in J} |\vec p_{i}|\,  |\vec p_{j}|  \,|\vec p_{k}| \left[2(1-\cos\theta_{ij})2(1-\cos\theta_{jk})2(1-\cos\theta_{ik})\right]^{\beta/2}\,, \nonumber
\end{align}
where $\vec p_i$ denotes the three-momenta of particle $i$. For massless particles, this definition is identical to that of \Eq{eq:ecf_def}, and so our perturbative analytics would be unchanged by using this definition or the definition of \Eq{eq:ecf_def}.\footnote{As will be discussed shortly, the differences in our analytic calculation due to hadron masses will arise through non-perturbative effects, namely the shape function.}  The definitions of \Eq{eq:ecf_def} and \Eq{eq:ecf_pscheme} differ for massive particles. In particular, the energy correlation functions as defined in \Eq{eq:ecf_pscheme} have the advantage that they vanish for low momentum or collimated particles regardless of whether these particles are massless or massive, which is not true of the definition in \Eq{eq:ecf_def}. Because of this, we expect that the energy correlation functions as defined in \Eq{eq:ecf_pscheme} are less sensitive to hadron mass effects and that kinematic restrictions on the energy correlation functions remain the same before and after hadronization, so that the phase space studied in \Sec{sec:power_counting} assuming massless particles is not significantly modified.

At LEP energies, hadronization will also have a larger effect on the $D_2$ spectrum than at $1$ TeV. However, a particularly important aspect of our all orders factorization theorem is that it isolates perturbative and non-perturbative physics contributions. In this section we will again implement non-perturbative effects into our analytic calculation using the shape function defined in \Eq{eq:shape_func}. There are two effects which determine how the shape function depends on the jet mass, $m_J$, and the center of mass energy, $Q$. First, for a fixed valued of $\Omega_D$, the 
shift in the first moment of the $D_2$ distribution was given in \Eq{eq:bkg_shift_np}, which we recall here for convenience, by
\begin{equation}\label{eq:LEP_shape}
\Delta_D=\frac{\Omega_D}{ E_J \left(  \frac{m_J}{E_J} \right)^4}\,.
\end{equation}
This has dependence on both $m_J$ and $Q$ (through $E_J$), and for the jets we consider at LEP, this is a considerably larger shift than for the $1$ TeV jets studied in \Sec{sec:Hadronization}. This scaling is a non-trivial prediction of our factorization framework, and we will see that it is well respected when we perform a comparison of our analytic results with Monte Carlo. Furthermore, the parameter $\Omega_D$ has a logarithmic dependence on a renormalization scale, $\Omega_D=\Omega_D(\mu)$, through renormalization group evolution \cite{Mateu:2012nk}, which is briefly reviewed in \App{app:shape_RGE}. However, this effect is small compared with the linear change in the first moment with $E_J$ for a fixed $m_J/ E_J$. A numerical estimate for the effect of the running of $\Omega_D(\mu)$ is given in \App{app:shape_RGE}. At the level of accuracy to which we work in this paper, we cannot probe this logarithmic running, although we will see that our results are consistent with it.

The definition of the energy correlation functions given in \Eq{eq:ecf_pscheme} also has an effect on the universality of the non-perturbative parameter $\Omega_D$, when hadron mass effects are included. Power corrections due to hadron mass effects are of order $\mathcal{O}(m_H/Q)$, where $m_H$ is a light hadron mass, and are therefore of the same order as the leading $\mathcal{O}(\Lambda_{\text{QCD}}/Q)$ power corrections. In the $p$-scheme definition of the energy correlation functions which we have chosen in \Eq{eq:ecf_pscheme}, it is no longer possible to extract the dependence on the angular exponent alpha from $\Omega_D$, as was done in \Eq{eq:alpha_np}. However, to the accuracy to which we work, we expect this to be a negligible effect, and furthermore, the case $\alpha=2$ is of most phenomenological interest, and is the case we have focused on exclusively in this paper. Furthermore, even in the presence of hadron mass effects, it is still possible to extract the parameter $\Omega_D$ from dijet event shapes in the same universality class \cite{Mateu:2012nk}. This exhibits the benefits of the factorization approach both for separating perturbative and non-perturbative effects, and for relating non-perturbative parameters to maintain predictivity.

One further distinction between the case of boosted $Z$ discrimination and the measurement of QCD jet shapes at the $Z$ pole is that while a tight mass cut is natural for boosted $Z$ discrimination, it is not natural in jet shape analyses. However, our shape function analysis, as derived in \Sec{sec:Hadronization}, is valid at a fixed jet mass (or correspondingly fixed value of $\ecf{2}{\beta}$). This is clear from both \Eq{eq:shape_function_eq1} and from the equation for the shift in the first moment in \Eq{eq:bkg_shift_np}. However, we emphasize that the non-perturbative parameter $\Omega_D$ is unique, and the scaling of the non-perturbative shift with the jet mass is fully determined. To achieve an analytic prediction for the non-perturbative $D_2$ spectrum inclusive over the jet mass mass, one must calculate the perturbative $D_2$ spectra differentially in the jet mass, convolve with a shape function for each value of the jet mass, and then integrate over the jet mass. While this is in principle straightforward, it is computationally intensive, and is beyond the scope of this paper. Instead, we will enforce a jet mass cut of $8< m_J <16$ GeV. This mass cut was chosen because it is near to the Sudakov peak of the jet mass distribution for this jet energy and the $m_J$ in this range are set by low scale, but still perturbative, emissions.

\begin{figure}
\begin{center}
\subfloat[]{\label{fig:D2_LEP1_a}
\includegraphics[width= 7.2cm]{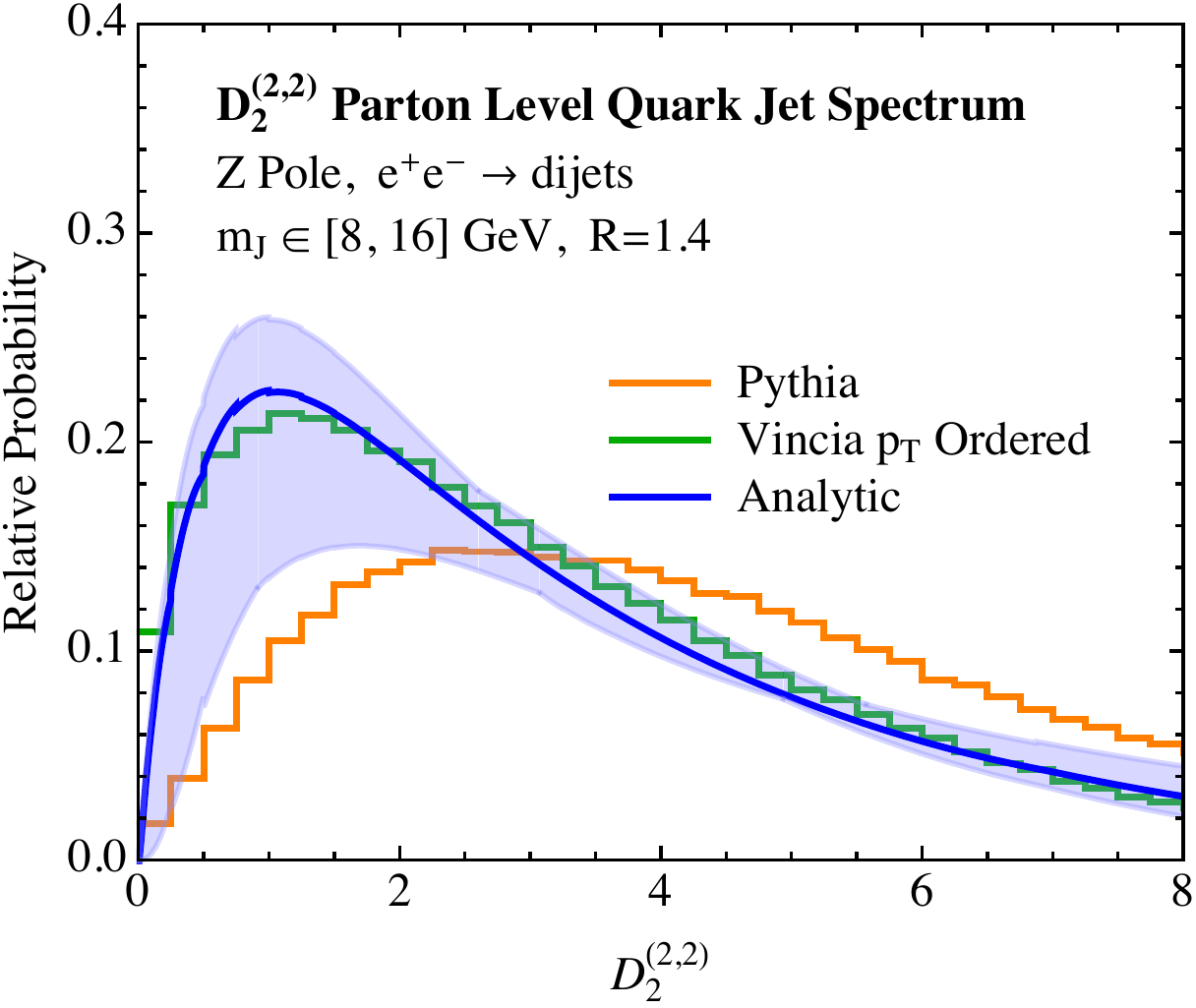}
}
\ 
\subfloat[]{\label{fig:D2_LEP1_b}
\includegraphics[width = 7.2cm]{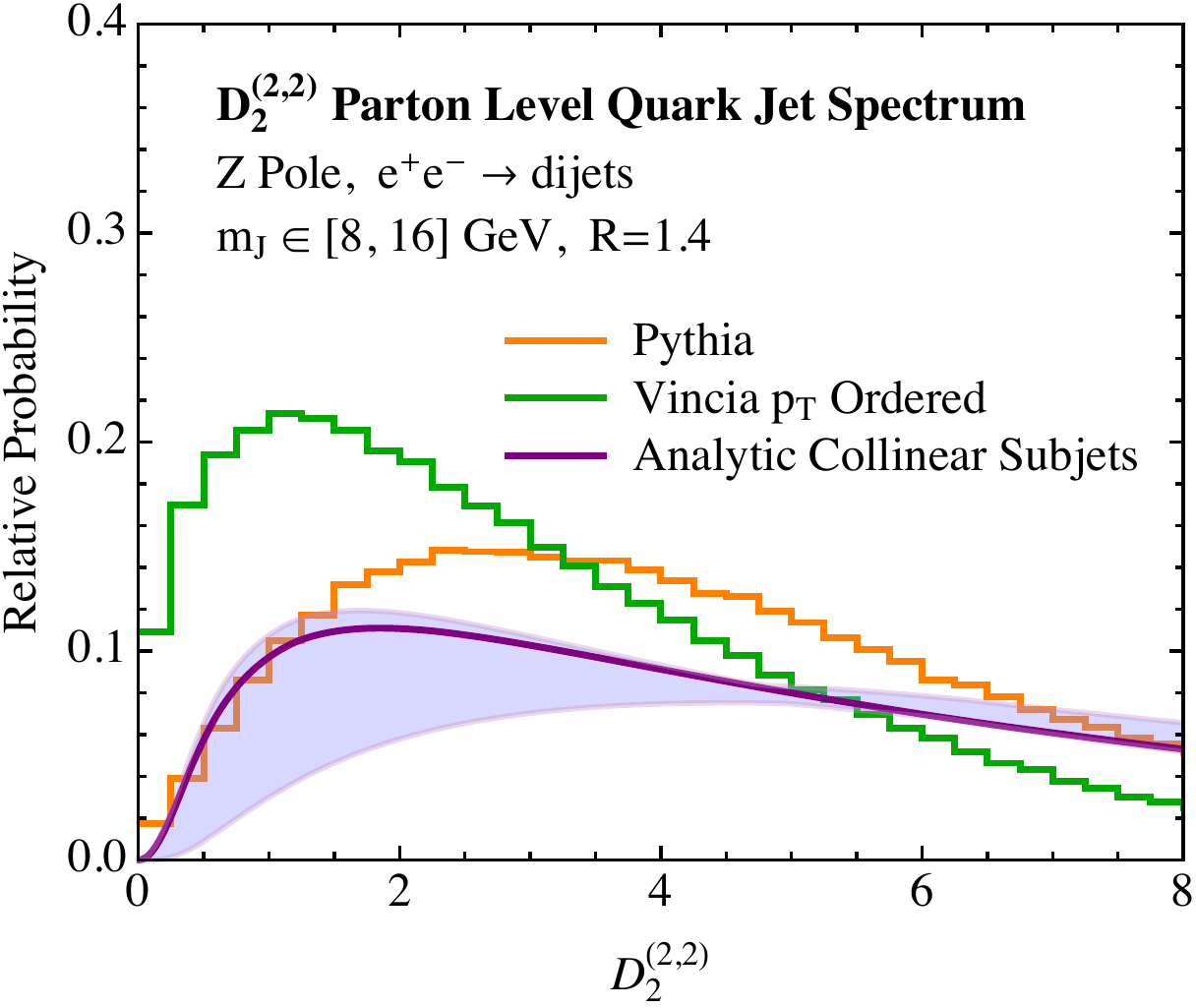}
}
\end{center}
\vspace{-0.2cm}
\caption{A comparison of the $D_2$ spectrum as measured on quark initiated jets at the $Z$ pole from the \pythia{} and $p_T$-ordered \vincia{} Monte Carlo generators to our analytic predictions. a) Comparison of our complete analytic calculation including both the soft subjet and collinear subjets region of phase space with the predictions of the Monte Carlo generators. b) Comparison of our analytic calculation including only the collinear subjets region of phase space compared with the predictions of the Monte Carlo generators.  
The pinch in the scale variations is a consequence of unit normalizing the distributions.
}
\label{fig:D2_LEP1}
\end{figure}

Similar to what we did in our numerical analysis at $1$ TeV, we begin in \Fig{fig:D2_LEP1} by comparing our analytic prediction for the $D_2$ spectrum with the distributions from parton level Monte Carlo. In \Fig{fig:D2_LEP1_a}, we show a comparison of our complete analytic calculation, including perturbative scale variations, along with Monte Carlo predictions from both  \pythia{} and $p_T$-ordered \vincia{}, which we take as representative of the different Monte Carlo generators. We use a jet radius of $R=1.4$ to approximate hemisphere jets. We find good agreement with the predictions of the \vincia{} Monte Carlo. It is important to emphasize, however, that at LEP energies, non-perturbative effects are large, and therefore a comparison with parton level Monte Carlo is difficult due to large uncertainties in the treatment of the shower cutoff. We also show, in \Fig{fig:D2_LEP1_b}, a comparison of our analytic prediction, including only the collinear subjets region of phase space, with both Monte Carlo generators. The difference between the analytic predictions in  \Fig{fig:D2_LEP1_a} and  \Fig{fig:D2_LEP1_b} emphasizes the large effect played by the soft subjet at LEP energies. Unfortunately, due to large hadronization corrections, the treatment in Monte Carlo of the soft subjet region is difficult to disentangle from the treatment of non-perturbative physics.

\begin{figure}
\begin{center}
\subfloat[]{\label{fig:D2_LEP_a}
\includegraphics[width= 7.2cm]{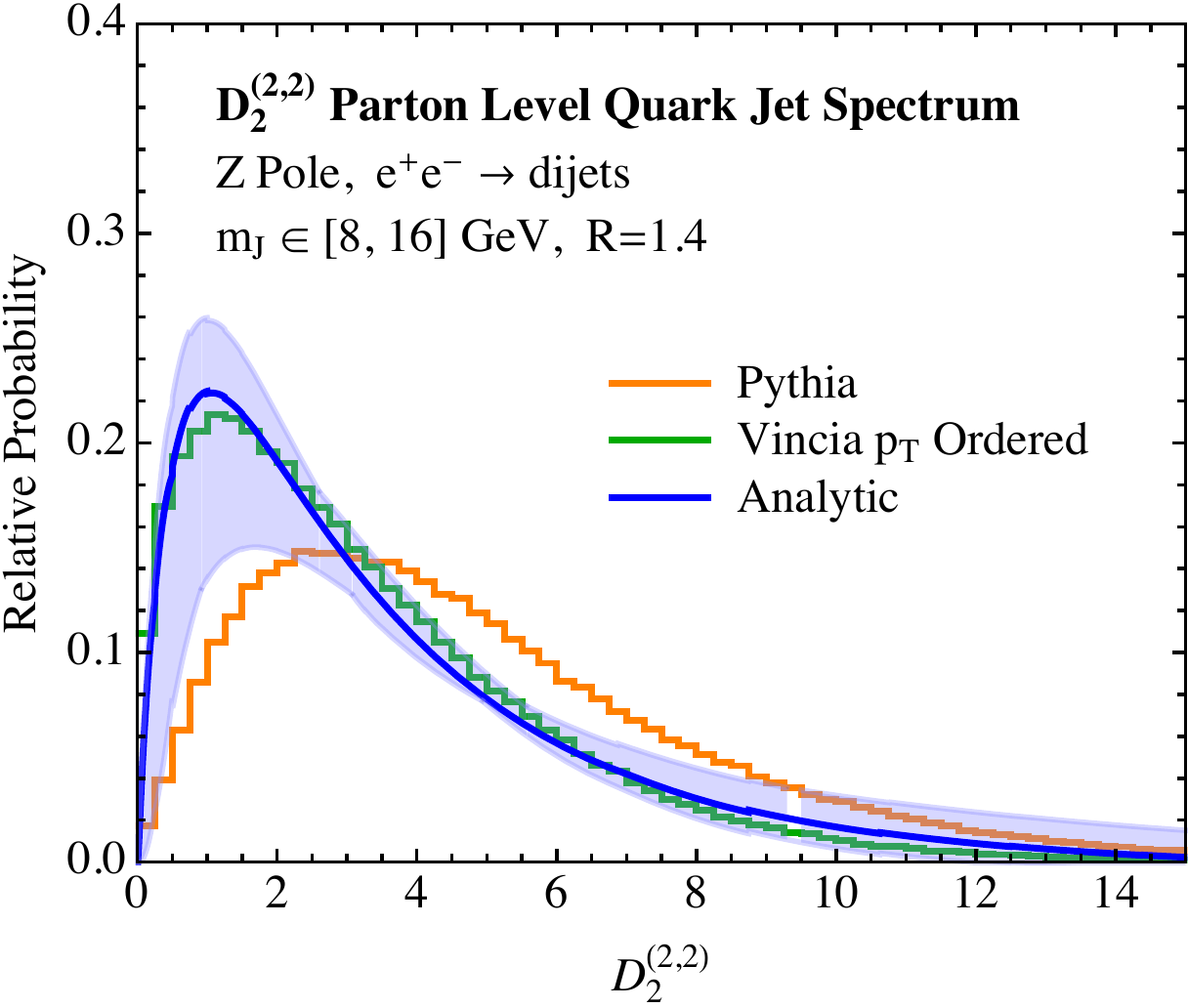}
}
\ 
\subfloat[]{\label{fig:D2_LEP_b}
\includegraphics[width = 7.2cm]{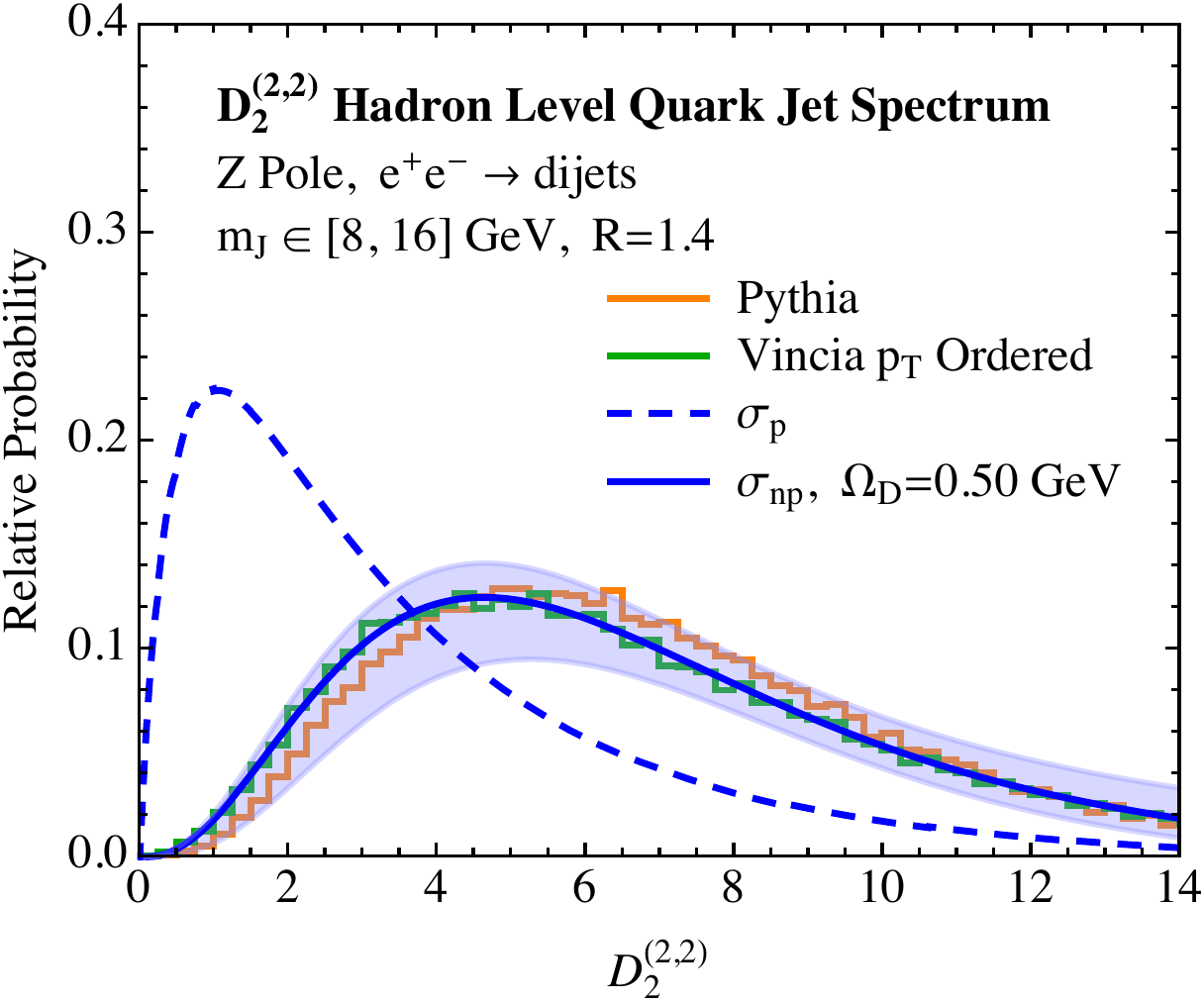}
}
\end{center}
\vspace{-0.2cm}
\caption{A comparison of the $D_2$ spectrum as measured on quark initiated jets at the $Z$ pole from the \pythia{} and $p_T$-ordered \vincia{} Monte Carlo generators to our analytic predictions. Results are shown both for parton level Monte Carlo compared with perturbative analytics in a), and for hadron level Monte Carlo compared with non-perturbative analytics in b). 
The pinch in the scale variations is a consequence of unit normalizing the distributions.
}
\label{fig:D2_LEP}
\end{figure}

In \Fig{fig:D2_LEP_b} we show our analytic prediction for the non-perturbative spectra using the shape function. An alternate view of the perturbative spectrum is shown in  \Fig{fig:D2_LEP_a} for reference, and to show the overall shape of the perturbative distribution.  We have used a valued of $\Omega_D=0.50$ GeV, which was obtained by fitting to the \vincia{} Monte Carlo.
There is considerable uncertainty on this value, probably of the order $\pm0.3$ GeV due to the wide mass window, which is probably slightly large for the na\"ive application of our shape function. Furthermore, as demonstrated in \Sec{sec:Hadronization}, there is some ambiguity in the value of $\Omega_D$, depending on whether it is extracted from hadron level \pythia{} or \vincia{}, which is of this same order. However, this value is consistent with $\Omega_D=0.34$ GeV as extracted from our analysis at $1$ TeV. Although it is expected that the logarithmic running of the $\Omega_D$ parameter will decrease its value slightly, this effect is expected to be small.  The amount by which it is expected to decrease depends on another non-perturbative parameter, but is estimated in \App{app:shape_RGE} that $\Omega_D$ should decrease by approximately $0.1$ GeV between our predictions at $1$ TeV and those at LEP energies.  This is an important consistency check on our results, but due to the large uncertainty, we cannot claim to probe this running over the scales that we have considered. The analytic perturbative spectrum is also shown for reference. Good overall agreement with both Monte Carlo generators is observed, and the discrepancy between the \pythia{} and \vincia{} generators which was present at parton level is reduced, although still non-negligible. As was discussed in \Sec{sec:Hadronization}, it could also be compensated for by a modification of the non-perturbative shape parameter. In particular, the effect of hadronization is well captured by non-perturbative shape function. Hadronization has a significantly larger effect on the $D_2$ observable at $Z$ pole energies than at $1$ TeV. This demonstrates the consistency of our implementation of the non-perturbative corrections through the shape function, which predicts the scaling of the shift in the first moment through \Eq{eq:LEP_shape}. 

Unlike for the $D_2$ distributions at $1$ TeV, where the effect of hadronization was well described only by a shift in the first moment, at LEP energies the hadronization also has a non-trivial effect on the shape of the distribution. This can clearly be seen by comparing the dashed perturbative spectrum and the non-perturbative results in \Fig{fig:D2_LEP_b}. While our factorization of non-perturbative effects in terms of a shape function is completely generic, it is only the first moment of the shape function which is universal, with the full non-perturbative shape function being in general observable dependent. However, the modification in the shape of the $D_2$ spectrum due to hadronization effects seems to be quite well captured by the shape function of \Eq{eq:shape_func}. In our plots we do not include any uncertainties due to the form of the non-perturbative shape function, despite the fact that they are the dominant effect throughout most of the hadronized distribution. More general shape functions, and a study of their associated uncertainties could be studied along the lines of \Ref{Ligeti:2008ac}, although this is beyond the scope of this paper, and could only be justified if the perturbative components of our calculation were computed to a higher level of accuracy.

Since the $D_2$ spectrum is sensitive to the emissions from the gluon subjet, it is sensitive to the radiation pattern generated by a gluon, and could potentially be used to improve the Monte Carlo description of gluons and the modeling of color coherence effects. In contrast to most observables which have been used for tuning Monte Carlos to LEP data, such as the jet mass which is set by a single emission, $D_2$ requires two emissions off of the initiating quark to be non-zero, and therefore can be used as a more detailed probe of the perturbative shower. Although non-perturbative effects play a large role for jets in this energy range, we have shown that our factorization theorem allows us to cleanly separate perturbative from non-perturbative effects, which could be useful when tuning Monte Carlo generators, allowing one to disentangle genuine perturbative effects which should be well described by the Monte Carlo shower, from effects which should be captured by the hadronization model. We believe that higher order calculations of QCD jet shapes sensitive to three particle correlations, such as $D_2$, and their use in Monte Carlo tunings is therefore well motivated.

For reference, in \App{app:twoemissionMC} we show a collection of $\ecf{2}{2}$ distributions measured at the $Z$ pole, at both parton and hadron level for both the \vincia{} and \pythia{} event generators. Unlike for the $D_2$ observable, the \vincia{} and \pythia{} generators agree both at parton and hadron level to an excellent degree. This is of course expected due to the fact that these Monte Carlos have been tuned to LEP event shapes, but further emphasizes the fact that $D_2$, and other observables sensitive to additional emissions, provide a more detailed probe of the perturbative shower.

\section{Looking Towards the LHC}\label{sec:LHC}

Throughout this paper, we have restricted our analysis to $e^+e^-$ colliders so that we could ignore subtleties with initial state radiation, pile-up and other features important at hadron colliders. However, it is precisely for including these effects that a rigorous factorization based approach to jet substructure, such as that presented in this paper, will prove most essential.  In this section, we discuss the extension to the LHC and in particular to what extent conclusions for $e^+e^-$ colliders holds for the LHC.

The energy correlation functions have a natural longitudinally-invariant generalization relevant for $pp$ colliders, which is given by \cite{Larkoski:2013eya,Larkoski:2014gra}
\begin{align}\label{eq:pp_def_ecf}
\ecf{2}{\beta} &= \frac{1}{p_{TJ}^2}\sum_{1\leq i<j\leq n_J} p_{Ti}p_{Tj} R_{ij}^\beta \ ,\nonumber \\
\ecf{3}{\beta} &= \frac{1}{p_{TJ}^3}\sum_{1\leq i<j<k\leq n_J} p_{Ti}p_{Tj}p_{Tk} R_{ij}^\beta R_{ik}^\beta R_{jk}^\beta \ .
\end{align}
Here $p_{TJ}$ is the transverse momentum of the jet with respect to the beam, $p_{Ti}$ is the transverse momentum of particle $i$, and $n_J$ is the number of particles contained in the jet.  The boost-invariant angle $R_{ij}^2 = (\phi_i-\phi_j)^2+(y_i-y_j)^2$ is defined as the Euclidean distance in the azimuth-rapidity plane. For central rapidity jets, which we will restrict ourselves to in this section, the power counting discussion of \Sec{sec:phase_space} is unmodified. Therefore, the same conclusions for the form of the optimal observable, $D_2$, as well as the range of angular exponents, apply. A simplified version of the $\Dobs{2}{\alpha, \beta}$ variable, restricted to have equal angular exponents $\alpha=\beta$, was used in \Ref{Larkoski:2014gra}, for jet substructure studies at the LHC.

It is in principle straightforward to extend the factorization theorems for $D_2$ to hadronic colliders, where $D_2$ is measured on a single jet in an exclusive $N$-jet event. Factorization theorems for exclusive $N$-jet production defined using $N$-jettiness \cite{Stewart:2009yx,Stewart:2010tn} or with a $p_T$-veto \cite{Liu:2012sz,Liu:2013hba} on additional radiation exist and could be combined with the factorization theorems of \Sec{sec:Fact} to describe the jet substructure. We now briefly discuss how each of these factorization theorems can be interfaced with the presence of additional eikonal lines, representing either additional jets or beam directions in $pp$ collisions.

Recall from \Sec{sec:ninja}, that the collinear subjets factorization theorem is formulated as a refactorization of the jet function for a particular jet in the $n$ direction,  and it is therefore insensitive to the global color structure of the event, seeing only the total color. Intuitively, the collinear-soft modes are boosted, and therefore all additional Wilson lines in the event are grouped in the $\bar n$ direction. Furthermore, the global soft modes, which resolve the global color structure of the event do not resolve the jet substructure. This property of the collinear subjets factorization theorem has the feature that it can be trivially combined with a factorization theorem with an arbitrary number of eikonal lines, without complicating the color structure. All that is then required, apart from the substructure components, is the addition of an additional measurement function in to the global soft function. Indeed, this extension has been discussed in detail in \Ref{Bauer:2011uc}. This same property is of course also true for the soft haze factorization theorem, as no additional Wilson lines are required to describe the jet substructure in the first place. 

However, for the soft subjet factorization theorem, the presence of additional Wilson lines does significantly complicate the factorization from a calculational perspective. In particular, since the subjet is soft, arising from a refactorization of the soft function, it is emitted coherently from the $N$-eikonal line structure as a whole, requiring a proper treatment of all color correlations, which becomes complicated with even a few additional Wilson lines. A conjectural proposal for the all orders soft subjet factorization theorem with $N$-eikonal lines was given in \Ref{Larkoski:2015zka}, where the soft subjet factorization theorem was first proposed and studied in the large $N_c$ limit. However, more work is required to understand its structure, and an efficient organization of the color correlations at finite $N_c$. Furthermore for the soft subjet factorization theorem, the final soft function has an additional eikonal line, since the jet substructure is resolved by the long wavelength global soft modes, further complicating the calculation (although there has recently been some progress in the computation of soft functions \cite{guido_talk,Boughezal:2015eha}). We emphasize however, that these are purely technical complications, and believe that the extension to a calculation of jet substructure in $pp$ would be well worthwhile for improving our understanding of analytic jet substructure. Furthermore, depending on the relevant boosts and jet radii, the techniques of this paper could be used to identify whether the soft subjet factorization theorem plays an important role, or could be formally neglected, simplifying the calculation in more complicated cases.

\begin{figure}
\begin{center}
\subfloat[]{\label{fig:D2_LHC_a}
\includegraphics[width= 7cm]{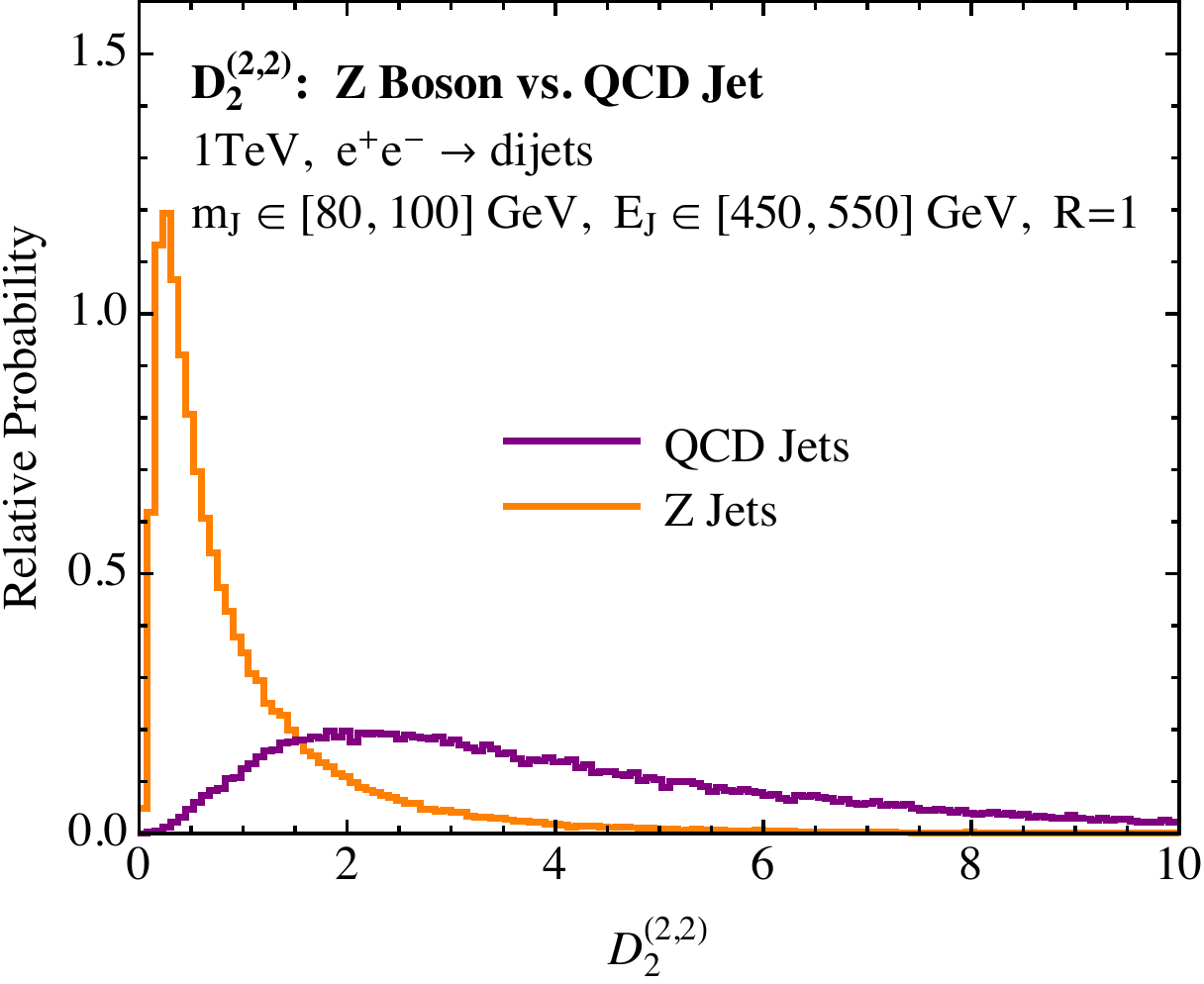}
}
\ 
\subfloat[]{\label{fig:D2_LHC_b}
\includegraphics[width = 7cm]{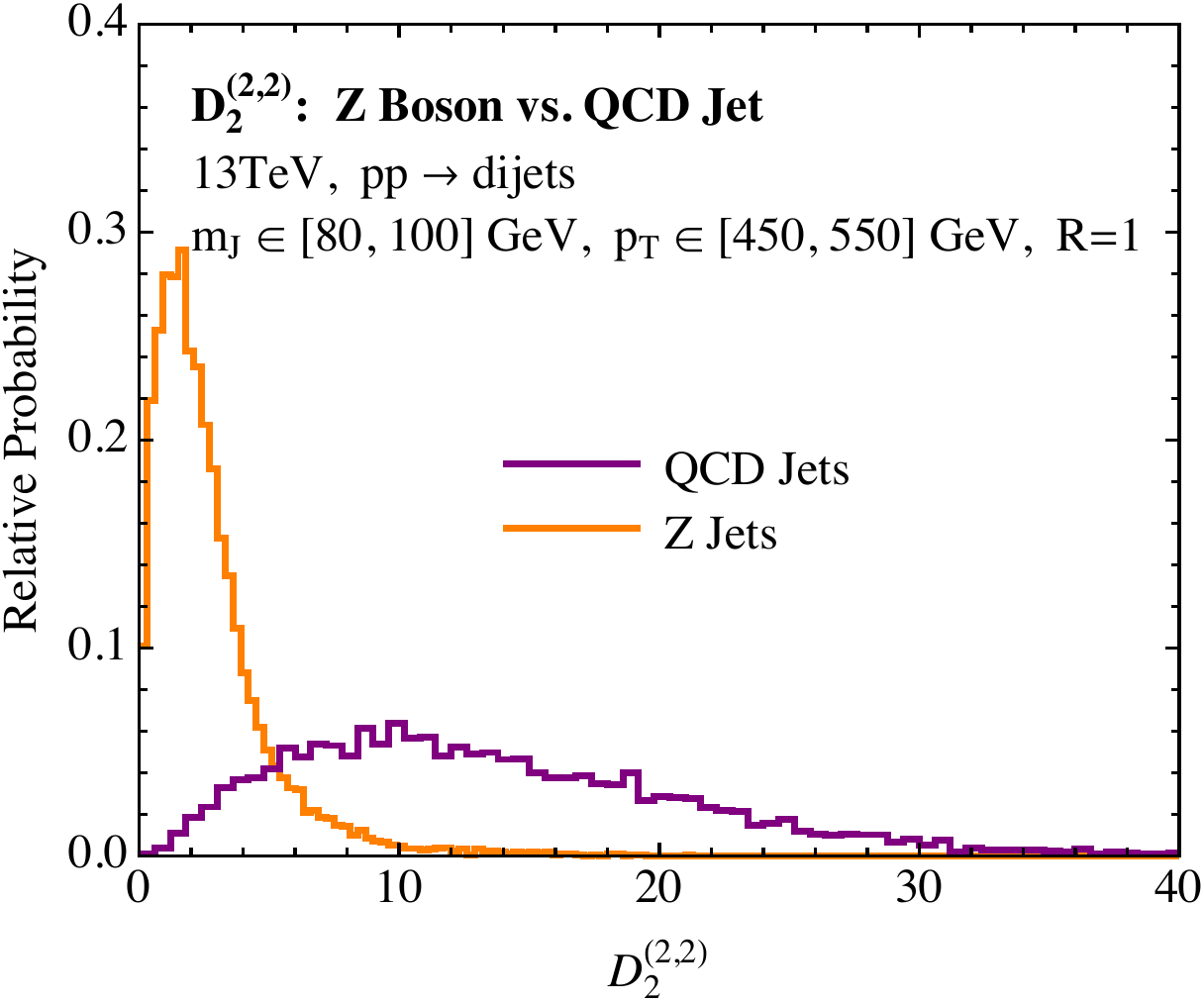}
}
\end{center}
\vspace{-0.2cm}
\caption{A comparison of the $\Dobs{2}{2,2}$ distributions for signal and background jets.  a) Distributions for $R=1$ jets at a 1 TeV $e^+e^-$ collider.  b) Distributions for $R=1$ jets at the 13 TeV LHC, for jets with transverse momenta in the range $p_T\in[450,550]$ GeV.
}
\label{fig:D2_LHC}
\end{figure}

For these reasons, a full calculation in $pp$ is well beyond the current scope of this initial investigation. We will instead restrict ourselves to a brief Monte Carlo study comparing the properties of $D_2$ in $e^+e^-$ and $pp$ to show that the distributions exhibit similar features. In \Fig{fig:D2_LHC} we compare the Monte Carlo predictions for $\Dobs{2}{2,2}$ as measured in $e^+e^-$ collisions, shown in \Fig{fig:D2_LHC_a}, and $pp$ collisions, shown in \Fig{fig:D2_LHC_b}.  For $e^+e^-$ collisions, the event selection is identical to earlier.  For $pp$ collisions, we generate background events from the parton-level process $pp \to q \bar q$ and signal events from $pp \to ZZ \to q\bar q q\bar q$ events, where $q$ denotes a massless quark, with \pythia{8.205} at the $13$ TeV LHC.\footnote{Since we only briefly mention the case of $pp$ colliders, we do not perform a systematic study of the variation of the $D_2$ distribution in $pp$ with different Monte Carlo generators, as we did for the case of $e^+e^-$. However, we believe that this is essential in any jet substructure study at $pp$, as we expect variations will be present, as in the $e^+e^-$ case. It would be particularly interesting to compare a $p_T$-ordered dipole-antenna shower, such as was recently implemented for $pp$ in \dire{} \cite{Hoche:2015sya}, with the \pythia{} and \herwigpp{} generators which are more commonly used in jet substructure studies at the LHC. 
} Jets are clustered with the anti-$k_T$ algorithm with radius $R=1.0$, and using the WTA recombination scheme, with a $p_T$ metric. We cut on the transverse momentum of the hardest jet, requiring $p_T \in[450,550]$ GeV, and on the jet mass requiring $m_J \in [80,100]$ GeV. These are chosen to be similar to the cuts on the jets for the case of $e^+e^-$, although they are of course not identical, and strict comparisons should not be made between the two cases. The shapes and general features of the $D_2$ distributions at the two colliders are very similar.  There is a relative scaling between the $D_2$ distributions in $e^+e^-$ and $pp$ due to the different observable definitions. The $e^+e^-$ definition uses the $1-\cos(\theta_{ij})$ measure of \Eq{eq:ecf_def}, while the $pp$ definition uses the boost invariant definition in terms of $R_{ij}$, as in \Eq{eq:pp_def_ecf}. Since the $\ecf{3}{\alpha}$ observable correlates particles of separation up to $2R$, where $R$ is the jet radius, for $\alpha=\beta=2$, this gives an expected factor of $4$ difference between the two cases, as is approximately observed in \Fig{fig:D2_LHC}.

The similar behavior of the $e^+e^-$ and $pp$ distributions suggests that a complete a calculation using our techniques would provide an excellent description of the $D_2$ distribution at a hadron collider, as we have found for $e^+e^-$. Such a calculation would also be interesting to better understand the effects of initial state radiation on the $D_2$ distribution. A simple setting where this calculation would be feasible, for example, would be to consider measuring the $D_2$ distribution on a jet recoiling against a color-singlet such as a $W$, $Z$ or $H$ boson, as was used in \Ref{Jouttenus:2013hs} to perform a NNLL calculation of the jet mass. Although the effects of non-global logarithms would need to be understood, and could play an important role, recent progress in this area suggests that this issue could be addressed, either by direct resummation of the NGLs \cite{Banfi:2002hw,Weigert:2003mm,Hatta:2013iba,Caron-Huot:2015bja,Larkoski:2015zka}, or through the use of jet grooming algorithms which remove NGLs \cite{Dasgupta:2013ihk,Dasgupta:2013via,Larkoski:2014wba}.  While it is truly uncorrelated with the jet, the effect of radiation from pile-up on $D_2$ could also be mitigated using similar jet grooming algorithms.

\section{Conclusions}\label{sec:conc}

In this paper we have presented a novel approach to the factorization of jet substructure observables, and applied it to the identification of two-prong substructure.  Instead of starting  with a given two-prong discriminant, we used the energy correlation functions as a basis of IRC safe observables to isolate the possible subjet configurations. We then studied the phase space defined by these IRC safe observables and proved all orders factorization theorems in each region of phase space.  This procedure naturally identified an observable, $D_2$, which we argued provided optimal discrimination power, and which preserved the factorization properties of the individual factorization theorems describing different regions of the phase space defined by our basis of observables. We showed that a factorized description of this observable could be obtained by merging the different factorization theorems, and introduced a novel zero bin procedure in factorization theorem space to implement this merging. An important benefit of this approach is that our factorization theorems are valid to all orders in $\alpha_s$ at leading power and therefore provide a systematically improvable description of $D_2$.

Using our factorized description of the $D_2$ observable, we presented a numerical study of our results at an $e^+e^-$ collider, for both the signal and background distributions, resulting in analytic boosted $Z$ boson versus massive QCD jet discrimination predictions. We compared with a variety of Monte Carlo generators, and demonstrated that the low $D_2$ region, where a hard two-prong substructure is resolved, is a sensitive probe of the Monte Carlo parton shower description.  We also studied the effect of non-perturbative corrections, showing that they can be well-described using a simple shape function, and related the single parameter of this shape function to a universal non-perturbative matrix element measured at LEP. This is vital for comparing our calculation with data. 

Because our calculation presents the first factorized description of a two-prong discriminant jet observable in both signal and background regions, there are a large number of directions for future study which are of great interest. First, our calculation was presented in the context of jets produced in $e^+e^-$ collisions. For applications at the LHC, where jet substructure plays a vital role, it is important to extend the calculation to jets produced at a $pp$ collider. The factorization theorem we presented straightforwardly generalizes to $pp$ colliders with only complications due to soft radiation from the beams and the more complicated color structure of the hard interaction. The treatment of both these effects are well-understood and their inclusion in a jet substructure calculation would allow the first precision comparisons of calculations with data.

An interesting potential application of our factorization theorems, and merging procedures, which describe in a more differential way the substructure of jets, is to improve jet shape based subtraction schemes for QCD calculations at NNLO and beyond. Quite recently, subtractions based on the $N$-jettiness observable \cite{Stewart:2010tn} have been used to perform NNLO calculations in QCD \cite{Boughezal:2015aha,Boughezal:2015dva,Gaunt:2015pea}. This allowed, in particular, the calculation of $W$, $H+1$ jet at NNLO  \cite{Boughezal:2015aha,Boughezal:2015dva} ($H+1$ jet at NNLO was also calculated using more traditional subtraction techniques in \cite{Boughezal:2015dra}). The use of more differential subtractions based on more differential factorization theorems would allow for more local, and potentially numerically more efficient subtractions.

It would also be interesting to apply our calculation approach to other observables. For example, the $N$-subjettiness observables \cite{Thaler:2010tr,Thaler:2011gf} are used extensively in jet substructure studies at the LHC, and it would be of significant phenomenological relevance to obtain a factorized description of these observables. The approach presented in this paper could also be extended to study more differential observables, such as those used for boosted top discrimination, which can resolve three subjets. A generalization of the $\Dobs{2}{\alpha, \beta}$ observable, $\Dobs{3}{\alpha, \beta, \gamma}$, which resolves three prong structure was introduced in \Ref{Larkoski:2014zma} (see also \Ref{Larkoski:2015yqa} where it was used for boosted top discrimination at a 100 TeV collider). The $\Dobs{3}{\alpha, \beta, \gamma}$ observable should exhibit similar factorization properties to that of $\Dobs{2}{\alpha, \beta}$, and hence should be calculable with similar techniques. A rigorous factorization will also prove essential in this case, allowing for the separation of perturbative and non-perturbative physics, as well as effects associated with the finite top width \cite{Fleming:2007qr,Fleming:2007xt}. More generally, we anticipate that the approach to the factorization of jet substructure observables presented in this paper will allow for the construction of more powerful jet substructure discriminants and will enable a more detailed analytic understanding of the substructure of high energy QCD jets.

\begin{acknowledgments}
We especially thank Bryan Webber and Andrzej Siodmok for alerting us to errors in the original \herwig analysis and its subsequent correction. We thank the referee for comments on the variation of the perturbative shower cutoff, and its effect on the $D_2$ distribution.  We thank Paolo Nason, Torbj\"orn Sj\"ostrand, and Peter Skands for discussions regarding the ordering of the parton shower in \pythia.  We thank Piotr Pietrulewicz, Stefan Prestel, Iain Stewart, Frank Tackmann, Jesse Thaler, and Wouter Waalewijn for helpful discussions. IM thanks Matthew Low and Lina Necib for discussions on Monte Carlo generation, and Jan Balewski for the use of, and extensive help with, the MIT computing cluster, on which the fixed order calculations of $D_2$ were performed. This work is supported by the U.S. Department of Energy (DOE) under grant Contract Numbers DE-SC00012567 and DE-SC0011090. D.N. is also supported by an MIT Pappalardo Fellowship.  I.M. is also supported by NSERC of Canada.
\end{acknowledgments}

\appendix

\section{Conventions and SCET Notation}

In the body of the text we have presented the required factorization theorems for studying the two-prong substructure of jets using the $D_2$ observable. Although all the factorization theorems were presented, only heuristic descriptions of the functions appearing in the factorization theorems were presented in an attempt to appeal to a broader audience, and so as to not distract the reader with technical complications. In these appendices, we give the operator definitions of the functions appearing in the factorization theorems of \Sec{sec:Fact}, and calculate the functions to one-loop accuracy. 

In this appendix we begin by summarizing some notation and conventions. The factorization theorems presented in this paper are formulated in the language of SCET \cite{Bauer:2000yr,Bauer:2001ct,Bauer:2001yt,Bauer:2002nz}. We assume that the reader has some familiarity with the subject, and will only define our particular notation, and review the definition for common SCET objects. We refer readers unfamiliar with SCET to the reviews \cite{iain_notes,Becher:2014oda}.

SCET is formulated as a multipole expansion in the momentum components along the jet directions. Since we take the jet directions to be lightlike, it is
convenient to work in terms of light-cone coordinates. We define two light-cone vectors
\begin{equation}
n^\mu = (1, \vec{n})
\,,\qquad
\bn^\mu = (1, -\vec{n})
\,,\end{equation}
with $\vec{n}$ a unit three-vector, which satisfy the relations $n^2 = \bn^2 = 0$ and  $n\cdot\bn = 2$.
We can then write any four-momentum $p$ as
\begin{equation} \label{eq:lightcone_dec}
p^\mu = \bn\sdt p\,\frac{n^\mu}{2} + n\sdt p\,\frac{\bn^\mu}{2} + p^\mu_{n\perp}
\,.\end{equation}
A particle in the $n$-collinear sector has momentum $p$ close to the $\vec{n}$ direction,
so that its momentum scales like
$(n\!\cdot\! p, \bn \!\cdot\! p, p_{n\perp}) \sim
\bn\!\cdot\! p$ $\,(\la^2,1,\la)$, with $\la \ll 1$ 
a small parameter. The parameter $\lambda$ is a generic substitute for the power counting parameters in the different factorization theorems presented in \Sec{sec:Fact}, and since our factorization theorems involve multiple scales, there are generically multiple distinct $\lambda$s.

In the effective field theory, the momentum of the particles in the $n$-collinear sector are multipole expanded, and written as
\begin{equation} \label{eq:label_dec}
p^\mu = \lp^\mu + k^\mu = \bn\sdt\lp\, \frac{n^\mu}{2} + \lp_{n\perp}^\mu + k^\mu\,,
\,\end{equation}
where $\bn\cdot\lp$ and $\lp_{n\perp}$ are large momentum
components, which label fields, while $k$ is a small residual momentum, suppressed by powers of $\lambda$. This gives rise to an effective theory expansion in powers of $\la$.

SCET fields for quarks and gluons in the $n$-collinear sector, $\xi_{n,\lp}(x)$ and
$A_{n,\lp}(x)$, are labeled by the lightlike vector of their collinear sector, $n$ and their large momentum $\lp$. We will write the fields in a mixed position space/momentum space notation, using position space for the residual momentum and momentum space for  the large momentum components. 
The residual momentum dependence can be extracted using the derivative operator $\img\partial^\mu \sim k$, while the large label momentum is obtained from the momentum label operator $\cP_n^\mu$.

Operators and matrix elements in SCET are constructed from collinearly gauge-invariant quark and gluon fields, defined as \cite{Bauer:2000yr,Bauer:2001ct}
\begin{align}
\chi_{n,\w}(x) &= \left[\delta(\w - \bnP_n)\, W_n^\dagger(x)\, \xi_n(x) \right]
\,,\\
\cB_{n,\w\perp}^\mu(x)
&= \frac{1}{g}\left[\delta(\w + \bnP_n)\, W_n^\dagger(x)\,\img D_{n\perp}^\mu W_n(x)\right]
\,.\end{align}
The $\perp$ derivative in the definition of the SCET fields is defined using the label momenta operator as
\begin{equation}
\img D_{n\perp}^\mu = \cP^\mu_{n\perp} + g A^\mu_{n\perp}\,,
\end{equation}
and
\begin{equation} 
W_n(x) = \left[\sum_\text{perms} \exp\left(-\frac{g}{\bnP_n}\,\bn\sdt A_n(x)\right)\right]\,,
\end{equation}
is a Wilson line of $n$-collinear gluons. 
We use the common convention that the label operators in the definition of the SCET fields only act inside the square brackets. Although the Wilson line $W_n(x)$ is a non-local operator, it is localized with respect to the residual position $x$, and we can therefore treat $\chi_{n,\w}(x)$ and $\cB_{n,\w}^\mu(x)$ as local quark and gluon fields when constructing operators. The operator definitions for jet functions in these appendices are given in terms of these collinear gauge invariant quark and gluon SCET fields.

Our operator definitions will also involve matrix elements of eikonal Wilson lines, which arise from the soft-collinear factorization through the BPS field redefinition at the Lagrangian level \cite{Bauer:2001yt}. The Wilson lines extend from the origin to infinity along the direction of a lightlike vector, $q$, specifying their directions. Explicitly
\begin{align}
S_q={\bf P} \exp \left( ig \int\limits_0^\infty ds\, q\cdot A(x+sq)    \right)\,.
\end{align}
Here $\bf P$ denotes path ordering, and $A$ is the appropriate gauge field for any sector which couples eikonally to a collinear sector with label $q$ (for example collinear-soft, soft, boundary soft), and the color representation has been suppressed. All Wilson lines are taken to be outgoing, since we consider the case of jet production from $e^+e^-$ collisions.

Throughout this paper we have considered the production of two jets, one of which has a possible two-prong substructure, in an $e^+e^-$ collider. This implies the presence of at most three Wilson lines in the soft or collinear soft function. With only three Wilson lines, all possible color structures can be written as a sum of color-singlet traces. In the more general case, with more than three Wilson lines, the soft function is a color matrix which must be traced against the hard functions, which are also matrices in color space, appearing in the factorization theorem for the cross section (see e.g.~\Refs{Ellis:2010rwa,Jouttenus:2011wh} for more details).

In \App{sec:ninja_app} through \App{sec:signal_app} we will give operator definitions for the functions appearing in the factorization theorems in terms of matrix elements of the SCET operators, $\chi_{n,\w}(x)$ and $\cB_{n,\w}^\mu(x)$, as well as products of soft Wilson lines. These matrix elements can be calculated using the leading power SCET Lagrangian, which can be found in \Refs{Bauer:2000yr,Bauer:2001ct,Bauer:2001yt,Bauer:2002nz}, or by using eikonal Feynman rules in the soft functions, and known results for the splitting functions to calculate the jet functions \cite{Ritzmann:2014mka}. We will use the latter approach, as it considerably simplifies the calculations at one-loop.

\section{One Loop Calculations of Collinear Subjets Functions}\label{sec:ninja_app}

In this appendix we collect the calculations relevant to the calculation in the collinear subjets region of phase space, and explicitly show the cancellation of anomalous dimensions. The calculation follows closely that of \Ref{Bauer:2011uc}, with the exception of the form of the measurement function. Nevertheless, the calculation is presented in detail, as the SCET$_+$ effective theory has not been widely used.

\subsection*{Kinematics and Notation}
For our general kinematic setup, we will denote by $Q$ the center of mass energy of the $e^+e^-$ collisions, so that $Q/2$ is the energy deposited in a hemisphere. i.e. the four-momenta of the two hemispheres are
\begin{align}
p_{\text{hemisphere}_1}=\left( \frac{Q}{2},\vec p_1   \right)\,, \qquad p_{\text{hemisphere}_2}=\left( \frac{Q}{2},-\vec p_1   \right)
\end{align}
so
\begin{align}
s=Q^2\,.
\end{align}
We will also denote the energy in a jet at intermediate stages of the calculation by $E_J$, but we will write our final results in terms of $Q$.

We work in the region where one  hemispherical jet splits into two hard subjets, assume the power counting $z\sim \frac{1}{2}$, with $z$ being the energy fraction of one of the jets. We further assume the power counting relations between the energy correlation functions valid in the collinear subjets region, as discussed in \Sec{sec:power_counting}. We adopt the following notation to describe the kinematics of the subjets
\begin{align}
&\text{Subjet a,b momenta: }& &p_a,p_b\\
&\text{Subjet a,b spatial directions: }& &\hat{n}_{a}, \hat{n}_b\\
&\text{Thrust axis: }& &\hat{n}=\frac{\hat{n}_{a}+ \hat{n}_b}{|\hat{n}_{a}+ \hat{n}_b|}\\
&\text{Light-cone vectors: }& &n=(1,\hat{n}),\,\bar{n}=(1,-\hat{n}),\nonumber\\
& & &n_{a,b}=(1,\hat{n}_{a,b}),\,\bar{n}_{a,b}=(1,-\hat{n}_{a,b})\,.
\end{align}
In the collinear soft region of phase space, we have $n_a \cdot n_b \ll 1$. When performing expansions, we can work to leading order in $n_a \cdot n_b$, and must use a consistent power counting. It is therefore useful to collect some kinematic relations between vectors which are valid at leading power. These will be useful for later evaluations of the measurement function and integrand at leading power.
These kinematics satisfy the following useful relations
\begin{align} \label{eq:cs_kinematics1}
n\cdot n_a=n\cdot n_b= \frac{n_a \cdot n_b}{4}\,
\end{align}
\begin{align}
\bar{n}\cdot n_a=\bar{n}\cdot n_b=2\,,
\end{align}
\begin{align}
n_{\perp a,b}\cdot \bar{n}_{\perp a,b}=-n_{\perp a,b}\cdot n_{\perp a,b}=\hat{n}_{\perp a,b}\cdot\hat{n}_{\perp a,b}=\frac{n_a\cdot n_b}{2}\,. \label{eq:cs_kinematics2}
\end{align}
For a particle with the power counting of collinear sector $a$ or $b$, we have the following simplified relations
\begin{align}\label{energy_in_diverse_coor}
p_a\sim\frac{1}{2}(\bar{n}\cdot p_a) n_a,&\quad p_b\sim\frac{1}{2}(\bar{n}\cdot p_b) n_b\,,\\
p_a^0\sim\frac{1}{2}(\bar{n}\cdot p_a),&\quad p_b^0\sim\frac{1}{2}(\bar{n}\cdot p_b)\,,
\end{align}
which are true to leading order in the power counting.
Finally, we label the energy fractions carried in each subjet by
\begin{align}\label{energy_fractions}
z_{a,b}&=\frac{2p_{a,b}^0}{Q}=\frac{\bar n\cdot p_{a,b}}{Q}\,,
\end{align}
where the second relation is true to leading power.

The value of $\ecf{2}{\alpha}$ is given to leading power by the subjet splitting
\begin{align}\label{eq:lp_e2}
\ecf{2}{\alpha}&=\frac{1}{E_J^2} E_a E_b \left( \frac{2p_a\cdot p_b}{E_a E_b}  \right)^{\alpha/2}   \\
&=2^{\alpha/2}  z_az_b\left(n_a\cdot n_b\right)^{\alpha/2}\,.
\end{align}

In the collinear soft region of phase space, the 3-point energy correlation function is dominated by the correlation  between two particles in different subjets, with a third collinear, soft, or collinear-soft particle. Depending on the identity of the third particle, the power counting of the observable is different. We begin by collecting expressions for the $\ecf{3}{\alpha}$ observable for a single soft, collinear-soft, or collinear emission, which will be required for the one-loop calculations.

For three emissions, with momenta $k_1, k_2, k_3$, the general expression for the three point energy correlation function is
\begin{align}
\ecf{3}{\alpha}&=\frac{1}{E_J^3}k_1^0 k_2^0 k_3^0 \left(\frac{2 k_1 \cdot k_2}{k_1^0 k_2^0}  \right)^{\alpha/2}     \left(\frac{2k_1\cdot k_3}{k_1^0 k_3^0}  \right)^{\alpha/2}      \left(\frac{2k_2 \cdot k_3}{k_2^0 k_3^0}  \right)^{\alpha/2}\,.
\end{align}
For an emission collinear with one of the subjets, where we have the splitting $p_{a,b}\rightarrow k_1+k_2$, we can write $\ecf{3}{\alpha}$ entirely in terms of $k_1\cdot k_2$, $n_a\cdot n_b$, and $\bar n_a \cdot k_{1,2}$, because there is a hierarchy between the opening angle of the dipole, and the opening angle of the splitting. At leading power it is given by
\begin{align}\label{eq:collinear_limits_eec3}
\ecf{3}{\alpha}&=_{k_1,k_2\parallel n_a}   2^{5\alpha/2}z_b (n_a\cdot n_b)^{\alpha}\left(\frac{k_1\cdot k_2}{Q^2}\right)^{\frac{\alpha}{2}}\left(\frac{\bar n_a\cdot k_1}{Q}\right)^{1-\frac{\alpha}{2}}\left(\frac{\bar n_a\cdot k_2}{Q}\right)^{1-\frac{\alpha}{2}}\,,\\
\ecf{3}{\alpha}&=_{k_1,k_2\parallel n_b} 2^{5\alpha/2}z_a (n_a\cdot n_b)^{\alpha}\left(\frac{k_1\cdot k_2}{Q^2}\right)^{\frac{\alpha}{2}}\left(\frac{\bar n_b\cdot k_1}{Q}\right)^{1-\frac{\alpha}{2}}\left(\frac{\bar n_b\cdot k_2}{Q}\right)^{1-\frac{\alpha}{2}}\,.
\end{align}

For a soft emission off of the dipole, with momentum $k$, which cannot resolve the opening angle of the dipole, we have
\begin{align}
n_a\cdot k\rightarrow n\cdot k\,, \qquad n_b\cdot k \rightarrow n\cdot k\,,
\end{align}
at leading power. We then find
\begin{align}\label{eq:soft_limits_eec3}
\ecf{3}{\alpha}&=   2^{3\alpha/2+1} z_a z_b (n_a \cdot n_b)^{\alpha/2}    \left(\frac{\bar{n}\cdot k+n\cdot k}{2Q}\right)^{1-\alpha}\left(\frac{n\cdot k}{Q}\right)^{\alpha}\,,
\end{align}
where we have used the full expression for the energy of the soft particle, as it is not boosted.

For a third collinear-soft emission $k$ off of the $p_{a,b}$ partons, for which there is no hierarchy between the opening angle of the dipole and the opening angle of the emission (i.e. a collinear soft emission), $\ecf{3}{\alpha}$ is given by
\begin{align}\label{eq:collinear_soft_limits_eec3}
\ecf{3}{\alpha}&=2^{3\alpha/2+1}z_az_b\left(n_a\cdot n_b\right)^{\alpha/2}\left(\frac{\bar{n}\cdot k}{2Q}\right)^{1-\alpha}\left(\frac{n_a\cdot k}{Q}\right)^{\alpha/2}\left(\frac{n_b\cdot k}{Q}\right)^{\alpha/2}\,.
\end{align}

For the SCET operators involved in the matching calculation, we follow the notation of \Ref{Bauer:2011uc}, defining
\begin{align}
\mathcal{O}_2=\bar \chi_n Y^\dagger_n \Gamma Y_{\bar n} \chi_{\bar n}\,,
\end{align}
which is the usual SCET operator for $e^+e^-\to$ dijets, and
\begin{align}
\mathcal{O}_3=\bar \chi_{n_a} \mathcal{B}_{\perp n_b}^{A} \left [ X^\dagger_{n_a} X_{n_b} T^A X_{n_b} V_{\bar n}    \right]_{ij}  \left[ Y^\dagger_n Y_{\bar n} \right]_{jk}    \Gamma  \left[ \chi_{\bar n} \right]_k\,,
\end{align}
which is the SCET$_+$ operator describing the production of the collinear subjets. Throughout this section, we will not be careful with the Dirac structure of the operators, as it is largely irrelevant to our discussion. With this in mind, we have not made the Lorentz indices explicit on the operators.
Here we have chosen to write the Wilson line corresponding to the gluon in the fundamental representation. Note that the two stage matching onto SCET$_+$ makes it clear that the partonic configuration in which the two collinear subjets are both quarks is power suppressed. In the operators $\mathcal{O}_2, \mathcal{O}_3$, we have used $Y$ to denote soft Wilson lines, and $X,V$ to denote collinear-soft Wilson lines. In the definitions of the factorized functions below, we will refer to all Wilson lines as $S$, as after factorization, no confusion can arise.

\subsection*{Definitions of Factorized Functions}

The functions appearing in the collinear subjets factorization theorem of \Eq{eq:NINJA_fact} have the following SCET operator definitions:
\begin{itemize}
\item Hard Matching Coefficient for Dijet Production
\begin{equation}
H\left(Q^2,\mu \right)=\left |C\left(Q^2,\mu\right) \right|^2\,,
\end{equation}
where $C\left(Q^2,\mu\right)$ is the Wilson coefficient obtained from matching the full theory QCD current $\bar \psi \Gamma \psi$ onto the SCET dijet operator $\bar \chi_n \Gamma \chi_{\bar n}$
\begin{align}
\langle q\bar q | \bar \psi \Gamma \psi |0\rangle=C\left(Q^2,\mu\right) \langle q\bar q | \mathcal{O}_2 |0\rangle\,.
\end{align}
When accounting for the Lorentz structure, there is a contraction with the leptonic tensor, which we have dropped for simplicity. See \Ref{Ellis:2010rwa} for a detailed discussion.
\item Hard Splitting Function:
\begin{align}
&H_2 \left(\ecf{2}{\alpha},z_a, \mu\right)= \left|C_2\left(\ecf{2}{\alpha},z_a, \mu\right)\right|^2\,,
\end{align}
where $C_2\left(\ecf{2}{\alpha},z_a, \mu\right)$ is the Wilson coefficient in the matching from $\mathcal{O}_2$ to $\mathcal{O}_3$, namely the relation between the following matrix elements
\begin{align}
\langle q\bar q g| \mathcal{O}_2 |0\rangle=C_2\left(\ecf{2}{\alpha},z_a, \mu\right) \langle q\bar q g| \mathcal{O}_3 |0\rangle\,.
\end{align}
\item Jet Function:
\begin{align}
J_{n_{a,b}}\Big(\ecf{3}{\alpha}\Big)&=\\
&\hspace{-0.5cm}\frac{(2\pi)^3}{C_F}\text{tr}\langle 0|\frac{\bar{n}\!\!\!\slash _{a,b}}{2}\chi_{n_{a,b}}(0) \delta(Q-\bar{n}_{a,b}\cdot{\mathcal P})\delta^{(2)}(\vec{{\mathcal P}}_{\perp})\delta\Big(\ecf{3}{\alpha}-\ecfop{3}{\alpha}\Big)\bar{\chi}_{n_{a,b}}(0)|0\rangle \nonumber
\end{align}
For simplicity, we have given the definition of the quark jet function. The gluon jet function is defined identically but with the SCET collinear invariant gluon field, $\cB_{n_{a,b},\perp}$, instead of the collinear invariant quark field.
\item Soft Function:
\begin{align}\label{eq:def_soft_function}
S_{n \, \bar n }\Big(\ecf{3}{\alpha};R\Big)&=\frac{1}{C_{A}}\text{tr}\langle 0|T\{S_{n } S_{\bar n }\}  \delta\Big(\ecf{3}{\alpha}-\Theta_{R}\ecfop{3}{\alpha}\Big)\bar{T}\{S_{n } S_{\bar n }\} |0\rangle
\end{align}
\item Collinear-Soft Function:
\begin{align}
\hspace{-1cm}S_{n_a \,n_b\,\bar{n}}\Big(\ecf{3}{\alpha}\Big)&=\text{tr}\langle 0|T\{S_{n_a } S_{n_b} S_{\bar{n}}\}\delta\Big(\ecf{3}{\alpha}-\ecfop{3}{\alpha}\Big)\bar{T}\{S_{n_a } S_{n_b} S_{\bar{n}}\} |0\rangle
\end{align}
\end{itemize}
In each of these definitions, we have defined an operator, $\ecfop{3}{\alpha}$, which measures the contribution to $\ecf{3}{\alpha}$ from final states, and must be appropriately expanded following the power counting of the sector on which it acts, as was shown explicitly in \Eq{eq:collinear_limits_eec3}, \Eq{eq:soft_limits_eec3}, and \Eq{eq:collinear_soft_limits_eec3}. These operators can be written in terms of the energy-momentum tensor of the full or effective theory \cite{Sveshnikov:1995vi,Korchemsky:1997sy,Lee:2006nr,Bauer:2008dt}, but we can simply view them as returning the value of $\ecf{3}{\alpha}$ as measured on a particular perturbative state. The soft function is also sensitive to the jet function definition, which is included through the operator $\Theta_R$. To simplify the notation, we have strictly speaking only defined in the in-jet contribution to the soft function. Additionally, we assume that some IRC safe observable is also measured in the out-of-jet region, although this will play little role in our discussion, so we have not made it explicit.

\subsection*{Hard Matching Coefficient for Dijet Production}

The hard matching coefficient for dijet production, $H(Q^2,\mu)$, appears in the factorization theorems in each region of phase space. $H(Q^2,\mu)$ is the well known hard function for the production of a $q \bar q$ pair in $e^+e^-$ annihilation. It is defined by
\begin{equation}\label{eq:hard_matching_coeff}
H\left(Q^2, \mu\right)= \left|C\left(Q^2,\mu\right)\right|^2\,,
\end{equation}
where $C\left(Q^2,\mu\right)$ is the Wilson coefficient obtained from matching the full theory QCD current $\bar \psi \Gamma \psi$ onto the SCET dijet operator $\bar \chi_n \Gamma \chi_{\bar n}$. This Wilson coefficient is well known (see, e.g., \Refs{Bauer:2003di,Manohar:2003vb,Ellis:2010rwa,Bauer:2011uc} ), and is given at one-loop by
\begin{equation}
C(Q^2,\mu)=1+\frac{\alpha_s(\mu)\, C_F}{4\pi}\left ( -\log^2\left[ \frac{-Q^2}{\mu^2}  \right]+3\log \left[ \frac{-Q^2}{\mu^2} \right]-8+\frac{\pi^2}{6}   \right )\,.
\end{equation}
The branch cut in the logarithms must be taken as $-Q^2 \to -Q^2-i\epsilon$. The hard function satisfies a multiplicative RGE, given by
\begin{equation}
\mu \frac{d}{d\mu}\ln H\left(Q^2,\mu\right)=2\text{Re}\left[ \gamma_{C}\left(Q^2,\mu\right)\right]\,,
\end{equation}
where $\gamma_{C}(Q^2,\mu)$ is the anomalous dimension for the Wilson coefficient, which is given to one-loop by
\begin{equation}\label{eq:hard_anom_dim}
\gamma_C(Q^2,\mu)=\frac{\alpha_s C_F}{4\pi}\left(  4\log\left[  \frac{-Q^2}{\mu^2} \right] -6  \right )\,.
\end{equation}

\subsection*{Hard Splitting Function}

The hard splitting function can be calculated using known results for the one-loop splitting functions \cite{Kosower:1999rx} or from the result for $e^+e^- \to 3$ jets \cite{Ellis:1980wv}. However, since at leading power the measurement of the 2-point energy correlation functions define the energy fractions and splitting angle, it is simplest to change variables in the results of \Ref{Bauer:2011uc}, where the hard splitting function matching was performed for jet mass. Using the notation $t=s_{q g}$, and $x=s_{q \bar q}/Q^2$, \Ref{Bauer:2011uc} gave the matching coefficient to one-loop as
{\small \begin{align} \label{eq:hardsplit_t}
H_2^{q\to qg}(t,x,\mu)&=Q^2 \frac{\alpha_s(\mu) C_F}{2\pi} \frac{1}{t} \frac{1+x^2}{1-x} \left \{ 1+\frac{\alpha_s(\mu)}{2\pi} \left[ \left( \frac{C_A}{2} -C_F \right) \left( 2 \log \frac{t}{\mu^2} \log\, x +\log^2 x +2\text{Li}_2(1-x)  \right) \right. \right. \nonumber \\
&\hspace{-1.5cm}\left. \left. -\frac{C_A}{2} \left (  \log^2 \frac{t}{\mu^2}-\frac{7\pi^2}{6}+2\,\log \frac{t}{\mu^2}\log (1-x) +\log^2(1-x)+2\text{Li}_2 (x) \right)+(C_A-C_F)\frac{1-x}{1+x^2} \right] \right\}\,.
\end{align}}
We can now perform a change of variables to rewrite this in terms of $\ecf{2}{\alpha}$, and the subjet energy fractions, using the leading power relation of \Eq{eq:lp_e2}, and the kinematic relations valid in the collinear subjets region of phase space. We find
\begin{equation}\label{eq:relate_to_frank}
t=\frac{Q^2}{2}\frac{(z_a z_b)^{1-2/\alpha} \left( \ecf{2}{\alpha}  \right)^{2/\alpha}  }{2}\,, \qquad x=z_q\,,
\end{equation}
and
{\small \begin{align}
H_2^{q\to qg}(\ecf{2}{\alpha},z_q,\mu)&= \frac{\alpha_s(\mu) C_F}{\alpha \pi} \frac{1}{  \ecf{2}{\alpha}      } \frac{1+z_q^2}{1-z_q}   
\\
& \hspace{-2.75cm}\times\left \{ 1+\frac{\alpha_s(\mu)}{2\pi} \left[ \left( \frac{C_A}{2} -C_F \right) \left( 2 \log\left( \frac{Q^2}{\mu^2} \frac{(z_az_b)^{1-2/\alpha}    \left( \ecf{2}{\alpha}  \right)^{2/\alpha}      }{4} \right) \log\, z_q +\log^2 z_q +2\text{Li}_2(1-z_q)  \right) \right. \right. \nonumber \\
&\hspace{-1.5cm}\left. \left. -\frac{C_A}{2} \left (  \log^2 \left( \frac{Q^2}{\mu^2} \frac{(z_az_b)^{1-2/\alpha}    \left( \ecf{2}{\alpha}  \right)^{2/\alpha}      }{4} \right)-\frac{7\pi^2}{6} \right. \right. \right.  \nonumber \\
& \hspace{-3cm}\left. \left. \left.+2\,\log \left( \frac{Q^2}{\mu^2} \frac{(z_az_b)^{1-2/\alpha}    \left( \ecf{2}{\alpha}  \right)^{2/\alpha}      }{4} \right)\log (1-z_q) +\log^2(1-z_q)+2\text{Li}_2 (z_q) \right)+(C_A-C_F)\frac{1-z_q}{1+z_q^2} \right] \right\}\,. \nonumber
\end{align}}
Note that the hard splitting function depends on the partons involved in the split, which in our case we have taken to be $q\to qg$, and therefore singled out $z_q$, which is the energy fraction of the quark jet (defined identically to $z_a$, $z_b$). Throughout the rest of this appendix, we will, whenever possible, write results in terms of $z_a$, and $z_b$ for generic partons, using general Casimirs. Since we consider the case $q\to qg$, we will calculate the jet functions for both quark and gluon jets, and therefore the results in this appendix are sufficient to treat general two-prong substructure, where the prongs are associated with generic partons by using the hard splitting function for other partonic splittings. 

For completeness, we also present the one-loop results for $g\to gg$ and $g\to q\bar q$ splittings.  While one-loop, and even two-loop, splitting helicity amplitudes exist in the literature \cite{Kosower:1999rx,Bern:2004cz,Badger:2004uk}, to our knowledge, the one-loop unpolarized splitting functions have not not been explicitly written down before.  Using the results from \Refs{Kosower:1999rx,Badger:2004uk}, the one-loop function for the $g\to gg$ splitting is
\begin{align} \label{eq:hardsplit_gg}
H_2^{g\to gg}(s_{gg},z,\mu)&= \frac{\alpha_s(\mu) C_A}{2\pi} \frac{1}{s_{gg}} \left(
\frac{z}{1-z}+\frac{1-z}{z}+z(1-z)
\right) \left \{ 1+\frac{\alpha_s(\mu)}{2\pi}N\left[
\log\frac{\mu^2}{s_{gg}}\log\left(
z(1-z)
\right)\right.\right.\nonumber\\
&
\hspace{-1.5cm}\left.\left.
-\frac{1}{2}\log^2\frac{s_{gg}}{\mu^2}-\frac{1}{2}\log^2\frac{z}{1-z}+\frac{5\pi^2}{12}+\left(
\frac{1}{3}-\frac{n_F}{3N}
\right)\frac{z(1-z)}{1+z^4+(1-z)^4}
\right]
\right\} \,.
\end{align}
Here, we have expressed the result in terms of the numbers of colors, $N$, of the gauge theory and number of active quarks, $n_F$.  Note that $C_A = N$.  The virtuality of the splitting is
\begin{equation}
s_{gg} = 2z_az_bE_J^2(n_a\cdot n_b) \,,
\end{equation}
where $a$ and $b$ denote the final-state gluons in the splitting.
Its anomalous dimension to one-loop is
\begin{align}
\gamma_{g\to gg}=\frac{\alpha_s(\mu)}{\pi}\left[
N\,\log\frac{s_{gg}}{\mu^2}+N\,\log\,z(1-z)-\frac{\beta_0}{2}
\right]\,.
\end{align}
For the one-loop result of the $g\to q\bar q$ splitting, we have
\begin{align}
 \label{eq:hardsplit_qq}
H_2^{g\to q\bar q}(s_{q\bar q},z,\mu)&= \frac{\alpha_s(\mu) n_F}{2\pi} \frac{1}{s_{q\bar q}} \left(
z^2+(1-z)^2
\right) \left \{ 1+\frac{\alpha_s(\mu)}{2\pi}\left[
N\, \log\frac{\mu^2}{s_{q\bar q}}\log(z(1-z))\right.\right.\nonumber\\
&
\hspace{-1.5cm}
+\frac{3}{2}\frac{1}{N}\log\frac{\mu^2}{s_{q\bar q}}-\frac{2n_F}{3}\log\frac{\mu^2}{s_{q\bar q}}+\frac{13}{6}N\log\frac{\mu^2}{s_{q\bar q}}+\frac{1}{2N}\log^2\frac{\mu^2}{s_{q\bar q}}
\nonumber \\
&
\hspace{-1.5cm}
\left.\left.
-\frac{1}{N}\frac{7\pi^2}{12}-N\frac{\pi^2}{6}-\frac{N}{2}\log^2\frac{z}{1-z}+\frac{40}{9}N-\frac{10}{9}n_F
\right]
\right\}\,.
\end{align}
Note that, in terms of the number of colors, $$C_F = \frac{N^2-1}{2N}\,.$$
Its anomalous dimension is 
\begin{align}
\gamma_{g\to q\bar q} = \frac{\alpha_s(\mu)}{\pi}\left[
\frac{1}{N}\log\frac{\mu^2}{s_{q\bar q}}+N\,\log(z(1-z))+\frac{\beta_0}{2}-3C_F
\right]\,.
\end{align}

%

\subsection*{Global Soft Function}
In this section we calculate the global soft function. The global soft modes can resolve the boundaries of the jet, so the jet algorithm constraint cannot be expanded. However, the soft modes do not resolve the dipole of the collinear splitting. The global soft function therefore has two Wilson lines in the $n$ and $\bar n$ directions. A general one-loop soft function can be written as
\begin{align}
S^{(1)}_{G}\left(\ecf{3}{\alpha}\right)&=\frac{1}{2}\sum_{i\neq j}\mathbf{T}_i\cdot\mathbf{T}_j \,S_{G,\,ij}^{(1)}\left(\ecf{3}{\alpha}\right)\,,
\end{align}
where $\mathbf{T}_i$ is the color generator of leg $i$ in the notation of \Refs{Catani:1996jh,Catani:1996vz}, and the sum runs over all pairs of legs. Here we have only the contribution from $i,j=n, \bar n$, but we still perform this extraction of the color structure to keep the results generic.

The one-loop integrand for the soft function is given by
{\small \begin{align} \label{eq:soft_integrand_cs}
S_{G,\,n \bar{n}}^{(1)}(\ecf{3}{\alpha})&=\\
&
\hspace{-2cm}
-g^2 \left( \frac{\mu^{2}  e^{\gamma_E }  }{4\pi} \right) ^{\epsilon}         \int[d^dk]_+\frac{2n\cdot \bar{n}}{n\cdot k\,k\cdot \bar{n}}\Theta\left(\text{tan}^2\frac{R}{2}-\frac{n\cdot k}{\bar{n}\cdot k}\right)\delta\left(\ecf{3}{\alpha}-N_S\left(\frac{\bar{n}\cdot k+n\cdot k}{2Q}\right)^{1-\alpha}\left(\frac{n\cdot k}{Q}\right)^{\alpha}\right) \nonumber
\end{align}}
with $d=4-2\epsilon$, and where here we have extracted the normalization factor
 \begin{align}
N_S&=2^{3\alpha/2+1} z_az_b\left(n_a\cdot n_b\right)^{\alpha/2}\,,
\end{align}
following the expression for the three point energy correlation function in the soft power counting, given in \Eq{eq:soft_limits_eec3}. The first $\Theta$-function in \Eq{eq:soft_integrand_cs} implements the jet algorithm constraint, which is simple for a single emission. To simplify notation, we also use the following shorthand for the measure for a positive energy, on-shell, collinear particle
\begin{align}\label{eq:pos_energy}
[d^dk]_+&=\frac{d^dk}{(2\pi)^d}2\pi\Theta(\nbar\cdot k)\delta(k^2)\,.
\end{align}
To perform this integral, it is convenient to make the change of variables
\begin{align}
\bar{n}\cdot k=v\,, \qquad
n\cdot k=v\, u\,,
\end{align}
which factorizes the jet algorithm constraint and the measurement function. The integrals can then be evaluated using standard techniques.
Performing all the integrals but the $u$ integral, and transforming to Laplace space, $\ecf{3}{\alpha}\rightarrow\ecflp{3}{\alpha}$, gives
 \begin{align}
\tilde{S}_{G,\,n \bar{n}}^{(1)}(\ecflp{3}{\alpha})=-\frac{g^2 e^{-\epsilon\gamma_E}\Gamma(-2\epsilon)}{(2\pi)^2\Gamma(1-\epsilon)}\Bigg(\frac{e^{\gamma_E}\mu\ecflp{3}{\alpha}N_S}{2^{1-\alpha}Q}\Bigg)^{2\epsilon}   \int_{0}^{\text{tan}^2\frac{R}{2}} \frac{du}{u^{1+\epsilon(1-2\alpha)}}(1+u)^{2\epsilon(1-\alpha)}\,. 
\end{align}
This can be integrated exactly in terms of hypergeometric functions, 
\begin{align}
 \int_{0}^{\text{tan}^2\frac{R}{2}} \frac{du}{u^{1+\epsilon(1-2\alpha)}}(1+u)^{2\epsilon(1-\alpha)}=   \frac{\Gamma(-\epsilon(1-2\alpha))}{\Gamma(1-\epsilon(1-2\alpha))}    &\\
 &\hspace{-8cm} \times\left (  \tan^2 \frac{R}{2} \right)^\epsilon    \left(   \frac{\tan^2 \frac{R}{2}}{1+\tan^2 \frac{R}{2}} \right)^{-2(1-\alpha)\epsilon}     \!\,_2F_1\left[1,-2(1-\alpha)\epsilon;1-(1-2\alpha)\epsilon;\frac{\text{tan}^2\frac{R}{2}}{1+\text{tan}^2\frac{R}{2}}\right]\,,\nonumber
\end{align}
where we have used both a Pfaff and an Euler transformation to extract the singular behavior from the hypergeometric function.
We therefore have
\begin{align}
\tilde{S}_{G,\,n \bar{n}}^{(1)}(\ecflp{3}{\alpha})=-\frac{\alpha_s}{\pi}\frac{e^{-\epsilon\gamma_E}\Gamma(-2\epsilon)}{\Gamma(1-\epsilon)}\Bigg(\frac{e^{\gamma_E}\mu\ecflp{3}{\alpha}N_S}{2^{1-\alpha}Q}  \tan \frac{R}{2}    \Bigg)^{2\epsilon}   \frac{\Gamma(-\epsilon(1-2\alpha))}{\Gamma(1-\epsilon(1-2\alpha))}    &\\
 &
 \hspace{-10cm} 
 \times   \left(   \frac{\tan^2 \frac{R}{2}}{1+\tan^2 \frac{R}{2}} \right)^{-2(1-\alpha)\epsilon}     \!\,_2F_1\left[1,-2(1-\alpha)\epsilon;1-(1-2\alpha)\epsilon;\frac{\text{tan}^2\frac{R}{2}}{1+\text{tan}^2\frac{R}{2}}\right]\,.\nonumber
\end{align}

Expanding in $\epsilon$ (throughout these appendices we use the {\tt HypExp} package \cite{Huber:2005yg,Huber:2007dx} for expansions of hypergeometric functions) and separating in divergent and finite pieces, we find
\begin{align}\label{eq:gsoft_final}
\tilde{S}_{G,\,n \bar{n}}^{(1)\text{div}}(\ecflp{3}{\alpha})&=  \frac{\alpha_s}{2\pi}\frac{1}{(2 \alpha -1) \epsilon ^2}+  \frac{\alpha_s}{\pi} \frac{ \log \left[\frac{e^{\gamma_E}\mu\ecflp{3}{\alpha}N_S}{2^{1-\alpha}Q}\right]}{ (2 \alpha -1) \epsilon } 
+\frac{\alpha_s}{2\pi}\frac{\log\left[\tan^2\frac{R}{2}\right]}{\epsilon}
\,, 
\end{align}
\begin{align}
\tilde{S}_{G,\,n \bar{n}}^{(1)\text{fin}}(\ecflp{3}{\alpha})&=   \frac{\alpha_s}{\pi}\Bigg\{
\frac{\log^2 \left[\frac{e^{\gamma_E}\mu\ecflp{3}{\alpha}N_S}{2^{1-\alpha}Q}\right]}{2\alpha-1}
+\log \left[\frac{e^{\gamma_E}\mu\ecflp{3}{\alpha}N_S}{2^{1-\alpha}Q}\right]\,\log\left[\tan^2\frac{R}{2}\right] \\
&
+\frac{\pi^2}{8(2\alpha-1)}+\frac{2\alpha-1}{4}\log^2\left[\tan^2\frac{R}{2}\right]+(\alpha-1)\text{Li}_2\left[
-\tan^2\frac{R}{2}
\right]
\Bigg\} \,, \nonumber
\end{align}
where $\text{Li}_2$ is the dilogarithm function.

\subsection*{Jet Function}\label{sec:NINJA_jet_calc}

To calculate the jet function, we use the approach of \Ref{Ritzmann:2014mka} and integrate the appropriate splitting functions against our measurement function. In the power counting of the jet function, we can expand the jet algorithm constraint
\begin{align}\label{eq:jet_expand}
\Theta\left(\text{tan}^2\frac{R}{2}-\frac{n\cdot k}{\bar{n}\cdot k}\right) \to 1\,.
\end{align}

The one-loop jet function in the $n_a$ direction is then given by 
\begin{equation}
J_{i,n_{a}}^{(1)}(Q_J,\ecf{3}{\alpha})=\int d\Phi_2^c\, \sigma_2^c\, \delta\Bigg(\ecf{3}{\alpha}-N_J\left(\frac{\bar{n}_{a}\cdot k_1}{Q}\right)^{1-\alpha/2}\left(\frac{\bar{n}_{a}\cdot k_2}{Q}\right)^{1-\alpha/2}\left(\frac{k_1\cdot k_2}{Q^2}\right)^{\alpha/2}\Bigg)\,.
\end{equation}
The two particle collinear phase space is given by \cite{Giele:1991vf}
\begin{equation}
d\Phi_2^c= 2(2\pi)^{3-2\epsilon} Q_J \left[  d^d k_1  \right]_+   \left[  d^d k_2  \right]_+    \delta (Q_J-\bar{n}_{a}\cdot k_1-\bar{n}_{a}\cdot k_2) \delta^{d-2}(k_{1\perp} +k_{2\perp})\,,
\end{equation}
and
\begin{equation}
\sigma_2^c=\left(   \frac{\mu^2 e^{\gamma_E}}{4\pi} \right)^{\epsilon} \frac{2g^2}{s} P_{i}(z)\,,
\end{equation}
 where
 \begin{equation}
 P_{q}(z)=C_F \left[  \frac{1+z^2}{1-z} -\epsilon(1-z) \right]\,,
 \end{equation}
 and
 \begin{equation}
 P_g=C_A \left[ \frac{z}{1-z}+\frac{1-z}{z}+z(1-z) \right]+\frac{n_f}{2} \left[ 1-\frac{2z(1-z)}{1-\epsilon} \right]\,,
 \end{equation}
 which includes both the $g\to gg$ and $g\to q\bar q$ contributions.
Explicitly, the integrand is then given by
\begin{align}
J_{i,n_{a}}^{(1)}(Q_J,\ecf{3}{\alpha})&=\left( \frac{\mu^2 e^{\gamma_E}}{4\pi} \right)^\epsilon 2(2\pi)^{3-2\epsilon}  Q_J 2g^2\int[d^dk_1]_+\int[d^dk_2]_+\frac{ P_{i}\left(\frac{\bar{n}_{a}\cdot k_1}{Q_J}\right)}{2k_1\cdot k_2}\\
&
\hspace{1cm}
\times
\delta(Q_J-\bar{n}_{a}\cdot k_1-\bar{n}_{a}\cdot k_2)\delta^{d-2}(\vec{k}_{1\perp}+\vec{k}_{2\perp})\nonumber\\
&
\hspace{1cm}
\times
\delta\Bigg(\ecf{3}{\alpha}-N_J\left(\frac{\bar{n}_{a}\cdot k_1}{Q}\right)^{1-\alpha/2}\left(\frac{\bar{n}_{a}\cdot k_2}{Q}\right)^{1-\alpha/2}\left(\frac{k_1\cdot k_2}{Q^2}\right)^{\alpha/2}\Bigg)\,,\nonumber
\end{align}
where we have extracted the normalization factor
 \begin{align}
N_J&= 2^{5\alpha/2} (n_a\cdot n_b)^{\alpha} z_b\,,
\end{align}
for simplicity, following the expression of \Eq{eq:collinear_limits_eec3} for the three point energy correlation function in the power counting for the emission of a single collinear particle. Furthermore, note that we have used $Q_J= z_a Q$ in this expression.

The integrals can be performed using standard techniques, and we find, after transforming to Laplace space, $\ecf{3}{\alpha}\rightarrow\ecflp{3}{\alpha}$, for the jet function in the $n_a$ direction
\begin{align}\label{eq:gjet_final}
\tilde{J}_{g,n_{a}}^{(1)}(Q_J,\ecflp{3}{\alpha})&=\frac{\alpha_s}{2\pi}C_A\Bigg(\frac{\alpha}{(\alpha -1) \epsilon^2}   +\frac{2 L_\alpha^{J,a}\left(  \ecflp{3}{\alpha}  \right) }{(\alpha-1) \epsilon}   +\frac{1}{ \epsilon} \frac{11C_A-2n_f}{6C_A}  \\
   &
    \hspace{-2cm}
    -\frac{\alpha  \pi^2}{12 (\alpha -1)}+\frac{\pi^2}{3 (\alpha -1) \alpha }-\frac{67 }{9 \alpha }+\frac{2\pi^2}{3 \alpha }+\frac{2 L_\alpha^{J,a}\left(  \ecflp{3}{\alpha}  \right)^2}{ (\alpha -1) \alpha }+\frac{11     L_\alpha^{J,a}\left(  \ecflp{3}{\alpha}  \right)}{3 \alpha }\nonumber \\
   &
   \hspace{-2cm}
   +\frac{67}{9}-\frac{2\pi^2}{3}-\frac{2 n_f L_\alpha^{J,a}\left(  \ecflp{3}{\alpha}  \right)    }{3 C_A \alpha }+\frac{13 n_f}{9 C_A \alpha }-\frac{23 n_f}{18 C_A }\Bigg)\,,\nonumber
\end{align}
 for gluon jets, and 
\begin{align}\label{eq:qjet_final}
\tilde{J}_{q,n_{a}}^{(1)}(Q_J,\ecflp{3}{\alpha})=\frac{\alpha_s}{2\pi}C_F\Bigg(&-\frac{\alpha}{\epsilon^2(1-\alpha)}+\frac{3}{2\epsilon}-\frac{2}{\epsilon(1-\alpha)}   L_\alpha^{J,a}\left(  \ecflp{3}{\alpha}  \right)\\
&
\hspace{1cm}
-\frac{2}{\alpha(1-\alpha)}L_\alpha^{J,a}\left(  \ecflp{3}{\alpha}  \right)^2+\frac{3}{\alpha}L_\alpha^{J,a}\left(  \ecflp{3}{\alpha}  \right)-\frac{\pi^2}{6\alpha}\nonumber\\
&
\hspace{1cm}
-\frac{\pi^2}{4\alpha(1-\alpha)}+\frac{3\pi^2(1-\alpha)}{4\alpha}+\frac{1}{2\alpha}-\frac{13(1-\alpha)}{2\alpha}\Bigg) \,, \nonumber
\end{align}
for quark jets respectively.  The jet function for the $n_b$ direction can be trivially found from $a\rightarrow b$.

Here we have used $L_\alpha^{J,a}\left(  \ecflp{3}{\alpha}  \right)$to denote the logarithm appearing in the jet functions. The argument of this logarithm depends on the subjet energy fraction. We indicate the specific logarithm for the subjet via the notation
\begin{align}\label{eq:jet_log}
L_\alpha^{J,a}\left(  \ecflp{3}{\alpha}  \right)&=\log\left[N_J\ecflp{3}{\alpha}e^{\gamma_E }\left(\frac{\mu}{\sqrt{2}Q}\right)^\alpha z_a^{2-\alpha}\right]\,.
\end{align}

\subsection*{Collinear-Soft Function}\label{sec:csoft_calc}
We now calculate the collinear-soft function. The collinear-soft modes couple eikonally to the collinear sector, and so the collinear-soft function has the one-loop form
\begin{align}
S^{(1)}_{c}\left(\ecf{3}{\alpha}\right)&=\frac{1}{2}\sum_{i\neq j}\mathbf{T}_i\cdot\mathbf{T}_j \,S_{c,\,ij}^{(1)}\left(\ecf{3}{\alpha}\right)\,,
\end{align}
where $\mathbf{T}_i$ is the color generator of leg $i$ in the notation of \Refs{Catani:1996jh,Catani:1996vz}, and the sum runs over all pairs of legs. Since the collinear-soft modes resolves the dipole from the collinear splitting, there are three Wilson lines, $n_a, n_b, \bar n$ to which the collinear-soft modes couple. We calculate separately the contributions arising from the pair of legs $n_a, n_b$, and from the pairs $n_{a,b}, \bar n$. In both cases the integral involves the jet algorithm constraint. In the power counting of the collinear-soft modes, this constraint can be expanded as
\begin{align}\label{eq:jet_expand2}
\Theta\left(\text{tan}^2\frac{R}{2}-\frac{n\cdot k}{\bar{n}\cdot k}\right) \to 1\,.
\end{align}
If this expansion was not performed, the contribution of the collinear soft modes sensitive to the jet radius $R$, would be removed by a soft zero bin subtraction.

\subsubsection*{$n_a$, $n_b$ Contribution:}

We begin by calculating the contribution from the emission between the $n_a$, $n_b$ eikonal lines. The integrand is given by
\begin{align}
S_{c,\,n_a n_b}^{(1)}(\ecf{3}{\alpha})&= \\
&\hspace{-2cm} -g^2   \left( \frac{\mu^{2}  e^{\gamma_E }  }{4\pi} \right)^{\epsilon}       \int[d^dk]_+\frac{2n_a\cdot n_b}{n_{a}\cdot k\,k\cdot n_b}\delta\left(\ecf{3}{\alpha}-N_{CS}\left(\frac{\bar{n}\cdot k}{2Q}\right)^{1-\alpha}\left(\frac{n_a\cdot k}{Q}\right)^{\alpha/2}\left(\frac{n_b\cdot k}{Q}\right)^{\alpha/2}\right)\,, \nonumber
\end{align}
where we have extracted the normalization factor
 \begin{align}\label{eq:ncs}
N_{CS}&=2^{3\alpha/2+1}z_az_b\left(n_a\cdot n_b\right)^{\alpha/2}\,,
\end{align}
for simplicity, following the expression of \Eq{eq:collinear_soft_limits_eec3} for the three point energy correlation function in the power counting for the emission of a single collinear-soft particle.

To perform the calculation, we go to the light-cone basis defined by $n,\bar{n}$. We then have
\begin{align}
n_a\cdot k&=\frac{n\cdot n_a}{2}\bar{n}\cdot k+\frac{\bar{n}\cdot n_a}{2}n\cdot k+k_\perp\cdot n_{a\perp}\nonumber\\
&=\frac{n\cdot n_a}{2}\bar{n}\cdot k+\frac{\bar{n}\cdot n_a}{2}n\cdot k-(\bar{n}\cdot k n\cdot k)^{1/2} |\hat{n}_{a\perp}|\text{cos }\theta \,,\\
n_b\cdot k&=\frac{n\cdot n_b}{2}\bar{n}\cdot k+\frac{\bar{n}\cdot n_b}{2}n\cdot k+k_\perp\cdot n_{b\perp}\nonumber\\
&=\frac{n\cdot n_b}{2}\bar{n}\cdot k+\frac{\bar{n}\cdot n_b}{2}n\cdot k+(\bar{n}\cdot k n\cdot k)^{1/2}|\hat{n}_{a\perp}|\text{cos }\theta \,,
\end{align}
where $\theta$ denotes the angle between the particle $k$ and the $n$ axis. In the above kinematic relations, we have made use of the fact that since $\hat{n}\sim \hat{n}_a+\hat{n}_b$, $k_\perp\cdot n_{b\perp}=-k_\perp\cdot n_{a\perp}$. Rewriting the integrand for a positive energy gluon in terms of $\theta$, we find
\begin{align}
\int[d^dk]_+&=\frac{1}{2^{4-2\epsilon}\pi^{\frac{5}{2}-\epsilon}\Gamma(\frac{1}{2}-\epsilon)}\int_0^\infty\frac{dn\cdot k}{n\cdot k^{\epsilon}}\int_0^\infty\frac{d\bar{n}\cdot k}{\bar{n}\cdot k^{\epsilon}}\int_0^\pi d\theta\,\text{sin}^{-2\epsilon}\,\theta \,, \\
&= c_\epsilon \int_0^\infty\frac{dn\cdot k}{n\cdot k^{\epsilon}}\int_0^\infty\frac{d\bar{n}\cdot k}{\bar{n}\cdot k^{\epsilon}}\int_0^\pi d\theta\,\text{sin}^{-2\epsilon}\,\theta \,,
\end{align}
for $d=4-2\epsilon$.
To simplify our expressions, we have extracted the following constant
\begin{align}
c_{\epsilon}&=\frac{1}{2^{4-2\epsilon}\pi^{\frac{5}{2}-\epsilon}\Gamma(\frac{1}{2}-\epsilon)}\,.
\end{align}

In the collinear soft region of phase space, we power count $n_a \cdot n_b \ll 1$. We can therefore work to leading power in $n_a\cdot n_b$ in the integrand. Using the relations of \Eq{eq:cs_kinematics1}- \Eq{eq:cs_kinematics2}, and expanding to leading power in $n_a\cdot n_b$, we have
\begin{align}
n_a\cdot k\,n_b\cdot k&=\left(n\cdot k+\frac{n_a\cdot n_b}{8} \bar{n}\cdot k\right)^2-\frac{n_a \cdot n_b}{2}(n\cdot k \, \bar{n}\cdot k)\,\text{cos}^2\theta\,.
\end{align}
Note that in our power counting, $n \cdot k \sim n_a \cdot n_b$, so that this expression scales homogeneously. To perform the integral, we make the change of variables
\begin{align}
\bar n \cdot k = v, \qquad n \cdot k =v w \left(  \frac{n_a \cdot n_b}{8}  \right)\,.
\end{align}
We then have
\begin{align}
n_a\cdot k\,n_b\cdot k&=  v^2 \left(  \frac{n_a \cdot n_b}{8}  \right)^2    \left[  (1+w)^2-4w\,\cos^2 \theta  \right]\\
&= v^2 \left(  \frac{n_a \cdot n_b}{8}  \right)^2    \left[  (1-w)^2+4w\,\sin^2 \theta  \right]\,.
\end{align}
The one loop expression for the collinear soft function can then be written
\begin{align}
S_{c,\,n_a n_b}^{(1)}(\ecf{3}{\alpha})&= \\
&\hspace{-2cm} -g^2   \left( \frac{\mu^{2}  e^{\gamma_E}  }{4\pi} \right)^{\epsilon}    16c_\epsilon   \left(   \frac{ n_a\cdot n_b   }{ 8   }     \right)^{-\epsilon}      \int_0^\infty\frac{dw}{w^{\epsilon}}             \int_0^\infty\frac{dv}{v^{1+2\epsilon}}         \int_0^\pi d\theta\,\text{sin}^{-2\epsilon}\,\theta          \frac{1}{   (1-w)^2+4w\,\sin^2 \theta     }    \nonumber \\
&      \times\delta  \left(\ecf{3}{\alpha}-  \frac{N_{CS} }{2^{1-\alpha}}\frac{v}{Q}           \left(\frac{n_a\cdot n_b}{8} \right)^{\alpha}   \left[   (1-w)^2+4w\,\sin^2 \theta  \right]^{\alpha/2}            \right)\,, \nonumber
\end{align}
The $v$ integral is straightforward. Transforming to Laplace space,  $\ecf{3}{\alpha}\rightarrow\ecflp{3}{\alpha}$, we find
\begin{align}
S_{c,\,n_a n_b}^{(1)}(\ecflp{3}{\alpha})&=  -g^2 \Gamma(-2\epsilon)     \left(    \frac{  \mu^2 N_{CS}^2 e^{\gamma_E}  (\ecflp{3}{\alpha})^2      \left( \frac{n_a \cdot n_b}{8}   \right)^{-1+2\alpha}  }{4\pi 4^{1-\alpha}  Q^2    }  \right)^{\epsilon} \\
&
\hspace{2cm}
\times16 c_\epsilon          \int_0^\infty\frac{dw}{w^{\epsilon}}    \int_0^\pi d\theta\,\text{sin}^{-2\epsilon}\,\theta     \left[ (1-w)^2+4w\,\sin^2 \theta   \right]^{-1+\alpha \epsilon}\,.   \nonumber
\end{align}
The $\theta$ integral can be performed exactly in terms of hypergeometric functions using
\begin{align}
\int_0^\pi d\theta\,\text{sin}^{-2\epsilon}\,\theta     \left[ (1-w)^2+4w\,\sin^2 \theta   \right]^{-1+\alpha \epsilon} &= \\
&\hspace{-4cm}    \frac{\Gamma[1/2-\epsilon] \Gamma[1/2]}{\Gamma[1-\epsilon]} (1-w)^{2(-1+\alpha\epsilon)}   \,_2F_1\left[1-\alpha \epsilon, 1/2-\epsilon, 1-\epsilon, -\frac{4w}{(1-w)^2}\right] \,, \nonumber
\end{align}
which can be rewritten using a Pfaff transformation as
\begin{align}
&\int_0^\pi d\theta\,\text{sin}^{-2\epsilon}\,\theta     \left[ (1-w)^2+4w\,\sin^2 \theta   \right]^{-1+\alpha \epsilon} =    \frac{\Gamma[1/2-\epsilon] \Gamma[1/2]}{\Gamma[1-\epsilon]}   \\
& \hspace{1cm} \times (1+w)^{-1+2\epsilon}  \left( (1-w)^2\right)^{-1/2-(1-\alpha)\epsilon}  \,_2F_1\left[ 1/2-\epsilon,-\epsilon+\alpha\epsilon, 1-\epsilon, \frac{4w}{(1+w)^2}\right] \,. \nonumber
\end{align}
The remaining integral in $w$ is given by
\begin{align}
S_{c,\,n_a n_b}^{(1)}(\ecflp{3}{\alpha})&=  -g^2 \Gamma(-2\epsilon)     \left(    \frac{  \mu^2 N_{CS}^2 e^{\gamma_E}  (\ecflp{3}{\alpha})^2      \left( \frac{n_a \cdot n_b}{8}   \right)^{-1+2\alpha}  }{4\pi 4^{1-\alpha}  Q^2    }  \right)^{\epsilon}   16 c_\epsilon     \frac{\Gamma[1/2-\epsilon] \Gamma[1/2]}{\Gamma[1-\epsilon]}  \nonumber \\
& \hspace{-1.5cm}          \int_0^\infty\frac{dw}{w^{\epsilon}}   (1+w)^{-1+2\epsilon}  \left( (1-w)^2\right)^{-1/2-(1-\alpha)\epsilon}  \,_2F_1\left[ 1/2-\epsilon,-\epsilon+\alpha\epsilon, 1-\epsilon, \frac{4w}{(1+w)^2}\right] \,. 
\end{align}
Re-mapping the integral to the unit interval, we have
\begin{align}
S_{c,\,n_a n_b}^{(1)}(\ecflp{3}{\alpha})&=  -g^2 \Gamma(-2\epsilon)     \left(    \frac{  \mu^2 N_{CS}^2 e^{\gamma_E}  (\ecflp{3}{\alpha})^2      \left( \frac{n_a \cdot n_b}{8}   \right)^{-1+2\alpha}  }{4\pi 4^{1-\alpha}  Q^2    }  \right)^{\epsilon}    16 c_\epsilon      \frac{\Gamma[1/2-\epsilon] \Gamma[1/2]}{\Gamma[1-\epsilon]} \nonumber\\
& \hspace{-2.5cm}           \int_0^1dw \Big(w^{-\epsilon}+w^{(1-2\alpha)\epsilon}\Big)  (1+w)^{-1+2\epsilon}  (1-w)^{-1-2(1-\alpha)\epsilon}  \,_2F_1\left[ 1/2-\epsilon,-\epsilon+\alpha\epsilon, 1-\epsilon, \frac{4w}{(1+w)^2}\right] \,. \nonumber
\end{align}
We could not perform this integral exactly, but it can be done as a Laurent expansion in $\epsilon$ by expanding the hypergeometric function as 
 \begin{align}
\,_2F_1\left[\frac{1}{2}-\epsilon,-(1-\alpha)\epsilon;1-\epsilon;\frac{4w}{(1+w)^2}\right]&=1-2(1-\alpha)\epsilon\,\text{ln}(1+w)+{\mathcal O}(\epsilon^2)\,,
\end{align}
which is valid for $0\leq w \leq 1$, and we have truncated the expansion at ${\mathcal O}(\epsilon^2)$ as we are only interested in the terms up to ${\mathcal O}(\epsilon^0)$ in the one-loop result.
We then have
\begin{align}
S_{c,\,n_a n_b}^{(1)}(\ecflp{3}{\alpha})&=  -g^2 \Gamma(-2\epsilon)     \left(    \frac{  \mu^2 N_{CS}^2 e^{\gamma_E}  (\ecflp{3}{\alpha})^2      \left( \frac{n_a \cdot n_b}{8}   \right)^{-1+2\alpha}  }{4\pi 4^{1-\alpha}  Q^2    }  \right)^{\epsilon}   16 c_\epsilon     \frac{\Gamma[1/2-\epsilon] \Gamma[1/2]}{\Gamma[1-\epsilon]} \nonumber  \\
& \hspace{-1cm}          \int_0^1dw \Big(w^{-\epsilon}+w^{(1-2\alpha)\epsilon}\Big) (1+w)^{-1+2\epsilon}  (1-w)^{-1-2(1-\alpha)\epsilon} \left( 1-2(1-\alpha)\epsilon\,\text{ln}(1+w)   \right)\,. 
\end{align}
For the remaining integral in $w$, we have
 \begin{align}
& \int_0^1dw \Big(w^{-\epsilon}+w^{(1-2\alpha)\epsilon}\Big)    (1+w)^{-1+2\epsilon}  (1-w)^{-1-2(1-\alpha)\epsilon}    \left( 1-2(1-\alpha)\epsilon\,\text{ln}(1+w)   \right)= \nonumber \\
& \int_0^1dw \Big(w^{-\epsilon}+w^{(1-2\alpha)\epsilon}\Big)    (1+w)^{-1+2\epsilon}  (1-w)^{-1-2(1-\alpha)\epsilon}     \\
&\hspace{1cm}  -2(1-\alpha)\epsilon   \int_0^1dw \Big(w^{-\epsilon}+w^{(1-2\alpha)\epsilon}\Big)    (1+w)^{-1+2\epsilon}  (1-w)^{-1-2(1-\alpha)\epsilon}  \,   \log(1+w)  \,. \nonumber
 \end{align}
 The first integral can be done in terms of hypergeometric functions, while the second can be done using plus functions (for a detailed discussion of their properties, see e.g. \cite{Ligeti:2008ac}), and applying the identity
 \begin{align}
 \frac{1}{z^{1+a\epsilon}}=-\frac{1}{a\epsilon}\delta(z) +\sum \limits_{i=0}^{\infty}   \frac{(-a\epsilon)^i}{i!} \mathcal{D}_i (z)\,,
 \end{align}
 with
 \begin{align}
 \mathcal{D}_i (z)=\left[  \frac{\log^i z}{z}  \right]_+\,.
 \end{align}
 We find 
  \begin{align}
 & \int_0^1dw \Big(w^{-\epsilon}+w^{(1-2\alpha)\epsilon}\Big)  \Big((1+w)^2\Big)^{-\frac{1}{2}+\epsilon}\Big((1-w)^2\Big)^{-\frac{1}{2}-(1-\alpha)\epsilon} \left( 1-2(1-\alpha)\epsilon\,\text{ln}(1+w)   \right) \nonumber \\
&\hspace{0cm}= \frac{\Gamma[2(\alpha-1) \epsilon] \Gamma[1-\epsilon]}{ \Gamma[1-3\epsilon+2\alpha \epsilon]} \,_2F_1[1-2\epsilon,1-\epsilon; 1-3\epsilon+2\alpha \epsilon;-1]       
\\    &+       \frac{\Gamma[2(\alpha-1) \epsilon] \Gamma[1+\epsilon-2\alpha \epsilon]}{ \Gamma[1-\epsilon]} \,_2F_1[1-2\epsilon,1+\epsilon-2\alpha\epsilon;1-\epsilon;-1] \nonumber\\
&+ 2^{2\epsilon} \log 2 -2(1-\alpha)\epsilon \left( \log^2 2 -\frac{\pi^2}{12}   \right)  \nonumber \\
&=\frac{1}{(2 \alpha -2) \epsilon }+\frac{\alpha 
   \log (2)}{\alpha -1}+\log (2)+\frac{\epsilon  \left(-\pi ^2 \alpha ^2+36 \alpha ^2 \log ^2(2)+3 \pi ^2 \alpha -24 \alpha 
   \log ^2(2)-2 \pi ^2\right)}{12 (\alpha -1)}   \, . \nonumber
 \end{align}
Therefore, in total, we have
 \begin{align}
S_{c,\,n_a n_b}^{(1)}(\ecflp{3}{\alpha})&=  -g^2 \Gamma(-2\epsilon)     \left(    \frac{  \mu^2 N_{CS}^2 e^{\gamma_E}  (\ecflp{3}{\alpha})^2      \left( \frac{n_a \cdot n_b}{8}   \right)^{-1+2\alpha}  }{4\pi 4^{1-\alpha}  Q^2    }  \right)^{\epsilon}   16 c_\epsilon     \frac{\Gamma[1/2-\epsilon] \Gamma[1/2]}{\Gamma[1-\epsilon]} \nonumber  \\
& \hspace{-1.75cm}   \left(  \frac{1}{(2 \alpha -2) \epsilon }+\frac{\alpha 
   \log (2)}{\alpha -1}+\log (2)+\frac{\epsilon  \left(-\pi ^2 \alpha ^2+36 \alpha ^2 \log ^2(2)+3 \pi ^2 \alpha -24 \alpha 
   \log ^2(2)-2 \pi ^2\right)}{12 (\alpha -1)}      \right)     \,. 
\end{align}
Expanding in $\epsilon$, and keeping only the divergent piece, as relevant for the anomalous dimensions, we find 
 \begin{align} \label{eq:csoft_nanb_final}
\tilde{S}_{c,\,n_a n_b}^{(1)\text{div}}(\ecflp{3}{\alpha})&= \frac{\alpha_s}{\pi}\frac{1}{  (\alpha -1) \epsilon ^2}+2\frac{\alpha_s}{\pi}\frac{ \left(2 \alpha  \log (2)+ \log \left[   \frac{  \mu N_{CS} e^{\gamma_E}  (\ecflp{3}{\alpha})      \left( \frac{n_a \cdot n_b}{8}   \right)^{-1/2+\alpha}  }{2^{1-\alpha}  Q    }  \right]- \log (2)\right)}{   (\alpha -1) \epsilon }\nonumber \\
&=\frac{\alpha_s}{\pi}\frac{1}{  (\alpha -1) \epsilon ^2}+2\frac{\alpha_s}{\pi}\frac{L_\alpha^{cs}}{   (\alpha -1) \epsilon }\,,
\end{align}
where
\begin{align}\label{eq:csoft_log_def}
L_\alpha^{cs}= \log \left(    \frac{  \mu N_{CS} e^{\gamma_E}  (\ecflp{3}{\alpha})      \left( n_a \cdot n_b   \right)^{-1/2+\alpha}  }{\sqrt{2}  Q    }  \right)\,.
\end{align}

\subsubsection*{$n_a,\bar n$ and $n_b,\bar n$ Contributions:}

We now calculate the $n_a, \bar n$ contribution to the collinear-soft function. The $n_b, \bar n$ contribution will be identical. The one-loop integrand is given by
\begin{align}
S_{c,\,n_a \bar{n}}^{(1)}(\ecf{3}{\alpha})&= \\
&\hspace{-2cm}-g^2  \left( \frac{\mu^{2}  e^{\gamma_E }  }{4\pi} \right)^{\epsilon}    \int[d^dk]_+\frac{2n_a\cdot \bar{n}}{n_{a}\cdot k\,k\cdot \bar{n}}\delta\left(\ecf{3}{\alpha}-N_{CS}\left(\frac{\bar{n}\cdot k}{2Q}\right)^{1-\alpha}\left(\frac{n_a\cdot k}{Q}\right)^{\alpha/2}\left(\frac{n_b\cdot k}{Q}\right)^{\alpha/2}\right)\,,\nonumber
\end{align}
where we have again extracted the normalization factor
\begin{align}
N_{CS}&= 2^{3\alpha/2+1} z_az_b\left(n_a\cdot n_b\right)^{\frac{\alpha}{2}}\,.
\end{align}

As with the $n_a \cdot n_b$ contribution, we expand the integrand to leading power in $n_a \cdot n_b$ using
\begin{align}
n_a\cdot k\,n_b\cdot k&=\left(n\cdot k+\frac{n_a\cdot n_b}{8} \bar{n}\cdot k\right)^2-\frac{n_a \cdot n_b}{2}(n\cdot k \, \bar{n}\cdot k)\,\text{cos}^2\theta\,, \\
n_a\cdot \bar n&=2 \,,\\
n_a \cdot k&=\frac{n_a \cdot n_b}{8} \bar n\cdot k   +n\cdot k -\left( n\cdot k \bar n \cdot k    \right)^{1/2}    \sqrt{\frac{n_a \cdot n_b}{2}} \cos \theta\,.
\end{align}
To perform the integral, it is again convenient to make the change of variables
\begin{align}
\bar n \cdot k = v, \qquad n \cdot k =v w \left(  \frac{n_a \cdot n_b}{8}  \right)\,.
\end{align}
We then have
\begin{align}
n_a\cdot k\,n_b\cdot k&= v^2 \left(  \frac{n_a \cdot n_b}{8}  \right)^2    \left[  (1-w)^2+4w\,\sin^2 \theta  \right]\,, \\
n_a\cdot k&=\frac{n_a \cdot n_b}{8} v  +v w \left(  \frac{n_a \cdot n_b}{8}  \right) -\left(v^2 w \left(  \frac{n_a \cdot n_b}{8}  \right)   \right)^{1/2}    \sqrt{\frac{n_a \cdot n_b}{2}} \cos \theta \nonumber \\
&=v \left(  \frac{n_a \cdot n_b}{8}  \right) \left(   1+w-2\sqrt{w} \cos \theta \right)\,.
\end{align}

The one-loop expression for the contribution to the collinear soft function can then be written
\begin{align}
S_{c,\,n_a \bar{n}}^{(1)}(\ecf{3}{\alpha})&= \\
&\hspace{-1cm} -g^2   \left( \frac{\mu^{2}  e^{\gamma_E}  }{4\pi} \right)^{\epsilon}   4 c_\epsilon \left( \frac{n_a \cdot n_b}{8} \right)^{-\epsilon}  \int_0^\infty\frac{dw}{w^{\epsilon}}             \int_0^\infty\frac{dv}{v^{1+2\epsilon}}         \int_0^\pi d\theta\,\text{sin}^{-2\epsilon}\,\theta          \frac{1}{   1+w-2\sqrt{w} \cos \theta     }    \nonumber \\
&      \delta  \left(\ecf{3}{\alpha}-  \frac{N_{CS} }{2^{1-\alpha}}\frac{v}{Q}           \left(\frac{n_a\cdot n_b}{8} \right)^{\alpha}   \left[   (1-w)^2+4w\,\sin^2 \theta  \right]^{\alpha/2}            \right)\,. \nonumber
\end{align}
This integral can be performed in a similar manner to the $n_a\cdot n_b$ integral. The $v$ integral is straightforward, after transforming to Laplace space $\ecf{3}{\alpha}\to \ecflp{3}{\alpha}$, we find
 \begin{align}
 S_{c,\,n_a \bar{n}}^{(1)}(\ecflp{3}{\alpha})&=  -g^2    4 c_\epsilon    \Gamma(-2\epsilon)     \left(    \frac{  \mu^2 N_{CS}^2 e^{\gamma_E}  (\ecflp{3}{\alpha})^2      \left( \frac{n_a \cdot n_b}{8}   \right)^{-1+2\alpha}  }{4\pi 4^{1-\alpha}  Q^2    }  \right)^{\epsilon}  \nonumber \\
&    \hspace{2cm}     \int_0^\infty\frac{dw}{w^{\epsilon}}                  \int_0^\pi d\theta\,\text{sin}^{-2\epsilon}\,\theta          \frac{\left[  (1-w)^2+4w\,\sin^2 \theta  \right]^{\alpha \epsilon}}{   1+w-2\sqrt{w} \cos \theta     } \,.
\end{align}
We now focus on the integral
\begin{align}
 \int_0^\infty\frac{dw}{w^{\epsilon}}                  \int_0^\pi d\theta\,\text{sin}^{-2\epsilon}\,\theta          \frac{\left[  (1-w)^2+4w\,\sin^2 \theta  \right]^{\alpha \epsilon}}{   1+w-2\sqrt{w} \cos \theta     }\,.
\end{align}
Remapping to the unit interval,  we find
\begin{align}
& \int_0^1 du   \,  \left[  u^{-\epsilon}+u^{-1+(1-2\alpha)\epsilon} \right]               \int_0^\pi d\theta\,\text{sin}^{-2\epsilon}\,\theta      \frac{\left[  (1-u)^2+4u\,\sin^2 \theta  \right]^{\alpha \epsilon}}{   1+u-2\sqrt{u} \cos \theta     }     \nonumber\\
&= \int_0^1 du   \, \left[  u^{-\epsilon}+u^{-1+(1-2\alpha)\epsilon} \right]      \nonumber \\
& \hspace{2cm} \times        \int_0^\pi d\theta\,\text{sin}^{-2\epsilon}\,\theta         \left[  (1-u)^2+4u\,\sin^2 \theta  \right]^{\alpha \epsilon-1}    (1+u+2\sqrt{u} \cos \theta )     
\end{align} 
The $\theta$ integral can be performed in terms of hypergeometric functions using
\begin{align}
&\int_0^\pi d\theta\,\text{sin}^{-2\epsilon}\,\theta     \left[ (1-u)^2+4u\,\sin^2 \theta   \right]^{-1+\alpha \epsilon} =    \frac{\Gamma[1/2-\epsilon] \Gamma[1/2]}{\Gamma[1-\epsilon]}   \\
& \hspace{1cm} \times (1+u)^{-1+2\epsilon}  \left( (1-u)^2\right)^{-1/2-(1-\alpha)\epsilon}  \,_2F_1\left[ 1/2-\epsilon,-\epsilon+\alpha\epsilon, 1-\epsilon, \frac{4u}{(1+u)^2}\right] \,, \nonumber
\end{align}
and
\begin{align}
 \int_0^\pi d\theta\,\text{sin}^{-2\epsilon}\,\theta         \left[  (1-u)^2+4u\,\sin^2 \theta  \right]^{\alpha \epsilon-1}     \cos \theta = 0\,,
\end{align}
by symmetry.

The hypergeometric function has the expansion
 \begin{align}
\,_2F_1\left[\frac{1}{2}-\epsilon,-(1-\alpha)\epsilon;1-\epsilon;\frac{4u}{(1+u)^2}\right]&=1-2(1-\alpha)\epsilon\,\text{ln}(1+u)+{\mathcal O}(\epsilon^2)\,,
\end{align}
which is valid for $0\leq u \leq 1$, 

The final $u$ integral is then
\begin{align}
&    \frac{\Gamma[1/2-\epsilon] \Gamma[1/2]}{\Gamma[1-\epsilon]} \int_0^1 du   \, \left[  u^{-\epsilon}+u^{-1+(1-2\alpha)\epsilon} \right]          (1+u)^{2\epsilon}  \left( (1-u)^2\right)^{-1/2-(1-\alpha)\epsilon} \nonumber \\
&  \hspace{8cm} \times    \left(  1-2(1-\alpha)\epsilon\,\text{ln}(1+u)  \right)\,.
\end{align}

We expect this integral to contribute both $\frac{1}{(1-\alpha) \epsilon}$ and $\frac{1}{(1-2\alpha)\epsilon}$ poles, unlike the $n_a n_b$ contribution, which are evident in the $u\to 1$ and $u\to 0$ limits respectively. We need to do the integral to $\mathcal{O}(\epsilon)$ to get the finite pieces, but only $\mathcal{O}(\epsilon^0)$ to get the anomalous dimensions, which is sufficient for now. We have
\begin{align}
&   = \frac{\Gamma[1/2-\epsilon] \Gamma[1/2]}{\Gamma[1-\epsilon]} \int_0^1 du   \,  u^{-\epsilon}         (1+u)^{2\epsilon}  \left( (1-u)^2\right)^{-1/2-(1-\alpha)\epsilon}  \nonumber \\
&   -2(1-\alpha)\epsilon\, \frac{\Gamma[1/2-\epsilon] \Gamma[1/2]}{\Gamma[1-\epsilon]} \int_0^1 du   \,  u^{-\epsilon}         (1+u)^{2\epsilon}  \left( (1-u)^2\right)^{-1/2-(1-\alpha)\epsilon} \log(1+u) \nonumber \\
&+ \frac{\Gamma[1/2-\epsilon] \Gamma[1/2]}{\Gamma[1-\epsilon]} \int_0^1 du   \,  u^{-1+(1-2\alpha)\epsilon}           (1+u)^{2\epsilon}  \left( (1-u)^2\right)^{-1/2-(1-\alpha)\epsilon}  \nonumber \\
&-2(1-\alpha)\epsilon\, \frac{\Gamma[1/2-\epsilon] \Gamma[1/2]}{\Gamma[1-\epsilon]} \int_0^1 du   \,  u^{-1+(1-2\alpha)\epsilon}           (1+u)^{2\epsilon}  \left( (1-u)^2\right)^{-1/2-(1-\alpha)\epsilon}  \log(1+u)   
\end{align}
This integral can be done systematically using $+$-functions, but to the order we need the result, it is easier to use subtractions, evaluate the log at the value of the singularity, and then perform the integral in terms of hypergeometric functions. The integral can be written
\begin{align}
&   = \frac{\Gamma[1/2-\epsilon] \Gamma[1/2]}{\Gamma[1-\epsilon]} \int_0^1 du   \,  u^{-\epsilon}         (1+u)^{2\epsilon}  \left( (1-u)^2\right)^{-1/2-(1-\alpha)\epsilon}   \\
&   -2(1-\alpha)\epsilon\, \frac{\Gamma[1/2-\epsilon] \Gamma[1/2]}{\Gamma[1-\epsilon]} \int_0^1 du   \,  u^{-\epsilon}         (1+u)^{2\epsilon}  \left( (1-u)^2\right)^{-1/2-(1-\alpha)\epsilon} \log(2) \nonumber \\
&+ \frac{\Gamma[1/2-\epsilon] \Gamma[1/2]}{\Gamma[1-\epsilon]} \int_0^1 du   \,  u^{-1+(1-2\alpha)\epsilon}           (1+u)^{2\epsilon}  \left( (1-u)^2\right)^{-1/2-(1-\alpha)\epsilon}  \nonumber \\
&-2(1-\alpha)\epsilon\, \frac{\Gamma[1/2-\epsilon] \Gamma[1/2]}{\Gamma[1-\epsilon]} \int_0^1 du   \,  u^{-1+(1-2\alpha)\epsilon}         (1+u)^{2\epsilon}     \left[  \left( (1-u)^2\right)^{-1/2-(1-\alpha)\epsilon} -1\right] \log(2)   \,, \nonumber
\end{align}
which gives
\begin{align}
&   = \frac{\Gamma[1/2-\epsilon] \Gamma[1/2]}{\Gamma[1-\epsilon]}                    \frac{ \Gamma[1-\epsilon] \Gamma[-2(1-\alpha)\epsilon]   }{  \Gamma[1-\epsilon-2(1-\alpha) \epsilon]   }\, _2F_1[-2\epsilon,1-\epsilon; 1-\epsilon-2(1-\alpha)\epsilon;-1]   \nonumber \\
&   -2(1-\alpha)\epsilon\, \frac{\Gamma[1/2-\epsilon] \Gamma[1/2]}{\Gamma[1-\epsilon]}  \log(2)      \frac{ \Gamma[1-\epsilon] \Gamma[-2(1-\alpha)\epsilon]   }{  \Gamma[1-\epsilon-2(1-\alpha) \epsilon]   }\, _2F_1[-2\epsilon,1-\epsilon; 1-\epsilon-2(1-\alpha)\epsilon;-1]        \nonumber \\
&+ \frac{\Gamma[1/2-\epsilon] \Gamma[1/2]}{\Gamma[1-\epsilon]}  \frac{ \Gamma[ (1-2\alpha)\epsilon] \Gamma[-2(1-\alpha)\epsilon]   }{  \Gamma[  (1-2\alpha)\epsilon-2(1-\alpha)\epsilon       ]   }\, _2F_1[ -2\epsilon,(1-2\alpha)\epsilon;(1-2\alpha)\epsilon-2(1-\alpha)\epsilon;-1]  \nonumber \\
&-2(1-\alpha)\epsilon\, \frac{\Gamma[1/2-\epsilon] \Gamma[1/2]}{\Gamma[1-\epsilon]}  \log(2) \nonumber \\
& \times \left(        \frac{ \Gamma[ (1-2\alpha)\epsilon] \Gamma[-2(1-\alpha)\epsilon]   }{  \Gamma[  (1-2\alpha)\epsilon-2(1-\alpha)\epsilon       ]   }\, _2F_1[ -2\epsilon,(1-2\alpha)\epsilon;(1-2\alpha)\epsilon-2(1-\alpha)\epsilon;-1]  \right. \nonumber \\
&\hspace{5.5cm}\left.  -  \frac{\Gamma[(1-2\alpha)\epsilon] }{\Gamma[1+(1-2\alpha)\epsilon]} \,_2 F_1[-2\epsilon,1;1+(1-2\alpha)\epsilon;-1]   \right) \,.
\end{align}
Expanding this to  $\mathcal{O}(\epsilon^0)$ gives
\begin{align}
=\frac{\pi }{(\alpha -1) \epsilon }-\frac{\pi }{(2 \alpha -1) \epsilon } -\frac{2\pi  \log (2)}{2 \alpha -1}+\frac{4 \pi  \log (2)}{\alpha -1}+2\pi  \log (2)
\end{align}
We then have
 \begin{align}
 S_{c,\,n_a \bar{n}}^{(1)}(\ecflp{3}{\alpha})&=  -g^2     4 c_\epsilon    \Gamma(-2\epsilon)     \left(    \frac{  \mu^2 N_{CS}^2 e^{\gamma_E}  (\ecflp{3}{\alpha})^2      \left( \frac{n_a \cdot n_b}{8}   \right)^{-1+2\alpha}  }{4\pi 4^{1-\alpha}  Q^2    }  \right)^{\epsilon}  \nonumber \\
&    \hspace{2cm}    \,\times    \left(\frac{\pi }{(\alpha -1) \epsilon }-\frac{\pi }{(2 \alpha -1) \epsilon } -\frac{2\pi  \log (2)}{2 \alpha -1}+\frac{4 \pi  \log (2)}{\alpha -1}+2\pi  \log (2)   \right)\,.
\end{align}
Extracting just the divergent pieces so as to get the anomalous dimensions, we find
 \begin{align}\label{eq:csoft_nbarna_final}
 S_{c,\,n_a \bar{n}}^{(1)}(\ecflp{3}{\alpha})&= \frac{\alpha _s}{2\pi  (\alpha -1) \epsilon ^2}-\frac{\alpha _s}{2\pi  (2 \alpha
   -1) \epsilon ^2}+\frac{ \log\left[   \frac{  \mu N_{CS} e^{\gamma_E}  (\ecflp{3}{\alpha})      \left( \frac{n_a \cdot n_b}{8}   \right)^{-1/2+\alpha}  }{2^{1-\alpha}  Q    }     \right] \alpha _s}{\pi  (\alpha -1) \epsilon }    \nonumber \\
   &        -\frac{ \log \left[   \frac{  \mu N_{CS} e^{\gamma_E}  (\ecflp{3}{\alpha})      \left( \frac{n_a \cdot n_b}{8}   \right)^{-1/2+\alpha}  }{2^{1-\alpha}  Q    }   \right] \alpha _s}{\pi  (2 \alpha
   -1) \epsilon }    +\frac{ \log (2) \alpha _s}{\pi  (\alpha -1) \epsilon }+\frac{\log (2) \alpha _s}{\pi  \epsilon }\,,
\end{align}
which can be simplified to
\begin{align}
 S_{c,\,n_a \bar{n}}^{(1)}(\ecflp{3}{\alpha})&=\frac{\alpha _s}{2\pi  (\alpha -1) \epsilon ^2}-\frac{\alpha _s}{2\pi  (2 \alpha -1) \epsilon ^2}\nonumber\\
 &\hspace{2cm} +\frac{\alpha_s}{\pi (\alpha-1) \epsilon} L_\alpha^{cs} \left( \ecflp{3}{\alpha}  \right)-\frac{\alpha_s}{\pi (2\alpha-1) \epsilon} L_\alpha^{cs} \left( \ecflp{3}{\alpha}  \right)\,,
\end{align}
where, as for the logarithm in the $n_a\, n_b$ contribution, \Eq{eq:csoft_log_def}, the logarithm that appears is
\begin{align}
L_\alpha^{cs} \left( \ecflp{3}{\alpha}  \right)
= \log \left(    \frac{  \mu N_{CS} e^{\gamma_E}  (\ecflp{3}{\alpha})      \left( n_a \cdot n_b   \right)^{-1/2+\alpha}  }{\sqrt{2}  Q    }  \right)\,.
\end{align}

The contribution from an emission between the $n_b$ and $\bar n$ Wilson lines is identical, so we have
\begin{align}\label{eq:csoft_log}
S_{c,\,n_b \bar{n}}^{(1)}  \left(\ecf{3}{\alpha}\right)=S_{c,\,n_a \bar{n}}^{(1)}   \left(\ecf{3}{\alpha} \right)\,.
\end{align}

Note that for both the $\bar n\, n_a$ and $\bar n\, n_b$ contributions, and unlike for the $n_a \, n_b$ contribution, we have $1/\epsilon$ contributions both of the soft form $1/(1-2\alpha)$, and of the collinear form, $1/(1-\alpha)$. This will be crucial to achieve the cancellation of anomalous dimensions, as required for the consistency of the collinear subjets factorization theorem.

It is interesting to note that this structure is very different than that which appeared for the case of the $N$-subjettiness observable in \Ref{Bauer:2011uc}. In this case only a single angular exponent appears throughout the calculation, unlike both the $1/(1-2\alpha)$ and $1/(1-\alpha)$ that we find here, and the divergent pieces of the $\bar n\, n_a$ and $\bar n\, n_b$ contributions vanish.

\subsection*{Cancellation of Anomalous Dimensions}\label{sec:bkg_cancel_anomdim}

We now review the renormalization group evolution of each of the functions in the factorization theorem, and show that sum of the anomalous dimensions vanishes, as required for renormalization group consistency.  

The hard function satisfies a multiplicative RGE, given by
\begin{equation}
\mu \frac{d}{d\mu}\log\, H(Q^2,\mu)=\gamma_H (Q^2,\mu)=2\text{Re}\left[ \gamma_C(Q^2,\mu)\right]\,,
\end{equation}
where
\begin{equation}
\gamma_C(Q^2,\mu)=\frac{\alpha_s C_F}{4\pi}\left(  4\log\left[  \frac{-Q^2}{\mu^2} \right] -6  \right )\,,
\end{equation}
is the anomalous dimension of the dijet Wilson coefficient. Explicitly
\begin{align}
\gamma_H (Q^2,\mu)=\frac{\alpha_s C_F}{2\pi}\left(  4\log\left[  \frac{Q^2}{\mu^2} \right] -6  \right )\,.
\end{align}

The anomalous dimension of the hard splitting function $H_2$ can be extracted from \Ref{Bauer:2011uc} by performing a change of variables. It satisfies a multiplicative RGE
 \begin{align}
\mu \frac{d}{d\mu}H_2(t,x,\mu)=\gamma_{H_2}(t,x,\mu) H_2(t,x,\mu)\,, 
\end{align}
with anomalous dimension
\begin{align}
\gamma_{H_2}(t,x,\mu)=\frac{\alpha_s(\mu)}{2\pi} \left [  2C_A \log \frac{t}{\mu^2} +4\left( C_F-\frac{C_A}{2}\right ) \log\, x +2C_A \log(1-x)-\beta_0 \right]\,.
\end{align}
Here $\beta_0$ is defined with the normalization
\begin{equation}
\beta_0=\frac{11 C_A}{3}-\frac{2n_f}{3 } \,.
\end{equation}
Converting to $\ecf{2}{\alpha}$ by performing the change of variables given in \Eq{eq:relate_to_frank}, we find
\begin{align}\label{eq:h2_anom}
\gamma_{H_2}\left(\ecf{2}{\alpha},z_q,\mu\right)&=\frac{\alpha_s(\mu)}{2\pi} \left [  2C_A \log\left( \frac{Q^2}{\mu^2} \frac{(z_az_b)^{1-2/\alpha}    \left( \ecf{2}{\alpha}  \right)^{2/\alpha}      }{4} \right) \right. \\
&\hspace{3cm}\left.+4\left( C_F-\frac{C_A}{2}\right ) \log\, z_q +2C_A \log(1-z_q)-\beta_0        \vphantom{\log\left( \frac{Q^2}{\mu^2} \frac{(z_az_b)^{1-2/\alpha}    \left( \ecf{2}{\alpha}  \right)^{2/\alpha}      }{4} \right)}        \right]\,. \nonumber
\end{align}
Since the anomalous dimensions of the jet, soft and collinear-soft functions are written in terms of $\ecflp{3}{\alpha}$, $z_a$, $z_b$, and $n_a\cdot n_b$, for demonstrating cancellation of anomalous dimensions, it is convenient to replace $\ecf{2}{\alpha}$ in \Eq{eq:h2_anom} with its leading power expression from \Eq{eq:lp_e2}. We then have
\begin{align}\label{eq:h2_anom2}
\gamma_{H_2}\left(\ecf{2}{\alpha},z_q,\mu\right)&=\frac{\alpha_s(\mu)}{2\pi} \left [  2C_A \log\left( \frac{Q^2}{\mu^2} \frac{ z_az_b\,    n_a \cdot n_b     }{2} \right) \right. \\
&\hspace{3cm}\left.+4\left( C_F-\frac{C_A}{2}\right ) \log\, z_q +2C_A \log(1-z_q)-\beta_0        \vphantom{ \log\left( \frac{Q^2}{\mu^2} \frac{ z_az_b\,    n_a \cdot n_b     }{2} \right)}        \right]\,. \nonumber
\end{align}
Note that $1-z_q=z_g$.

The jet functions satisfy multiplicative RGEs in Laplace space (they satisfy convolutional RGEs in $\ecf{3}{\alpha}$, see \Ref{Ellis:2010rwa} for a detailed discussion)
\begin{align}
\mu\frac{d}{d\mu}\log\,\tilde{J}_{g,q\,n}    \left(Q_J,\ecflp{3}{\alpha}\right)&=\gamma_{g,q}^{\alpha}     \left(Q_J,\ecflp{3}{\alpha} \right)\,,
\end{align}
where the one-loop anomalous dimension is determined from \Eqs{eq:gjet_final}{eq:qjet_final}, and is given by
\begin{align}\label{eq:jet_anoms}
\gamma_{g,q}^{\alpha}\left(Q_J,\ecflp{3}{\alpha} \right)&=-2\frac{\alpha_s}{\pi}\frac{C_{g,q}}{(1-\alpha)}L_\alpha^{J,a}\left(  \ecflp{3}{\alpha}  \right)+\gamma_{g,q}\,,
\end{align}
where the logarithm $L_\alpha^{J,a}\left(  \ecflp{3}{\alpha}  \right)$ was defined in \Eq{eq:jet_log}, and is given by
\begin{align} 
L_\alpha^{J,a}\left(  \ecflp{3}{\alpha}  \right)&=\log\left[N_J\ecflp{3}{\alpha}e^{\gamma_E }\left(\frac{\mu}{\sqrt{2}Q}\right)^\alpha z_a^{2-\alpha}\right]\,.
\end{align}
Here $C_{g,q}$ is the appropriate Casimir ($C_A$ for gluon jets and $C_F$ for quark jets), and with $\gamma_{g,q}$ the standard functions
 \begin{align}
\gamma_{q}= \frac{3\alpha_s C_F}{2\pi}\,, \qquad
\gamma_{g}=\frac{\alpha_s}{\pi}\frac{11C_A- 2n_f}{6}\,.
\end{align}
For subjet $b$, we simply have $a\to b$.

Similarly, the soft function satisfies a multiplicative RGE in Laplace space 
\begin{align}
\mu\frac{d}{d\mu}\log\,\tilde{S}_{G}    \left(\ecflp{3}{\alpha}\right)&=\gamma_G     \left(\ecflp{3}{\alpha}    \right)\,,
\end{align}
with one-loop anomalous dimension determined by \Eq{eq:gsoft_final}, and given by
\begin{align}
\gamma_G    \left(\ecflp{3}{\alpha}\right)&=\frac{-2\alpha_s}{\pi(1-2\alpha)}{\mathbf T}_n\cdot{\mathbf T}_{\bar{n}} \,L_\alpha^{G} \left(\ecflp{3}{\alpha}\right)  \,.
\end{align}
Here the logarithm is given by
\begin{align}
L_\alpha^{G}\left(\ecflp{3}{\alpha}\right)&=\log\left[\frac{e^{\gamma_E}\mu\ecflp{3}{\alpha}N_S}{2^{1-\alpha}Q}\right]-\frac{(1-2\alpha)}{2}\log\left[\text{tan}^2\frac{R}{2}\right]\,.
\end{align}
Finally, the collinear soft function satisfies a multiplicative RGE in Laplace space
\begin{align}
\mu\frac{d}{d\mu}\log\,S^{}_{c}     \left(\ecflp{3}{\alpha}\right)&=\gamma_{cs}     \left(\ecflp{3}{\alpha}\right)\,,
\end{align}
with the one-loop anomalous dimension determined by \Eqs{eq:csoft_nanb_final}{eq:csoft_nbarna_final}
\begin{align}
\gamma_{cs}   \left(\ecflp{3}{\alpha}\right)&={\mathbf T}_a\cdot{\mathbf T}_b\gamma_{ab}    \left(\ecflp{3}{\alpha}\right)+{\mathbf T}_a\cdot{\mathbf T}_{\bar{n}}\gamma_{a\bar{n}}\left(\ecflp{3}{\alpha}\right)+{\mathbf T}_{\bar{n}}\cdot{\mathbf T}_b\gamma_{\bar{n}b}  \left(\ecflp{3}{\alpha}\right)\,,
\end{align}
where 
\begin{align}
\gamma_{ab}   \left (\ecflp{3}{\alpha}\right )&=     \frac{-4\alpha_s}{\pi(1-\alpha)}L_\alpha^{cs}\left(\ecflp{3}{\alpha}\right)\,,\\
\gamma_{a\bar{n}}    \left(\ecflp{3}{\alpha}\right)&=\gamma_{b\bar{n}} \left(\ecflp{3}{\alpha}\right)=\frac{-2\alpha_s}{\pi(1-\alpha)}L_\alpha^{cs}\left(\ecflp{3}{\alpha}\right)+\frac{2\alpha_s}{\pi(1-2\alpha)}L_\alpha^{cs}\left(\ecflp{3}{\alpha}\right)\,.
\end{align}
The argument of the logarithm appearing in the collinear soft function, was defined in \Eq{eq:csoft_log_def}, and is given by
\begin{align}
L_\alpha^{cs}= \log \left(    \frac{  \mu N_{CS} e^{\gamma_E}  (\ecflp{3}{\alpha})      \left( n_a \cdot n_b   \right)^{-1/2+\alpha}  }{\sqrt{2}  Q    }  \right)\,.
\end{align}

We can now explicitly check the cancellation of anomalous dimensions. We consider the particular partonic subprocess $e^+e^- \to \bar q q \to \bar q q g$ for which we have explicitly given the hard splitting function, in which case the color algebra can be simplified and written entirely in terms of Casimirs using the color conservation relations
\begin{align}
{\mathbf T}_n={\mathbf T}_q+{\mathbf T}_g\,,\\
{\mathbf T}_n+{\mathbf T}_{\bar n}=0\,.
\end{align}
We then have
\begin{align}
{\mathbf T}_n\cdot{\mathbf T}_{\bar{n}}&=-C_F\,,\\
{\mathbf T}_q\cdot{\mathbf T}_{g}&=-\frac{C_A}{2}\,,\\
{\mathbf T}_q\cdot{\mathbf T}_{\bar{n}}&=\frac{C_A}{2}-C_F\,,\\
{\mathbf T}_g\cdot{\mathbf T}_{\bar{n}}&=-\frac{C_A}{2}\,, \\
{\mathbf T}_g\cdot{\mathbf T}_{g}&=C_A\,, \\
{\mathbf T}_n\cdot{\mathbf T}_{n}&=C_F\,.
\end{align}
However, for most of the cancellation of the anomalous dimensions, it will be convenient to work in the abstract color notation, so as not to need to use relations between the color Casimirs.

The independence of the total cross section under renormalization group evolution implies the following relation between anomalous dimensions
 \begin{align}\label{eq:sum_anom_zero}
\gamma_H \left(Q^2,\mu  \right)+\gamma_{H_2}\left(\ecf{2}{\alpha},z_q,\mu\right)+\gamma_{g}^{\alpha}\left(\ecflp{3}{\alpha}\right)+\gamma_{q}^{\alpha}\left(\ecflp{3}{\alpha}\right)+\gamma_G\left(\ecflp{3}{\alpha}\right)+\gamma_{cs}\left(\ecflp{3}{\alpha}\right)\sim0\,,
\end{align}
where the $\sim$ means up to a term corresponding to the measurement of the jet in the $\bar n$ direction, and the out-of-jet contribution to the soft function, which is independent of the $\ecflp{3}{\alpha}$ measurement, and the kinematics of the substructure, namely $n_a\cdot n_b$, $z_a$, and $z_b$. We will make this relation precise shortly.

We now show explicitly that this cancellation occurs, and how it arises, which provides a non-trivial cross-check on the collinear-subjets factorization theorem. Substituting in the expressions above, we find
\begin{align}
&\sum_\gamma =\gamma_H \left(Q^2,\mu  \right)+\gamma_{H_2}\left(\ecf{2}{\alpha},z_q,\mu\right) \nonumber \\
&+  \left[   -    {\mathbf T}_a \cdot{\mathbf T}_{b}  \frac{4\alpha_s L_\alpha^{cs}\left(\ecflp{3}{\alpha}\right)}{\pi (1-\alpha)}     \right. \nonumber \\
&\left. \hspace{0.0cm}-2 {\mathbf T}_a \cdot{\mathbf T}_{\bar n}  \left( \frac{\alpha_s L_\alpha^{cs}\left(\ecflp{3}{\alpha}\right)}{\pi (1-\alpha)}-\frac{\alpha_s L_\alpha^{cs}\left(\ecflp{3}{\alpha}\right)}{\pi (1-2\alpha)}   \right)         -2 {\mathbf T}_b \cdot{\mathbf T}_{\bar n}  \left( \frac{\alpha_s L_\alpha^{cs}\left(\ecflp{3}{\alpha}\right)}{\pi (1-\alpha)}-\frac{\alpha_s L_\alpha^{cs}\left(\ecflp{3}{\alpha}\right)}{\pi (1-2\alpha)}   \right)  \right] \nonumber \\
&- \left[      {\mathbf T}_n \cdot{\mathbf T}_{\bar n}   \frac{2\alpha_s L_\alpha^{G}\left(\ecflp{3}{\alpha}\right)}{\pi (1-2\alpha)}   \right]        -   C_A   \frac{2\alpha_s L_\alpha^{g}\left(  \ecflp{3}{\alpha}  \right)}{\pi (1-\alpha)} +\gamma_g        - C_F   \frac{2\alpha_s L_\alpha^{q}\left(  \ecflp{3}{\alpha}  \right)}{\pi (1-\alpha)} +\gamma_q \,.
\end{align}
To make manifest the separate cancellations, we use the color conservation relation ${\mathbf T}_n={\mathbf T}_a+{\mathbf T}_b$ in the soft anomalous dimension, and ${\mathbf T}_{\bar n}=-{\mathbf T}_a-{\mathbf T}_b$ in the $1/(1-\alpha)$ pieces of the collinear soft anomalous dimensions. Grouping together collinear like terms ($1/(1-\alpha)$) and soft like terms ($1/(1-2\alpha)$), we then have
\begin{align}
&\sum_\gamma =\gamma_H \left(Q^2,\mu  \right)+\gamma_{H_2}\left(\ecf{2}{\alpha},z_q,\mu\right) \nonumber \\
&-  \left[      ({\mathbf T}_a+{\mathbf T}_b )\cdot{\mathbf T}_{\bar n}   \frac{2\alpha_s L_\alpha^{G}\left(\ecflp{3}{\alpha}\right)}{\pi (1-2\alpha)}   \right]+ \left[ {\mathbf T}_a \cdot{\mathbf T}_{\bar n}  \frac{2\alpha_s L_\alpha^{cs}\left(\ecflp{3}{\alpha}\right)}{\pi (1-2\alpha)}           + {\mathbf T}_b \cdot{\mathbf T}_{\bar n} \frac{2\alpha_s L_\alpha^{cs}\left(\ecflp{3}{\alpha}\right)}{\pi (1-2\alpha)}   \right] \nonumber \\
&-       {\mathbf T}_a \cdot{\mathbf T}_{b}  \frac{4\alpha_s L_\alpha^{cs}\left(\ecflp{3}{\alpha}\right)}{\pi (1-\alpha)} - \left[  {\mathbf T}_a \cdot   (-{\mathbf T}_a-{\mathbf T}_b)  \frac{2\alpha_s  L_\alpha^{cs}\left(\ecflp{3}{\alpha}\right) }{\pi (1-\alpha)}           + {\mathbf T}_b \cdot(-{\mathbf T}_a-{\mathbf T}_b)  \frac{2\alpha_s  L_\alpha^{cs}\left(\ecflp{3}{\alpha}\right)}{\pi (1-\alpha)}   \right] \nonumber \\
&-      C_A   \frac{2\alpha_s L_\alpha^{g}\left(  \ecflp{3}{\alpha}  \right)}{\pi (1-\alpha)} +\gamma_g  -     C_F   \frac{2\alpha_s L_\alpha^{q}\left(  \ecflp{3}{\alpha}  \right)}{\pi (1-\alpha)} +\gamma_q \,.
\end{align}
Since all the logs are linear in the $\ecflp{3}{\alpha}$, we immediately see that the color conservation relations have led to the cancellation of the $\ecflp{3}{\alpha}$ dependence in the soft like pieces between the $\bar nn_b$ and $\bar nn_a$ contributions to the collinear soft function with the global soft contribution, and the cancellation between the collinear like pieces involve all three contributions to the collinear soft function, as well as the jet functions. This nontrivial cancellation supports the validity of the collinear subjets factorization theorem.

It is also straightforward to check that the dependence on $\ecf{2}{\alpha}$ as well as on the jet energy fractions also cancels, although this is more tedious to perform step by step. We therefore simply quote the summed result of the anomalous dimensions, to make clear the meaning of the equivalence relation in \Eq{eq:sum_anom_zero}. We have
\begin{align}
&\gamma_H \left(Q^2,\mu  \right)+\gamma_{H_2}\left(\ecf{2}{\alpha},z_q,\mu\right)+\gamma_{g}^{\alpha}\left(\ecflp{3}{\alpha}\right)+\gamma_{q}^{\alpha}\left(\ecflp{3}{\alpha}\right)+\gamma_G\left(\ecflp{3}{\alpha}\right)+\gamma_{cs}\left(\ecflp{3}{\alpha}\right)= \nonumber \\
&\hspace{2cm}-\frac{3\alpha_s C_F}{2\pi}-\frac{  \alpha_s  C_F    \log \left[ \tan^2 \frac{R}{2}   \right] }{\pi} -\frac{  \alpha_s  C_F   \log \frac{\mu^2}{Q^2}}{\pi}\,.
\end{align}
These remaining terms are exactly those expected to cancel against the out-of-jet contribution; see, e.g., \Ref{Ellis:2010rwa} for a detailed discussion. 

The out-of-jet jet function is then given by the unmeasured jet function of \Ref{Ellis:2010rwa}
\begin{align}
\mu\frac{d}{d\mu}\ln J_{oj }(R_B) &= \frac{2\alpha_s C_F}{\pi} \log\left[    \frac{\mu}{ Q \tan \frac{ R_B}{2} }   \right]+\frac{3\alpha_s C_F}{2\pi}\,,
\end{align}
where here $R_B$ is the radius of the recoiling jet. For simplicity, throughout this paper, we have taken $R_B=R$.

The out-of-jet contribution to the soft function has a pure cusp anomalous dimension  \cite{Ellis:2010rwa}
\begin{align}
\mu\frac{d}{d\mu}\ln S_{oj }(R_B) &= \frac{2\alpha_s C_F}{\pi} \log\left[  \tan^2 \frac{ R}{2}  \right]   \,.
\end{align}

\section{One Loop Calculations of Soft Subjet Functions}\label{sec:softjet_app}
In this appendix we give the operator definitions and one-loop results for the functions appearing in the factorization theorem of \Eq{fact_inclusive_form_1} for the soft subjet region of phase space. The factorization theorem in the soft subjet region of phase space was first presented in \Ref{Larkoski:2015zka}, where all functions were calculated to one-loop, and a detailed discussion of the structure of the required zero bin subtractions was given. This calculation was performed with a broadening axis cone algorithm, however it was argued in \Sec{sec:soft_jet} that to leading power, the factorization theorem is identical in the case of an anti-$k_T$ algorithm. Because of this, in this appendix we give only the final results for the one-loop anomalous dimensions, and the tree level matching for the soft subjet production, as are required for the resummation considered in this paper. The interested reader is referred to \Ref{Larkoski:2015zka} for the detailed calculation, as well as a discussion of the intricate zero bin structure of the factorization theorem, which is only briefly mentioned in this appendix.         

\subsection*{Definitions of Factorized Functions}
The functions appearing in the soft subjet factorization theorem of \Eq{fact_inclusive_form_1} have the following SCET operator definitions:
\begin{itemize}
\item Hard Matching Coefficient for Dijet Production
\begin{equation}
H(Q^2,\mu)=|C(Q^2,\mu)|^2\,,
\end{equation}
where $C\left(Q^2,\mu\right)$ is the Wilson coefficient obtained from matching the full theory QCD current $\bar \psi \Gamma \psi$ onto the SCET dijet operator $\bar \chi_n \Gamma \chi_{\bar n}$
\begin{align}
\langle q\bar q | \bar \psi \Gamma \psi |0\rangle=C\left(Q^2,\mu\right) \langle q\bar q | \mathcal{O}_2 |0\rangle\,.
\end{align}
As before, we have neglected the contraction with the Leptonic tensor.
\item Soft Subjet Jet Function:
{\small\begin{align}
&J_{\sja }\Big(\ecf{3}{\alpha}\Big)=\\
& \hspace{.25cm}
\frac{(2\pi)^3}{C_A}\text{tr}\langle 0|\mathcal{B}_{\perp_{\sja}}^{\mu}(0)\Theta_{O}(B)\delta(Q_{SJ}-\sjabar \cdot{\mathcal P})\delta^{(2)}(\vec{{\mathcal P}}_{\perp_{SJ}})\delta\Big(\ecf{3}{\alpha}-\Theta_{FJ}\ecfop{3}{\alpha}\big|_{SJ}\Big)\,\mathcal{B}_{\perp_{\sja}\mu}(0)|0\rangle \nonumber
\end{align}}
\item Jet Function:
{\small\begin{align}
\hspace{-1cm}
J_{n}\Big(\ecf{3}{\alpha}\Big)&=\frac{(2\pi)^3}{C_F}\text{tr}\langle 0|\frac{\bar{n}\!\!\!\slash}{2}\chi_{n}(0) \Theta_{O}(B)\delta(Q-\bar{n}\cdot{\mathcal P})\delta^{(2)}(\vec{{\mathcal P}}_{\perp})\delta\Big(\ecf{3}{\alpha}-\Theta_{FJ}\ecfop{3}{\alpha}\big|_{HJ}\Big)\bar{\chi}_n(0)|0\rangle
\end{align}}
\item Boundary Soft Function:
\begin{align}
S_{\sja \,\sjabar }\Big(\ecf{3}{\alpha};R\Big)&=\frac{1}{C_{A}}\text{tr}\langle 0|T\{S_{\sja } S_{\sjabar }\} \Theta_{O}(B) \delta\Big(\ecf{3}{\alpha}-\Theta_{FJ}\ecfop{3}{\alpha}\big|_{BS}\Big)\bar{T}\{S_{\sja } S_{\sjabar }\} |0\rangle
\end{align}
\item Soft Subjet Soft Function:
\begin{align}
\hspace{-1cm}S_{\sja \,n\,\bar{n}}\Big(\ecf{3}{\alpha},B;R\Big)&=\text{tr}\langle 0|T\{S_{\sja } S_{n} S_{\bar{n}}\}\Theta_{O}(B)\delta\Big(\ecf{3}{\alpha}-\Theta_{FJ}\ecfop{3}{\alpha}\big|_{S}\Big)\bar{T}\{S_{\sja } S_{n} S_{\bar{n}}\} |0\rangle
\end{align}
\end{itemize}

The definitions of these functions include measurement operators, which when acting on the final state, return the value of a given observable. The operator $\ecfop{3}{\alpha}$ measures the contribution to $\ecf{3}{\alpha}$ from final states, and must be appropriately expanded following the power counting of the sector on which it acts. Expressions for the expansions in the power counting of the different sectors will be given shortly, after kinematic notation has been set up. The operators $\Theta_{FJ}$, and $\Theta_{O}$ constrain the measured radiation to be in the jet or out of the jet, respectively, and will be defined shortly. 

\subsection*{Kinematics and Notation}

For our general kinematic setup, we will denote by $Q$ the center of mass energy of the $e^+e^-$ collisions, so that $Q/2$ is the energy deposited in a hemisphere. i.e. the four-momenta of the two hemispheres are
\begin{align}
p_{\text{hemisphere}_1}=\left( \frac{Q}{2},\vec p_1   \right)\,, \qquad p_{\text{hemisphere}_2}=\left( \frac{Q}{2},-\vec p_1   \right)
\end{align}
so
\begin{align}
s=Q^2\,.
\end{align}

We are now interested in the regime where there is a wide angle soft subjet carrying a small energy fraction, and an energetic subjet, carrying the majority of the energy fraction. We will label the lightcone directions of the energetic subjet by $n,\bar n$, and the lightcone directions of the soft subjet as $n_{sj}, \bar n_{sj}$. We will use the variable $z_{sj}$ to label the energy fraction of the soft subjet, namely
\begin{align}
E_{sj}=z_{sj} \frac{Q}{2}\,, \qquad z_{sj} \ll 1\,.
\end{align}

In this region of phase space, to leading power the value of the two point energy correlation function is set by the two subjets, and is given by
\begin{align}
\ecf{2}{\alpha}= 2^{\alpha/2} z_{sj} \left(  n\cdot n_{sj} \right)^{\alpha/2}   \,.
\end{align}

The action of the measurement function $\ecfop{3}{\alpha}$ on a arbitrary state for each of the factorized sectors contributing to the 3-point energy correlation function measurement is given by
\begin{align}\label{eq:action_measurements}
\ecfop{3}{\alpha}\big|_{SJ}\Big|X_{sj}\Big\rangle&=\sum_{k_i,k_j\in X_{sj}} N_{SJ} \frac{\sjabar\cdot k_i}{Q}\frac{\sjabar\cdot k_j}{Q}\left(\frac{k_i\cdot k_j}{\sjabar\cdot k_i\sjabar\cdot k_j}\right)^{\frac{\alpha}{2}}\Big|X_{sj}\Big\rangle\,,\\
\ecfop{3}{\alpha}\big|_{HJ}\Big|X_{hj}\Big\rangle&=\sum_{k_i,k_j\in X_{hj}} N_{HJ}\frac{\nbar\cdot k_i}{Q}\frac{\nbar\cdot k_j}{Q}\left(\frac{k_i\cdot k_j}{\nbar\cdot k_i\nbar\cdot k_j}\right)^{\frac{\alpha}{2}}\Big|X_{hj}\Big\rangle\,,\\
\ecfop{3}{\alpha}\big|_{BS}\Big|X_{bs}\Big\rangle&=\sum_{k\in X_{bs}}N_{BS} \frac{\sjabar\cdot k}{Q}\left(\frac{\sja\cdot k}{\sjabar\cdot k}\right)^{\frac{\alpha}{2}}\Big|X_{bs}\Big\rangle\,,\\
\ecfop{3}{\alpha}\big|_{S}\Big|X_{s}\Big\rangle&=\sum_{k\in X_{s}}N_S\frac{k^0}{Q}\left(\frac{\sja\cdot k}{k^0}\frac{n\cdot k}{k^0}\right)^{\frac{\alpha}{2}}\Big|X_{s}\Big\rangle\,,
\end{align}
where, for simplicity, we have extracted the normalization factors
\begin{alignat}{2}\label{eq:norm_factors}
N_{SJ}&=2^{5\alpha/2}(n\cdot\sja)^{\alpha}\,, & \qquad
N_{HJ}&=2^{5\alpha/2}z_{sj}(n\cdot\sja)^{\alpha}\,,\\
N_{BS}&=2^{2\alpha}z_{sj}(n\cdot\sja)^{\alpha}\,,& \qquad
N_{S}&=2^{1+3\alpha/2} z_{sj}(n\cdot\sja)^{\alpha/2}\,.
\end{alignat}%

These expressions follow from properly expanding the definition of the energy correlation function measurements in the power counting of each of the sectors. Note that on the jet sectors, the 3-point correlation measurement becomes an effective 2-point correlation measurement, since the 2-point energy correlation function is set by the initial splitting of the subjet.

The in-jet restriction, $\Theta_{FJ}$, is given by
\begin{align}
\Theta_{FJ}(k)&=\Theta\left(\tan ^2\frac{R}{2}-\frac{n\cdot k}{\bar{n}\cdot k}\right)\,.
\end{align}
The jet restriction must also be expanded following the power counting of the given sector. We will see that this is actually quite subtle for the soft subjet modes, since the angle between the soft subjet axis and the boundary of the jet has a non-trivial power counting. In particular, the expansion of $\Theta_{FJ}(k)$ is different for the soft subjet jet and boundary soft modes, and will demonstrate the necessity of performing the complete factorization of the soft subjet dynamics into jet and boundary soft modes. Finally, since we are considering the case where the out-of-jet scale $B$ is much less than the in-jet scale, the operator
$$
\Theta_{O}(B)
$$
must also be included in the definition of the soft subjet functions. This operators vetoes out-of-jet radiation above the scale $B$. The explicit expression for $\Theta_{O}(B)$ expanded in the power counting of each of the factorized sectors can be found in \Ref{Larkoski:2015zka}.

\subsection*{Hard Matching Coefficient for Dijet Production}

The hard matching coefficient for dijet production, $H(Q^2,\mu)$, is identical to that for the collinear subjets factorization theorem by hard-collinear-soft factorization,  and is given in
\Eq{eq:hard_matching_coeff}.


\subsection*{Hard Matching for Soft Subjet Production}
The hard matching coefficient $H^{sj}(\sje,\sjtheta)$ is determined by the finite parts of the logarithm of the soft matrix element for a single soft state
\begin{align}
H^{sj}(\sje,\sja)&=\text{tr}\langle 0|T\{S_nS_{\bar{n}}\}|sj \rangle\langle sj|\bar{T}\{S_nS_{\bar{n}}\}|0\rangle_{\text{fin}}\,.
\end{align}
The virtual corrections of the effective theory cancel the IR divergences of this matrix element, giving a finite matching coefficient. This matrix element can be calculated from the square of the soft gluon current \cite{Berends:1988zn,Catani:2000pi}, which is known to two loop order \cite{Duhr:2013msa,Li:2013lsa}. The tree level hard matching coefficient for the soft subjet production is given by
\begin{align}
H^{sj(\text{tree})}_{n\bar{n}}(\sje,\sja)&=\frac{\alpha_s C_F}{\pi\sje}\frac{n\cdot \bar{n}}{n\cdot\sja\,\sja\cdot\bar{n}}\,.
\end{align}
The results of \Ref{Catani:2000pi} can be used to determine the soft subjet production matching from an arbitrary number of hard jets at one loop.

\subsection*{Anomalous Dimensions}\label{app:Anom_Dim}

In this section we collect the one-loop anomalous dimensions for all the functions calculated in this appendix. The two hard functions satisfy multiplicative renormalization group equations. For the dijet production hard function, we have
\begin{equation}
\mu \frac{d}{d\mu}\ln H(Q^2,\mu)=\gamma_H (Q^2,\mu)=2\text{Re}\left[ \gamma_C(Q^2,\mu)\right]\,.
\end{equation}
Explicitly
\begin{align}
\gamma_H (Q^2,\mu)=\frac{\alpha_s C_F}{2\pi}\left(  4\log\left[  \frac{Q^2}{\mu^2} \right] -6  \right )\,.
\end{align}

For the soft subjet production hard function, we have
\begin{equation}
\mu\frac{d}{d\mu}\ln H^{sj}_{n\bar{n}}(\sje,\sja,\mu) =-\frac{\alpha_s C_A}{\pi}\ln \Bigg[\frac{2\mu^2\bar{n}\cdot n}{Q^2 z_{sj}^2 n\cdot\sja\,\sja\cdot\bar{n}}\Bigg]-\frac{\alpha_s}{2\pi}\beta_0\,.
\end{equation}

The jet, boundary soft, and global soft functions satisfy multiplicative renormalization group equations in Laplace space, where the Laplace conjugate variable to $\ecf{3}{\alpha}$ will be denoted $\eeclp{3}{\alpha}$.

The jet function for the soft subjet satisfies the RGE
\begin{align}
\mu\frac{d}{d\mu}\ln J_{\sja }\Big(\eeclp{3}{\alpha}\Big) &=-4 \frac{\alpha_s C_A}{2\pi (1-\alpha)} \log\left[  2^{-\alpha/2} \eeclp{3}{\alpha}   e^{\gamma_E}   z_{sj}^2 \left( \frac{\mu}{z_{sj} Q} \right)^\alpha N_{SJ}  \right]+\frac{\alpha_s}{2\pi}\beta_0\,,
\end{align}
where the normalization factor $N_{SJ}$ was defined in  \Eq{eq:norm_factors}. We have assumed that the soft subjet is a gluon jet, as it is this case that exhibits the soft singularity of QCD.

The jet function for the hard subjet, which we have assumed to be a quark jet, satisfies the RGE
\begin{align}
\mu\frac{d}{d\mu}\ln J_{hj }\Big(\eeclp{3}{\alpha}\Big) &=-4 \frac{\alpha_s C_F}{2\pi (1-\alpha)} \log\left[  2^{-\alpha/2} \eeclp{3}{\alpha}   e^{\gamma_E}   \left( \frac{\mu}{ Q} \right)^\alpha N_{HJ}  \right]+\frac{3\alpha_s C_F}{2\pi}\,.
\end{align}
where the normalization factor $N_{HJ}$ was defined in  \Eq{eq:norm_factors}.

Since the soft subjet factorization theorem is sensitive to the boundary of the jet, it is also necessary to include out-of-jet contributions. We assume that nothing is measured on the recoiling jet. The out-of-jet jet function is then given by the unmeasured jet function of \Ref{Ellis:2010rwa}
\begin{align}
\mu\frac{d}{d\mu}\ln J_{oj }(R_B) &= \frac{2\alpha_s C_F}{\pi} \log\left[    \frac{\mu}{ Q \tan \frac{ R_B}{2} }   \right]+\frac{3\alpha_s C_F}{2\pi}\,,
\end{align}
where here $R_B$ is the radius of the recoiling jet. For simplicity, throughout this paper, we have taken $R_B=R$.

The boundary soft function, satisfies the RGE
\begin{align}
\mu\frac{d}{d\mu}\ln  S_{\sja \,\sjabar }\Big(\eeclp{3}{\alpha};R\Big)&=\frac{\alpha_s C_A}{\pi (1-\alpha)} \log \left[  \frac{\mu}{Q}   \eeclp{3}{\alpha} e^{\gamma_E} 2^{1-\alpha}N_{BS}     \left(  \frac{\bar n \cdot \sja}{n \cdot \sja}   \tan^4 \frac{R}{2}    \right)^{\frac{-(1-\alpha)}{2}}   \right. \nonumber\\
&\hspace{4cm}    \left.     \left(   1-   \frac{n \cdot \sja}{  \bar n \cdot \sja  \tan^2 \frac{R}{2}  }     \right)^{-(1-\alpha)}      \right]\,.
\end{align}
where the normalization factor $N_{BS}$ was defined in  \Eq{eq:norm_factors}.

For the soft function, it is necessary to perform a refactorization into in-jet and out-of-jet contributions along the lines of \Ref{Ellis:2010rwa}. This is particularly important in the present case, since as was discussed in detail in \Ref{Larkoski:2015zka}, the out-of-jet contribution to the soft function is sensitive to the large logarithm, $\log\left[  \tan^2 \frac{R}{2} -   \tan^2 \frac{\theta_{sj} }{2}   \right]$, but due to zero bin subtractions, the in-jet contribution to the soft function does not exhibit such a sensitivity.

The in-jet anomalous dimension has both $C_A$ and $C_F$ contributions. It is given by
\begin{align}
\gamma_{GS}^{(\text{in})}=& -\left(   \frac{C_A}{2}-C_F  \right)       \left(  \frac{2 \alpha_s}{\pi (1-\alpha)}  \log[T]  -\frac{2\alpha_s}{\pi}  \log\left[       \frac{2\tan \frac{R}{2}}{  \tan \frac{\theta_{sj} }{2}  } \right]           \right) \nonumber\\
&-\left(   \frac{-C_A}{2} \right)       \left(  \frac{3 \alpha_s}{\pi (1-\alpha)}  \log[T]  +\frac{2\alpha_s}{\pi}  \log\left[    \frac{1}{2\bar n \cdot \sja}   \frac{\tan \frac{R}{2}}{  \tan \frac{\theta_{sj} }{2}  } \right]     \right)\nonumber \\
&-\left(   \frac{-C_A}{2} \right)       \left(  \frac{ \alpha_s}{\pi (1-\alpha)}  \log[T]  -\frac{2\alpha_s}{\pi}  \log\left[       \bar n \cdot \sja \right]   \right) \\
&=(C_A+2C_F)\frac{\alpha_s}{\pi (1-\alpha)} \log[T]+\frac{C_A \alpha_s}{\pi} \log \left[    \frac{   \tan^2 \frac{R}{2}     }{  (\bar n \cdot \sja)^2 \tan^2 \frac{\theta_{sj} }{2}        }     \right] \nonumber\\
&\hspace{4cm}- \frac{2C_F \alpha_s}{\pi}   \log \left[  \frac{2\tan\frac{R}{2}}{   \tan \frac{\theta_{sj} }{2}  }   \right]\,,
\end{align}
where in the first equality we have separated the contributions from a gluon between the three different Wilson lines, and to simplify the expression we have extracted the argument of the logs
\begin{align}
T&= e^{\gamma_E} N_S\frac{ \eeclp{3}{\alpha}\, \mu }{Q \,\text{tan}^{1-\alpha }\frac{\sjtheta}{2}}\Big(\frac{n\cdot \sja}{2}\Big)^{\alpha/2}\,.
\end{align}
We choose the canonical scale for the in-jet soft function by minimizing the arguments of the $C_A$ log. Namely, we rewrite the anomalous dimension as
\begin{align}
\gamma_{GS}^{(\text{in})}=&  \frac{   (C_A +2C_F) \alpha_s    }{\pi (1-\alpha)   }      \log \left[   T  \left( \frac{    \tan \frac{R}{2}    }{      (\bar n \cdot \sja) \tan \frac{\theta_{sj} }{2}    }  \right)^{2(1-\alpha)} \right]    \nonumber \\
&\hspace{2cm} + \frac{  2C_F \alpha_s     }{ \pi(1-\alpha)        }     \log \left[   \left(  \frac{    \tan \frac{\theta_{sj} }{2}   } { 2 \tan \frac{R}{2}    }   \right)^{1-\alpha}       \left(  \frac{  \tan \frac{R}{2}    } {  (\bar n \cdot \sja) \tan \frac{\theta_{sj} }{2}    }   \right)^{-2(1-\alpha)}   \right]\,.
\end{align}
The argument of the second logarithm is formally an $\mathcal{O}(1)$ number in the soft subjet region of phase space, and is treated as the non-cusp anomalous dimension. The argument of the first logarithm is used to set the scale.

The out-of-jet anomalous dimension is purely non-cusp, and is given by
\begin{align}
\gamma_{GS}^{(\text{out})}=&  -\left(   \frac{C_A}{2}-C_F  \right)\frac{2\alpha_s}{\pi} \log  \left[  \tan \frac{R}{2} \tan \frac{R_B }{2}         \right]    \\
&\hspace{0cm} -\frac{\alpha_s C_A}{2\pi}  \log \left[ \frac{ \tan^2 \frac{R}{2}        }{  \tan^2 \frac{R}{2}  -    \tan^2 \frac{\theta_{sj} }{2}      }    \right]         -\frac{\alpha_s C_A}{2\pi}  \log \left[   \frac{ 1            }{    \tan^2 \frac{R_B}{2}    \left(   \tan^2 \frac{R}{2} -   \tan^2 \frac{\theta_{sj} }{2}     \right)           }     \right]\,.   \nonumber
\end{align}
The natural scale for the out-of-jet soft function is
\begin{align}
\mu_{\text{out}}=\frac{2 n\cdot \sja B}{\tan \frac{\theta_{sj} }{2}}\,,
\end{align}
where $B$ is the out-of-jet scale. We set $B=Q\left( \ecf{2}{\alpha}\right)^2$ as discussed in \Sec{sec:soft_jet}.

For consistency of our soft subjet factorization theorem, the sum of the anomalous dimensions listed above should cancel.  Indeed, one can explicitly check that the anomalous dimensions satisfy the consistency condition
\begin{align}
&\mu \frac{d}{d\mu}\ln H(Q^2,\mu)+\mu\frac{d}{d\mu}\ln H^{sj}_{n\bar{n}}(\sje,\sja,\mu)+\mu\frac{d}{d\mu}\ln J_{\sja }\Big(\eeclp{3}{\alpha}\Big)  +    \mu\frac{d}{d\mu}\ln J_{hj }\Big(\eeclp{3}{\alpha}\Big)    \nonumber \\
&+\mu\frac{d}{d\mu}\ln J_{oj }(R_B)    +\mu\frac{d}{d\mu}\ln  S_{\sja \,\sjabar }\Big(\eeclp{3}{\alpha};R\Big)+\mu\frac{d}{d\mu}\ln  S_{\sja \,n\,\bar{n}}\Big(\eeclp{3}{\alpha},B;R, R_B\Big) = 0 \,.
\end{align}
This cancellation is highly non-trivial, involving intricate cancellations between a large number of scales, providing support for the structure of our factorization theorem at the one-loop level. Some further details on the structure of the cancellations, particularly on the dependence of the angle between the soft subjet axis and the boundary, are discussed in \Ref{Larkoski:2015zka}.


\section{Soft Subjet Collinear Zero Bin}\label{sec:soft_subjet_cbin}

In this appendix we summarize the one-loop anomalous dimensions, and required tree level matrix elements for the calculation of the collinear zero bin of the soft subjet factorization theorem, which is required to interpolate between the collinear subjets and soft subjets factorization theorem. Although all the ingredients in this appendix can be obtained straightforwardly from \App{app:Anom_Dim} using the standard zero bin procedure \cite{Manohar:2006nz}, we explicitly summarize the results here for completeness. 

To perform the zero-bin, all anomalous dimensions and matrix elements of the soft subjet factorization theorem are written in terms of $\ecf{2}{\alpha}$ and $z_{sj}$, and then the limit
\begin{align}
\frac{\ecf{2}{\alpha}}{z_{sj}} \to 0
\end{align}
is taken. We will therefore write the anomalous dimensions and matrix elements in this section in terms of $\ecf{2}{\alpha}$, $z_{sj}$, and $\eeclp{3}{\alpha}$. To keep the notation as simple as possible, we will use only a tilde to denote a collinear zero binned matrix element or anomalous dimension, e.g. $\gamma_{GS}^{(\text{in})} \to \tilde \gamma_{GS}^{(\text{in})}$.

\subsection*{Hard Matching for Soft Subjet Production}

The collinear binned hard matching coefficient for soft subjet production is given at tree level by
\begin{align}
\tilde H^{sj(\text{tree})}_{n\bar{n}}(\sje, \ecf{2}{\alpha})&=    \frac{\alpha_s   C_F}{\pi} \frac{2}{\alpha}       \frac{1   }{  \sje \ecf{2}{\alpha}   }\,.
\end{align}

\subsection*{Anomalous Dimensions}
Since the renormalization group evolution of all functions in the zero bin is identical to in the soft subjet factorization theorem, here we simply list the results for the zero binned one-loop anomalous dimensions:
\begin{align}
 \gamma_{ \tilde H}     &=\frac{\alpha_s C_F}{2\pi}\left(  4\log\left[  \frac{Q^2}{\mu^2} \right] -6  \right )   \,, \\
 \gamma_{\tilde H^{sj}_{n\bar{n}}   } &= -\frac{  2C_A \alpha_s  }{   \pi } \log \left[ \frac{   2\mu      } {       Q      z_{sj}^{1-1/\alpha}  \left(      \ecf{2}{\alpha}  \right)^{1/\alpha}    }     \right]   -\frac{\alpha_s}{2\pi} \beta_0   \,, \\
 \gamma_{\tilde J_{oj}}&=   \frac{2\alpha_s C_F}{\pi} \log\left[    \frac{\mu}{ Q \tan \frac{ R_B}{2} }   \right]+\frac{3\alpha_s C_F}{2\pi}  \,, \\
\tilde \gamma_{GS}^{(\text{out})}&=  \frac{C_A \alpha_s}{2 \pi}  \log\left[  \tan^2 \frac{R}{2} \tan^2 \frac{R_B}{2}  \right]  -\frac{\left(   \frac{C_A}{2}-C_F \right)   \alpha_s}{\pi} \log\left[  \tan^2\frac{R}{2} \tan^2 \frac{R_B}{2}  \right]  \,, \\
 \gamma_{\tilde J_{hj}}&=  -4 \frac{\alpha_s C_F}{2\pi (1-\alpha)} \log\left[  2^{-\alpha/2} \eeclp{3}{\alpha}   e^{\gamma_E}   \left( \frac{\mu}{ Q} \right)^\alpha N_{HJ}  \right]+\frac{3\alpha_s C_F}{2\pi} \,, \\
 \gamma_{\tilde J_{sj}}&= -4 \frac{\alpha_s C_A}{2\pi (1-\alpha)} \log\left[  2^{-\alpha/2} \eeclp{3}{\alpha}   e^{\gamma_E}   z_{sj}^2 \left( \frac{\mu}{z_{sj} Q} \right)^\alpha N_{SJ}  \right]+\frac{\alpha_s}{2\pi}\beta_0  \,, \\
 \tilde \gamma_{GS}^{(\text{in})}&= \frac{C_F \alpha_s}{\pi (1-\alpha)} \log \left[ 2^{1-\alpha} \tan^{-3(1-\alpha)} \frac{R}{2} \left(  \frac{2^{-\alpha } \ecf{2}{\alpha}  }{z_{sj}}  \right)^{-3+3/\alpha}  \right]  \nonumber \\
 & +     \frac{(C_A+2C_F) \alpha_s}{\pi (1-\alpha)} \log \left[  \frac{2^{-1+4\alpha}   \mu \eeclp{3}{\alpha}   e^{\gamma_E}  z_{sj}   }{  Q }  \tan^{2(1-\alpha)} \frac{R}{2}   \left(  \frac{2^{-\alpha } \ecf{2}{\alpha}  }{z_{sj}}  \right)^{5-3/\alpha}        \right]   \,, \\
 \gamma_{ \tilde S_{\sja \,\sjabar }}&=  \frac{C_A \alpha_s}{\pi(1-\alpha)}     \log  \left[ \frac{\mu \eeclp{3}{\alpha}   e^{\gamma_E} 2^{1+2\alpha}  \tan^{2(\alpha-1)} \frac{R}{2}    }{Q} \left(  \frac{2^{-\alpha } \ecf{2}{\alpha}  }{z_{sj}}  \right)^{1+1/\alpha}       \right] \,.
\end{align}

As for the soft subjet anomalous dimensions, one can check that the zero binned anomalous dimensions satisfy the consistency relation
\begin{align}
&\mu \frac{d}{d\mu}\ln \tilde H(Q^2,\mu)+\mu\frac{d}{d\mu}\ln  \tilde H^{sj}_{n\bar{n}}(\sje,\sja,\mu)+\mu\frac{d}{d\mu}\ln  \tilde J_{\sja }\Big(\eeclp{3}{\alpha}\Big)  +    \mu\frac{d}{d\mu}\ln  \tilde J_{hj }\Big(\eeclp{3}{\alpha}\Big)    \nonumber \\
&+\mu\frac{d}{d\mu}\ln  \tilde J_{oj }(R_B)    +\mu\frac{d}{d\mu}\ln   \tilde S_{\sja \,\sjabar }\Big(\eeclp{3}{\alpha};R\Big)+\mu\frac{d}{d\mu}\ln   \tilde S_{\sja \,n\,\bar{n}}\Big(\eeclp{3}{\alpha},B;R, R_B\Big) = 0 \,,
\end{align}
as required for the consistency of the factorization theorem.


\section{One Loop Calculations of Signal Factorization Theorem}\label{sec:signal_app}

In this section we give the operator definitions, and one-loop results for the functions appearing in the factorization theorem of \Eq{eq:signal_fact} for the signal contribution from $Z\to q\bar q$. These are formulated in the SCET$_+$ effective theory of \Ref{Bauer:2011uc}, in an attempt to have a consistent approach to factorization for both the signal and background distributions. In the collinear subjets region of phase space the two are identical (including identical power counting for the modes) up to the absence of global soft modes for the signal distribution. Alternatively, the factorization theorem for the signal region can be formulated by boosting the factorization theorems for appropriately chosen $e^+e^-$ event shapes, as was considered in \Ref{Feige:2012vc}. While this approach is less in the spirit of developing effective field theory descriptions for jet substructure that was pursued in this paper, it has the potential advantage of being easily able to relate to higher order known results for event shapes.

\subsection*{Definitions of Factorized Functions}

The functions appearing in the collinear subjets factorization theorem of \Eq{eq:NINJA_fact} have the following SCET operator definitions:
\begin{itemize}
\item Hard Matching Coefficient:
\begin{align}
H_Z\left(Q^2\right)=\left |C_Z\left(Q^2\right) \right|^2\,,
\end{align}
where $C_Z\left(Q^2\right)$ is the matrix element for the process $e^+e^- \to ZZ$, and also includes the leptonic decay of one of the $Z$ bosons. Since we use the narrow width approximation, flat polarization distributions for the $Z$, and normalize our distributions to unity, it will play no role in our calculation.
\item Jet Functions:
\begin{align}
J_{n_{a,b}}\Big(\ecf{3}{\alpha}\Big)&=\\
&\hspace{-1cm}\frac{(2\pi)^3}{C_F}\text{tr}\langle 0|\frac{\bar{n}\!\!\!\slash _{a,b}}{2}\chi_{n_{a,b}}(0) \delta(Q-\bar{n}_{a,b}\cdot{\mathcal P})\delta^{(2)}(\vec{{\mathcal P}}_{\perp})\delta\Big(\ecf{3}{\alpha}-  \ecfop{3}{\alpha} \Big)\bar{\chi}_{n_{a,b}}(0)|0\rangle \nonumber
\end{align}
\item Collinear-Soft Function:
\begin{align}
\hspace{-1cm} S_{c, \,n_a n_b} \Big(\ecf{3}{\alpha}\Big)&=\text{tr}\langle 0|T\{ S_{n_a } S_{n_b} \}\delta\Big(\ecf{3}{\alpha}-\ecfop{3}{\alpha}\Big)\bar{T}\{ S_{n_a } S_{n_b} \} |0\rangle
\end{align}
\end{itemize}
As in \App{sec:ninja_app} and \App{sec:softjet_app}, the operator, $\ecfop{3}{\alpha}$, measures the contribution to $\ecf{3}{\alpha}$ from final states, and must be appropriately expanded following the power counting of the sector on which it acts. Since the power counting is identical as for the collinear subjets factorization theorem, the expansions are given in \Eq{eq:collinear_limits_eec3}, and \Eq{eq:collinear_soft_limits_eec3}. In the collinear subjets region that we consider for the signal, all modes are boosted, and so there is no dependence on the jet algorithm at leading power.

\subsection*{Hard Matching Coefficient}

The hard matching coefficient for the process $e^+e^-\to ZZ$, with one $Z$ decaying leptonically, $H_Z(Q^2)$, does not carry an SCET anomalous dimension (hence we have dropped the $\mu$ dependence), as it is colorless. Because we work in the narrow width approximation, at a fixed $Q^2$, and consider only normalized distributions, it is therefore irrelevant to our discussion.

\subsection*{Matrix Element for $Z \to q\bar q$ Decay}
The anomalous dimension for the $Z\to q\bar q$ splitting function appearing in the factorization theorem of \Eq{eq:NINJA_fact} is the same as that for the SCET quark bilinear operator, which was given in \Eq{eq:hard_anom_dim}, but evaluated at the appropriately boosted scale. 

For simplicity, in this paper we do not account for spin correlations, and assume a flat profile in the polarization of the $Z$ boson. The tree level $Z\to q\bar q$ matrix element is well known and first calculated in \Ref{Altarelli:1979ub}. The full matrix element is known to two loops \cite{Matsuura:1988sm}.

The anomalous dimension depends only on the color structure, and is therefore the same as the anomalous dimension for the hard matrix element for $e^+e^-\to q\bar q$, namely
\begin{align}
\gamma_{H_Z}=1+\frac{\alpha_s C_F}{2\pi} \left (  -8 +\frac{\pi^2}{6} -\log^2 \left[  \frac{\mu_H^2}{\mu^2}  \right]     +3 \log \left[  \frac{\mu_H^2}{\mu^2}  \right]  \right)\,.
\end{align}
Here $\mu_H$ is the scale of the splitting. It is essential for the cancellation of anomalous dimensions that the scale $\mu_H$ is equal to the invariant mass of the jet. In terms of the energy correlation functions, this is given by
\begin{align}
m_J^2&=\frac{Q^2 \left[ z (1-z) \right]^{1-2/\alpha}  \left(   \ecf{2}{\alpha} \right)^{2/\alpha}    }{4} \nonumber \\
&=\frac{Q^2 z (1-z) n_a \cdot n_b   }{2}\,.
\end{align}
The necessity for the appearance of the jet mass as the scale in the anomalous dimension is due to the fact that it is a Lorentz invariant quantity, and as has been discussed in \Ref{Feige:2012vc}, the factorization theorem for the case of the boosted boson can be obtained by boosting an $e^+e^-$ event shape, where it is of course known that the scale $Q^2$ of the off-shell $Z$, or $\gamma$ is the scale appearing in the hard anomalous dimension.

\subsection*{Jet Functions}
The jet functions appearing in the signal factorization theorem are identical to the quark (and antiquark) jet functions calculated in \App{sec:NINJA_jet_calc} for the collinear subjets region of phase space.  This is because the power counting is identical in the two cases and the jet functions are only sensitive to the color of the jet that they describe. Therefore we do not repeat them here.

\subsection*{Collinear-Soft Function}
The power counting for the signal is identical to the power counting for the collinear subjet region for the QCD background. However, the collinear-soft function contains only Wilson lines along the collinear subjet directions. The collinear-soft function for the QCD background was calculated in pairs of dipoles in \App{sec:csoft_calc}, and therefore the contribution from a collinear-soft exchange between the $n_a$ and $n_b$ Wilson lines can simply be extracted from that calculation. The result for this contribution is given by
 \begin{align}
S_{c,\,n_a n_b}^{(1)}(\ecflp{3}{\alpha})&=  -g^2 \Gamma(-2\epsilon)     \left(    \frac{  \mu^2 N_{CS}^2 e^{\gamma_E}  (\ecflp{3}{\alpha})^2      \left( \frac{n_a \cdot n_b}{8}   \right)^{-1+2\alpha}  }{4\pi 4^{1-\alpha}  Q^2    }  \right)^{\epsilon}   16 c_\epsilon     \frac{\Gamma[1/2-\epsilon] \Gamma[1/2]}{\Gamma[1-\epsilon]} \nonumber  \\
& \hspace{-1cm}   \left(  \frac{1}{(2 \alpha -2) \epsilon }+\frac{\alpha 
   \log (2)}{\alpha -1}+\log (2)+\frac{\epsilon  \left(-\pi ^2 \alpha ^2+36 \alpha ^2 \log ^2(2)+3 \pi ^2 \alpha -24 \alpha 
   \log ^2(2)-2 \pi ^2\right)}{12 (\alpha -1)}      \right)     \,,
\end{align}
where we recall that the normalization factor is given by
 \begin{align}
N_{CS}&=2^{3\alpha/2+1}z_az_b\left(n_a\cdot n_b\right)^{\alpha/2}\,,
\end{align}
as defined in \Eq{eq:ncs}. Also note that we have factored out the color generators, so that the collinear-soft function is defined as
\begin{align}
S^{(1)}_{c}\left(\ecf{3}{\alpha}\right)&=\frac{1}{2}\sum_{i\neq j}\mathbf{T}_i\cdot\mathbf{T}_j S_{c,\,ij}^{(1)} \left(\ecf{3}{\alpha}\right)\,,
\end{align}
which is the generic form of the collinear-soft (or soft) function to one-loop.

Expanding in $\epsilon$, and keeping only the divergent piece, as relevant for the anomalous dimensions, we find 
 \begin{align} 
\tilde{S}_{c,\,n_a n_b}^{(1)\text{div}}(\ecflp{3}{\alpha})&= \frac{\alpha_s}{\pi}\frac{1}{  (\alpha -1) \epsilon ^2}+2\frac{\alpha_s}{\pi}\frac{ \left(2 \alpha  \log (2)+ \log \left[   \frac{  \mu N_{CS} e^{\gamma_E}  (\ecflp{3}{\alpha})      \left( \frac{n_a \cdot n_b}{8}   \right)^{-1/2+\alpha}  }{2^{1-\alpha}  Q    }  \right]- \log (2)\right)}{   (\alpha -1) \epsilon }\nonumber \\
&=\frac{\alpha_s}{\pi}\frac{1}{  (\alpha -1) \epsilon ^2}+2\frac{\alpha_s}{\pi}\frac{L_\alpha^{cs}}{   (\alpha -1) \epsilon }\,,
\end{align}
where
\begin{align}
L_\alpha^{cs}= \log \left(    \frac{  \mu N_{CS} e^{\gamma_E}  (\ecflp{3}{\alpha})      \left( n_a \cdot n_b   \right)^{-1/2+\alpha}  }{\sqrt{2}  Q    }  \right)\,.
\end{align}
Since there is no global-soft function the cancellation of anomalous dimensions, to be discussed shortly, requires that only $1/(1-\alpha)$ contributions appear in the collinear soft function, as is observed.

\subsection*{Cancellation of Anomalous Dimensions}\label{sec:signal_cancel_anom}

It is also interesting to explicitly check the cancellation of anomalous dimensions for the signal factorization theorem as formulated in SCET$_+$ to further confirm the cancellation mechanism which took place for the background distribution. The functions appearing in the signal factorization theorem obey identical evolution equations to those for the background distribution, which were explicitly given in \App{sec:bkg_cancel_anomdim}, so we do not repeat them here.

The independence of the total cross section under renormalization group evolution implies the following relation between anomalous dimensions
 \begin{align} \label{eq:consistency_signal}
\gamma_{H_Z}  +\gamma_{q}^{\alpha}\left(\ecflp{3}{\alpha}\right)+\gamma_{\bar q}^{\alpha}\left(\ecflp{3}{\alpha}\right)+\gamma_{cs}\left(\ecflp{3}{\alpha}\right)= 0\,.
\end{align}
Here $\gamma_{H_Z} $ is the anomalous dimension of the $Z\to q\bar q$ matrix element,  $\gamma_{q}^{\alpha}\left(\ecflp{3}{\alpha}\right)$ and $\gamma_{\bar q}^{\alpha}\left(\ecflp{3}{\alpha}\right)$ are the anomalous dimensions of the quark and antiquark jet functions and  $\gamma_{cs}\left(\ecflp{3}{\alpha}\right)$ is the anomalous dimension of the collinear soft function.

For the case of $Z\to q \bar q$, we have the color conservation relation
\begin{align}\label{eq:color_cons_signal}
{\mathbf T}_q+{\mathbf T}_{\bar q}&=0\,.
\end{align}
The explicit values of the relevant Casimirs are
\begin{align}
{\mathbf T}_q\cdot{\mathbf T}_{q}=C_F\,, \qquad {\mathbf T}_{\bar q}\cdot{\mathbf T}_{\bar q}=C_F\,, \qquad {\mathbf T}_q\cdot{\mathbf T}_{\bar q}=-C_F\,,
\end{align}
however, for most of the cancellation of the anomalous dimensions, it will be convenient to work in the abstract color notation.

Substituting the explicit expressions for the anomalous dimensions into the consistency relation of \Eq{eq:consistency_signal}, we find
\begin{align}\label{eq:casimirs_signal}
&\sum_\gamma =\gamma_{H_Z} +  \left[   -    {\mathbf T}_q \cdot{\mathbf T}_{\bar q}  \frac{4\alpha_s L_\alpha^{cs}\left(\ecflp{3}{\alpha}\right)}{\pi (1-\alpha)}     \right] \nonumber \\
&  \hspace{4cm}     -   C_F   \frac{2\alpha_s L_\alpha^{g}\left(  \ecflp{3}{\alpha}  \right)}{\pi (1-\alpha)} +\gamma_q        - C_F   \frac{2\alpha_s L_\alpha^{q}\left(  \ecflp{3}{\alpha}  \right)}{\pi (1-\alpha)} +\gamma_q \,,
\end{align}
where $L_\alpha^{g}\left(  \ecflp{3}{\alpha}  \right)$ and $L_\alpha^{q}\left(  \ecflp{3}{\alpha}  \right)$, were defined in \Eq{eq:jet_anoms}.

As expected, all contributions are collinear in nature, having a $1/(1-\alpha)$ dependence, and using the color conservation relation of \Eq{eq:color_cons_signal} along with the explicit expressions for the Casimirs of \Eq{eq:casimirs_signal}, we immediately see the cancellation of the $\ecflp{3}{\alpha}$ dependence. It is also straightforward to check the cancellation of the remaining dependencies. It is a nice consistency check on the calculation that the cancellation occurs in exactly the same way as for the background cancellation, namely between the $ {\mathbf T}_q \cdot{\mathbf T}_{\bar q}$ contribution and the jet functions. It is important to emphasize that the cancellation only occurs if the scale of the splitting is given by the invariant mass of the jet, as expected from boosting $e^+e^-$ event shapes.

\section{Soft Haze Factorization Theorem}\label{app:softhaze}

For completeness, we list the operator definitions of the functions appearing in the soft haze factorization theorems. We also give the explicit forms of the measurement operators expanded in the appropriate kinematics. 

The quark jet functions are given as:
\begin{align}
J_{n}\Big(\ecf{2}{\alpha}\Big)&= \nonumber\\
&\hspace{-0.5cm}\frac{(2\pi)^3}{C_F}\text{tr}\langle 0|\frac{\bar{n}\!\!\!\slash _{a,b}}{2}\chi_{n_{a,b}}(0) \delta(Q-\bar{n}_{a,b}\cdot{\mathcal P})\delta^{(2)}(\vec{{\mathcal P}}_{\perp})\delta\Big(\ecf{2}{\alpha}-\ecfop{2}{\alpha}\Big)\bar{\chi}_{n_{a,b}}(0)|0\rangle\,.
\end{align}
The gluon jet functions are similarly defined. The soft functions appearing in the factorization theorems \eqref{eq:fact_soft_haze} and \eqref{eq:fact_soft_haze2} are: 
\begin{align}\label{eq:def_soft_function_SH_1}
S_{n \, \bar n }\Big(\ecf{2}{\beta},\ecf{2}{\alpha},\ecf{3}{\alpha};R\Big)&=\frac{1}{C_{A}}\text{tr}\langle 0|T\{S_{n } S_{\bar n }\}  \delta\Big(\ecf{2}{\beta}-\Theta_{R}\ecfop{2}{\beta}\Big)\delta\Big(\ecf{2}{\alpha}-\Theta_{R}\ecfop{2}{\alpha}\Big)\nonumber\\
&\qquad\qquad\delta\Big(\ecf{3}{\alpha}-\Theta_{R}\ecfop{3}{\alpha}\Big)\bar{T}\{S_{n } S_{\bar n }\} |0\rangle\,,\\
\label{eq:def_soft_function_SH_2}
S_{n \, \bar n }\Big(\ecf{2}{\alpha},\ecf{3}{\alpha};R\Big)&=\frac{1}{C_{A}}\text{tr}\langle 0|T\{S_{n } S_{\bar n }\}  \delta\Big(\ecf{2}{\alpha}-\Theta_{R}\ecfop{2}{\alpha}\Big)\delta\Big(\ecf{3}{\alpha}-\Theta_{R}\ecfop{3}{\alpha}\Big)\bar{T}\{S_{n } S_{\bar n }\} |0\rangle\,.
\end{align}

The action of the energy correlation functions on the collinear and soft haze states are given as:
\begin{align}
\ecfop{2}{\alpha}\big|_{C}\Big|X_{n}\Big\rangle&=\sum_{k,p\in X_{n}}\frac{\nbar\cdot k}{Q}\frac{\nbar\cdot p}{Q}\left(\frac{8\,p\cdot k}{\nbar\cdot p\,\nbar\cdot k}\right)^{\frac{\alpha}{2}}\Big|X_{n}\Big\rangle\,,\\
\ecfop{2}{\alpha}\big|_{SH}\Big|X_{s}\Big\rangle&=\sum_{k\in X_{s}}2\frac{k^0}{Q}\left(\frac{2 n\cdot k}{k^0}\right)^{\frac{\alpha}{2}}\Big|X_{s}\Big\rangle\,,\\
\ecfop{3}{\alpha}\big|_{SH}\Big|X_{s}\Big\rangle&=\sum_{k,p\in X_{s}}4\frac{k^0}{Q}\frac{p^0}{Q}\left(\frac{2 n\cdot k}{k^0}\frac{2 n\cdot p}{p^0}\frac{2 p\cdot k}{p^0k^0}\right)^{\frac{\alpha}{2}}\Big|X_{s}\Big\rangle\,.
\end{align}

\section{Summary of Canonical Scales}\label{sec:canonical_merging_scales}

As many of our factorization theorems involve a large number of scales, in this section we summarize for convenience the scales used in the resummation. Unless otherwise indicated, all scales are taken to be the canonical scales of the logarithms appearing in the factorization theorems.

When performing the numerical resummation, we perform the renormalization group evolution in Laplace space, and compute the cumulative distribution. We then perform the scale setting at the level of the cumulative distribution and numerically differentiate to derive the differential $D_2$ spectrum. While this is formally equivalent to scale setting in the differential distribution when working to all orders in perturbation theory, differences between scale setting in the differential and cumulative distribution arise when working to fixed order in perturbation theory \cite{Almeida:2014uva}. We have not investigated the size of the effect that this has on our $D_2$ distributions. We utilized only two loop running of $\alpha_s$, to be consistent with the Monte Carlos, and avoided the Landau pole by freezing out the running coupling at a specific $\mu_{Landau}\sim 1$ GeV.

Throughout this appendix we will use $z_q$ and $z_g$ to denote the energy fractions of the quark and gluon subjets, respectively. For simplicity, we restrict to the case $\alpha=\beta$. Finally, we estimate the soft out-of-jet radiation scale to be:
\begin{align}
B\approx Q\Big(\ecf{2}{\alpha}\Big)^2
\end{align}
This is consistent with the jet algorithm constraint given by \Eq{eq:out_jet_estimate}.

\subsection*{Collinear Subjets}

We take the canonical scales for the functions appearing in the collinear subjets factorization theorem as
\begin{align}
\mu_H&=Q\,, \\
\mu_{H_2}&=\frac{Q \left( \ecf{2}{\alpha} \right)^{1/\alpha}     z_q^{\frac{1}{2}-  \frac{1}{\alpha}}    z_g^{\frac{1}{2}-  \frac{1}{\alpha}}    }{2}\,,\\
\mu_{J_g}&=\frac{ e^{-\gamma_E/\alpha} Q \left( \ecf{2}{\alpha} \right)^{-2/\alpha}    \left( \ecflp{3}{\alpha} \right)^{1/\alpha}           z_q^{ \frac{1}{\alpha}}    z_g   }{2}\,,\\
\mu_{J_q}&=\frac{ e^{-\gamma_E/\alpha} Q \left( \ecf{2}{\alpha} \right)^{-2/\alpha}    \left( \ecflp{3}{\alpha} \right)^{1/\alpha}           z_q    z_g^{ \frac{1}{\alpha}}   }{2}\,,\\
\mu_{CS}&=\frac{ e^{-\gamma_E} Q \left( \ecf{2}{\alpha} \right)^{-3+1/\alpha}     \ecflp{3}{\alpha}           z_q^{ 2-\frac{1}{\alpha}}     z_g^{2- \frac{1}{\alpha}}   }{2} \,,\\
\mu_{S}^{(\text{in})}&= \frac{ 4^{-\alpha} e^{-\gamma_E} Q \left( \ecf{2}{\alpha} \right)^{-1}     \ecflp{3}{\alpha}       \tan^2\frac{R}{2}       }{2} \,, \\
\mu_{S}^{(\text{out})}&=B
\end{align}
where the scales are indexed by the name of the associated function in the factorization theorem.

\subsection*{Soft Subjets}
We take the canonical scales for the functions appearing in the soft subjets factorization theorem as
\begin{align}
\mu_H&=Q\,, \\
\mu_{H_{sj}}&=\frac{   Q  \left(   \ecf{2}{\alpha} \right)^{1/\alpha}  z_{sj}^{(\alpha-1)/\alpha}   \sqrt{4- \left(\ecf{2}{\alpha} \right)^{2/\alpha}  z_{sj}^{-2/\alpha}  }      }{         4           }\,, \\
\mu_{S_{n_{sj} \bar n_{sj}  }}&=   2^{-\alpha}  e^{-\gamma_E}  Q   \tan^{2(1-\alpha)}\frac{R}{2}    \left(\ecf{2}{\alpha}     \right)^{-(1+\alpha)/\alpha}   \eeclp{3}{\alpha}   \left(  z_{sj}  \right)^{1/\alpha}  \nonumber \\
&  \hspace{0.5cm}     \left( 1-\frac{1}{4}  \left(\ecf{2}{\alpha}     \right)^{2/\alpha}  z_{sj}^{-2/\alpha}    \right)^{(1-\alpha)/2} 
 \left(  \frac{ \left( 1+\tan^{2}\frac{R}{2} \right) \left(   \ecf{2}{\alpha} \right)^{2/\alpha}   -4 \tan^{2}\frac{R}{2} z_{sj}^{2/\alpha}    }{  \tan^{2}\frac{R}{2}   \left(  \left(   \ecf{2}{\alpha} \right)^{2/\alpha}   -4z_{sj}^{2/\alpha}   \right)       }   \right)^{1-\alpha}  \,, \\
\mu^{(\text{in})}_{S_{n_{sj} n \bar n  }}&= 2^{-2+\alpha}  e^{-\gamma_E}  Q   \tan^{2(\alpha-1)}\frac{R}{2}    \left(\ecf{2}{\alpha}     \right)^{-5+3/\alpha}   \eeclp{3}{\alpha}   \left(  z_{sj}  \right)^{4-3/\alpha}  \nonumber \\
&\hspace{4cm}     \left( 1-\frac{1}{4}  \left(\ecf{2}{\alpha}     \right)^{2/\alpha}  z_{sj}^{-2/\alpha}    \right)^{(1-\alpha)/2}   \,, \\
\mu^{(\text{out})}_{S_{n_{sj} n \bar n  }}&= \frac{2 n\cdot \sja B}{\tan \frac{\theta_{sj} }{2}}   \,, \\
\mu_{J_{hj}}&=   \frac{ Q e^{-\gamma_E/\alpha}    \left(   \ecf{2}{\alpha} \right)^{-2/\alpha}     \left( \eeclp{3}{\alpha}   \right)^{1/\alpha}   z_{sj}^{1/\alpha}  }{   2     }    \,, \\
\mu_{J_{\bar n}}&= Q \tan \frac{ R_B}{2}   \,, \\
\mu_{J_{n_{sj}}}&=\frac{ Q e^{-\gamma_E/\alpha}    \left(   \ecf{2}{\alpha} \right)^{-2/\alpha}     \left( \eeclp{3}{\alpha}   \right)^{1/\alpha}   z_{sj}  }{   2     }  \,.
\end{align}

\subsection*{Soft Subjet Collinear Zero Bin}

We take the canonical scales for the functions appearing in the collinear zero bin of the soft subjets factorization theorem as
\begin{align}
\mu_{\tilde H}&=Q\,, \\
\mu_{\tilde H_{sj}}&=   \frac{ Q  \left(  \ecf{2}{\alpha} \right)^{1/\alpha}  z_{sj}^{(\alpha-1)/\alpha} }{2}\,, \\
\mu_{\tilde S_{n_{sj} \bar n_{sj}  }}&=  \frac{   2^{-\alpha}Q e^{-\gamma_E} \tan^{2(1-\alpha)} \frac{R}{2} \eeclp{3}{\alpha} z_{sj}^{1/\alpha}       }{    \left(  \ecf{2}{\alpha} \right)^{(1+\alpha)/\alpha}         }   \,, \\
\mu^{(\text{in})}_{S_{n_{sj} n \bar n  }}&=    \frac{   2^{-2+\alpha} Q e^{-\gamma_E} \tan^{2(\alpha-1)} \frac{R}{2} \eeclp{3}{\alpha} z_{sj}^{4-3/\alpha}       }{    \left(  \ecf{2}{\alpha} \right)^{5-3/\alpha}         }   \,, \\
\mu^{(\text{out})}_{\tilde S_{n_{sj} n \bar n  }}&= \frac{2 n\cdot \sja B}{\tan \frac{\theta_{sj} }{2}}    \,, \\
\mu_{\tilde J_{hj}}&= \frac{ e^{-\gamma_E/\alpha}  Q \left(\eeclp{3}{\alpha} \right)^{1/\alpha}  z_{sj}^{1/\alpha}  }{  2   \left( \ecf{2}{\alpha} \right)^{2/\alpha}    } \,, \\
\mu_{\tilde J_{oj}}&= Q \tan \frac{ R_B}{2}   \,, \\
\mu_{\tilde J_{n_{sj}}}&= \frac{ e^{-\gamma_E/\alpha}  Q \left(\eeclp{3}{\alpha} \right)^{1/\alpha}  z_{sj}  }{   2    \left( \ecf{2}{\alpha} \right)^{2/\alpha}    }\,.
\end{align}

\subsection*{Scale Variation}
Here we list all the variations that went into the scale uncertainties of the QCD background calculations. Any common scale between the soft subjet factorization and its collinear bin are always varied together. Hence we will only discuss variations of the soft subjet and collinear subjets. It is important to note that $\mu^{(\text{out})}_{S_{n_{sj} n \bar n  }}$ of the soft subjet is not exactly the same as the $\mu^{(\text{out})}_{S}$ of the collinear factorization. The extra angular factor improves cancellation with the soft subjet collinear zero bin in the collinear region of the phase space. In the soft subjet region, the angular factor becomes an $O(1)$ number. Given the arbitrariness of the out-of-jet scale setting, we included several different schemes.
\begin{itemize}
\item Splitting scales $\mu_{H_2}$ and $\mu_{H_{sj}}$ from half to twice canonical.
\item $\mu_{Landau}$ where the running of the coupling is frozen from $0.5$ GeV to $1.5$ GeV, canonical is $1$ GeV.
\item All in-jet soft scales $\mu^{(\text{in})}_{S_{n_{sj} n \bar n  }}, \mu_{\tilde S_{n_{sj} \bar n_{sj}  }}, \mu_{CS}$, and $\mu_{S}$ from half to twice canonical. This included the scales in the collinear factorization and soft subjet factorization being varied together, and independently.
\item All out-of-jet soft scales $\mu^{(\text{out})}_{S_{n_{sj} n \bar n  }}, \mu_{S}^{(out)}$ from half to twice canonical. This included the scales in the collinear factorization and soft subjet factorization being varied together, and independently.
\item Soft subjet out-of-jet soft scale $\mu^{(\text{out})}_{S_{n_{sj} n \bar n  }}= Q z_{sj}^2$ from half to twice canonical. Also in this scheme the splitting scales were varied from half to twice canonical, and $\mu_{Landau}$ from $0.5$ GeV to $1.5$ GeV.
\item Soft subjet out-of-jet soft scale $\mu^{(\text{out})}_{S_{n_{sj} n \bar n  }}= \mu_{S}^{(out)}$ from half to twice canonical. Also in this scheme the splitting scales were varied from half to twice canonical, and $\mu_{Landau}$ from $0.5$ GeV to $1.5$ GeV.
\end{itemize}

The final uncertainty bands were taken as the envolope of these variations. Though these variations do not cover all perturbative functions that can be varied, we believe that they are representative of NLL uncertainties.


\section{Renormalization Group Evolution of the Shape Function}\label{app:shape_RGE}

In this appendix we briefly summarize some of the properties of the non-perturbative shape function used in the analysis of the $D_2$ observable, including hadron mass effects, so as to ensure that the level of renormalization group evolution of the parameter $\Omega_D$ is consistent with our results at both $1$ TeV and $91$ GeV, as discussed in \Secs{sec:Hadronization}{sec:LEP}, respectively. There we found that the value of $\Omega_D$ was approximately equal at the two energies, to within our uncertainties. As in the text, we assume that the dominant non-perturbative corrections arise from the global soft modes of the collinear subjets factorization theorem, so that we are working with a soft function with Wilson lines only along the $n$ and $\bar n$ directions. We follow closely the formalism originally developed in \Ref{Mateu:2012nk}.

In \Ref{Mateu:2012nk} it was shown that for dijet observables which can be written in terms of the rapidity $y$ and the transverse velocity $r$, defined as
\begin{align}
r=\frac{p_\perp}{\sqrt{p_\perp^2 +m_H^2  }}\,,
\end{align}
where $m_H$ is a light hadron mass, have a leading power correction that is universal, for event shapes with the same $r$ dependence. Furthermore, the leading power corrections can be written as an integral over an $r$ dependent power correction,
\begin{align}\label{eq:omega_int}
\Omega_D=\int\limits_0^1 dr g(r)\, \Omega_D(r),
\end{align}
where $g(r)$ is a function of $r$ which depends only on the definition of the event shape (see \Ref{Mateu:2012nk}), and $\Omega_D(r)$ exhibits a multiplicative renormalization group evolution in $r$, which is independent of $y$. In particular, for $\Omega_D$, we have
\begin{align}
\mu \frac{d}{d\mu} \Omega_D (r,\mu) =\gamma_{\Omega_D}(r,\mu)  \Omega_D (r,\mu)=\left( -\frac{\alpha_s C_A}{\pi} \log(1-r^2)\Omega_D (r,\mu) \right)\,,
\end{align}
to one loop accuracy \cite{Mateu:2012nk}. This renormalization group equation can be solved exactly for each $r$, however, the computation of $\Omega_D$ using \Eq{eq:omega_int} requires knowledge of the exact $r$ dependence of $\Omega_D (r,\mu)$ at a particular scale. However, it was shown that to order $\alpha_s$, only a single non-perturbative parameter is required to described the evolution, so that one can write
\begin{align}
\Omega_D (\mu) =\Omega_D(\mu_0)   +   \frac{\alpha_s (\mu_0) C_A}{\pi}   \log \left(    \frac{\mu}{\mu_0} \right)  \Omega_D^{\text{ln}}(\mu_0)\,,
\end{align}
where apart from the non-perturbative parameter $\Omega_D(\mu_0)$ evaluated at a particular scale, we have also had to introduce the non-perturbative parameter $\Omega_D^{\text{ln}}(\mu_0)$, which captures the logarithmic running (hence the notation).

The additional non-perturbative parameter $\Omega_D^{\text{ln}}(\mu_0)$ is not well constrained in the literature, and therefore as a simple estimate to make sure that the values used for $\Omega_D$ at both LEP energies and at $1$ TeV are consistent, we consider the estimate $\Omega_D^{\text{ln}}(\mu_0)=\Omega_D(\mu_0)$. Making this approximation, we find the difference between the values of $\Omega_D$ as relevant for LEP and our $1$ TeV analysis to differ by $\lesssim 0.1$, with the value at LEP being lower. This is small compared to our uncertainties, and compared to the scaling in the shift of the first moment with $E_J$ and $m_J$. However, it is an important check that the values of $\Omega_D$ that we use are consistent with each other in our different analyses, and could be important in analyses for which jets are probed over large energy ranges.

\section{Comparison of MC Generators for Single Emission Observables}\label{app:twoemissionMC}

Throughout this paper, we have extensively compared different Monte Carlo generators both at parton and hadron level for the observable $D_2$, which is set by two emissions off the initiating quark. We found significant differences between different Monte Carlo generators, and as compared with our analytic calculation, particularly at parton level. After hadronization, differences remained but these were quantitative differences, not differences in the shapes of distributions. For reference, in this appendix we compare the Monte Carlo generators used in this paper, at both parton and hadron level for an observable set by a single emission off of the initiating parton, namely the jet mass. Observables set by a single emission have been extensively studied in the literature, and are well understood. There exist automated codes for their resummation to NNLL \cite{Banfi:2004yd,Banfi:2014sua}, and they have been extensively used to tune Monte Carlo generators. We therefore expect to see much better agreement than for the $D_2$ observable, demonstrating that $D_2$ is a more differential probe of the perturbative shower structure.\footnote{Differences between Monte Carlo generators for single emission observable can also be accentuated by departing from jet mass, and considering angularities, or energy correlation functions, or differences between quark and gluon jets, for which limited data from LEP can be used for tuning \cite{Larkoski:2014pca,Sakaki:2015iya}.}

\begin{figure}
\begin{center}
\subfloat[]{\label{fig:D2_mass_a}
\includegraphics[width= 7.2cm]{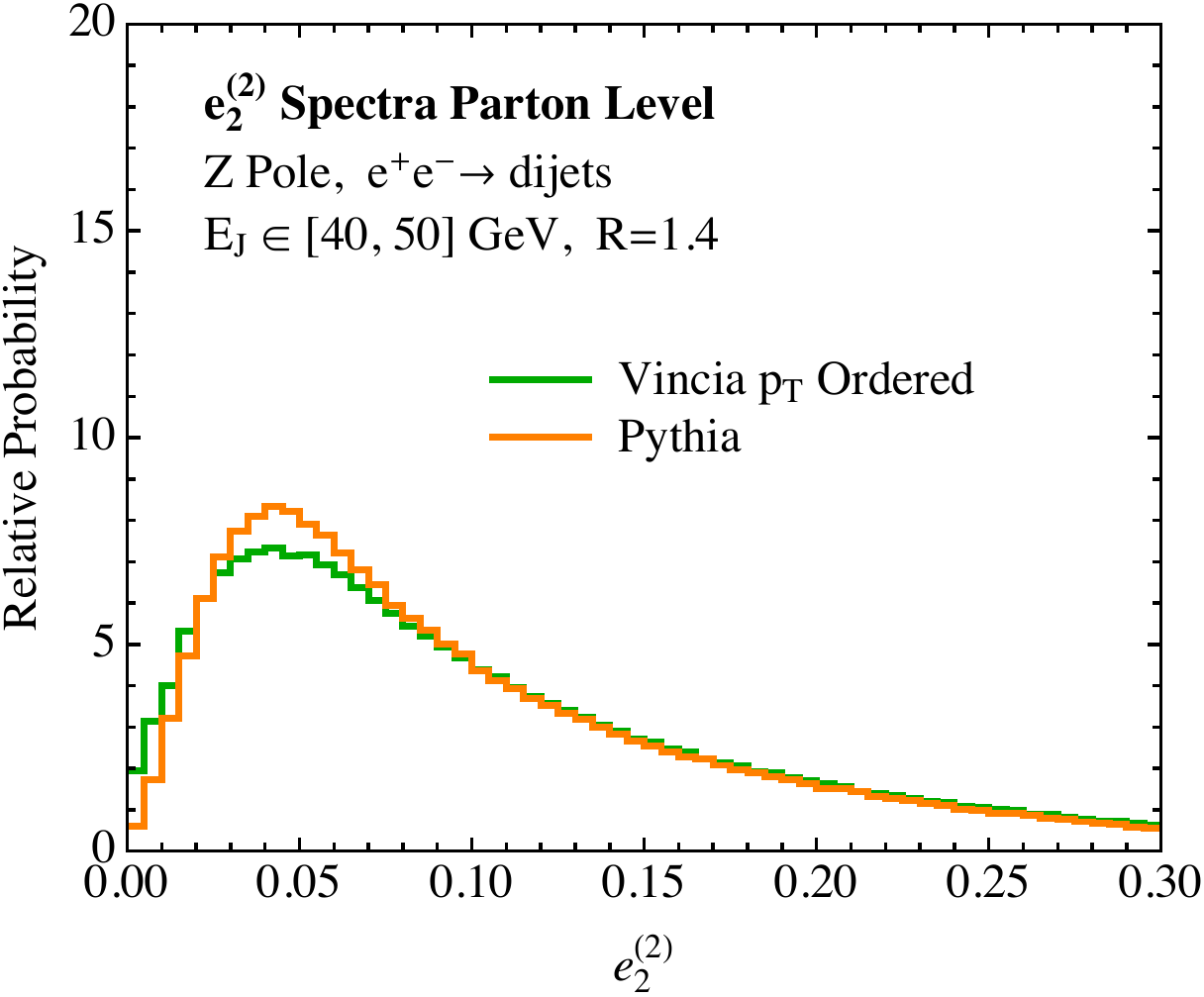}
}
\ 
\subfloat[]{\label{fig:D2_mass_b}
\includegraphics[width = 7.2cm]{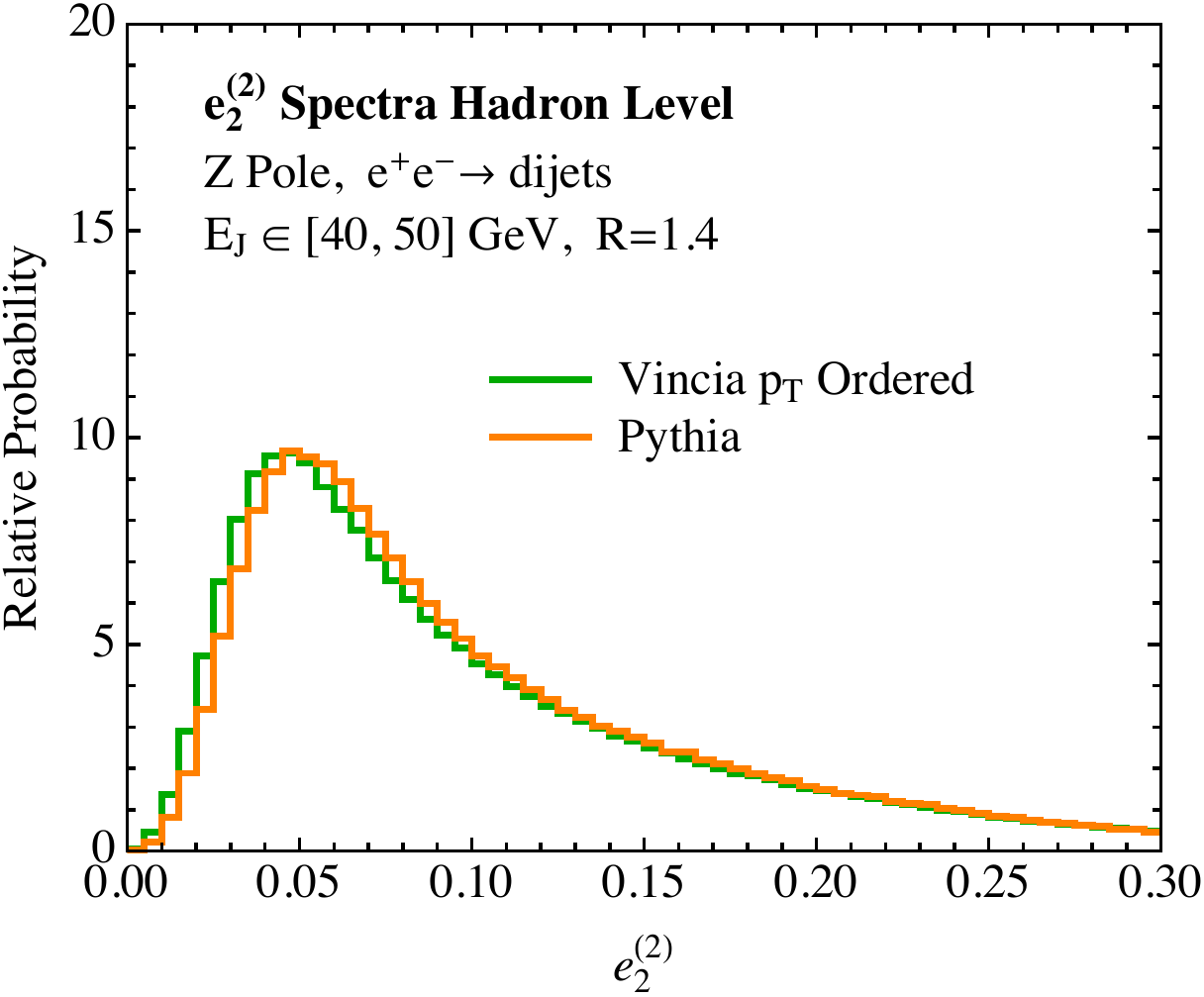}
}
\end{center}
\vspace{-0.2cm}
\caption{A comparison of the $\ecf{2}{2}$ spectrum as measured on quark initiated jets at the $Z$ pole from the \pythia{} and $p_T$-ordered \vincia{} Monte Carlo generators. Results are shown both for parton level Monte Carlo in a), and for hadron level Monte Carlo in b). 
}
\label{fig:D2_mass}
\end{figure}

In \Fig{fig:D2_mass} we compare the $\ecf{2}{2}$ spectra both at parton and hadron level for the \pythia{} and \vincia{} event generators at the $Z$ pole. We choose to the use $\ecf{2}{2}$ instead of the jet mass, as it is dimensionless. The level of agreement should be contrasted with \Fig{fig:D2_LEP} for the $D_2$ observable at the $Z$ pole, with and without hadronization. In particular, for the $\ecf{2}{2}$ observable, there is excellent agreement in the distributions at parton level, which is not true for $D_2$. For $D_2$, the disagreement is largely remedied by hadronization, while for $\ecf{2}{2}$, the level of disagreement before and after hadronization is much more similar. This supports our claim that the $D_2$ observable provides a more differential probe of the perturbative shower in particular, and could be used to improve its description.

\begin{figure}
\begin{center}
\subfloat[]{\label{fig:D2_mass_1TeV_a}
\includegraphics[width= 7.2cm]{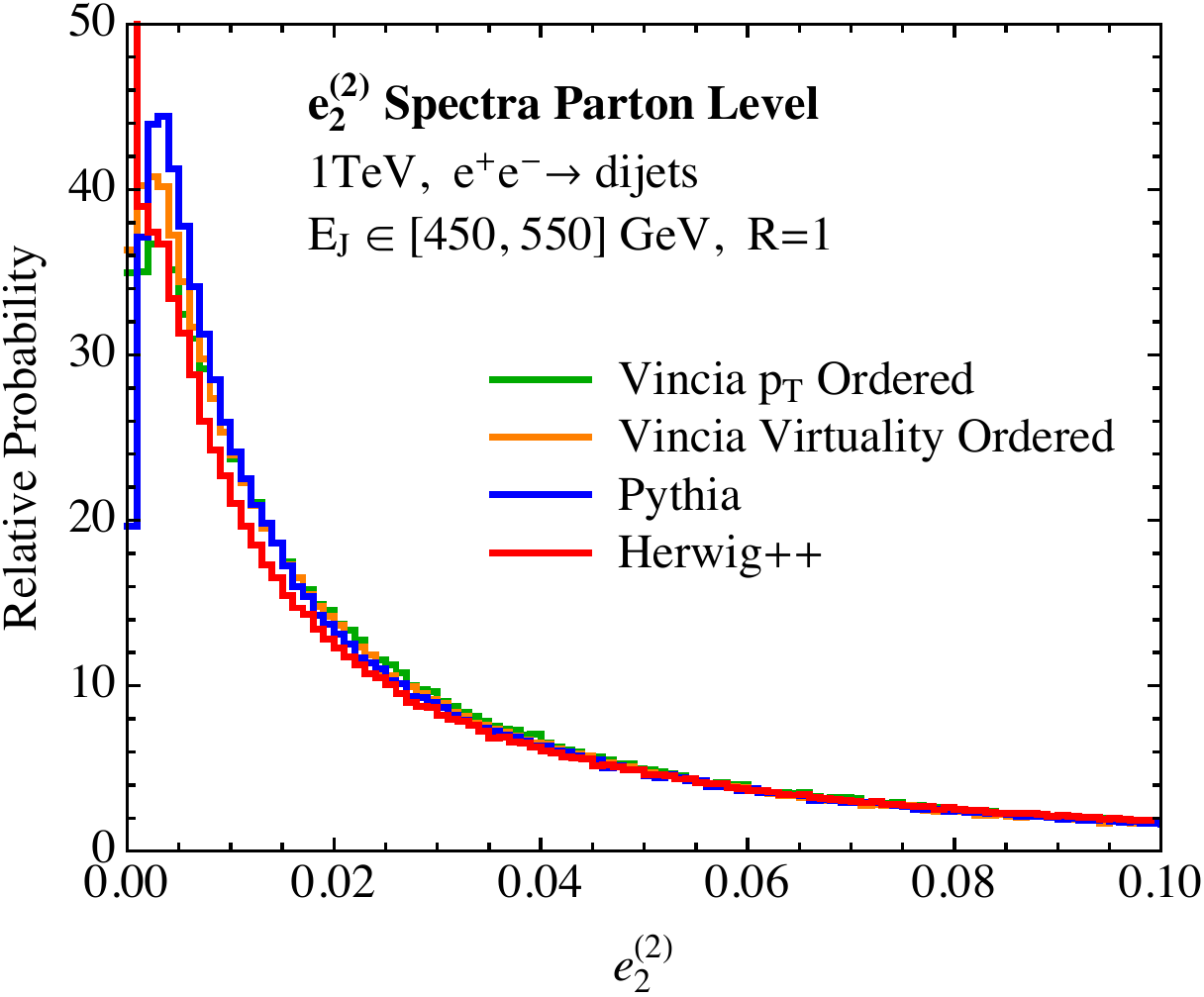}
}
\ 
\subfloat[]{\label{fig:D2_mass_1TeV_b}
\includegraphics[width = 7.2cm]{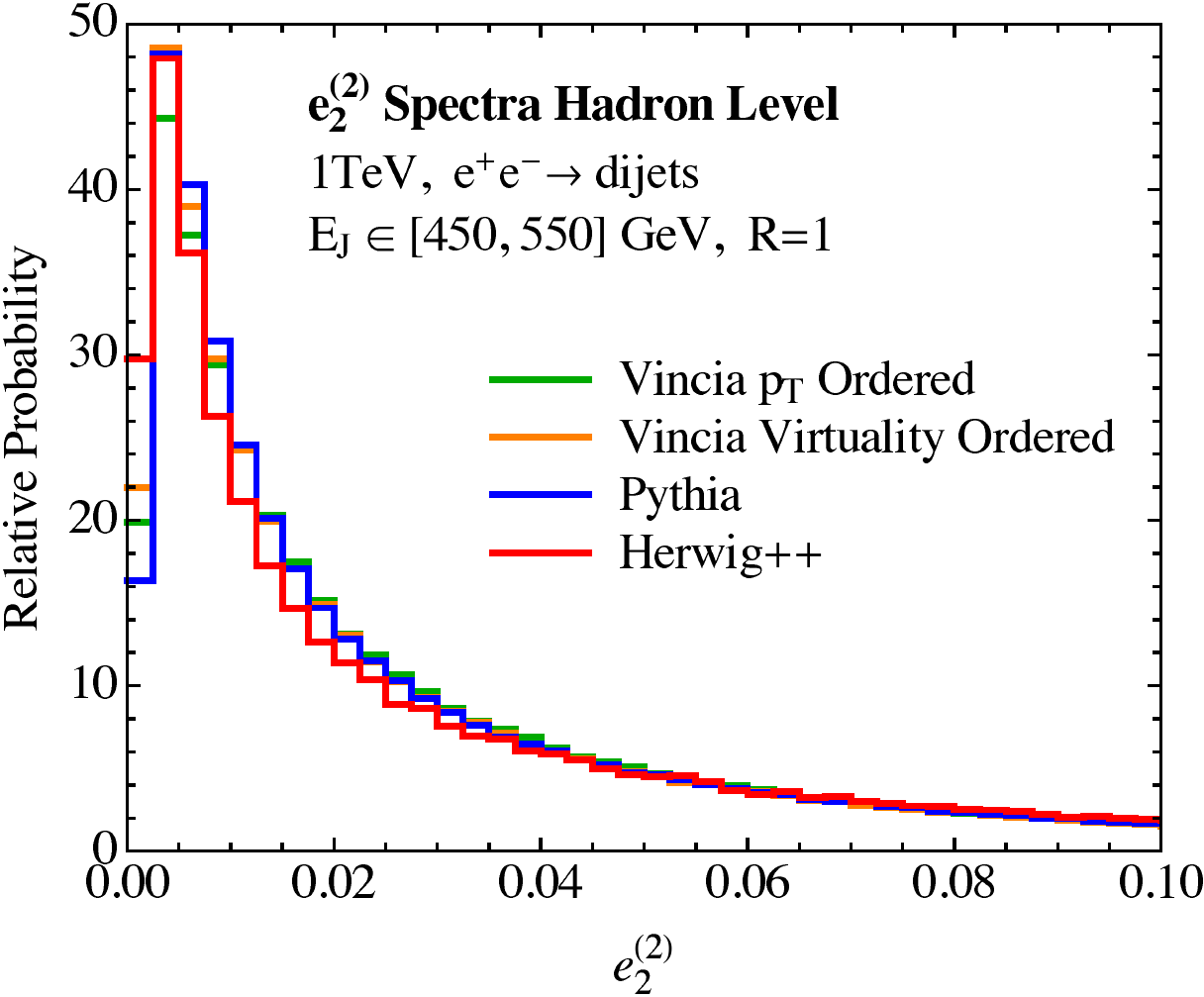}
}
\end{center}
\vspace{-0.2cm}
\caption{A comparison of the $\ecf{2}{2}$ spectrum as measured on quark initiated jets at a center of mass energy of $1$ TeV from the \pythia{}, $p_T$-ordered \vincia{}, virtuality ordered \vincia{}, and \herwigpp{} Monte Carlo generators. Results are shown both for parton level Monte Carlo in a), and for hadron level Monte Carlo in b).  
}
\label{fig:D2_mass_1TeV}
\end{figure}

In \Fig{fig:D2_mass_1TeV} we compare the $\ecf{2}{2}$ spectra both at parton and hadron level for the \pythia{}, $p_T$-ordered \vincia{}, virtuality ordered \vincia{}, and \herwigpp{} event generators at a center of mass energy of $1$ TeV and jet radius $R=1$, as was used for the majority of numerical comparisons with analytic calculations throughout the paper. The level of agreement in \Fig{fig:D2_mass_1TeV} should be compared with that for the $D_2$ spectra throughout \Sec{sec:results}. In particular, it is interesting to compare the level of agreement observed for the partonic  $\ecf{2}{2}$ spectra as compared with the partonic $D_2$ spectra in \Fig{fig:MC_compare}. There is still some difference between the \herwigpp{} spectrum at parton level and those of \vincia{} and \pythia{}, however, this is to be expected, as these Monte Carlos have different hadronization models and the comparison at parton level should be taken with caution. At hadron level, all Monte Carlos also agree well for the $\ecf{2}{2}$ spectra.

\begin{figure}
\begin{center}
\subfloat[]{\label{fig:D2_mass_500GeV_cut}
\includegraphics[width= 7.2cm]{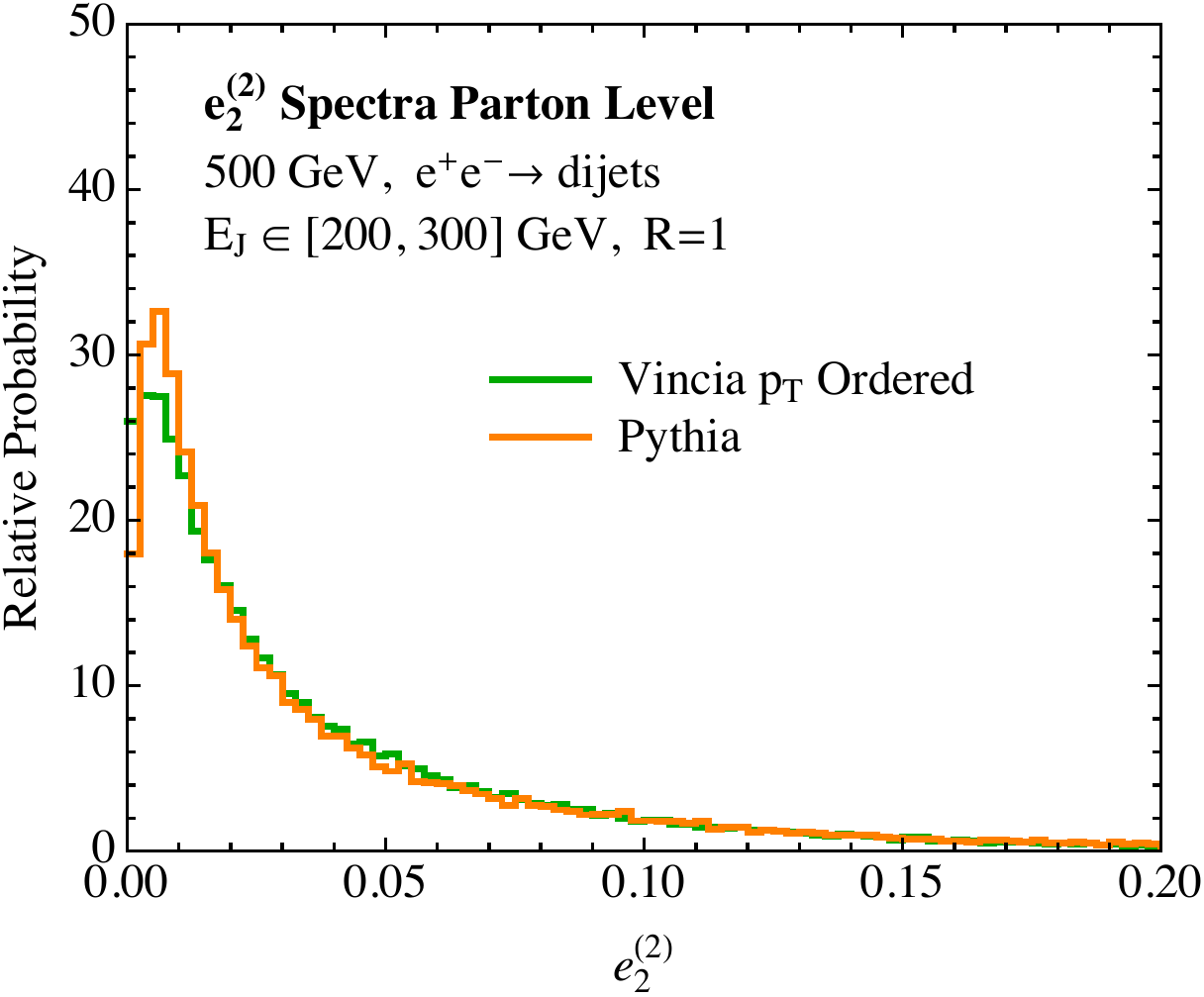}
}
\ 
\subfloat[]{\label{fig:D2_mass_2TeV_cut}
\includegraphics[width = 7.2cm]{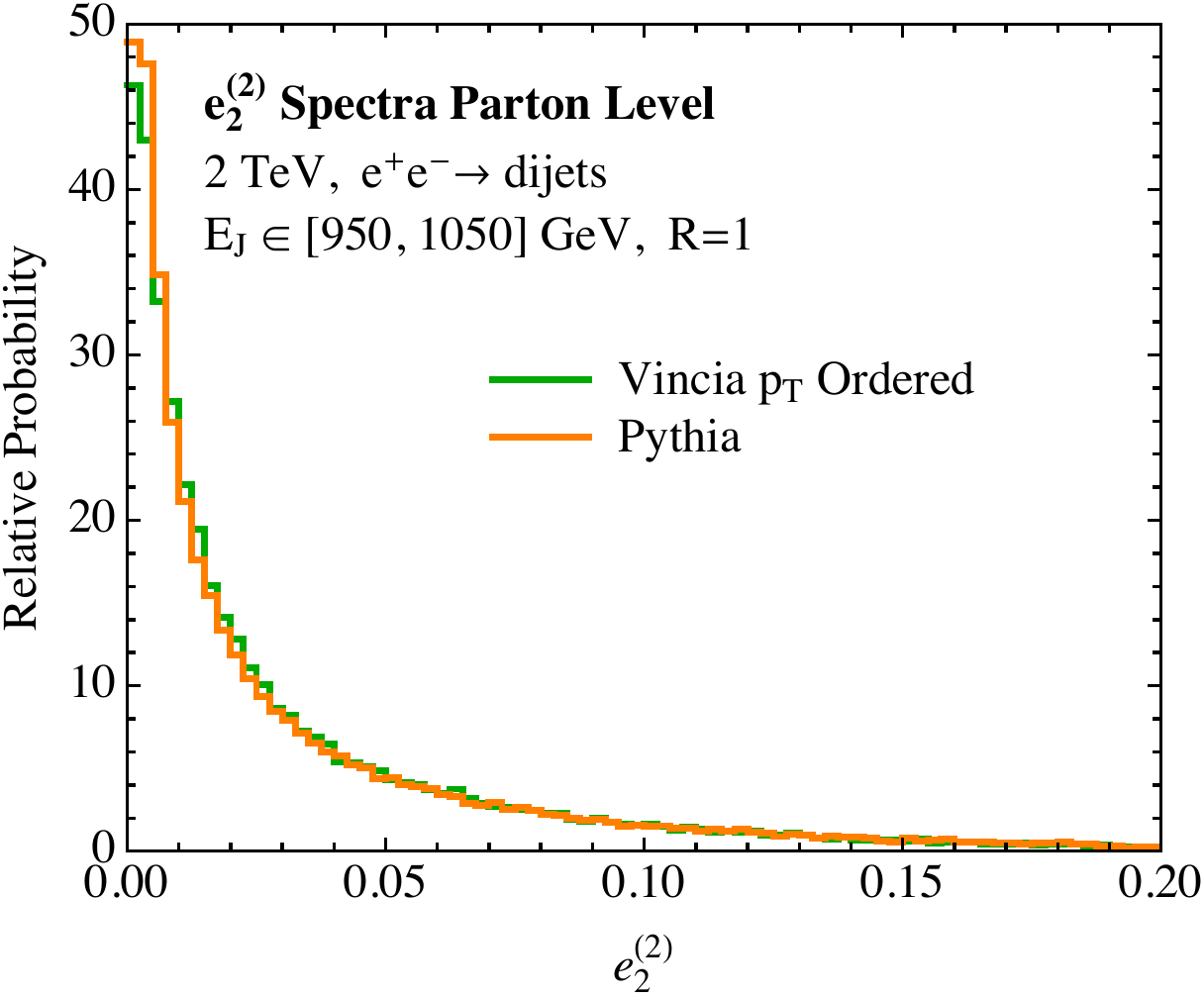}
}
\end{center}
\vspace{-0.2cm}
\caption{A comparison of the $\ecf{2}{2}$ spectrum as measured on quark initiated jets at a center of mass energy of $500$ GeV in a). and $2$ TeV in b). Results are shown for both the \pythia{}, and $p_T$-ordered \vincia{} Monte Carlo generators at parton level.
}
\label{fig:D2_mass_cut_500and2000}
\end{figure}

For completeness, in this appendix we will also include parton level plots of the $\ecf{2}{2}$ distributions for the other parameter ranges that were explored in detail in the text. In \Fig{fig:D2_mass_cut_500and2000} we show the $\ecf{2}{2}$ distributions at a center of mass energy of $500$ GeV and $2$ TeV, the two energies considered in the text. Only the \pythia{} and $p_T$-ordered \vincia{} generators are considered. The level of agreement between the different generators for $\ecf{2}{2}$ should be compared with the level of agreement for the $D_2$ spectra at these two energies, shown in \Fig{fig:E_dependence}. While for the $D_2$ observable, there was a significant discrepancy between the two generators at $2$ TeV, even in the general shape of the distribution, for $\ecf{2}{2}$, the distributions from the two generators agree quite well both at $500$ GeV and $2$ TeV. In particular, they exhibit a similar peak position and shape of the distributions.

\begin{figure}
\begin{center}
\subfloat[]{\label{fig:dep_app_1}
\includegraphics[width= 7.2cm]{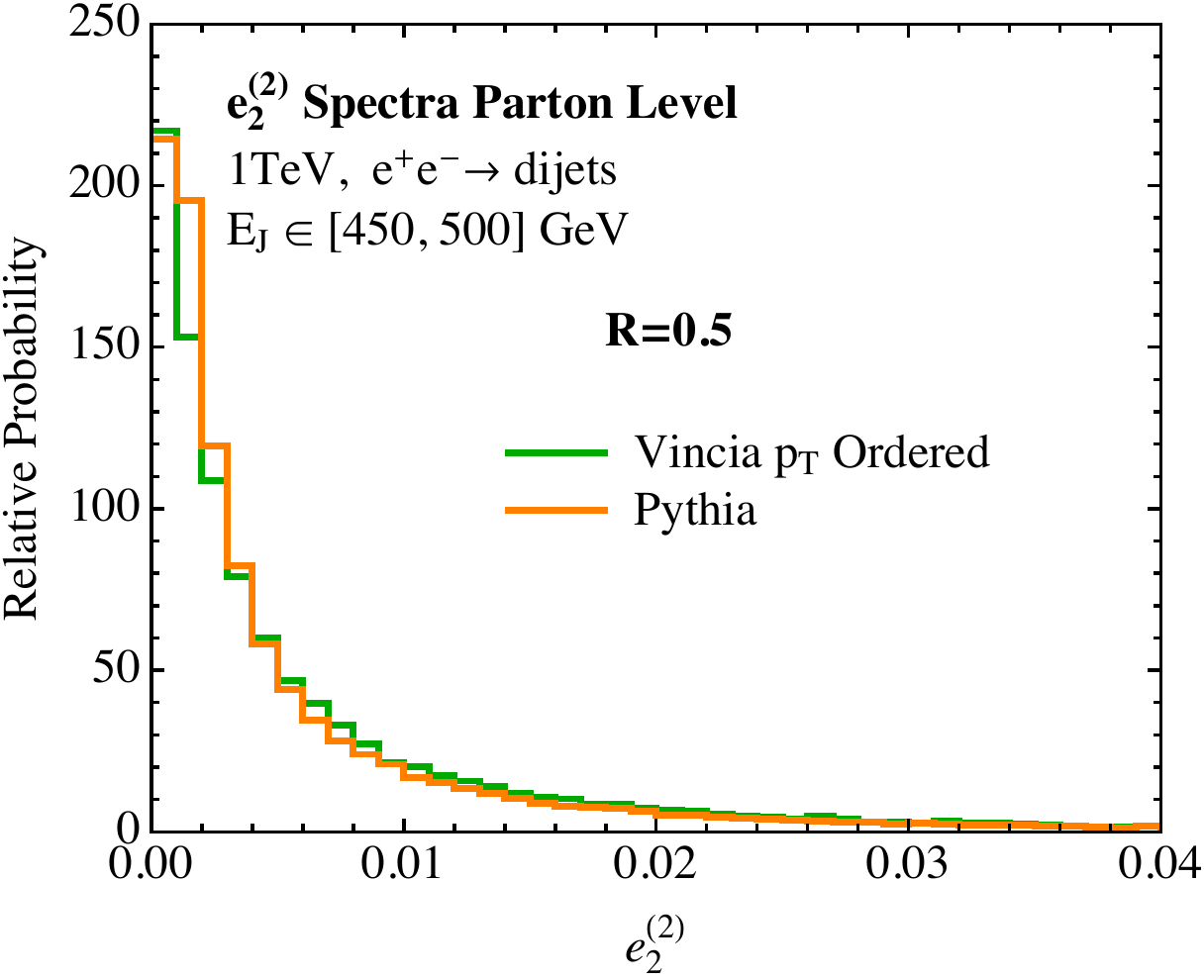}
}
\ 
\subfloat[]{\label{fig:dep_app_2}
\includegraphics[width = 7.2cm]{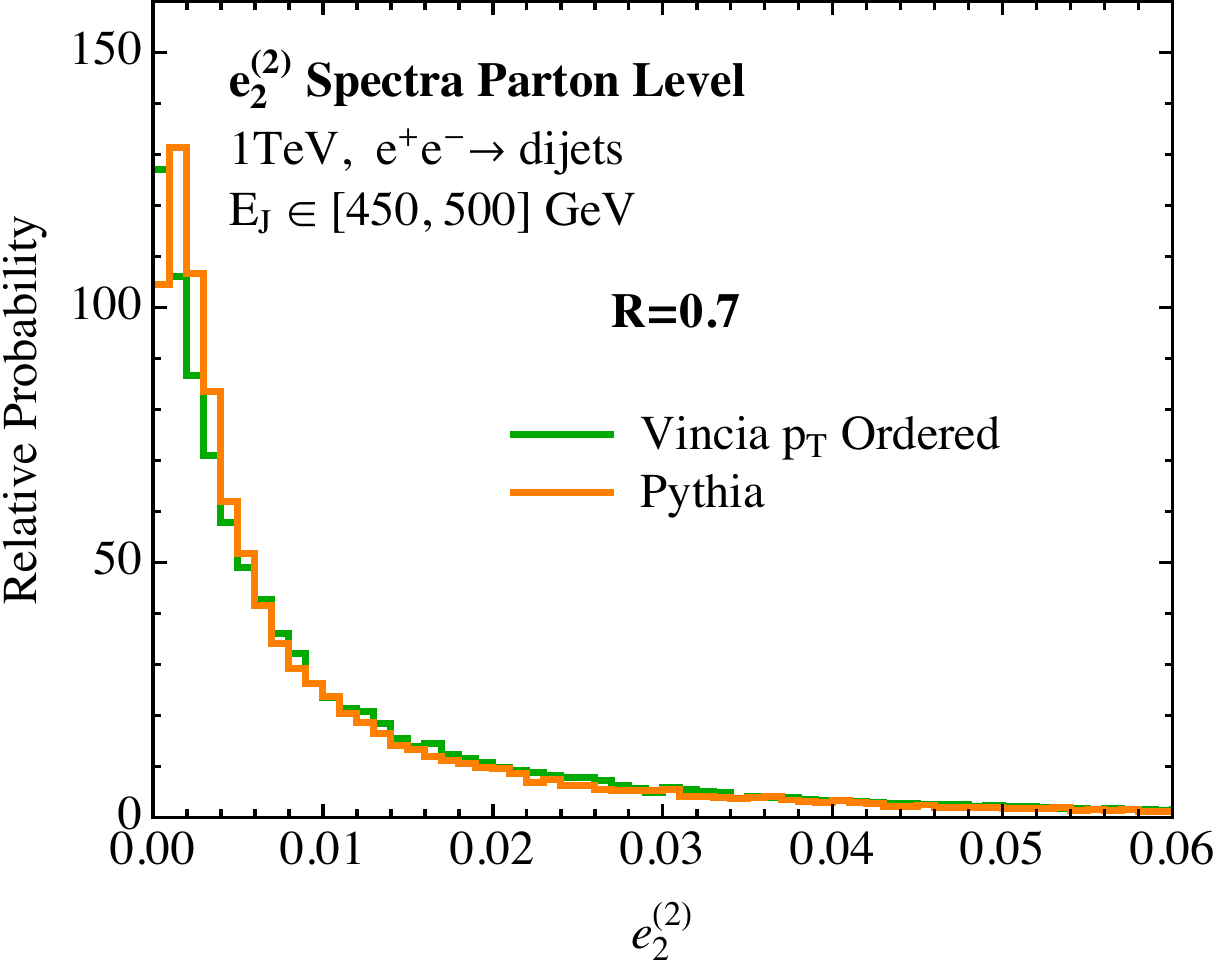}
}\\
\subfloat[]{\label{fig:dep_app_3}
\includegraphics[width= 7.2cm]{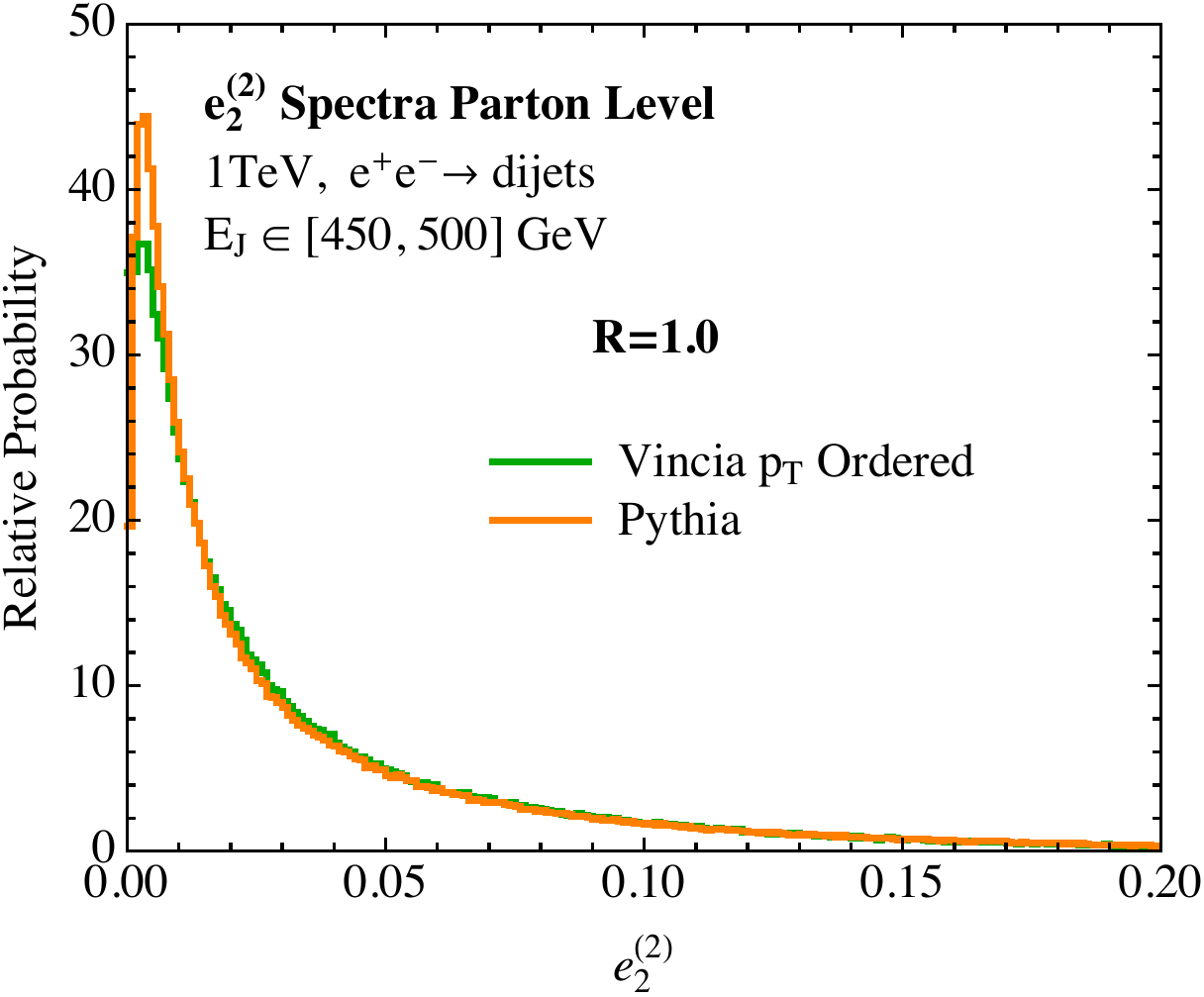}
}
\ 
\subfloat[]{\label{fig:Rdep_app_4}
\includegraphics[width = 7.2cm]{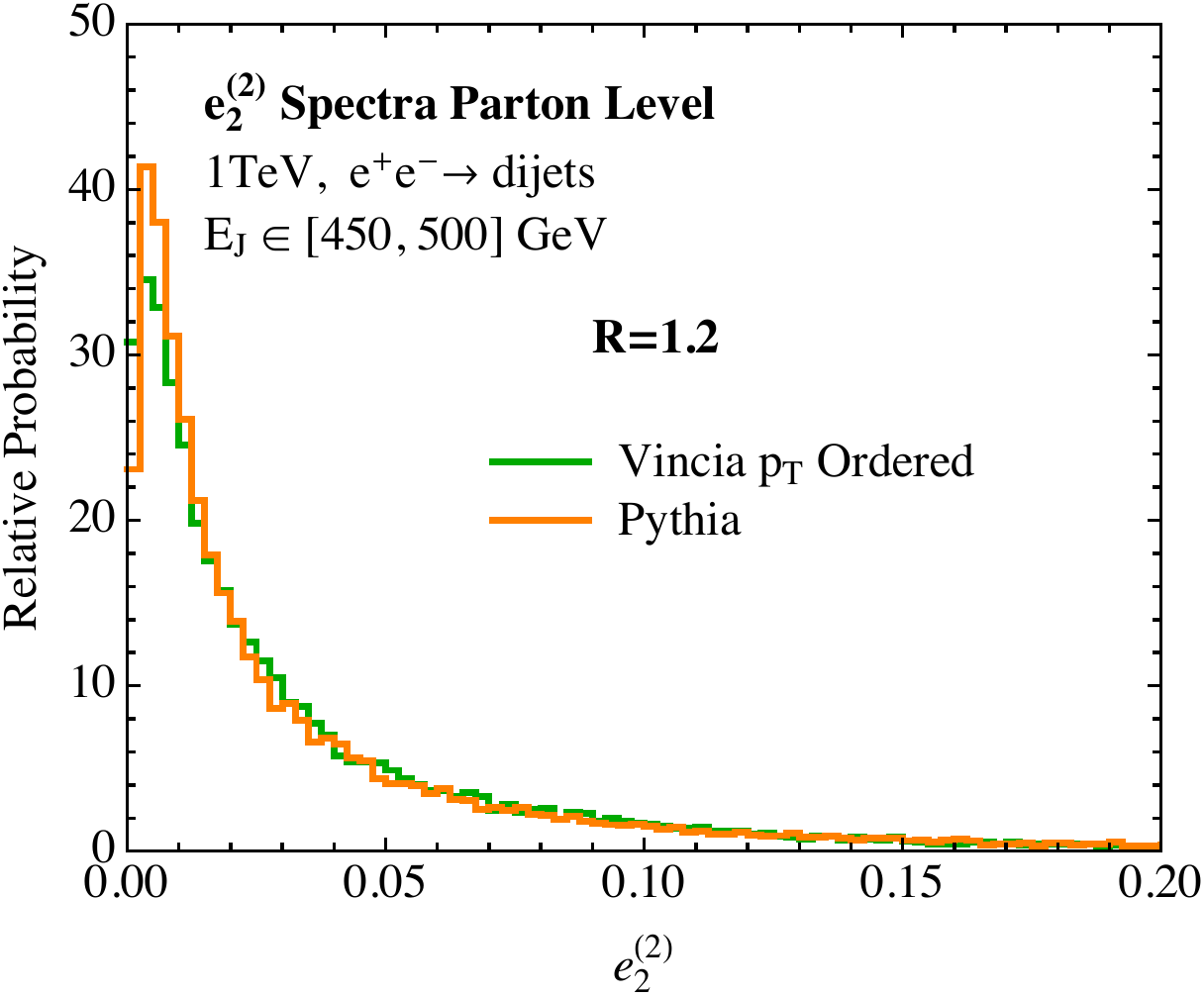}
}
\end{center}
\vspace{-0.2cm}
\caption{A comparison of the $\ecf{2}{2}$ spectrum as measured on quark initiated jets for different $R$ values at a center of mass energy of $1$ TeV from the \pythia{}, and $p_T$-ordered \vincia{} Monte Carlo generators at parton level. Results are shown $R=0.5,0.7,1.0, 1.2$ in a).-d). respectively.}
\label{fig:appendix_Rdependences}
\end{figure}

\begin{figure}
\begin{center}
\subfloat[]{
\includegraphics[width= 7.2cm]{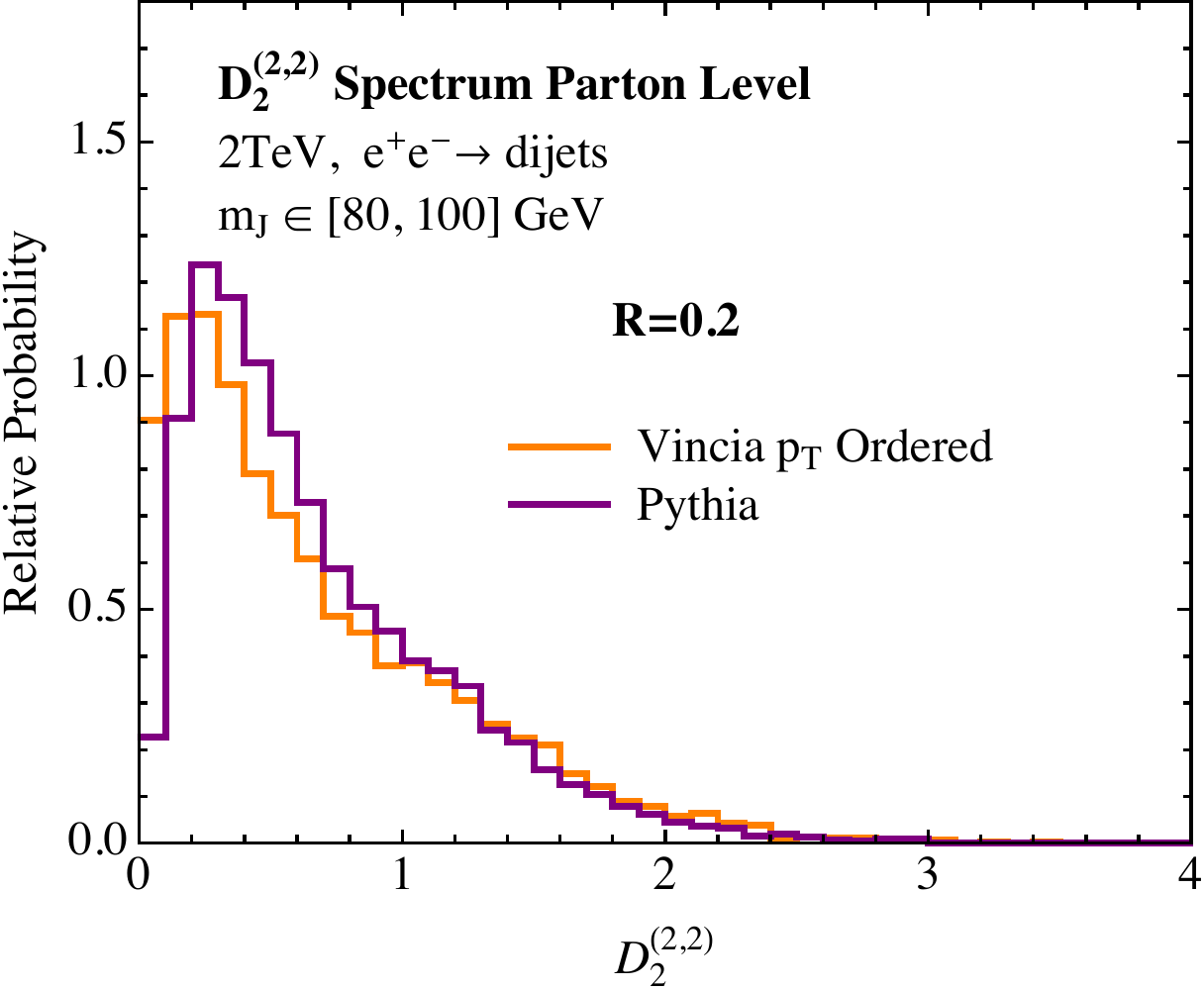}
}
\ 
\subfloat[]{
\includegraphics[width = 7.2cm]{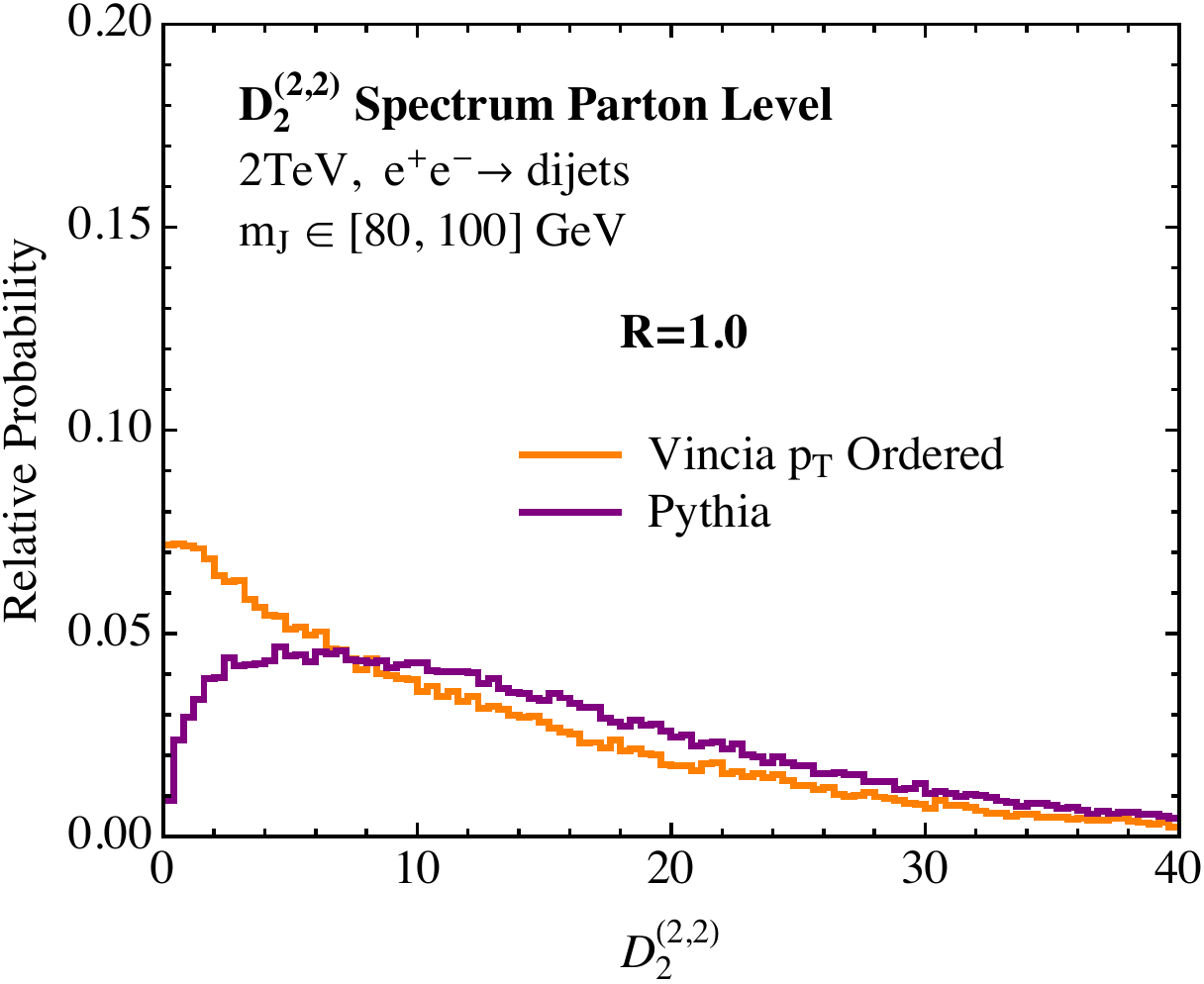}
}
\end{center}
\vspace{-0.2cm}
\caption{A comparison of the $D_2$ spectrum as measured on quark initiated jets at a center of mass energy of $2$ TeV from the \pythia{}, $p_T$-ordered \vincia{} Monte Carlo generators at parton level. A jet radius of $R=0.2$ is used in a) and $R=1.0$ is used in b).  
}
\label{fig:D2_app_R021}
\end{figure}

In \Fig{fig:appendix_Rdependences}, we consider the $R$ dependence of the parton level $\ecf{2}{2}$ distributions as measured in \pythia{} and $p_T$-ordered \vincia{}, as was considered in \Fig{fig:R_dependence} in the text for the $D_2$ observable. Unlike for the $D_2$ distributions, we see good agreement at parton level over the entire range of $R$. To conclude our discussion of $R$ dependence at parton level, we also include in \Fig{fig:D2_app_R021} a comparison of the parton level $D_2$ spectra as measured in in \pythia{} and $p_T$-ordered \vincia{} at $2$ TeV, with $R=0.2$ and $R=1.0$. As was referenced in \Sec{sec:jet_energy}, while poor agreement between the two generators is seen for $R=1$, comparably good agreement is seen at $R=0.2$. 
We view the ability to perform analytic calculations of observables which are sensitive to the substructure of the jet in this manner as an opportunity to improve the perturbative description of the QCD shower as implemented in Monte Carlo generators.

\bibliography{D2}

\providecommand{\href}[2]{#2}\begingroup\raggedright\begin{thebibliography}{100}

\bibitem{Abdesselam:2010pt}
A.~Abdesselam, E.~B. Kuutmann, U.~Bitenc, G.~Brooijmans, J.~Butterworth,
  et~al., {\it {Boosted objects: A Probe of beyond the Standard Model
  physics}},  {\em Eur.Phys.J.} {\bf C71} (2011) 1661,
  [\href{http://arxiv.org/abs/1012.5412}{{\tt arXiv:1012.5412}}].

\bibitem{Altheimer:2012mn}
A.~Altheimer, S.~Arora, L.~Asquith, G.~Brooijmans, J.~Butterworth, et~al., {\it
  {Jet Substructure at the Tevatron and LHC: New results, new tools, new
  benchmarks}},  {\em J.Phys.} {\bf G39} (2012) 063001,
  [\href{http://arxiv.org/abs/1201.0008}{{\tt arXiv:1201.0008}}].

\bibitem{Altheimer:2013yza}
A.~Altheimer, A.~Arce, L.~Asquith, J.~Backus~Mayes, E.~Bergeaas~Kuutmann,
  et~al., {\it {Boosted objects and jet substructure at the LHC. Report of
  BOOST2012, held at IFIC Valencia, 23rd-27th of July 2012}},  {\em
  Eur.Phys.J.} {\bf C74} (2014) 2792,
  [\href{http://arxiv.org/abs/1311.2708}{{\tt arXiv:1311.2708}}].

\bibitem{Adams:2015hiv}
D.~Adams, A.~Arce, L.~Asquith, M.~Backovic, T.~Barillari, et~al., {\it {Towards
  an Understanding of the Correlations in Jet Substructure}},
  \href{http://arxiv.org/abs/1504.00679}{{\tt arXiv:1504.00679}}.

\bibitem{CMS:2011xsa}
{\bf CMS} Collaboration, {\it {Jet Substructure Algorithms}},  Tech. Rep.
  CMS-PAS-JME-10-013, 2011.

\bibitem{Miller:2011qg}
{\bf ATLAS} Collaboration, D.~W. Miller, {\it {Jet substructure in ATLAS}},
  Tech. Rep. ATL-PHYS-PROC-2011-142, 2011.

\bibitem{Chatrchyan:2012mec}
{\bf CMS} Collaboration, S.~Chatrchyan et~al., {\it {Shape, transverse size,
  and charged hadron multiplicity of jets in pp collisions at 7 TeV}},  {\em
  JHEP} {\bf 1206} (2012) 160, [\href{http://arxiv.org/abs/1204.3170}{{\tt
  arXiv:1204.3170}}].

\bibitem{ATLAS:2012jla}
{\bf ATLAS} Collaboration, {\it {Studies of the impact and mitigation of
  pile-up on large-$R$ and groomed jets in ATLAS at $\sqrt{s}=7$ TeV}},  Tech.
  Rep. ATLAS-CONF-2012-066, ATLAS-COM-CONF-2012-097, 2012.

\bibitem{Aad:2012meb}
{\bf ATLAS} Collaboration, G.~Aad et~al., {\it {ATLAS measurements of the
  properties of jets for boosted particle searches}},  {\em Phys.Rev.} {\bf
  D86} (2012) 072006, [\href{http://arxiv.org/abs/1206.5369}{{\tt
  arXiv:1206.5369}}].

\bibitem{ATLAS:2012kla}
{\bf ATLAS} Collaboration, {\it {Performance of large-R jets and jet
  substructure reconstruction with the ATLAS detector}},  Tech. Rep.
  ATLAS-CONF-2012-065, ATLAS-COM-CONF-2012-095, 2012.

\bibitem{ATLAS:2012am}
{\bf ATLAS} Collaboration, G.~Aad et~al., {\it {Jet mass and substructure of
  inclusive jets in $\sqrt{s}=7$ TeV $pp$ collisions with the ATLAS
  experiment}},  {\em JHEP} {\bf 1205} (2012) 128,
  [\href{http://arxiv.org/abs/1203.4606}{{\tt arXiv:1203.4606}}].

\bibitem{Aad:2013gja}
{\bf ATLAS} Collaboration, G.~Aad et~al., {\it {Performance of jet substructure
  techniques for large-$R$ jets in proton-proton collisions at $\sqrt{s}$ = 7
  TeV using the ATLAS detector}},  {\em JHEP} {\bf 1309} (2013) 076,
  [\href{http://arxiv.org/abs/1306.4945}{{\tt arXiv:1306.4945}}].

\bibitem{Aad:2013fba}
{\bf ATLAS} Collaboration, G.~Aad et~al., {\it {Measurement of jet shapes in
  top pair events at sqrt(s) = 7 TeV using the ATLAS detector}},
  \href{http://arxiv.org/abs/1307.5749}{{\tt arXiv:1307.5749}}.

\bibitem{TheATLAScollaboration:2013tia}
{\bf ATLAS} Collaboration, {\it {Performance and Validation of Q-Jets at the
  ATLAS Detector in pp Collisions at $\sqrt{s}$=8 TeV in 2012}},  Tech. Rep.
  ATLAS-CONF-2013-087, ATLAS-COM-CONF-2013-099, 2013.

\bibitem{TheATLAScollaboration:2013sia}
{\bf ATLAS} Collaboration, {\it {Jet Charge Studies with the ATLAS Detector
  Using $\sqrt{s} = 8$ TeV Proton-Proton Collision Data}},  Tech. Rep.
  ATLAS-CONF-2013-086, ATLAS-COM-CONF-2013-101, 2013.

\bibitem{TheATLAScollaboration:2013ria}
{\bf ATLAS} Collaboration, {\it {Performance of pile-up subtraction for jet
  shapes}},  Tech. Rep. ATLAS-CONF-2013-085, ATLAS-COM-CONF-2013-100, 2013.

\bibitem{TheATLAScollaboration:2013pia}
{\bf ATLAS} Collaboration, {\it {Pile-up subtraction and suppression for jets
  in ATLAS}},  Tech. Rep. ATLAS-CONF-2013-083, ATLAS-COM-CONF-2013-097, 2013.

\bibitem{CMS:2013uea}
{\bf CMS} Collaboration, {\it {Identifying Hadronically Decaying Vector Bosons
  Merged into a Single Jet}},  Tech. Rep. CMS-PAS-JME-13-006, 2013.

\bibitem{CMS:2013kfa}
{\bf CMS} Collaboration, {\it {Performance of quark/gluon discrimination in 8
  TeV pp data}},  Tech. Rep. CMS-PAS-JME-13-002, 2013.

\bibitem{CMS:2013wea}
{\bf CMS} Collaboration, {\it {Pileup Jet Identification}},  Tech. Rep.
  CMS-PAS-JME-13-005, 2013.

\bibitem{CMS-PAS-JME-10-013}
{\bf CMS} Collaboration, {\it Jet substructure algorithms},  Tech. Rep.
  CMS-PAS-JME-10-013, CERN, Geneva, 2011.

\bibitem{CMS-PAS-QCD-10-041}
{\bf CMS} Collaboration, {\it Measurement of the subjet multiplicity in dijet
  events from proton-proton collisions at sqrt(s) = 7 tev},  Tech. Rep.
  CMS-PAS-QCD-10-041, CERN, Geneva, 2010.

\bibitem{Aad:2014gea}
{\bf ATLAS} Collaboration, G.~Aad et~al., {\it {Light-quark and gluon jet
  discrimination in pp collisions at $\sqrt{s}$ = 7 TeV with the ATLAS
  detector}},  \href{http://arxiv.org/abs/1405.6583}{{\tt arXiv:1405.6583}}.

\bibitem{LOCH:2014lla}
{\bf ATLAS} Collaboration, P.~Loch, {\it {Studies of jet shapes and jet
  substructure in proton-proton collisions at $\sqrt{s} =$ 7 TeV with ATLAS}},
  {\em PoS} {\bf EPS-HEP2013} (2013) 442.

\bibitem{CMS:2014fya}
{\bf CMS} Collaboration, {\it {Boosted Top Jet Tagging at CMS}},  Tech. Rep.
  CMS-PAS-JME-13-007, 2014.

\bibitem{CMS:2014joa}
{\bf CMS} Collaboration, {\it {V Tagging Observables and Correlations}},  Tech.
  Rep. CMS-PAS-JME-14-002, 2014.

\bibitem{Aad:2014haa}
{\bf ATLAS} Collaboration, G.~Aad et~al., {\it {Measurement of the
  cross-section of high transverse momentum vector bosons reconstructed as
  single jets and studies of jet substructure in $pp$ collisions at
  ${\sqrt{s}}$ = 7 TeV with the ATLAS detector}},  {\em New J.Phys.} {\bf 16}
  (2014), no.~11 113013, [\href{http://arxiv.org/abs/1407.0800}{{\tt
  arXiv:1407.0800}}].

\bibitem{CMS:2011bqa}
{\bf CMS} Collaboration, {\it {Search for BSM ttbar Production in the Boosted
  All-Hadronic Final State}},  Tech. Rep. CMS-PAS-EXO-11-006, 2011.

\bibitem{Fleischmann:2013woa}
{\bf ATLAS, CMS} Collaboration, S.~Fleischmann, {\it {Boosted top quark
  techniques and searches for $t\bar{t}$ resonances at the LHC}},  {\em
  J.Phys.Conf.Ser.} {\bf 452} (2013), no.~1 012034.

\bibitem{Pilot:2013bla}
{\bf ATLAS, CMS} Collaboration, J.~Pilot, {\it {Boosted Top Quarks, Top Pair
  Resonances, and Top Partner Searches at the LHC}},  {\em EPJ Web Conf.} {\bf
  60} (2013) 09003.

\bibitem{TheATLAScollaboration:2013qia}
{\bf ATLAS} Collaboration, {\it {Performance of boosted top quark
  identification in 2012 ATLAS data}},  Tech. Rep. ATLAS-CONF-2013-084,
  ATLAS-COM-CONF-2013-074, 2013.

\bibitem{Chatrchyan:2012ku}
{\bf CMS} Collaboration, S.~Chatrchyan et~al., {\it {Search for Anomalous
  $t\bar{t}$ Production in the Highly-Boosted All-Hadronic Final State}},  {\em
  JHEP} {\bf 1209} (2012) 029, [\href{http://arxiv.org/abs/1204.2488}{{\tt
  arXiv:1204.2488}}].

\bibitem{Chatrchyan:2012sn}
{\bf CMS} Collaboration, S.~Chatrchyan et~al., {\it {Search for a Higgs boson
  in the decay channel $H$ to ZZ(*) to $q$ qbar $\ell^-$ l+ in $pp$ collisions
  at $\sqrt{s}=7$ TeV}},  {\em JHEP} {\bf 1204} (2012) 036,
  [\href{http://arxiv.org/abs/1202.1416}{{\tt arXiv:1202.1416}}].

\bibitem{CMS:2013cda}
{\bf CMS} Collaboration, {\it {Search for a Standard Model-like Higgs boson
  decaying into WW to l nu qqbar in pp collisions at sqrt s = 8 TeV}},  Tech.
  Rep. CMS-PAS-HIG-13-008, 2013.

\bibitem{CMS:2014afa}
{\bf CMS} Collaboration, {\it {Search for pair-produced vector-like quarks of
  charge -1/3 decaying to bH using boosted Higgs jet-tagging in pp collisions
  at sqrt(s) = 8 TeV}},  Tech. Rep. CMS-PAS-B2G-14-001, 2014.

\bibitem{CMS:2014aka}
{\bf CMS} Collaboration, {\it {Search for top-Higgs resonances in all-hadronic
  final states using jet substructure methods}},  Tech. Rep.
  CMS-PAS-B2G-14-002, 2014.

\bibitem{Khachatryan:2015axa}
{\bf CMS} Collaboration, V.~Khachatryan et~al., {\it {Search for vector-like T
  quarks decaying to top quarks and Higgs bosons in the all-hadronic channel
  using jet substructure}},  \href{http://arxiv.org/abs/1503.01952}{{\tt
  arXiv:1503.01952}}.

\bibitem{CMS:1900uua}
{\bf CMS} Collaboration, C.~Collaboration, {\it {Search for pair-produced
  vector-like top quark partners decaying to bW in the fully hadronic channel
  using jet substructure at 8 TeV}}, .

\bibitem{Khachatryan:2015bma}
{\bf CMS} Collaboration, V.~Khachatryan et~al., {\it {Search for a massive
  resonance decaying into a Higgs boson and a W or Z boson in hadronic final
  states in proton-proton collisions at sqrt(s) = 8 TeV}},
  \href{http://arxiv.org/abs/1506.01443}{{\tt arXiv:1506.01443}}.

\bibitem{Aad:2015owa}
{\bf ATLAS} Collaboration, G.~Aad et~al., {\it {Search for high-mass diboson
  resonances with boson-tagged jets in proton-proton collisions at $\sqrt{s}$ =
  8 TeV with the ATLAS detector}},  \href{http://arxiv.org/abs/1506.00962}{{\tt
  arXiv:1506.00962}}.

\bibitem{Feige:2012vc}
I.~Feige, M.~D. Schwartz, I.~W. Stewart, and J.~Thaler, {\it {Precision Jet
  Substructure from Boosted Event Shapes}},  {\em Phys.Rev.Lett.} {\bf 109}
  (2012) 092001, [\href{http://arxiv.org/abs/1204.3898}{{\tt
  arXiv:1204.3898}}].

\bibitem{Field:2012rw}
M.~Field, G.~Gur-Ari, D.~A. Kosower, L.~Mannelli, and G.~Perez, {\it
  {Three-Prong Distribution of Massive Narrow QCD Jets}},  {\em Phys.Rev.} {\bf
  D87} (2013), no.~9 094013, [\href{http://arxiv.org/abs/1212.2106}{{\tt
  arXiv:1212.2106}}].

\bibitem{Dasgupta:2013ihk}
M.~Dasgupta, A.~Fregoso, S.~Marzani, and G.~P. Salam, {\it {Towards an
  understanding of jet substructure}},  {\em JHEP} {\bf 1309} (2013) 029,
  [\href{http://arxiv.org/abs/1307.0007}{{\tt arXiv:1307.0007}}].

\bibitem{Dasgupta:2013via}
M.~Dasgupta, A.~Fregoso, S.~Marzani, and A.~Powling, {\it {Jet substructure
  with analytical methods}},  {\em Eur.Phys.J.} {\bf C73} (2013), no.~11 2623,
  [\href{http://arxiv.org/abs/1307.0013}{{\tt arXiv:1307.0013}}].

\bibitem{Larkoski:2014pca}
A.~J. Larkoski, J.~Thaler, and W.~J. Waalewijn, {\it {Gaining (Mutual)
  Information about Quark/Gluon Discrimination}},  {\em JHEP} {\bf 1411} (2014)
  129, [\href{http://arxiv.org/abs/1408.3122}{{\tt arXiv:1408.3122}}].

\bibitem{Dasgupta:2015yua}
M.~Dasgupta, A.~Powling, and A.~Siodmok, {\it {On jet substructure methods for
  signal jets}},  \href{http://arxiv.org/abs/1503.01088}{{\tt
  arXiv:1503.01088}}.

\bibitem{Seymour:1997kj}
M.~Seymour, {\it {Jet shapes in hadron collisions: Higher orders, resummation
  and hadronization}},  {\em Nucl.Phys.} {\bf B513} (1998) 269--300,
  [\href{http://arxiv.org/abs/hep-ph/9707338}{{\tt hep-ph/9707338}}].

\bibitem{Li:2011hy}
H.-n. Li, Z.~Li, and C.-P. Yuan, {\it {QCD resummation for jet substructures}},
   {\em Phys.Rev.Lett.} {\bf 107} (2011) 152001,
  [\href{http://arxiv.org/abs/1107.4535}{{\tt arXiv:1107.4535}}].

\bibitem{Larkoski:2012eh}
A.~J. Larkoski, {\it {QCD Analysis of the Scale-Invariance of Jets}},  {\em
  Phys.Rev.} {\bf D86} (2012) 054004,
  [\href{http://arxiv.org/abs/1207.1437}{{\tt arXiv:1207.1437}}].

\bibitem{Jankowiak:2012na}
M.~Jankowiak and A.~J. Larkoski, {\it {Angular Scaling in Jets}},  {\em JHEP}
  {\bf 1204} (2012) 039, [\href{http://arxiv.org/abs/1201.2688}{{\tt
  arXiv:1201.2688}}].

\bibitem{Chien:2014nsa}
Y.-T. Chien and I.~Vitev, {\it {Jet Shape Resummation Using Soft-Collinear
  Effective Theory}},  {\em JHEP} {\bf 1412} (2014) 061,
  [\href{http://arxiv.org/abs/1405.4293}{{\tt arXiv:1405.4293}}].

\bibitem{Chien:2014zna}
Y.-T. Chien, {\it {Resummation of Jet Shapes and Extracting Properties of the
  Quark-Gluon Plasma}},  {\em Int.J.Mod.Phys.Conf.Ser.} {\bf 37} (2015)
  1560047, [\href{http://arxiv.org/abs/1411.0741}{{\tt arXiv:1411.0741}}].

\bibitem{Isaacson:2015fra}
J.~Isaacson, H.-n. Li, Z.~Li, and C.~P. Yuan, {\it {Factorization for
  substructures of boosted Higgs jets}},
  \href{http://arxiv.org/abs/1505.06368}{{\tt arXiv:1505.06368}}.

\bibitem{Krohn:2012fg}
D.~Krohn, M.~D. Schwartz, T.~Lin, and W.~J. Waalewijn, {\it {Jet Charge at the
  LHC}},  {\em Phys.Rev.Lett.} {\bf 110} (2013), no.~21 212001,
  [\href{http://arxiv.org/abs/1209.2421}{{\tt arXiv:1209.2421}}].

\bibitem{Waalewijn:2012sv}
W.~J. Waalewijn, {\it {Calculating the Charge of a Jet}},  {\em Phys.Rev.} {\bf
  D86} (2012) 094030, [\href{http://arxiv.org/abs/1209.3019}{{\tt
  arXiv:1209.3019}}].

\bibitem{Bertolini:2015pka}
D.~Bertolini, J.~Thaler, and J.~R. Walsh, {\it {The First Calculation of
  Fractional Jets}},  {\em JHEP} {\bf 1505} (2015) 008,
  [\href{http://arxiv.org/abs/1501.01965}{{\tt arXiv:1501.01965}}].

\bibitem{Bhattacherjee:2015psa}
B.~Bhattacherjee, S.~Mukhopadhyay, M.~M. Nojiri, Y.~Sakaki, and B.~R. Webber,
  {\it {Associated jet and subjet rates in light-quark and gluon jet
  discrimination}},  {\em JHEP} {\bf 1504} (2015) 131,
  [\href{http://arxiv.org/abs/1501.04794}{{\tt arXiv:1501.04794}}].

\bibitem{Catani:1991bd}
S.~Catani, G.~Turnock, and B.~Webber, {\it {Heavy jet mass distribution in e+
  e- annihilation}},  {\em Phys.Lett.} {\bf B272} (1991) 368--372.

\bibitem{Chien:2010kc}
Y.-T. Chien and M.~D. Schwartz, {\it {Resummation of heavy jet mass and
  comparison to LEP data}},  {\em JHEP} {\bf 1008} (2010) 058,
  [\href{http://arxiv.org/abs/1005.1644}{{\tt arXiv:1005.1644}}].

\bibitem{Chien:2012ur}
Y.-T. Chien, R.~Kelley, M.~D. Schwartz, and H.~X. Zhu, {\it {Resummation of Jet
  Mass at Hadron Colliders}},  {\em Phys.Rev.} {\bf D87} (2013) 014010,
  [\href{http://arxiv.org/abs/1208.0010}{{\tt arXiv:1208.0010}}].

\bibitem{Dasgupta:2012hg}
M.~Dasgupta, K.~Khelifa-Kerfa, S.~Marzani, and M.~Spannowsky, {\it {On jet mass
  distributions in Z+jet and dijet processes at the LHC}},  {\em JHEP} {\bf
  1210} (2012) 126, [\href{http://arxiv.org/abs/1207.1640}{{\tt
  arXiv:1207.1640}}].

\bibitem{Jouttenus:2013hs}
T.~T. Jouttenus, I.~W. Stewart, F.~J. Tackmann, and W.~J. Waalewijn, {\it {Jet
  mass spectra in Higgs boson plus one jet at next-to-next-to-leading
  logarithmic order}},  {\em Phys.Rev.} {\bf D88} (2013), no.~5 054031,
  [\href{http://arxiv.org/abs/1302.0846}{{\tt arXiv:1302.0846}}].

\bibitem{Thaler:2010tr}
J.~Thaler and K.~Van~Tilburg, {\it {Identifying Boosted Objects with
  N-subjettiness}},  {\em JHEP} {\bf 1103} (2011) 015,
  [\href{http://arxiv.org/abs/1011.2268}{{\tt arXiv:1011.2268}}].

\bibitem{Thaler:2011gf}
J.~Thaler and K.~Van~Tilburg, {\it {Maximizing Boosted Top Identification by
  Minimizing N-subjettiness}},  {\em JHEP} {\bf 1202} (2012) 093,
  [\href{http://arxiv.org/abs/1108.2701}{{\tt arXiv:1108.2701}}].

\bibitem{Larkoski:2013eya}
A.~J. Larkoski, G.~P. Salam, and J.~Thaler, {\it {Energy Correlation Functions
  for Jet Substructure}},  {\em JHEP} {\bf 1306} (2013) 108,
  [\href{http://arxiv.org/abs/1305.0007}{{\tt arXiv:1305.0007}}].

\bibitem{Larkoski:2014gra}
A.~J. Larkoski, I.~Moult, and D.~Neill, {\it {Power Counting to Better Jet
  Observables}},  {\em JHEP} {\bf 1412} (2014) 009,
  [\href{http://arxiv.org/abs/1409.6298}{{\tt arXiv:1409.6298}}].

\bibitem{Larkoski:2014zma}
A.~J. Larkoski, I.~Moult, and D.~Neill, {\it {Building a Better Boosted Top
  Tagger}},  {\em Phys.Rev.} {\bf D91} (2015), no.~3 034035,
  [\href{http://arxiv.org/abs/1411.0665}{{\tt arXiv:1411.0665}}].

\bibitem{Almeida:2008yp}
L.~G. Almeida, S.~J. Lee, G.~Perez, G.~F. Sterman, I.~Sung, et~al., {\it
  {Substructure of high-$p_T$ Jets at the LHC}},  {\em Phys.Rev.} {\bf D79}
  (2009) 074017, [\href{http://arxiv.org/abs/0807.0234}{{\tt
  arXiv:0807.0234}}].

\bibitem{Soyez:2012hv}
G.~Soyez, G.~P. Salam, J.~Kim, S.~Dutta, and M.~Cacciari, {\it {Pileup
  subtraction for jet shapes}},  {\em Phys.Rev.Lett.} {\bf 110} (2013), no.~16
  162001, [\href{http://arxiv.org/abs/1211.2811}{{\tt arXiv:1211.2811}}].

\bibitem{Larkoski:2013paa}
A.~J. Larkoski and J.~Thaler, {\it {Unsafe but Calculable: Ratios of
  Angularities in Perturbative QCD}},  {\em JHEP} {\bf 1309} (2013) 137,
  [\href{http://arxiv.org/abs/1307.1699}{{\tt arXiv:1307.1699}}].

\bibitem{Larkoski:2014wba}
A.~J. Larkoski, S.~Marzani, G.~Soyez, and J.~Thaler, {\it {Soft Drop}},  {\em
  JHEP} {\bf 1405} (2014) 146, [\href{http://arxiv.org/abs/1402.2657}{{\tt
  arXiv:1402.2657}}].

\bibitem{Larkoski:2014bia}
A.~J. Larkoski and J.~Thaler, {\it {Aspects of jets at 100 TeV}},  {\em
  Phys.Rev.} {\bf D90} (2014), no.~3 034010,
  [\href{http://arxiv.org/abs/1406.7011}{{\tt arXiv:1406.7011}}].

\bibitem{Larkoski:2015lea}
A.~J. Larkoski, S.~Marzani, and J.~Thaler, {\it {Sudakov Safety in Perturbative
  QCD}},  {\em Phys.Rev.} {\bf D91} (2015), no.~11 111501,
  [\href{http://arxiv.org/abs/1502.01719}{{\tt arXiv:1502.01719}}].

\bibitem{Larkoski:2014tva}
A.~J. Larkoski, I.~Moult, and D.~Neill, {\it {Toward Multi-Differential Cross
  Sections: Measuring Two Angularities on a Single Jet}},  {\em JHEP} {\bf
  1409} (2014) 046, [\href{http://arxiv.org/abs/1401.4458}{{\tt
  arXiv:1401.4458}}].

\bibitem{Procura:2014cba}
M.~Procura, W.~J. Waalewijn, and L.~Zeune, {\it {Resummation of
  Double-Differential Cross Sections and Fully-Unintegrated Parton Distribution
  Functions}},  {\em JHEP} {\bf 1502} (2015) 117,
  [\href{http://arxiv.org/abs/1410.6483}{{\tt arXiv:1410.6483}}].

\bibitem{Larkoski:2015zka}
A.~J. Larkoski, I.~Moult, and D.~Neill, {\it {Non-Global Logarithms,
  Factorization, and the Soft Substructure of Jets}},
  \href{http://arxiv.org/abs/1501.04596}{{\tt arXiv:1501.04596}}.

\bibitem{Bauer:2011uc}
C.~W. Bauer, F.~J. Tackmann, J.~R. Walsh, and S.~Zuberi, {\it {Factorization
  and Resummation for Dijet Invariant Mass Spectra}},  {\em Phys.Rev.} {\bf
  D85} (2012) 074006, [\href{http://arxiv.org/abs/1106.6047}{{\tt
  arXiv:1106.6047}}].

\bibitem{Dasgupta:2001sh}
M.~Dasgupta and G.~Salam, {\it {Resummation of nonglobal QCD observables}},
  {\em Phys.Lett.} {\bf B512} (2001) 323--330,
  [\href{http://arxiv.org/abs/hep-ph/0104277}{{\tt hep-ph/0104277}}].

\bibitem{Bauer:2000yr}
C.~W. Bauer, S.~Fleming, D.~Pirjol, and I.~W. Stewart, {\it {An Effective field
  theory for collinear and soft gluons: Heavy to light decays}},  {\em
  Phys.Rev.} {\bf D63} (2001) 114020,
  [\href{http://arxiv.org/abs/hep-ph/0011336}{{\tt hep-ph/0011336}}].

\bibitem{Bauer:2001ct}
C.~W. Bauer and I.~W. Stewart, {\it {Invariant operators in collinear effective
  theory}},  {\em Phys.Lett.} {\bf B516} (2001) 134--142,
  [\href{http://arxiv.org/abs/hep-ph/0107001}{{\tt hep-ph/0107001}}].

\bibitem{Bauer:2001yt}
C.~W. Bauer, D.~Pirjol, and I.~W. Stewart, {\it {Soft collinear factorization
  in effective field theory}},  {\em Phys.Rev.} {\bf D65} (2002) 054022,
  [\href{http://arxiv.org/abs/hep-ph/0109045}{{\tt hep-ph/0109045}}].

\bibitem{Bauer:2002nz}
C.~W. Bauer, S.~Fleming, D.~Pirjol, I.~Z. Rothstein, and I.~W. Stewart, {\it
  {Hard scattering factorization from effective field theory}},  {\em
  Phys.Rev.} {\bf D66} (2002) 014017,
  [\href{http://arxiv.org/abs/hep-ph/0202088}{{\tt hep-ph/0202088}}].

\bibitem{Korchemsky:1999kt}
G.~P. Korchemsky and G.~F. Sterman, {\it {Power corrections to event shapes and
  factorization}},  {\em Nucl.Phys.} {\bf B555} (1999) 335--351,
  [\href{http://arxiv.org/abs/hep-ph/9902341}{{\tt hep-ph/9902341}}].

\bibitem{Korchemsky:2000kp}
G.~Korchemsky and S.~Tafat, {\it {On power corrections to the event shape
  distributions in QCD}},  {\em JHEP} {\bf 0010} (2000) 010,
  [\href{http://arxiv.org/abs/hep-ph/0007005}{{\tt hep-ph/0007005}}].

\bibitem{Giele:2007di}
W.~T. Giele, D.~A. Kosower, and P.~Z. Skands, {\it {A simple shower and
  matching algorithm}},  {\em Phys.Rev.} {\bf D78} (2008) 014026,
  [\href{http://arxiv.org/abs/0707.3652}{{\tt arXiv:0707.3652}}].

\bibitem{Giele:2011cb}
W.~Giele, D.~Kosower, and P.~Skands, {\it {Higher-Order Corrections to Timelike
  Jets}},  {\em Phys.Rev.} {\bf D84} (2011) 054003,
  [\href{http://arxiv.org/abs/1102.2126}{{\tt arXiv:1102.2126}}].

\bibitem{GehrmannDeRidder:2011dm}
A.~Gehrmann-De~Ridder, M.~Ritzmann, and P.~Z. Skands, {\it {Timelike
  Dipole-Antenna Showers with Massive Fermions}},  {\em Phys.Rev.} {\bf D85}
  (2012) 014013, [\href{http://arxiv.org/abs/1108.6172}{{\tt
  arXiv:1108.6172}}].

\bibitem{Ritzmann:2012ca}
M.~Ritzmann, D.~Kosower, and P.~Skands, {\it {Antenna Showers with Hadronic
  Initial States}},  {\em Phys.Lett.} {\bf B718} (2013) 1345--1350,
  [\href{http://arxiv.org/abs/1210.6345}{{\tt arXiv:1210.6345}}].

\bibitem{Hartgring:2013jma}
L.~Hartgring, E.~Laenen, and P.~Skands, {\it {Antenna Showers with One-Loop
  Matrix Elements}},  {\em JHEP} {\bf 1310} (2013) 127,
  [\href{http://arxiv.org/abs/1303.4974}{{\tt arXiv:1303.4974}}].

\bibitem{Larkoski:2013yi}
A.~J. Larkoski, J.~J. Lopez-Villarejo, and P.~Skands, {\it {Helicity-Dependent
  Showers and Matching with VINCIA}},  {\em Phys.Rev.} {\bf D87} (2013), no.~5
  054033, [\href{http://arxiv.org/abs/1301.0933}{{\tt arXiv:1301.0933}}].

\bibitem{Salam:2001bd}
G.~Salam and D.~Wicke, {\it {Hadron masses and power corrections to event
  shapes}},  {\em JHEP} {\bf 0105} (2001) 061,
  [\href{http://arxiv.org/abs/hep-ph/0102343}{{\tt hep-ph/0102343}}].

\bibitem{Mateu:2012nk}
V.~Mateu, I.~W. Stewart, and J.~Thaler, {\it {Power Corrections to Event Shapes
  with Mass-Dependent Operators}},  {\em Phys.Rev.} {\bf D87} (2013), no.~1
  014025, [\href{http://arxiv.org/abs/1209.3781}{{\tt arXiv:1209.3781}}].

\bibitem{Cacciari:2011ma}
M.~Cacciari, G.~P. Salam, and G.~Soyez, {\it {FastJet User Manual}},  {\em
  Eur.Phys.J.} {\bf C72} (2012) 1896,
  [\href{http://arxiv.org/abs/1111.6097}{{\tt arXiv:1111.6097}}].

\bibitem{fjcontrib}
``Fastjet contrib.'' \url{http://fastjet.hepforge.org/contrib/}.

\bibitem{Stewart:2010tn}
I.~W. Stewart, F.~J. Tackmann, and W.~J. Waalewijn, {\it {N-Jettiness: An
  Inclusive Event Shape to Veto Jets}},  {\em Phys.Rev.Lett.} {\bf 105} (2010)
  092002, [\href{http://arxiv.org/abs/1004.2489}{{\tt arXiv:1004.2489}}].

\bibitem{Brandt:1978zm}
S.~Brandt and H.~Dahmen, {\it {Axes and Scalar Measures of Two-Jet and
  Three-Jet Events}},  {\em Z.Phys.} {\bf C1} (1979) 61.

\bibitem{Sjostrand:2006za}
T.~Sjostrand, S.~Mrenna, and P.~Z. Skands, {\it {PYTHIA 6.4 Physics and
  Manual}},  {\em JHEP} {\bf 0605} (2006) 026,
  [\href{http://arxiv.org/abs/hep-ph/0603175}{{\tt hep-ph/0603175}}].

\bibitem{Sjostrand:2007gs}
T.~Sjostrand, S.~Mrenna, and P.~Z. Skands, {\it {A Brief Introduction to PYTHIA
  8.1}},  {\em Comput.Phys.Commun.} {\bf 178} (2008) 852--867,
  [\href{http://arxiv.org/abs/0710.3820}{{\tt arXiv:0710.3820}}].

\bibitem{Cacciari:2008gp}
M.~Cacciari, G.~P. Salam, and G.~Soyez, {\it {The Anti-k(t) jet clustering
  algorithm}},  {\em JHEP} {\bf 0804} (2008) 063,
  [\href{http://arxiv.org/abs/0802.1189}{{\tt arXiv:0802.1189}}].

\bibitem{Larkoski:2014uqa}
A.~J. Larkoski, D.~Neill, and J.~Thaler, {\it {Jet Shapes with the Broadening
  Axis}},  {\em JHEP} {\bf 1404} (2014) 017,
  [\href{http://arxiv.org/abs/1401.2158}{{\tt arXiv:1401.2158}}].

\bibitem{Catani:1993hr}
S.~Catani, Y.~L. Dokshitzer, M.~Seymour, and B.~Webber, {\it {Longitudinally
  invariant $K_t$ clustering algorithms for hadron hadron collisions}},  {\em
  Nucl.Phys.} {\bf B406} (1993) 187--224.

\bibitem{Ellis:1993tq}
S.~D. Ellis and D.~E. Soper, {\it {Successive combination jet algorithm for
  hadron collisions}},  {\em Phys.Rev.} {\bf D48} (1993) 3160--3166,
  [\href{http://arxiv.org/abs/hep-ph/9305266}{{\tt hep-ph/9305266}}].

\bibitem{Dokshitzer:1997in}
Y.~L. Dokshitzer, G.~Leder, S.~Moretti, and B.~Webber, {\it {Better jet
  clustering algorithms}},  {\em JHEP} {\bf 9708} (1997) 001,
  [\href{http://arxiv.org/abs/hep-ph/9707323}{{\tt hep-ph/9707323}}].

\bibitem{Wobisch:1998wt}
M.~Wobisch and T.~Wengler, {\it {Hadronization corrections to jet
  cross-sections in deep inelastic scattering}},
  \href{http://arxiv.org/abs/hep-ph/9907280}{{\tt hep-ph/9907280}}.

\bibitem{Wobisch:2000dk}
M.~Wobisch, {\it {Measurement and QCD analysis of jet cross-sections in deep
  inelastic positron proton collisions at $\sqrt{s} = 300$~GeV}},  2000.

\bibitem{Appleby:2002ke}
R.~Appleby and M.~Seymour, {\it {Nonglobal logarithms in interjet energy flow
  with kt clustering requirement}},  {\em JHEP} {\bf 0212} (2002) 063,
  [\href{http://arxiv.org/abs/hep-ph/0211426}{{\tt hep-ph/0211426}}].

\bibitem{Banfi:2005gj}
A.~Banfi and M.~Dasgupta, {\it {Problems in resumming interjet energy flows
  with $k_t$ clustering}},  {\em Phys.Lett.} {\bf B628} (2005) 49--56,
  [\href{http://arxiv.org/abs/hep-ph/0508159}{{\tt hep-ph/0508159}}].

\bibitem{Banfi:2010pa}
A.~Banfi, M.~Dasgupta, K.~Khelifa-Kerfa, and S.~Marzani, {\it {Non-global
  logarithms and jet algorithms in high-pT jet shapes}},  {\em JHEP} {\bf 1008}
  (2010) 064, [\href{http://arxiv.org/abs/1004.3483}{{\tt arXiv:1004.3483}}].

\bibitem{Kelley:2012kj}
R.~Kelley, J.~R. Walsh, and S.~Zuberi, {\it {Abelian Non-Global Logarithms from
  Soft Gluon Clustering}},  {\em JHEP} {\bf 1209} (2012) 117,
  [\href{http://arxiv.org/abs/1202.2361}{{\tt arXiv:1202.2361}}].

\bibitem{Ellis:2010rwa}
S.~D. Ellis, C.~K. Vermilion, J.~R. Walsh, A.~Hornig, and C.~Lee, {\it {Jet
  Shapes and Jet Algorithms in SCET}},  {\em JHEP} {\bf 1011} (2010) 101,
  [\href{http://arxiv.org/abs/1001.0014}{{\tt arXiv:1001.0014}}].

\bibitem{Fleming:2007xt}
S.~Fleming, A.~H. Hoang, S.~Mantry, and I.~W. Stewart, {\it {Top Jets in the
  Peak Region: Factorization Analysis with NLL Resummation}},  {\em Phys.Rev.}
  {\bf D77} (2008) 114003, [\href{http://arxiv.org/abs/0711.2079}{{\tt
  arXiv:0711.2079}}].

\bibitem{Kelley:2011aa}
R.~Kelley, M.~D. Schwartz, R.~M. Schabinger, and H.~X. Zhu, {\it {Jet Mass with
  a Jet Veto at Two Loops and the Universality of Non-Global Structure}},  {\em
  Phys.Rev.} {\bf D86} (2012) 054017,
  [\href{http://arxiv.org/abs/1112.3343}{{\tt arXiv:1112.3343}}].

\bibitem{Hornig:2011tg}
A.~Hornig, C.~Lee, J.~R. Walsh, and S.~Zuberi, {\it {Double Non-Global
  Logarithms In-N-Out of Jets}},  {\em JHEP} {\bf 1201} (2012) 149,
  [\href{http://arxiv.org/abs/1110.0004}{{\tt arXiv:1110.0004}}].

\bibitem{Nagy:1997yn}
Z.~Nagy and Z.~Trocsanyi, {\it {Next-to-leading order calculation of four jet
  shape variables}},  {\em Phys.Rev.Lett.} {\bf 79} (1997) 3604--3607,
  [\href{http://arxiv.org/abs/hep-ph/9707309}{{\tt hep-ph/9707309}}].

\bibitem{Nagy:1998bb}
Z.~Nagy and Z.~Trocsanyi, {\it {Next-to-leading order calculation of four jet
  observables in electron positron annihilation}},  {\em Phys.Rev.} {\bf D59}
  (1999) 014020, [\href{http://arxiv.org/abs/hep-ph/9806317}{{\tt
  hep-ph/9806317}}].

\bibitem{Nagy:2001fj}
Z.~Nagy, {\it {Three jet cross-sections in hadron hadron collisions at
  next-to-leading order}},  {\em Phys.Rev.Lett.} {\bf 88} (2002) 122003,
  [\href{http://arxiv.org/abs/hep-ph/0110315}{{\tt hep-ph/0110315}}].

\bibitem{Nagy:2001xb}
Z.~Nagy and Z.~Trocsanyi, {\it {Multijet cross-sections in deep inelastic
  scattering at next-to-leading order}},  {\em Phys.Rev.Lett.} {\bf 87} (2001)
  082001, [\href{http://arxiv.org/abs/hep-ph/0104315}{{\tt hep-ph/0104315}}].

\bibitem{Nagy:2003tz}
Z.~Nagy, {\it {Next-to-leading order calculation of three jet observables in
  hadron hadron collision}},  {\em Phys.Rev.} {\bf D68} (2003) 094002,
  [\href{http://arxiv.org/abs/hep-ph/0307268}{{\tt hep-ph/0307268}}].

\bibitem{Manohar:2006nz}
A.~V. Manohar and I.~W. Stewart, {\it {The Zero-Bin and Mode Factorization in
  Quantum Field Theory}},  {\em Phys.Rev.} {\bf D76} (2007) 074002,
  [\href{http://arxiv.org/abs/hep-ph/0605001}{{\tt hep-ph/0605001}}].

\bibitem{Pietrulewicz:2016nwo}
P.~Pietrulewicz, F.~J. Tackmann, and W.~J. Waalewijn, {\it {Factorization and
  Resummation for Generic Hierarchies between Jets}},
  \href{http://arxiv.org/abs/1601.05088}{{\tt arXiv:1601.05088}}.

\bibitem{Fischer:2014bja}
N.~Fischer, S.~Gieseke, S.~Pl{\"a}tzer, and P.~Skands, {\it {Revisiting
  radiation patterns in $e^+e^-$ collisions}},  {\em Eur.Phys.J.} {\bf C74}
  (2014), no.~4 2831, [\href{http://arxiv.org/abs/1402.3186}{{\tt
  arXiv:1402.3186}}].

\bibitem{Fischer:2015pqa}
{\bf OPAL} Collaboration, N.~Fischer, S.~Gieseke, S.~Kluth, S.~Pl{\"a}tzer, and
  P.~Skands, {\it {Measurement of observables sensitive to coherence effects in
  hadronic Z decays with the OPAL detector at LEP}},
  \href{http://arxiv.org/abs/1505.01636}{{\tt arXiv:1505.01636}}.

\bibitem{Buckley:2011ms}
A.~Buckley, J.~Butterworth, S.~Gieseke, D.~Grellscheid, S.~Hoche, et~al., {\it
  {General-purpose event generators for LHC physics}},  {\em Phys.Rept.} {\bf
  504} (2011) 145--233, [\href{http://arxiv.org/abs/1101.2599}{{\tt
  arXiv:1101.2599}}].

\bibitem{Skands:2012ts}
P.~Skands, {\it {Introduction to QCD}},
  \href{http://arxiv.org/abs/1207.2389}{{\tt arXiv:1207.2389}}.

\bibitem{Seymour:2013ega}
M.~H. Seymour and M.~Marx, {\it {Monte Carlo Event Generators}},
  \href{http://arxiv.org/abs/1304.6677}{{\tt arXiv:1304.6677}}.

\bibitem{Gieseke:2013eva}
S.~Gieseke, {\it {Simulation of jets at colliders}},  {\em
  Prog.Part.Nucl.Phys.} {\bf 72} (2013) 155--205.

\bibitem{Hoche:2014rga}
S.~H{\"o}che, {\it {Introduction to parton-shower event generators}},
  \href{http://arxiv.org/abs/1411.4085}{{\tt arXiv:1411.4085}}.

\bibitem{Gleisberg:2003xi}
T.~Gleisberg, S.~Hoeche, F.~Krauss, A.~Schalicke, S.~Schumann, et~al., {\it
  {SHERPA 1. alpha: A Proof of concept version}},  {\em JHEP} {\bf 0402} (2004)
  056, [\href{http://arxiv.org/abs/hep-ph/0311263}{{\tt hep-ph/0311263}}].

\bibitem{Gleisberg:2008ta}
T.~Gleisberg, S.~Hoeche, F.~Krauss, M.~Schonherr, S.~Schumann, et~al., {\it
  {Event generation with SHERPA 1.1}},  {\em JHEP} {\bf 0902} (2009) 007,
  [\href{http://arxiv.org/abs/0811.4622}{{\tt arXiv:0811.4622}}].

\bibitem{Lonnblad:1992tz}
L.~Lonnblad, {\it {ARIADNE version 4: A Program for simulation of QCD cascades
  implementing the color dipole model}},  {\em Comput.Phys.Commun.} {\bf 71}
  (1992) 15--31.

\bibitem{Hoche:2015sya}
S.~H{\"o}che and S.~Prestel, {\it {The midpoint between dipole and parton
  showers}},  \href{http://arxiv.org/abs/1506.05057}{{\tt arXiv:1506.05057}}.

\bibitem{Marchesini:1991ch}
G.~Marchesini, B.~Webber, G.~Abbiendi, I.~Knowles, M.~Seymour, et~al., {\it
  {HERWIG: A Monte Carlo event generator for simulating hadron emission
  reactions with interfering gluons. Version 5.1 - April 1991}},  {\em
  Comput.Phys.Commun.} {\bf 67} (1992) 465--508.

\bibitem{Corcella:2000bw}
G.~Corcella, I.~Knowles, G.~Marchesini, S.~Moretti, K.~Odagiri, et~al., {\it
  {HERWIG 6: An Event generator for hadron emission reactions with interfering
  gluons (including supersymmetric processes)}},  {\em JHEP} {\bf 0101} (2001)
  010, [\href{http://arxiv.org/abs/hep-ph/0011363}{{\tt hep-ph/0011363}}].

\bibitem{Corcella:2002jc}
G.~Corcella, I.~Knowles, G.~Marchesini, S.~Moretti, K.~Odagiri, et~al., {\it
  {HERWIG 6.5 release note}},  \href{http://arxiv.org/abs/hep-ph/0210213}{{\tt
  hep-ph/0210213}}.

\bibitem{Bahr:2008pv}
M.~Bahr, S.~Gieseke, M.~Gigg, D.~Grellscheid, K.~Hamilton, et~al., {\it
  {Herwig++ Physics and Manual}},  {\em Eur.Phys.J.} {\bf C58} (2008) 639--707,
  [\href{http://arxiv.org/abs/0803.0883}{{\tt arXiv:0803.0883}}].

\bibitem{Platzer:2011bc}
S.~Platzer and S.~Gieseke, {\it {Dipole Showers and Automated NLO Matching in
  Herwig++}},  {\em Eur.Phys.J.} {\bf C72} (2012) 2187,
  [\href{http://arxiv.org/abs/1109.6256}{{\tt arXiv:1109.6256}}].

\bibitem{Catani:1990rr}
S.~Catani, B.~R. Webber, and G.~Marchesini, {\it {QCD coherent branching and
  semiinclusive processes at large x}},  {\em Nucl. Phys.} {\bf B349} (1991)
  635--654.

\bibitem{Dokshitzer:1995ev}
Y.~L. Dokshitzer, V.~A. Khoze, and S.~I. Troian, {\it {Specific features of
  heavy quark production. LPHD approach to heavy particle spectra}},  {\em
  Phys. Rev.} {\bf D53} (1996) 89--119,
  [\href{http://arxiv.org/abs/hep-ph/9506425}{{\tt hep-ph/9506425}}].

\bibitem{Alwall:2014hca}
J.~Alwall, R.~Frederix, S.~Frixione, V.~Hirschi, F.~Maltoni, et~al., {\it {The
  automated computation of tree-level and next-to-leading order differential
  cross sections, and their matching to parton shower simulations}},  {\em
  JHEP} {\bf 1407} (2014) 079, [\href{http://arxiv.org/abs/1405.0301}{{\tt
  arXiv:1405.0301}}].

\bibitem{Dasgupta:2002bw}
M.~Dasgupta and G.~P. Salam, {\it {Accounting for coherence in interjet E(t)
  flow: A Case study}},  {\em JHEP} {\bf 0203} (2002) 017,
  [\href{http://arxiv.org/abs/hep-ph/0203009}{{\tt hep-ph/0203009}}].

\bibitem{Tackmann:2012bt}
F.~J. Tackmann, J.~R. Walsh, and S.~Zuberi, {\it {Resummation Properties of Jet
  Vetoes at the LHC}},  {\em Phys.Rev.} {\bf D86} (2012) 053011,
  [\href{http://arxiv.org/abs/1206.4312}{{\tt arXiv:1206.4312}}].

\bibitem{Dasgupta:2014yra}
M.~Dasgupta, F.~Dreyer, G.~P. Salam, and G.~Soyez, {\it {Small-radius jets to
  all orders in QCD}},  {\em JHEP} {\bf 1504} (2015) 039,
  [\href{http://arxiv.org/abs/1411.5182}{{\tt arXiv:1411.5182}}].

\bibitem{Bosch:2004th}
S.~Bosch, B.~Lange, M.~Neubert, and G.~Paz, {\it {Factorization and shape
  function effects in inclusive B meson decays}},  {\em Nucl.Phys.} {\bf B699}
  (2004) 335--386, [\href{http://arxiv.org/abs/hep-ph/0402094}{{\tt
  hep-ph/0402094}}].

\bibitem{Hoang:2007vb}
A.~H. Hoang and I.~W. Stewart, {\it {Designing gapped soft functions for jet
  production}},  {\em Phys.Lett.} {\bf B660} (2008) 483--493,
  [\href{http://arxiv.org/abs/0709.3519}{{\tt arXiv:0709.3519}}].

\bibitem{Ligeti:2008ac}
Z.~Ligeti, I.~W. Stewart, and F.~J. Tackmann, {\it {Treating the b quark
  distribution function with reliable uncertainties}},  {\em Phys.Rev.} {\bf
  D78} (2008) 114014, [\href{http://arxiv.org/abs/0807.1926}{{\tt
  arXiv:0807.1926}}].

\bibitem{Akhoury:1995sp}
R.~Akhoury and V.~I. Zakharov, {\it {On the universality of the leading, 1/Q
  power corrections in QCD}},  {\em Phys.Lett.} {\bf B357} (1995) 646--652,
  [\href{http://arxiv.org/abs/hep-ph/9504248}{{\tt hep-ph/9504248}}].

\bibitem{Dokshitzer:1995zt}
Y.~L. Dokshitzer and B.~Webber, {\it {Calculation of power corrections to
  hadronic event shapes}},  {\em Phys.Lett.} {\bf B352} (1995) 451--455,
  [\href{http://arxiv.org/abs/hep-ph/9504219}{{\tt hep-ph/9504219}}].

\bibitem{Lee:2006fn}
C.~Lee and G.~F. Sterman, {\it {Universality of nonperturbative effects in
  event shapes}},  {\em eConf} {\bf C0601121} (2006) A001,
  [\href{http://arxiv.org/abs/hep-ph/0603066}{{\tt hep-ph/0603066}}].

\bibitem{Lee:2007jr}
C.~Lee, {\it {Universal nonperturbative effects in event shapes from
  soft-collinear effective theory}},  {\em Mod.Phys.Lett.} {\bf A22} (2007)
  835--851, [\href{http://arxiv.org/abs/hep-ph/0703030}{{\tt hep-ph/0703030}}].

\bibitem{Stewart:2014nna}
I.~W. Stewart, F.~J. Tackmann, and W.~J. Waalewijn, {\it {Dissecting Soft
  Radiation with Factorization}},  {\em Phys.Rev.Lett.} {\bf 114} (2015), no.~9
  092001, [\href{http://arxiv.org/abs/1405.6722}{{\tt arXiv:1405.6722}}].

\bibitem{Beneke:1998ui}
M.~Beneke, {\it {Renormalons}},  {\em Phys.Rept.} {\bf 317} (1999) 1--142,
  [\href{http://arxiv.org/abs/hep-ph/9807443}{{\tt hep-ph/9807443}}].

\bibitem{Gardi:2000yh}
E.~Gardi, {\it {Perturbative and nonperturbative aspects of moments of the
  thrust distribution in e+ e- annihilation}},  {\em JHEP} {\bf 0004} (2000)
  030, [\href{http://arxiv.org/abs/hep-ph/0003179}{{\tt hep-ph/0003179}}].

\bibitem{Hornig:2009vb}
A.~Hornig, C.~Lee, and G.~Ovanesyan, {\it {Effective Predictions of Event
  Shapes: Factorized, Resummed, and Gapped Angularity Distributions}},  {\em
  JHEP} {\bf 0905} (2009) 122, [\href{http://arxiv.org/abs/0901.3780}{{\tt
  arXiv:0901.3780}}].

\bibitem{Achard:2004sv}
{\bf L3} Collaboration, P.~Achard et~al., {\it {Studies of hadronic event
  structure in $e^{+} e^{-}$ annihilation from 30-GeV to 209-GeV with the L3
  detector}},  {\em Phys.Rept.} {\bf 399} (2004) 71--174,
  [\href{http://arxiv.org/abs/hep-ex/0406049}{{\tt hep-ex/0406049}}].

\bibitem{Gehrmann:2009eh}
T.~Gehrmann, M.~Jaquier, and G.~Luisoni, {\it {Hadronization effects in event
  shape moments}},  {\em Eur.Phys.J.} {\bf C67} (2010) 57--72,
  [\href{http://arxiv.org/abs/0911.2422}{{\tt arXiv:0911.2422}}].

\bibitem{Abbate:2010xh}
R.~Abbate, M.~Fickinger, A.~H. Hoang, V.~Mateu, and I.~W. Stewart, {\it {Thrust
  at $N^3LL$ with Power Corrections and a Precision Global Fit for
  alphas(mZ)}},  {\em Phys.Rev.} {\bf D83} (2011) 074021,
  [\href{http://arxiv.org/abs/1006.3080}{{\tt arXiv:1006.3080}}].

\bibitem{Abbate:2012jh}
R.~Abbate, M.~Fickinger, A.~H. Hoang, V.~Mateu, and I.~W. Stewart, {\it
  {Precision Thrust Cumulant Moments at $N^3$LL}},  {\em Phys.Rev.} {\bf D86}
  (2012) 094002, [\href{http://arxiv.org/abs/1204.5746}{{\tt
  arXiv:1204.5746}}].

\bibitem{Hoang:2014wka}
A.~H. Hoang, D.~W. Kolodrubetz, V.~Mateu, and I.~W. Stewart, {\it
  {$C$-parameter distribution at N$^3$LL′ including power corrections}},
  {\em Phys.Rev.} {\bf D91} (2015), no.~9 094017,
  [\href{http://arxiv.org/abs/1411.6633}{{\tt arXiv:1411.6633}}].

\bibitem{Hoang:2015hka}
A.~H. Hoang, D.~W. Kolodrubetz, V.~Mateu, and I.~W. Stewart, {\it {Precise
  determination of $\alpha_s$ from the $C$-parameter distribution}},  {\em
  Phys.Rev.} {\bf D91} (2015), no.~9 094018,
  [\href{http://arxiv.org/abs/1501.04111}{{\tt arXiv:1501.04111}}].

\bibitem{Stewart:2009yx}
I.~W. Stewart, F.~J. Tackmann, and W.~J. Waalewijn, {\it {Factorization at the
  LHC: From PDFs to Initial State Jets}},  {\em Phys.Rev.} {\bf D81} (2010)
  094035, [\href{http://arxiv.org/abs/0910.0467}{{\tt arXiv:0910.0467}}].

\bibitem{Liu:2012sz}
X.~Liu and F.~Petriello, {\it {Resummation of jet-veto logarithms in hadronic
  processes containing jets}},  {\em Phys.Rev.} {\bf D87} (2013), no.~1 014018,
  [\href{http://arxiv.org/abs/1210.1906}{{\tt arXiv:1210.1906}}].

\bibitem{Liu:2013hba}
X.~Liu and F.~Petriello, {\it {Reducing theoretical uncertainties for exclusive
  Higgs-boson plus one-jet production at the LHC}},  {\em Phys.Rev.} {\bf D87}
  (2013), no.~9 094027, [\href{http://arxiv.org/abs/1303.4405}{{\tt
  arXiv:1303.4405}}].

\bibitem{guido_talk}
J.~Talbert, {\it {Automated Calculations of Dijet Soft Functions}},  {\em SCET
  Conference} (2015).

\bibitem{Boughezal:2015eha}
R.~Boughezal, X.~Liu, and F.~Petriello, {\it {$N$-jettiness soft function at
  next-to-next-to-leading order}},  {\em Phys.Rev.} {\bf D91} (2015), no.~9
  094035, [\href{http://arxiv.org/abs/1504.02540}{{\tt arXiv:1504.02540}}].

\bibitem{Banfi:2002hw}
A.~Banfi, G.~Marchesini, and G.~Smye, {\it {Away from jet energy flow}},  {\em
  JHEP} {\bf 0208} (2002) 006, [\href{http://arxiv.org/abs/hep-ph/0206076}{{\tt
  hep-ph/0206076}}].

\bibitem{Weigert:2003mm}
H.~Weigert, {\it {Nonglobal jet evolution at finite N(c)}},  {\em Nucl.Phys.}
  {\bf B685} (2004) 321--350, [\href{http://arxiv.org/abs/hep-ph/0312050}{{\tt
  hep-ph/0312050}}].

\bibitem{Hatta:2013iba}
Y.~Hatta and T.~Ueda, {\it {Resummation of non-global logarithms at finite
  $N_c$}},  {\em Nucl.Phys.} {\bf B874} (2013) 808--820,
  [\href{http://arxiv.org/abs/1304.6930}{{\tt arXiv:1304.6930}}].

\bibitem{Caron-Huot:2015bja}
S.~Caron-Huot, {\it {Resummation of non-global logarithms and the BFKL
  equation}},  \href{http://arxiv.org/abs/1501.03754}{{\tt arXiv:1501.03754}}.

\bibitem{Boughezal:2015aha}
R.~Boughezal, C.~Focke, W.~Giele, X.~Liu, and F.~Petriello, {\it {Higgs boson
  production in association with a jet using jettiness subtraction}},  {\em
  Phys.Lett.} {\bf B748} (2015) 5--8,
  [\href{http://arxiv.org/abs/1505.03893}{{\tt arXiv:1505.03893}}].

\bibitem{Boughezal:2015dva}
R.~Boughezal, C.~Focke, X.~Liu, and F.~Petriello, {\it {$W$-boson production in
  association with a jet at next-to-next-to-leading order in perturbative
  QCD}},  \href{http://arxiv.org/abs/1504.02131}{{\tt arXiv:1504.02131}}.

\bibitem{Gaunt:2015pea}
J.~Gaunt, M.~Stahlhofen, F.~J. Tackmann, and J.~R. Walsh, {\it {N-jettiness
  Subtractions for NNLO QCD Calculations}},
  \href{http://arxiv.org/abs/1505.04794}{{\tt arXiv:1505.04794}}.

\bibitem{Boughezal:2015dra}
R.~Boughezal, F.~Caola, K.~Melnikov, F.~Petriello, and M.~Schulze, {\it {Higgs
  Boson Production in Association with a Jet at Next-to-Next-to-Leading
  Order}},  \href{http://arxiv.org/abs/1504.07922}{{\tt arXiv:1504.07922}}.

\bibitem{Larkoski:2015yqa}
A.~J. Larkoski, F.~Maltoni, and M.~Selvaggi, {\it {Tracking down hyper-boosted
  top quarks}},  {\em JHEP} {\bf 1506} (2015) 032,
  [\href{http://arxiv.org/abs/1503.03347}{{\tt arXiv:1503.03347}}].

\bibitem{Fleming:2007qr}
S.~Fleming, A.~H. Hoang, S.~Mantry, and I.~W. Stewart, {\it {Jets from massive
  unstable particles: Top-mass determination}},  {\em Phys.Rev.} {\bf D77}
  (2008) 074010, [\href{http://arxiv.org/abs/hep-ph/0703207}{{\tt
  hep-ph/0703207}}].

\bibitem{iain_notes}
I.~W. Stewart and C.~W. Bauer, ``Lectures on the soft-collinear effective
  theory.''
  \url{http://ocw.mit.edu/courses/physics/8-851-effective-field-theory-spring-2013/lecture-notes/MIT8_851S13_scetnotes.pdf}.

\bibitem{Becher:2014oda}
T.~Becher, A.~Broggio, and A.~Ferroglia, {\it {Introduction to Soft-Collinear
  Effective Theory}},  \href{http://arxiv.org/abs/1410.1892}{{\tt
  arXiv:1410.1892}}.

\bibitem{Jouttenus:2011wh}
T.~T. Jouttenus, I.~W. Stewart, F.~J. Tackmann, and W.~J. Waalewijn, {\it {The
  Soft Function for Exclusive N-Jet Production at Hadron Colliders}},  {\em
  Phys.Rev.} {\bf D83} (2011) 114030,
  [\href{http://arxiv.org/abs/1102.4344}{{\tt arXiv:1102.4344}}].

\bibitem{Ritzmann:2014mka}
M.~Ritzmann and W.~J. Waalewijn, {\it {Fragmentation in Jets at NNLO}},  {\em
  Phys.Rev.} {\bf D90} (2014) 054029,
  [\href{http://arxiv.org/abs/1407.3272}{{\tt arXiv:1407.3272}}].

\bibitem{Sveshnikov:1995vi}
N.~Sveshnikov and F.~Tkachov, {\it {Jets and quantum field theory}},  {\em
  Phys.Lett.} {\bf B382} (1996) 403--408,
  [\href{http://arxiv.org/abs/hep-ph/9512370}{{\tt hep-ph/9512370}}].

\bibitem{Korchemsky:1997sy}
G.~P. Korchemsky, G.~Oderda, and G.~F. Sterman, {\it {Power corrections and
  nonlocal operators}},  {\em AIP Conf.Proc.} {\bf 407} (1997) 988,
  [\href{http://arxiv.org/abs/hep-ph/9708346}{{\tt hep-ph/9708346}}].

\bibitem{Lee:2006nr}
C.~Lee and G.~F. Sterman, {\it {Momentum Flow Correlations from Event Shapes:
  Factorized Soft Gluons and Soft-Collinear Effective Theory}},  {\em
  Phys.Rev.} {\bf D75} (2007) 014022,
  [\href{http://arxiv.org/abs/hep-ph/0611061}{{\tt hep-ph/0611061}}].

\bibitem{Bauer:2008dt}
C.~W. Bauer, S.~P. Fleming, C.~Lee, and G.~F. Sterman, {\it {Factorization of
  e+e- Event Shape Distributions with Hadronic Final States in Soft Collinear
  Effective Theory}},  {\em Phys.Rev.} {\bf D78} (2008) 034027,
  [\href{http://arxiv.org/abs/0801.4569}{{\tt arXiv:0801.4569}}].

\bibitem{Bauer:2003di}
C.~W. Bauer, C.~Lee, A.~V. Manohar, and M.~B. Wise, {\it {Enhanced
  nonperturbative effects in Z decays to hadrons}},  {\em Phys.Rev.} {\bf D70}
  (2004) 034014, [\href{http://arxiv.org/abs/hep-ph/0309278}{{\tt
  hep-ph/0309278}}].

\bibitem{Manohar:2003vb}
A.~V. Manohar, {\it {Deep inelastic scattering as $x \to 1$ using soft
  collinear effective theory}},  {\em Phys.Rev.} {\bf D68} (2003) 114019,
  [\href{http://arxiv.org/abs/hep-ph/0309176}{{\tt hep-ph/0309176}}].

\bibitem{Kosower:1999rx}
D.~A. Kosower and P.~Uwer, {\it {One loop splitting amplitudes in gauge
  theory}},  {\em Nucl.Phys.} {\bf B563} (1999) 477--505,
  [\href{http://arxiv.org/abs/hep-ph/9903515}{{\tt hep-ph/9903515}}].

\bibitem{Ellis:1980wv}
R.~K. Ellis, D.~Ross, and A.~Terrano, {\it {The Perturbative Calculation of Jet
  Structure in e+ e- Annihilation}},  {\em Nucl.Phys.} {\bf B178} (1981) 421.

\bibitem{Bern:2004cz}
Z.~Bern, L.~J. Dixon, and D.~A. Kosower, {\it {Two-loop g ---> gg splitting
  amplitudes in QCD}},  {\em JHEP} {\bf 08} (2004) 012,
  [\href{http://arxiv.org/abs/hep-ph/0404293}{{\tt hep-ph/0404293}}].

\bibitem{Badger:2004uk}
S.~Badger and E.~N. Glover, {\it {Two loop splitting functions in QCD}},  {\em
  JHEP} {\bf 0407} (2004) 040, [\href{http://arxiv.org/abs/hep-ph/0405236}{{\tt
  hep-ph/0405236}}].

\bibitem{Catani:1996jh}
S.~Catani and M.~Seymour, {\it {The Dipole formalism for the calculation of QCD
  jet cross-sections at next-to-leading order}},  {\em Phys.Lett.} {\bf B378}
  (1996) 287--301, [\href{http://arxiv.org/abs/hep-ph/9602277}{{\tt
  hep-ph/9602277}}].

\bibitem{Catani:1996vz}
S.~Catani and M.~Seymour, {\it {A General algorithm for calculating jet
  cross-sections in NLO QCD}},  {\em Nucl.Phys.} {\bf B485} (1997) 291--419,
  [\href{http://arxiv.org/abs/hep-ph/9605323}{{\tt hep-ph/9605323}}].

\bibitem{Huber:2005yg}
T.~Huber and D.~Maitre, {\it {HypExp: A Mathematica package for expanding
  hypergeometric functions around integer-valued parameters}},  {\em
  Comput.Phys.Commun.} {\bf 175} (2006) 122--144,
  [\href{http://arxiv.org/abs/hep-ph/0507094}{{\tt hep-ph/0507094}}].

\bibitem{Huber:2007dx}
T.~Huber and D.~Maitre, {\it {HypExp 2, Expanding Hypergeometric Functions
  about Half-Integer Parameters}},  {\em Comput.Phys.Commun.} {\bf 178} (2008)
  755--776, [\href{http://arxiv.org/abs/0708.2443}{{\tt arXiv:0708.2443}}].

\bibitem{Giele:1991vf}
W.~Giele and E.~N. Glover, {\it {Higher order corrections to jet cross-sections
  in e+ e- annihilation}},  {\em Phys.Rev.} {\bf D46} (1992) 1980--2010.

\bibitem{Berends:1988zn}
F.~A. Berends and W.~Giele, {\it {Multiple Soft Gluon Radiation in Parton
  Processes}},  {\em Nucl.Phys.} {\bf B313} (1989) 595.

\bibitem{Catani:2000pi}
S.~Catani and M.~Grazzini, {\it {The soft gluon current at one loop order}},
  {\em Nucl.Phys.} {\bf B591} (2000) 435--454,
  [\href{http://arxiv.org/abs/hep-ph/0007142}{{\tt hep-ph/0007142}}].

\bibitem{Duhr:2013msa}
C.~Duhr and T.~Gehrmann, {\it {The two-loop soft current in dimensional
  regularization}},  {\em Phys.Lett.} {\bf B727} (2013) 452--455,
  [\href{http://arxiv.org/abs/1309.4393}{{\tt arXiv:1309.4393}}].

\bibitem{Li:2013lsa}
Y.~Li and H.~X. Zhu, {\it {Single soft gluon emission at two loops}},  {\em
  JHEP} {\bf 1311} (2013) 080, [\href{http://arxiv.org/abs/1309.4391}{{\tt
  arXiv:1309.4391}}].

\bibitem{Altarelli:1979ub}
G.~Altarelli, R.~K. Ellis, and G.~Martinelli, {\it {Large Perturbative
  Corrections to the Drell-Yan Process in QCD}},  {\em Nucl.Phys.} {\bf B157}
  (1979) 461.

\bibitem{Matsuura:1988sm}
T.~Matsuura, S.~van~der Marck, and W.~van Neerven, {\it {The Calculation of the
  Second Order Soft and Virtual Contributions to the Drell-Yan Cross-Section}},
   {\em Nucl.Phys.} {\bf B319} (1989) 570.

\bibitem{Almeida:2014uva}
L.~G. Almeida, S.~D. Ellis, C.~Lee, G.~Sterman, I.~Sung, et~al., {\it
  {Comparing and counting logs in direct and effective methods of QCD
  resummation}},  {\em JHEP} {\bf 1404} (2014) 174,
  [\href{http://arxiv.org/abs/1401.4460}{{\tt arXiv:1401.4460}}].

\bibitem{Banfi:2004yd}
A.~Banfi, G.~P. Salam, and G.~Zanderighi, {\it {Principles of general
  final-state resummation and automated implementation}},  {\em JHEP} {\bf
  0503} (2005) 073, [\href{http://arxiv.org/abs/hep-ph/0407286}{{\tt
  hep-ph/0407286}}].

\bibitem{Banfi:2014sua}
A.~Banfi, H.~McAslan, P.~F. Monni, and G.~Zanderighi, {\it {A general method
  for the resummation of event-shape distributions in $e^{+} e^{−}$
  annihilation}},  {\em JHEP} {\bf 1505} (2015) 102,
  [\href{http://arxiv.org/abs/1412.2126}{{\tt arXiv:1412.2126}}].

\bibitem{Sakaki:2015iya}
Y.~Sakaki, {\it {Evolution variable dependence of jet substructure}},
  \href{http://arxiv.org/abs/1506.04811}{{\tt arXiv:1506.04811}}.

\end{thebibliography}\endgroup

\end{document}